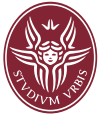

# Finite Size Corrections to Disordered Systems
## mean field results and applications to finite dimensional models

Scuola di dottorato Vito Volterra

Dottorato di Ricerca in Fisica – XXVII Ciclo

Candidate

Carlo Lucibello
ID number 1100021

Thesis Advisors

Prof. Giorgio Parisi
Prof. Federico Ricci-Tersenghi

A thesis submitted in partial fulfillment of the requirements
for the degree of Doctor of Philosophy in Physics

December 2014

Thesis defended on 29th January 2015
in front of a Board of Examiners composed by:

Prof. Gaetano Senatore (chairman)
Prof. Silvio Franz
Prof. Riccardo Zecchina

**Finite Size Corrections to Disordered Systems**
Ph.D. thesis. Sapienza – University of Rome

The research leading to these results has received funding from the European Research Council under the European Union's Seventh Framework Programme (FP7/2007-2013) / ERC grant agreement No. 247328 (**CriPheRaSy**).

This thesis has been typeset by L<sup>A</sup>T<sub>E</sub>X and the Sapthesis class.

Version: February 19, 2015

Author's email: carlo.lucibello@gmail.com

*Alla mia famiglia*

# Contents

# Part I

# Introduction

# Chapter 1

# General Introduction

*Disordered Systems* is a broad, well established and actively investigated branch of statistical physics. The disordered systems we consider in the context of this thesis are those whose Hamiltonian encloses some kind of randomness, usually in the form of random couplings, fields or topologies. While this definition is quite general, it does not encompass the whole discipline, since it leaves aside systems with non-random Hamiltonian where the disorder is self-induced [7], such as structural glasses.

Most prominent examples of disordered systems are spin glass models, introduced by Edwards and Anderson [8] and simplified by Sherrington and Kirkpatrick [9]. The exact solution for the latter mean field modl produced by Parisi in 1979 [10] and clarified a few years later [11–13], showed a surprisingly rich phenomenology. In such systems the order parameter is a function describing the decomposition of the Gibbs measure into an exponential number of pure states, organized in an ultrametric structure [13].

The analytical techniques developed in those years to solve the mean field spin glass, namely the replica and the cavity method, successively refined to deal with diluted systems [14, 15], proved to be highly effective and general tools. In the last thirty years, concepts, analytical techniques, numerical tools and people as well from disordered systems found application to a plethora of other scientific fields: most noticeably supercooled liquids [16, 17] and combinatorial optimization [18–24], but also inference [25–27], protein motion [28], signal processing [29], immunology [30, 31], neural networks [32–35], metal-insulator transition [36, 37], quantum algorithms [38], epidemic spreading [39], game theory [40, 41] photonics [42], biological networks [43], random matrices [44, 45], finance [46, 47] and random interfaces [48].

Also from the mathematical perspective, the effort to rigorously prove the results obtained through heuristic physics methods lead to the development of new techniques [49, 50] and ideas [51, 52].

Another paradigmatic model of disordered system, much discussed along this thesis, is the Random Field Ising Model (RFIM). A renormalization group analysis shows that disorder is a relevant direction for the renormalization flow, and that the phase transition is controlled by a zero temperature fixed point [53]. Recently this fixed point was showed to be in the same universality class of a critical point displayed by glass-forming liquids [54, 55]. Another interesting feature of this model, still under investigation, is the breaking down, in finite dimension, of the dimensional reduction property that characterizes its mean field version [56, 57].

One of the heuristic techniques used to deal with disordered systems, the Cavity Method, relies on some explicit factorization assumptions, valid in fully connected and diluted random graphs only in the thermodynamic limit. In these cases, in the non-glassy, replica symmetric phase, the asymptotic free energy takes the name of Bethe free energy. In particular, in diluted models, the validity of the Bethe approximation relies on locally tree-likeness, that is on the asymptotic absence of short loops. Departure from the Bethe free energy is observed in diluted systems when the system size is finite. One of the aims of this thesis is to connect the finite size corrections to the presence of (few) short loops. The perspective in studying this mean field models with finite connectivity, is to apply the same formalism to deal with the many short loops one finds in finite dimensional lattices. The basic step for a perturbation theory around the Bethe approximation in finite dimension are in fact taken in Chapter 8.

A large part of this thesis deals with disordered Ising models, such as RFIMs and spin-glasses, although most of the techniques presented can be generalized to others models. The Matching Problem (MP), a combinatorial optimization problem with a long standing tradition in the statistical physics community [18, 58–62], is the other big player of this essay.

The material presented is organized as follows:

- Part I : **Introduction**

    – Chapter 1 : **General introduction**

      This very same Chapter. It contains a brief introduction to the scientific field, states aims and scopes of the thesis, and explains the organization of the material with a short overview of each Chapter.

    – Chapter 2 : **Preliminaries**

      The main concepts and tools recurring throughout the thesis are introduced: the notion of random graphs; disorder systems and the main tools that physicists developed to cope with them, the replica and the cavity methods; combinatorial optimization problems, as seen from a physicist viewpoint; a brief overview of the problem of finite size corrections.

    – Chapter 3 : **The Replicated Transfer Matrix**

      This is the first original contribution of this thesis. We set up an analytical framework to characterize the properties of one-dimensional disordered systems. The formalism is based on the spectral theory of the Replicated Transfer Matrix, though most results can be rederived using a probabilist approach. The main application of the formalism are the computation of many types of correlation functions in diluted systems and the computations of free energies of closed and open chains embedded in locally tree-like graphs. This last result will be much used in successive Chapters. This Chapter is based on the work we published in Ref. [1].

- Part II : **Perturbative Finite Size Corrections**

  – Chapter 4 : **Finite Size Corrections On Random Graphs**

    In diluted random graph ensembles the average number of short loops is finite also when the number of nodes in the graph goes to infinity. Therefore, in this systems, the free energy in the thermodynamic limit is that of an infinite tree graph. The first finite size corrections, though, resents the presence of these sparse (simple) loops. The computation of finite size corrections in diluted random graphs is an analytically manageable calculation where the additional contributions of loops to the Bethe free energy can be clearly highlighted. This is done in three steps: we express the replicated partition function as an integral over a certain order parameter; we perform the saddle point computation, which gives the leading order Bethe contribution; we compute the Gaussian fluctuations around the saddle point. It is then easy to show that the Gaussian fluctuations can be expressed as a linear combination of free energies of open and closed chains. We carry on this program in two different graph ensembles, the Erdős-Rényi and the Random Regular Graph ensembles. We published the contents of this Chapter in Refs. [2] and [3].

- Part III : **Non-Perturbative Finite Size Corrections**

  Here we discuss two types of finite size corrections that do not arise from the presence of loops. The techniques we employ are specific to the problems involved.

  – Chapter 5 : **The Random Field Ising Model**

    We consider magnetic systems with random external fields. In the ferromagnetic phase the fluctuations of the field causes a free energy difference among the up and down states. Minimization of the free energy on a sample-by-sample basis leads to an anomalous $O(1/\sqrt{N})$ subleading scaling for the average free energy. We show how to compute the coefficient of this correction using a variant of the replica method with $m + n$ replicas constrained to stay in the two different states. The computation can be carried out exactly in diluted and fully connected systems. This is the content of Ref. [4].

  – Chapter 6 : **The Euclidean Assignment Problem**

    We investigate the scaling behaviour of the average cost in the Euclidean Assignment Problem, also known as th uclidan Bipartite Matching Problem. We show how the difference in the number of blue and red points in a small region of space is a source for a transport field. This is formalized in a Poisson-like equation that yields a surprisingly ample set of predictions for the leading and subleading behaviour and coefficients of the average cost, for each dimension $d$. These results have been published in Ref. [5].

- Part IV : **Beyond the Bethe Approximation In Finite Dimension**

  We address two different problems in the Euclidean space, where diagrammatic terms corresponding to simple loops give the first order contribution in perturbative expansions around the Bethe approximation.

  – Chapter 7 : **The Euclidean Matching Problem**

    We first consider the Euclidean (monopartite) Matching Problem. We write a replicated action containing the correlations among distances between any subset of points. Then we keep only the terms corresponding to polygons and we compute them perturbatively.

  – Chapter 8 : **The Large M Expansion**

    Here we propose a very general method to perform a perturbative expansion around the Bethe approximation for finite dimensional systems, whether disordered or not. The expansion involves an integer parameter $M$. Varying $M$ between infinity and one we interpolate between a random regular graph and the original lattice. The first order in the $1/M$ expansion corresponds to the contributions from simple loops.

  In both cases we asses the formal equivalence between these contributions and the first order finite size corrections to mean field models. The material presented in this chapter has not been published yet and both problems are currently under investigation. Therefore these results have to be considered incomplete and not polished.

- Part V : **Conclusions**

  We finally discuss the results obtained in the previous chapters and highlight some directions for future investigation.

The computation of finite size corrections is the common link between Part II and III, while Part IV is the natural follow up of Chapter 4 regarding the characterization of loop contributions to the free energy. The discussion over the replicated transfer matrix of Chapter 3 will be exploited in several part of the thesis, specifically in Chapters 4, 7 and 8. An alternative arrangement of the material would see Chapter 4 along with Chapter 8 as a unique discussion about loops expansions, and Chapters 6 and 7 as a Part dedicated to Euclidean matching problems.

A reasoned history of my doctoral studies would go as follows. At the beginning of the first year one of my advisors, Giorgio Parisi, set me to work on the Euclidean and the random link matching problems. He was convinced that the term corresponding to correlations in triangles in the Euclidean space was the same one arising in the finite size corrections in the random link model. I am glad he was right. After a few months I was hijacked by Giorgio himself and pointed toward the study of finite size corrections on random graphs. In fact the idea there was the same we where prosecuting in the study of the matching problem: to relate the finite size corrections to the presence of short simple loops. Some time later I discovered that this was also a preparatory study, aiming at shaping up analytical and numerical tools, in order to tackle the problem of the large $M$ expansion on lattices. In the middle of the second year, the need to better understand the properties of chains in random graphs, steamed in a discussion between Flaviano Morone, Tommaso Rizzo and me, that led to the work on the replicated transfer matrix. At the end of the second year we started working on the bipartite matching problem. The aim was to characterize the anomalous leading scaling of the cost in very low dimensions. It turned out that the analytical framework we set up predicts in high dimension an anomalous subleading scaling, so that I can subsume also this work under the umbrella of "Finite Size Corrections". The presence of $O(1/\sqrt{N})$ corrections in the RFIM instead, came as an unexpected surprise from numerical simulations at zero temperature on Erdős-Rényi graphs. Fortunately we had the analytical technology to deal with it. I devoted most part of the third year to finishing and polishing all these projects. In the last part of the year I blew the dust off the replica calculations I did on the matching problem in the first year, and did many new ones, to produce, along with Giorgio and Gabriele Sicuro, the contents of Chapter 7. Also, according to our long thought plan, we are finally focusing more and more on the large $M$ expansion. We are confident that, once the formalism is polished, it will provide a useful tool to investigate finite dimensional systems.

# Chapter 2

# Preliminaries

## 2.1 Graph Theory

### 2.1.1 Basic notions

In this paragraph we introduce the concept of graph, a simple mathematical structure that encompasses encompasses and broadens the commonly known lattices. Most statistical physics models we address in this thesis are defined on a (random) graph. *Graph Theory* is a well established subject in mathematics, here we review its basics aspects.

A *graph* $G = (V, E)$ consists in a *vertex* set $V$, usually taken as a subset of the natural numbers $V = \{1, 2, \ldots, N\}$, such that $|V| = N$, and an *edge* set $E$, a collection of unordered pairs of vertices. We will denote with $i, j, \ldots$ the elements of $V$, and with $(i, j)$ and edge between vertices $i$ and $j$. Elements of $E$ the form $(i, i)$ are called *self-loops*. Multiple instances of the same edge $(i, j)$ are called *multi-edges*. A graph without self-loops and multi-edges called *simple*. In what follows when we write graph we will always mean simple graph. It is common use to call vertices also *nodes*, and to call edges also *links*. We will use both terminologies indistinctly.

The *neighbourhood* of a vertex $i$ is the set of vertexes that have am edge in common with $i$, and it is denoted as $\partial i \equiv \{j \in V : (i, j) \in E\}$. The *degree* (or *connectivity*) of a vertex $i$ is the number of its neighbours, $\deg(i) \equiv |\partial i|$, and it is sometimes denoted as $k_i$. A vertex with degree zero is called an *isolated* vertex. A vertex with degree one is called a *leaf* or a *dangling node*. The *residual degree* of a node $i$ is $k'_i = k_i - 1$. The *degree distribution* $p_G(k)$ is the frequency of the node degrees, namely

$$p_G(k) = \frac{1}{N} \sum_{i \in V} \mathbb{I}(k_i = k), \qquad (2.1)$$

and the *mean degree* $z_G$ is given by

$$z_G \equiv \frac{1}{N} \sum_{i \in V} k_i = \sum_{k \geq 0} p_G(k)\, k. \qquad (2.2)$$

A *path* of *length* $\ell$ in a graph is a sequence of vertices $w = (i_0, i_1, \ldots, i_\ell)$ such that $(i_m, i_{m+i}) \in E$. The path is *closed* if $i_0 = i_\ell$, and it is *open* otherwise. If no vertex, except for the extremities, is crossed more then once the path is called *simple*. A closed path can also be called *loop* or *closed chain*. An open path is also called an *open chain*. The *distance* between two nodes is the length of the shortest path joining them. A graph is *connected* if there is a path joining any two nodes. The *d-neighbourhood* of a vertex $i$ is the set of vertexes at a distance at most $d$ from $i$. A graph is called *complete* or *fully connected* if any pair of nodes is connected by an edge. A graph is *bipartite* if $V$ can be partitioned in two set such that there is no edge joining vertices in the same set. A *subgraph* of a graph $G$ is a graph $G' = (V', E')$ such that $V' \subset V$ and $E' \subset E$.

The set $E$ can also be taken to be composed of ordered pairs of vertices. in this case the graph is called *oriented* and the elements of $E$ are called *oriented edges* or *arcs*, and written in the form $(i \to j)$. Analogous definitions to the ones given above apply for directed graphs.

### 2.1.2 Random Graphs

**Graph Ensembles**

It is often the case, especially when dealing with complex networks such as the ones appearing in finance and biology [63, 64], that a particular graph is considered just as a contingent realization of an ample class of them. The mathematical structure corresponding to this concept is that of random graph ensemble [65]. A *random graph ensemble* $\mathbb{G} = (\mathcal{G}, \mathbb{P})$ is a set of graphs $\mathcal{G}$ and a probability law $\mathbb{P}$ defined over it. The average value of some observable $A(G)$ over the ensemble is then given by

$$\mathbb{E}[A] \equiv \sum_{G \in \mathcal{G}} \mathbb{P}[G] \, A(G). \qquad (2.3)$$

As an example, the *average degree distribution* $\{p_k\}$ is given by and the (ensemble) average degree $z$ are defined as

$$p_k \equiv \mathbb{E}[\, p_G(k) \,] \qquad z \equiv \mathbb{E}_k \, k, \qquad (2.4)$$

where $\mathbb{E}_k$ denotes expectations over $p_k$.

We shall now introduce some random graph ensembles commonly used in statistical physics and other disciplines.

- $\mathbb{G}_{\text{ER}}(N, p)$: the Erdős-Rényi (ER) ensemble.

  Historically, this is the first random graph ensemble ever proposed, marking the beginning of random graph theory [66]. It is also the most simple one. Each graph of the ensemble has $N$ vertices and can be sampled as follow: start with a graph with $N$ vertices and no edges, then for each of the $\binom{n}{2}$ pairs of vertices independently add an edge with probability $p$. Calling $M$ the number of edges corresponding to a graph $G$, the law of the ER ensemble depends only on $M$ and is thus given by

  $$\mathbb{P}[G] = p^M (1-p)^{\binom{N}{2} - M}. \qquad (2.5)$$

  If $p$ is scaled as $p = \frac{c}{N}$, in order for the nodes to have finite connectivities, taking the limit $N \to +\infty$ we have that

  $$p_k \sim \frac{e^{-c} c^k}{k!}, \qquad (2.6)$$

that is, the degree is Poisson distributed with mean $c$, that is $z = c$ for large $N$.

In order to sample a graph from the ER ensemble it is convenient first to extract $M$ from the appropriate binomial distribution, then to assign each of the $M$ edges to a pair of nodes. This procedure requires $O(N)$ operations, while choosing each potential edge with probability $p$ would require $O(N^2)$ operations.

- $\mathbb{G}_{\text{RRG}}(N, z)$: the Random Regular Graph (RRG) ensemble.

  The measure of this ensemble is uniform over all the graphs with $N$ vertices where each vertex has connectivity $z$ [67]. Graph with this property are called regular. Notice that in order for the ensemble to be non-empty $Nz$ has to be an even number. Sampling of graphs in the RRG ensemble is more involved than in the ER ensemble.

  An exact sampling method was proposed by Bollobás [68], and goes under the name of configuration model, although we reserve this name to the more general ensemble and generative procedure we present next. Here we resume Bollobá s' procedure. Start with a set of $Nz$ elements, called *stubs* or *half-edges*, and partition them in $N$ subset containing $z$ stubs each and representing the nodes of the graph. Pair the stubs at random (they are even), that is choose uniformly at random a matching among them (see also Sec. sec:prel-matching). Construct the graph associated to this particular pairing (sometimes called "configuration"). Repeat the pairing procedure until the resulting graph has no self-loops and multi-edges. It is easy to show that this contruction induces a law that is uniform over all regular graphs. It is very slow though, and it is exponentially slower as the degree increases, as we will discuss in Paragraph 2.1.2, due to the fact that the probability of obtain a non-simple graph after a matching is finite.

  An alternative method, proposed in Ref. [69], is less computationally expensive at the expense of provably yielding a uniform measure only in the infinite graph limit. The method is a slight and reasonable variation to Bollobás's one. One starts matching pairs of stubs one by one at random, and if at a certain step a self-loop or a multi-edge is produced, only this last matching is discarded and a new one is proposed, instead of starting over the entire matching procedure. If during the procedure no feasible pairing remains available the graph is discarded and the procedure restarts with a new set of unpaired stubs.

  This is the method we use throughout the thesis to generate random regular graphs. Since we deal with finite size correction we took care to check that this heuristic that does not produce biases in our numerical estimates to the relevant (i.e. $O(1/N)$) order.

- $\mathbb{G}_{\text{CONF}}(N, \{k_i\}_{i=1}^N)$: the configuration model.

  This graph ensemble has uniform measure over all the graphs of $N$ nodes where the node $i$ has degree $k_i$. Notice that this definition is non-trivial only if $\sum_i k_i$ is even. The degree distribution in this ensemble is the same for all graphs and $\mathbb{G}_{\text{CONF}}$ is sometimes named arbitrary degree distribution ensemble.

Both the exact and the approximate sampling procedure for this ensembles are the straightforward generalizations of the one given for the random regular graph ensemble. On the other hand $\mathbb{G}_{\mathrm{RRG}}$ is just a particular instance of the $\mathbb{G}_{\mathrm{CONF}}$ ensemble.

Alternative definition can be considered for this ensemble, for example vertex degrees could be independently generated from a distribution $\{p_k\}$ instead of being fixed to $\{k_i\}$.

Random graph theory, and statistical physics as well, is often concerned with the properties of ensemble in the limit of large $N$, that we will call the *large graph limit*, therefore we will assume the existence of a prescription for the scaling with $N$ of the parameters defining the ensembles, such as $p$ in $\mathbb{G}_{\mathrm{ER}}(N,p)$ and $\{k_i\}$ in $\mathbb{G}_{\mathrm{CONF}}(N,\{k_i\})$. In what follows we will be interested mainly in *sparse random graph ensembles*, that is ensembles with finite average degree $z$ in the limit $N \uparrow \infty$. Therefore, for instance, we will assume the edge probability $p$ in $\mathbb{G}_{\mathrm{ER}}$ to scale as $p = \frac{c}{N}$ to achieve sparsity.

Beyond sparsity, a property possessed by finite dimensional lattices as well, some peculiar features characterize the ensembles we presented above. One of them, very convenient for their analytical investigation, is the property of being *locally tree-like*. We will give a more precise definition of locally tree-likeness in Paragraph 2.1.2, the basic idea though is that a ball of finite size centered on a randomly chosen node does not contain any loop with a probability that goes to one as $N$ goes to infinity. In this sense the random graph is locally a tree. As discussed in Paragraph 2.1.2, loops of finite length are rare in random graphs.

Since we mainly discuss the large graph limit, let us define for convenience the symbol $\sim$ as

$$f(N) \sim g(N) \qquad \Longleftrightarrow \qquad f(N) = g(N) + o(1) \quad \text{as} \quad N \uparrow \infty. \qquad (2.7)$$

Hereafter we will refer to $\mathbb{G}$ as to one of the ensembles above in the large $N$ limit.

Asymptotic properties depended only on the degree distribution $\{p_k\}$, due to the lack of additional constraints in the definition of the ensembles. For example it can be shown that the ensemble $\mathbb{G}_{\mathrm{ER}}$ is asymptotically equivalent to $\mathbb{G}_{\mathrm{ER}}$ with Poissonian distributed $\{k_i\}$. More refined random topologies where utter local or global properties can be devised (e.g. assortativity [63]), their analysis is usually more involved though.

Let us define the random variable *edge perspective degree* $\tilde{k}$, of a node chosen as follows: choose uniformly at random one of its edges from a random graph $G$ in $\mathbb{G}$ and let $\tilde{k}$ be the residual degree (i.e. $\tilde{k} = k - 1$) of one of its extremities. A little thought shows that the probability $\tilde{p}_{\tilde{k}}$ of obtaining a certain residual degree $\tilde{k}$ is proportional to $\tilde{k} + 1$ and to the number of nodes having degree $\tilde{k} + 1$. Therefore, for the edge perspective degree distribution $\{\tilde{p}_{\tilde{k}}\}$, and its average edge perspective degree $\tilde{z}$, that we also call residual degree distribution and average residual degree, we have

$$\tilde{p}_{\tilde{k}} \sim \frac{1}{z} p_{\tilde{k}+1} (\tilde{k} + 1) \qquad \tilde{z} \equiv \mathbb{E}_{\tilde{k}} \tilde{k}, \qquad (2.8)$$

where $\mathbb{E}_{\tilde{k}}$ denotes expectations over $\tilde{p}_{\tilde{k}}$. For Poissonian random graph $\tilde{z} = z$, while for RRGs $\tilde{z} = z - 1$ trivially.

Sparse graph ensembles $\mathbb{G}$ enjoy some powerful factorization properties due to the large graph limit. For example, if we choose uniformly and independently at random a $n$ nodes $i_1, \ldots, i_n$, their joint degree distribution factorizes as follows:

$$\mathbb{P}[\deg(i_1) = k_1, \ldots, \deg(i_n) = k_n] \sim \prod_{m=1}^{n} \mathbb{P}[\deg(i_m) = k_m]. \qquad (2.9)$$

Also the joint degree distribution of nodes in a finite neighbourhood of a given node factorizes. The local structure of finite neighbourhoods will be discussed in Paragraph 2.1.2.

The features we discussed have very important consequences to the analytical treatment of sparse random graph ensembles, in fact they are the among basic assumption on which the Cavity Method relies. We point the reader to Ref. [15] for a more in-depth discussion of all these arguments.

**Giant Component**

An outstanding feature of our sparse random graphs ensembles is the existence of a sharp threshold, depending on the first two moments of the degree distribution, and in particular only on the average residual degree $\tilde{z}$, above which a *giant connected component* arises. For some values of the parameters defining a graph ensembles in fact, the almost sure presence of a unique connected component, which contains a finite fraction of the nodes, can be rigorously proven [70]. Below that threshold, it is possible to show that *w.h.p.* only tree-like components exist, thus almost any graph sampled from $\mathbb{G}$ is a *forest*.

We will give an heuristic argument, which holds to for large graphs, to explain this phenomena. This the first *cavity argument* we present in this thesis, and it goes as follows: consider a uniformly chosen vertex $i$ from the random graph $G$ and call $m$ the probability it belongs to a giant component. Notice that according to this definition $mN$ would be the average size of the giant component. The node $i$ is not in the giant component if no one of its neighbours is in the giant component once the node $i$ has been removed from the graph, that is once a *cavity* is formed. We call $m_{\text{cav}}$ this probability for one of the neighbours to belong to the giant component in the cavity graph. In the large graph limit we can write

$$1 - m = \mathbb{E}_k (1 - m_{\text{cav}})^k, \qquad (2.10)$$

where in the r.h.s. we exploited the factorization of the joint probability distribution in the cavity graph. We can then write a self-consistent equation for $m_{\text{cav}}$ reiterating previous argument:

$$1 - m_{\text{cav}} = \mathbb{E}_{\tilde{k}} (1 - m_{\text{cav}})^{\tilde{k}}, \qquad (2.11)$$

where the residual degree $\tilde{k}$ is considered. Last equation always admits the solution $m_{\text{cav}} = 0$. It turns out, as it can be easily seen expanding Eq. (2.11) for small $m_{\text{cav}}$, that a second solution exists for $\tilde{z} > 1$, and disappears continuously for $\tilde{z} < 1$. Therefore the threshold condition for the emergence of a giant component is given by

$$\tilde{z}_c = 1. \qquad (2.12)$$

**Simple loops distribution**

As already stated, and as discussed more in depth in the next Paragraph, sparse random graphs have the local structure of a tree in the large $N$ limit. Still a finite number of loops with finite lengths, therefore containing a finite number of nodes although a null fraction of them, survives the limit. These loops are simple, that is non-intersecting. It can be proven [67, 68, 71] that, given a set of non-negative integers $\{r_\ell\}_{\ell=3}^{+\infty}$, the joint distribution of the number of simple loops for large $N$ is given by

$$\mathbb{P}[\{r_\ell \text{ simple loops of length } \ell\}_{\ell=3}^{+\infty}] \sim \prod_{\ell \geq 3} \frac{e^{-\lambda_\ell}\lambda_\ell^{r_\ell}}{r_\ell!}. \tag{2.13}$$

Therefore the numbers of loops of lengths $\ell$ are Poissonian independent random variables, with mean

$$\lambda_\ell \equiv \frac{\tilde{z}^\ell}{2\ell}. \tag{2.14}$$

A simple heuristic argument can be given to understand the mean value $\frac{\tilde{z}^\ell}{2\ell}$. In the setting of the configuration model consider the probability of two randomly chosen nodes, $i_0$ and $i_1$, to be connected by and edge. Calling $k_0$ and $k_1$ the degree of the two nodes, to the leading order in $1/N$ each of the $k_0$ stubs of $i_0$ tries to connect independently at random with one of the $k_1$ stubs of $i_1$ over the $zN$ total stubs. Therefore the probability of the edge is given by

$$\begin{aligned}\mathbb{P}[(i_0, i_1) \in E] &= \sum_{k_0,k_1} p_{k_0} p_{k_1} k_0 \frac{k_1}{zN} + O\left(\frac{1}{N^2}\right) \\ &= \frac{z}{N} + O\left(\frac{1}{N^2}\right).\end{aligned} \tag{2.15}$$

Now consider the probability of the existence of a randomly chosen open path $(i_0, i_1, i_2)$ of length two. We have a factor $k_0$ accounting for the outgoing stubs from $i_0$, then a factor $k_1/zN$ to connect to $i_1$, then $k_1 - 1$ ways to depart from $i_1$ and a factor $k_2/zN$ to connect to $i_2$. In the end

$$\begin{aligned}\mathbb{P}[\{(i_0, i_1), (i_1, i_2)\} \subset E] &= \sum_{k_0,k_1,k_2} p_{k_0} p_{k_1} p_{k_2} k_0 \frac{k_1(k_1-1)}{zN} \frac{k_2}{zN} + O\left(\frac{1}{N^3}\right) \\ &= \frac{z\tilde{z}}{N^2} + O\left(\frac{1}{N^3}\right).\end{aligned} \tag{2.16}$$

Extending the above reasoning to the probability of existence of an open chain of length $\ell$ we straightforwardly obtain

$$\mathbb{P}[\text{open chain of length } \ell] = \frac{z\tilde{z}^{\ell-1}}{N^\ell} + O\left(\frac{1}{N^{\ell+1}}\right). \tag{2.17}$$

The number of different paths is given by the way of choosing $\ell + 1$ nodes, $N!/(N-\ell-1)!$ divided by a factor two accounting for path reversal symmetry, therefore

$$\mathbb{E}[\# \text{ open chains of length } \ell] = \frac{z\tilde{z}^{\ell-1}}{2}N + O(1). \tag{2.18}$$

The computation of the probability of existence of a certain closed path

$$\mathbb{P}[\text{ closed chain of length } \ell\,] \sim \frac{\tilde{z}^\ell}{N^\ell} + O\left(\frac{1}{N^{\ell+1}}\right) \tag{2.19}$$

Sine a simple loop of length $\ell$ contains $\ell$ nodes and has $2\ell$ has a symmetry factor, we finally have

$$\mathbb{E}[\,\#\text{ closed chains of length } \ell\,] \sim \frac{\tilde{z}^\ell}{2\ell}, \tag{2.20}$$

where we have recovered the results expected from Eqs. (2.13) and (2.14). The factorized Poissonian form of the loops' joint distribution is due to the asymptotic independence of this rare events.

Also in this framework an important result regarding the configuration model can be simply understood. In the large $N$ limit, the probability that the graph induced by a random pairing is simple is given by [67]

$$P[\text{pairing is simple}] \sim e^{-\frac{\tilde{z}}{2} - \frac{\tilde{z}^2}{4}}. \tag{2.21}$$

This formula can be seen as an extension of Eq. (2.13) to loops of length $\ell = 1$ and $\ell = 2$ (self-loops and multi-edges), and in this sense it is just the probability for the graph to contain $r_1 = 0$ and $r_2 = 0$ of them.

Notice that at finite $N$ the maximum length of simple loops is limited by $N$ itself. Moreover probabilistic arguments such as the one leading to Eq. (2.13) fail as soon as $\tilde{z}^\ell \approx N$, that is for lengths greater then $\ell \approx \frac{\ln N}{\ln z}$.

**Local weak convergence**

Here we define one of the most important characteristic of sparse random graph ensembles, that of being locally tree-like, in a sounder mathematical framework. In these ensembles in fact almost any finite neighbourhood of a node is isomorphic to a random tree in the large graph limit. To accurately define this property we have to go through some definitions. We will follow the analytical framework proposed by Aldous and Steele [50], since it provided very useful in turning Cavity Method predictions to rigorous results.

Let $G_*$ be the class of graphs with a countable (also infinite) number of nodes and a distinguished node that we call *root*. We will define a notion of convergence of elements in $G_*$ through which we will infer a topology on $G_*$, so that we get to use all the instruments from weak convergence theory. For $G \in G_*$ let us call $B_l(G)$ the $l$-neighbourhood of its root.

**Definition 1** (Convergence in $G_*$). *Let $\{G_n\}$ be an infinite sequence of elements of $G_*$. We say that $\{G_n\}$ converges to $G_\infty$ if for each integer $l$ there is a $n_l$ such that $B_l(G_n)$ is isomorphic to $B_l(G_\infty)$ for each $n > n_l$.*

It can be shown that it is possible to provide $G_*$ with a topology compatible with this notion of convergence, such that it becomes a complete separable metric space. As a consequence all the tool of weak convergence theory apply and we denote weak convergence of measure on $G_*$ as $\mu_n \xrightarrow{d} \mu$.

Let $\mathbb{G}_N$ be one of the ensemble of paragraph 2.1.2. Let $\{G_N\}$ be a random sequence of graphs in $G_*$ sampled from $\mathbb{G}_N$, with root selected uniformly at random. Denote with $\{\mu_N\}$ the corresponding sequence of laws. We call $\{T_l\}$ a random sequence of trees of increasing depth $l$ that we define inductively. $T_0$ has a single node $i_0$. $T_1$ is constructed from $T_0$ extracting $k_0$ from $\{p_k\}$, adding $k_0$ nodes and linking them with $i_0$. We call these last nodes *children* of $i_0$. We build $T_2$ extracting independently for each children $i_m$ at the first level $\tilde{k}_m$ new children, sampled from $\{p_k\}$, at the second level and linking them with $i_m$. Each successive level is constructed analogously, leaves of the preceding level sprouting independently a new offspring according to an edge-perspective degree-distribution $\{\tilde{p}_k\}$.

Let us call $T$ the random limiting object of $\{T_l\}$ and $\mu$ its law. Graphs sampled from the ensemble $\mathbb{G}_N$ are then said to be *locally tree-like*, as it can be proved that $\mu_N \xrightarrow{d} \mu$ (i.e. they are locally isomorphic to a random tree). With a little abuse of notation we could state this as $B_l(\mathbb{G}_N) \stackrel{\mathrm{d}}{=} B_l(T) \stackrel{\mathrm{d}}{=} T_l$ for large $N$.

### 2.1.3 Factor Graphs

In order to introduce the Cavity Method, an useful analytical tool to deal with diluted disordered systems, it's convenient to make use of the factor graph mathematical structure (it's not mandatory though, we could have resorted to ordinary graphs with a little more effort and clumsiness). Factor graphs allows us to model a great variety of inferential statistic problems [15] [72] and are the natural set for the Belief Propagation (BP) algorithm we are going to introduce in the next paragraph. They are used in statistical physics as a graph representation of the geometrical structure (defined by their interactions) imposed on a set of variables.

A *factor graph* is an ordered triple $G = (V, F, E)$, where $V$ is the set of *variable nodes* (without loss of generality we take it to be $[N]$) and $F$ is the set of *function nodes* (for us will be $F = [M]$). $E$ is a set containing ordered couple $(i, r)$ such that $i \in V$ and $r \in F$, so that $E \subset V \times F$. We will generically call nodes elements of both $V$ and $G$. To easy the notation we will use the letters $i, j, ...$ for variable nodes (that we shall also call variable nodes) and $r, s, ...$ for function nodes (f-nodes). In summations or productories where dummy indices $i$ or $r$ are present without further specifications, they are intended to run over the sets $V$ and $F$ respectively. Factor graphs can be subsumed in the contest of ordinary Graph Theory as bipartite graphs and from Graph Theory we shall import definitions and results. We call *neighbourhood* of a variable node the set $\partial i = \{r \in F : (i, r) \in E\}$ and neighbours its elements. The *degree* of a node is the number of its neighbours. We write unambiguously $i \in r$ to mean in $i \in \partial r$. We call *v-neighbourhood* of $i$ the set of variable nodes having a common neighbour with $i$. Specular definitions hold for f-nodes. A *path* between nodes is a sequence of edge $P = (e_1, e_2, ..., e_p)$ such that two consecutive edges share a common node and it *length* is the number of edge it contains ($p$ for $P$). The *distance* between two nodes is the length of shortest path such that the first and the last edge contain respectively the two nodes. With this definition neighbours of a node are simply nodes at distance one from it and we can define an $l$-neighbourhood of a node as the set of nodes at distance at most $l$ from it. A *loop* is a path that start and ends with the same node and a *tree* (we will see the importance of this definition) is a factor graph without loops.

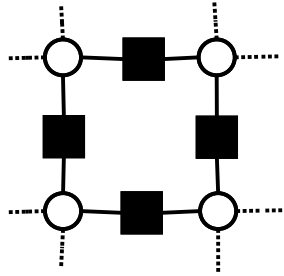

**Fig. 2.1.** Visual representation of the factor graph associated to the bidimensional Ising model. Black squares are function nodes, white circles are variable nodes.

Having gone through the definitions, we want to relate the abstract mathematical structure with actual (physical) problems. To do so we extend the factor graph definition to that of *graphical model*, that is a factor graph with variables $x_i$ associated to its variable nodes $i$, taking values in a set $X$ (that we shall suppose to be finite), and real-valued function $\psi_r$ associated to each f-node $r$ and taking as argument the variables associated to its neighbours. If we write $x_r = \{x_i : i \in r\}$ for the argument of $\psi_r$ and $\mathbf{x}$ for the entire set of variables associated to $G$, for a generic graphical model (that we shall also call $G$ for convenience) we can express the probability density distribution as

$$\mu_G(\mathbf{x}) = \frac{\prod_{r=1}^M \psi_r(x_r)}{Z_G}. \tag{2.22}$$

$Z_G$ is a normalization factor that statistical physicists will interpret as the partition function. Vice-versa to each probability density of type (2.22) (that is every one, but the corresponding factor graph could be trivial in some cases) we can naturally assign a graphical model. We arrived then to a probabilistic interpretation of a factor graph structure, where f-nodes represent 'interaction' among variable nodes neighbours. As an example, in a Ising spin model (see Fig. 2.1) the interaction term would be

$$\psi_{ij}(x_i, x_j) = e^{\beta J x_i x_j}. \tag{2.23}$$

The utility of factor graphs has been discovered by statistical physics community thanks to the success of Cavity Method in solving computational problems (random K-Sat for example) that couldn't be modelled on ordinary (non bipartite) graphs . Sometimes we will use the term graph instead of factor graph when the context allows no ambiguity.

## 2.2 Spin glasses and Random Fields

Archetypal example of disordered systems are spin glasses and random fields models. Here we will briefly describe their phenomenology. For a broader discussion we point out to the reader Refs. [13] and [73]. Both models present disorder in the form of some kind of randomness encoded in the Hamiltonian: random couplings for spin glasses, and random external magnetic fields for random field models. This kind of disorder, that can be thought as to correspond to very slow (frozen) degrees of freedom, goes under the name of *quenched disorder*. The free energy of this system depends on the particular realization of the disorder. One is then usually interested in the average properties of this system, the most relevant one being the average free energy

$$f \equiv \mathbb{E} f_{\text{sample}} \equiv -\frac{1}{\beta N} \mathbb{E} \ln Z_{\text{sample}}, \tag{2.24}$$

where $E$ is the average over the disorder. Some observables $O$ in disordered systems, most noticeably the free energy, have the *self-averaging* property, that is the relation

$$\lim_{N \to +\infty} P(|O - \mathbb{E}[O]| > \epsilon) = 0, \tag{2.25}$$

holds, therefore almost any realization of the system has the same value of $O$ in the thermodynamic limit.

Spin glasses are statistical mechanics models of magnetic systems presenting competing interactions, at odd with the purely ferromagnetic usual ones. In this thesis we will deal with the Ising spin glass, where the dynamical variables are binary (Ising) spins $\{\sigma_i\}_{i=1}^N$ taking values in $\{-1, +1\}$. Although binary variables provide a drastic simplification to real systems, Ising models are widely recognized to be able to catch in many cases most qualitative aspects of real systems. Generalizations to vectorial spins go under the name of Heisenberg models. Given a graph $G = (V, E)$ as the underlying topology of the model, we associate to each edge an interaction $J_{ij}$ extracted independently at random according to some distribution $P(J)$. The single sample, $\{J_{ij}\}$-dependent Hamiltonian is then given by

$$\mathcal{H}_J[\{\sigma\}] = -\sum_{(i,j)} J_{ij} \sigma_i \sigma_j - h \sum_i \sigma_i \tag{2.26}$$

and the corresponding (random) partition function is

$$Z_J = \sum_{\{\sigma\}} e^{-\beta \mathcal{H}_J[\{\sigma\}]} \tag{2.27}$$

If $P(J)$ has support only on positive reals the model is called a disordered ferromagnetic model, while if $J_{ij}$ can take also negative values we call it a spin glass. We call *frustration* the presence, in a given sample, of simple loops such that the product of the signs of $J_{ij}$ along the path is $-1$, meaning that it is not possible to satisfy each bond, i.e. to find a configuration $\{\sigma_i\}$ such that $\sigma_i J_{ij} \sigma_j = 1$ along the frustrated loop. Frustration is the distinguishing feature of spin glasses, and the cause of some of their peculiar features, such as degeneracy of ground states and computational hardness.

The correct solution of the fully-connected spin glass model (the Sherrington-Kirkpatrick (SK) model) was obtained by Parisi [10] within the replica method, through a peculiar scheme to escape the symmetry of the action under replica permutation. This approach, called (Full) Replica Symmetry Breaking, beside yielding the correct value for the average free energy, hinted to a deep probabilistic structure of the SK models, clarified in the following years. Below a certain temperature $T_c$ the system enters a disordered glassy phase, characterized by the slowing down of the dynamics, slow decay of correlation functions, presence of an exponential number of pure states. The complexity of states, ergodicity breaking, ultrametricity, stochastic stability, slowing down of the dynamics,became central concept to the analysis of the SK model. We point to Refs. [13] and [52] for a detailed discussion of the argument. Finite dimensional spin glass on the other hand are a much open problem. Two competing theories, one inspired to the solution of the SK model, the other, the droplet theory, to domain based arguments, depicts two very different scenarios for the thermodynamic phase under $T_c$. Replica symmetry breaking advocates support the existence of a spin glass phase with an exponential number of pure states [74], while the droplet community claims that only two symmetry related pure states exist [75]. Monte Carlo numerical simulations, plagued by the extreme slowing down of the dynamics, have not given yet conclusive results, despite the intense effort of the last years.

Another class of disordered systems is that of magnetic systems in a random external field. The prototype of these systems is the Random Field Ising Model (RFIM), defined by the Hamiltonian

$$\mathcal{H}_h[\{\sigma\}] = -J \sum_{(i,j)} \sigma_i \sigma_j - \sum_i h_i \sigma_i, \qquad (2.28)$$

with $J > 0$ and $\{h_i\}$ independent identically distributed random variables $\mathbb{E}\, h_i = 0$, $\mathbb{E}\, h_i h_j = \delta_{ij} \Delta^2$. The control parameters for the system are the temperature $T$ and the ratio $r = \frac{\Delta}{J}$. A simple argument, due to Imry and Ma, shows that the lower critical dimension for the system is $d_l = 2$. For $d \geq 2$ in fact there is a critical line $r_c(T)$ in the $r - T$ space separating the paramagnetic phase above from a ferromagnetic phase below. For pure ferromagnetic systems instead the lower critical dimension is $d_l = 2$. While there was some argument in favour of a spin glass phase for some values of the parameters [76,77], this possibility has been recently excluded [78]. Another interesting conjecture, based on a supersymmetry argument [56], concern the equivalence to all order of perturbation theory of random field systems in dimension $d$ with pure systems in dimension $d - 2$. This equivalence is called dimensional reduction, and it would predict a lower critical dimension $d_l = 3$ at odd with the rigorous result $d_l = 3$. Therefore dimensional reduction breaks down at least at low dimension. Recently some authors, using a functional renormalization group approach, proposed the value $d^* \approx 5.1$ as the lower critical dimension for the validity of dimensional reduction [79]. In the RFIM disorder is a relevant direction for the renormalization group flow, and the phase transition is controlled by the zero temperature fixed point [53]. In-depth discussions of the RFIM can be found in Refs. [80] and [73].

## 2.3 Replica Method

Replica method is a powerful analytical tool to address systems with quenched disorder [13]. The necessity of its use steams from the difficulty in averaging the random partition function $Z$ as argument of the logarithm in

$$f = -\frac{1}{\beta N} \mathbb{E} \ln Z. \qquad (2.29)$$

Its main ingredient is the mathematical identity

$$\lim_{n \to 0} \frac{Z^n - 1}{n} = \ln Z, \qquad (2.30)$$

such that we can write

$$f = \lim_{n \to 0} -\frac{1}{n\beta N} \ln \mathbb{E} Z^n. \qquad (2.31)$$

Computation of $\mathbb{E}Z^n$ is then performed for integer $n$, since this case is mathematically tractable. One therefore introduces $n$ copies of the system, called *replicas*, having the same disordered in the Hamiltonian. Computation is then carried out, usually introducing an order parameter and performing a steepest descent evaluation point in the thermodynamic limit ($N \uparrow \infty$). The $n \downarrow 0$ limit is finally taken, yielding the average thermodynamic free energy. There are some faulty steps in this approach, preventing it from being rigorous: analytical continuation to $n = 0$, which is not an accumulation point; exchange of the limits $n \downarrow 0$ and $N \uparrow \infty$; saddle point evaluation is always performed in a restricted subspace, defined through some symmetry related *ansatz*), since minimization of the action in the full parametric space is infeasible. Nonetheless replica method has proved to be an highly effective method and some of its predictions where proved to be right (sometimes decades later), while none to be wrong.

Throughout this thesis we will give numerous examples and applications of the replica method , namely in Chapters 3,4,5,7 and 8, therefore we will not go to great detail in this paragraph.

As an example of the method, the replicated partition function of the Sherrington-Kirkpatrick spin glass model, after an Hubbard-Stratonovich transformation and summation over the the replicated spin configurations, for large $N$ is given by

$$\mathbb{E}Z^n \sim \int \prod_{a<b}^n \mathrm{d}q_{ab}\ e^{-NS[\{q\}]} \qquad (2.32)$$

where the replicated action is given by

$$S[\{q\}] = -n\frac{\beta^2}{4} + \frac{\beta^2}{2}\sum_{a<b}^n q_{ab}^2 - \ln \sum_{\sigma^1,\dots,\sigma^n} e^{-\beta \sum_{a<b} q_{ab}\sigma^a \sigma^b}. \qquad (2.33)$$

The order parameter here is the $n \times n$ zero diagonal symmetric matrix $\{q_{ab}\}$, for a total of $n(n-1)/2$ degrees of freedom. In order to make the analytical continuation to real $n$, a explicitly $n$-dependent ansatz for $\{q_{ab}\}$ has to be proposed. The most simple one is to the Replica Symmetric (RS) ansatz, and assumes independence of the saddle point value of $q_{ab}$ from its indexes, that is $q_{ab} = q$ for all $a,b, a \neq b$. The RS ansatz is usually exact on fully-connected or locally-tree like topologies in the high temperature phase. In spin glass systems though, below a certain temperature $T_c$ a more complicated ansatz that goes under the name Replica Symmetry Breaking (RSB) is needed. This low temperature phase is called the spin glass phase. In this thesis we will deal only with RS systems, therefore we point the reader to Refs. [13] and [15] for an in-depth discussion of the broad phenomenology of the spin glass phase.

As a final note we want to discuss the equivalence between to alternative description for the order parameter in Ising model on diluted systems. One can choose as a set of order parameters the multioverlaps $\{q_{a_1\ldots a_p}\}$, with $p \geq 1$, expressing correlation among replicas according to

$$
\begin{aligned}
q_{a_1} &= \frac{1}{N} \sum_i \mathbb{E}\langle \sigma_i^{a_1} \rangle \\
q_{a_1 a_2} &= \frac{1}{N} \sum_i \mathbb{E}\langle \sigma_i^{a_1} \sigma_i^{a_2} \rangle \\
&\ldots \\
q_{a_1\ldots a_p} &= \frac{1}{N} \sum_i \mathbb{E}\langle \sigma_i^{a_1} \ldots \sigma_i^{a_p} \rangle \\
&\ldots
\end{aligned}
\tag{2.34}
$$

where $\{s_i^1\},\ldots,\{\sigma_i^n\}$ are the thermally independent replicas and $\langle \bullet \rangle$ is the thermal average. In fully-connected systems only the first two moment $q_a$ and $q_{ab}$ are needed to have a complete description of the system, since due the central limit theorem effective fields acting on replicas are Gaussian distributed. An example of multioverlap description is given in the matching problem of Chapter 7. Although the problem is set on a fully-connected graph it acts effectively as a random diluted graph, since only a small fraction of costs contributed to the relevant configurations.

The other possible description of replicated diluted system, and the on we employ most often during the thesis, is the one given by the order parameter $\rho(\sigma^1, \sigma^2, \ldots, \sigma^n)$, a real or complex valued function of the $2^n = |\{-1, 1\}^n|$ replicated spin configurations. Throughout the thesis we will use the notation $\rho(\sigma) \equiv \rho(\sigma^1,\ldots,\sigma^n)$ and we will refer to $\rho(\sigma)$ also as a vector with $2^n$ components. The function $\rho(\sigma)$ is related to the multioverlaps $\{q_{a_1\ldots a_p}\}$ by

$$
q_{a_1\ldots a_p} = \sum_{\sigma^{a_1},\ldots,\sigma^{a_p}} \rho(\sigma)\, \sigma^{a_1} \ldots \sigma^{a_p}.
\tag{2.35}
$$

The Replica Symmetric ansatz for this two sets of order parameters corresponds, with a little abuse of notation, to $\rho(s^1,\ldots,\sigma^n) \equiv \rho(\sum_{a=1}^n \sigma^a)$ and $q_{a_1\ldots a_p} \equiv c_p$.

## 2.4 Cavity Method

### 2.4.1 Belief Propagation

Cavity Method has been developed in the context of mean field glass theory [13] as an alternative to the Replica Trick approach and turned out to be very effective in tackling statistical models on locally tree-like structures. It is a generalization of the long known Bethe-Peierls approximation and it's main idea, used in different forms and in different scenarios [13] [15], is that properties of a system of large size do not change too much if the system is increased by a single element. Following [15] we will take an algorithmic approach introducing the Cavity Method through Belief Propagation, a message passing algorithm on single instances of the problem, and we will deal with random instances in the next paragraph. Belief Propagation (BP) is a message passing algorithm on a generic graphical model $(G, \{\psi_r\})$. It gives an estimate of $\mu$ probability distribution marginals that is correct for tree factor graphs and has proved to be fairly accurate on some models with long (or even short) loops and short range correlations. If BP equations admit a unique fixed point, BP fixed point messages can be used to calculate the so called Bethe approximation to free energy.

Let's define $\mu$ distribution marginals as

$$\begin{aligned}
\mu_i(x_i) &= \sum_{x_{V \setminus i}} \mu(\mathbf{x}) & i \in V; \\
\mu_r(x_r) &= \sum_{x_{V \setminus \partial r}} \mu(\mathbf{x}) & r \in F; \\
\mu_A(x_A) &= \sum_{x_{V \setminus A}} \mu(\mathbf{x}) & A \subset V.
\end{aligned} \quad (2.36)$$

We then introduce the aforementioned *messages*, that is we assign to each $(i, r) \in E$ two probability distributions, $\hat{\nu}^{(t)}_{r \to i}(x_i)$ and $\nu^{(t)}_{i \to r}(x_i)$, over the space $X$, where index $t$ denotes $t$-th iteration of the BP algorithm we are going to define.

BP update rules are local, messages at time $t+1$ at each edge are calculated from messages at time $t$ on adjacent edges. We write the BP update rules, also known as *sum-product* rules as

$$\begin{aligned}
\hat{\nu}^{(t)}_{r \to i}(x_i) &\cong \sum_{x_{r \setminus i}} \psi_r(x_r) \prod_{j \in r \setminus i} \nu^{(t-1)}_{j \to r}(x_j), \\
\nu^{(t)}_{i \to r}(x_i) &\cong \prod_{s \in i \setminus r} \hat{\nu}^{(t)}_{s \to i}(x_i).
\end{aligned} \quad (2.37)$$

Symbol $\cong$ denotes equivalence up to a constant factor given by the normalization condition.

Starting from a given initial condition $\{\nu^{(0)}_{i \to r}\}$, that could be the uniform distribution over $X$ (i.e. $\nu^{(0)}_{i \to r}(x_i) = \frac{1}{|X|}$), one could ask himself if Eq. (2.37) leads to some limit distributions at finite $t$ or as $t \uparrow \infty$. As a complementary point of view we can search for solution of

$$\begin{aligned}
\hat{\nu}_{r \to i}(x_i) &\cong \sum_{x_{r \setminus i}} \psi_r(x_r) \prod_{j \in r \setminus i} \nu_{j \to r}(x_j), \\
\nu_{i \to r}(x_i) &\cong \prod_{s \in i \setminus r} \hat{\nu}_{s \to i}(x_i),
\end{aligned} \quad (2.38)$$

fixed points of dynamic (2.37), and study their stability and their attraction basin. We notice that a probability distribution over a finite set of $|X|$ elements can be parametrized as a point of the $|X|$-dimensional simplex, so distribution limits are well defined using equivalent norm in these spaces.

The *BP estimate* of marginal $\mu_i$ after $t$ iterations is given by

$$\nu^{(t)}_i(x_i) \cong \prod_{r \in i} \hat{\nu}^{(t)}_{r \to i}(x_i) \quad (2.39)$$

A first taste of the consistence and utility of these definitions is given by the following fundamental theorem:

**Teorema 1** (BP is exact on trees). *Let $(\mathcal{G}, \{\psi_r\})$ be a tree-like graphical model with diameter $t_*$. Then:*

1. *For any initial condition BP messages converge to the unique fixed point in $t_* + 1$ iterations at most.*

2. *Fixed point messages yield an exact estimate of marginals, $\nu_i^*(x_i) = \mu_i(x_i) \quad \forall i \in V$.*

*Proof.* In tree-like graphical models BP it's just a clever way to sum over variables in the partition function, beginning from leaf and down to the root. Let $\mathcal{T}_{i \to r}$ denote the connected component including variable $i$ once removed $(i, r)$ from $G$. We shall call $t_*(i \to r)$ the depth of $\mathcal{T}_{i \to r}$.

We are going to prove that, for each number of iteration $t > t_*(i \to r)$, message $\nu_{i \to r}^{(t)}$ converges to the marginal distribution of $i$ in the graphical model $\mathcal{T}_{i \to r}$. It follows that for each $t > t_*$ every marginal is exactly computed. Proof proceeds by induction on $t_*(i \to r)$. The first step is trivial: $\mathcal{T}_{i \to r}$ is composed by the single node $i$, $t_*(i \to r) = 0$ and its marginal distribution in $\mathcal{T}_{i \to r}$ given by (2.39) is equal to $\nu_{i \to r}^{(t)} = \frac{1}{|\mathcal{X}|} \quad \forall t > 0$ in the original model thanks to (2.37). Let our assumption hold true for $t_*(i \to a) \leq \tau$ and let's prove it for $t_*(i \to r) = \tau + 1$. Nodes $j \in s \setminus i$, with $s \in i \setminus r$, are the roots of subtrees with depth $\tau$ at most, therefore we can use our inductive hypothesis and for $t > \tau$ messages $\{\nu_{j \to s}^{(t)}\}$ correspond to marginals on respective subtrees. Putting in $\{\nu_{j \to s}^{(t)}\}$(2.37) and thanks to (2.39) they grant the exact marginal of $i$ in $\mathcal{T}_{i \to r}$, which is the message $\nu_{i \to r}^{(t+1)}$ in the original model. This completes the proof. □

Since $t_* < N$, once fixed the root $i$, assuming the time to the neighbours of a given node $a$ to be of order $O(deg_a)$ and taking the factor graph to be a uniformly sparse tree, we found an algorithm of complexity $O(N)$ to exactly compute the marginal $\mu_i$ iterating BP equations only for messages descending from leafs to the root. This is a huge leap forward compared to the naive approach of summing over $x_{V \setminus i}$ variables which takes $O(|X|^{N-1})$ operations. Moreover even on factor graphs with loops often BP estimates get to be fairly accurate, and especially so when length of loops are of order $O(\log(N))$, as in the ensemble we are going to study, and no long range correlations arise. Even in same cases where many short loops exists, such as in the two index assignment problem, BP has been proven to yield exact results [81].

Preceding theorem could be generalized as follows, still on tree factor graphs: given $A \subset V$, let $F_A = \{r \in F : \partial r \subseteq A\}$ and $\partial A = \{r \in F : \partial r \cap A \neq \emptyset\} \setminus F_A$. Then

$$\mu_A(x_A) \cong \prod_{r \in F_A} \psi_r(x_r) \prod_{r \in \partial A} \hat{\nu}_{r \to i(r)}^*(x_{i(r)}), \qquad (2.40)$$

where $i(r)$ is the unique variable node of $A$ neighbour of the the f-node $r \in \partial A$.

Update rule (2.37) can be often simplified taking into account the symmetries and the structure of the considered model. When the alphabet $X$ is composed by two letters, that is variables can take only two values, probability distribution over $X$ ( and messages in particular) can be parametrized by a single real number. For a given parametrization Eq. (2.37) can be recast as equations involving multivariable real functions, much more analytically and numerically amenable. A typical choice for Ising-like systems ($X = \{-1, 1\}$) is

$$\hat{h}_{r \to i}^{(t)} = \frac{1}{2\beta} \log \frac{\hat{\nu}_{r \to i}^{(t)}(1)}{\hat{\nu}_{r \to i}^{(t)}(-1)}, \qquad (2.41)$$

and the new messages $\{\hat{h}_{r \to i}^{(t)}\}$ acts as effective magnetic field on spins.

On a tree-like factor graph ensemble $\mathbb{G}$ we are mainly interested in the distribution of messages over the whole graph, that we denote with $P(\hat{\nu})$. In the large graph limit we can associate to BP equations (2.37) a distributional recursion rule

$$P^{(t)}(\hat{\nu}) = \mathbb{E} \int \delta\big(\hat{\nu} - \hat{f}_R(\{\hat{\nu}_{sj}\})\big) \prod_{j=1}^{C} \prod_{s=1}^{D_j} P^{(t-1)}(\hat{\nu}_{sj}) \mathrm{d}\hat{\nu}_{sj} \qquad (2.42)$$

that goes under the name of Density Evolution (DE). In last equation expectation is taken over random f-node residual degree $C$, random independent variable node redsidual degrees $\{D_j\}$ and a compatibility function label $R$. In fact, in random graphical models, each f-node has an associated compatibility function randomly chosen among a certain collection (e.g. in spin glass models the interaction term $J_{ij}$ is a random number ). Independence of messages coming from different subtrees is exploited factorizing the incoming messages joint probability distribution. Once sampled $C$ and $\{D_j\}$, $\mathcal{M}(X)$-valued function $\hat{f}_R$ corresponds to BP updated rules (we use only factor node to variable node messages for the sake of concision)

$$\hat{f}_R(\{\hat{\nu}_{sj}\})(x) \cong \sum_{\{x_j\}} \psi_R(x, \{x_j\}) \prod_{j=1}^{C} \prod_{s=1}^{D_j} \hat{\nu}_{sj}(x_j). \qquad (2.43)$$

Fixed points of Density Evolution take a crucial role in Cavity Method as we will soon elucidate.

### 2.4.2 Replica Symmetry

Let's turn our attention to fixed point BP equations

$$\hat{\nu}_{r \to i}(x_i) \cong \sum_{x_{r \setminus i}} \psi_r(x_r) \prod_{j \in r \setminus i} \prod_{s \in j \setminus r} \hat{\nu}_{s \to j}(x_j). \qquad (2.44)$$

Cavity Method in its simplest form, the Replica Symmetric (RS) Cavity Method, assumes that (2.44) have only one solution[1] or a symmetry related few (as in the Ising model in the ferromagnetic phase, we shall omit this case though). In physical terms this scenario corresponds to a single pure state for the system: the Gibbs measure $\mu$ is not decomposable in a convex linear combination of independent measures (when this assumptions fails to be true, and an exponential number of pure states arises, we have to resort to Replica Symmetry Breaking). What we are establishing is a one-to-one mapping of the BP (or DE) fixed points to the pure states of the system. Linkage to usual thermodynamic functions is done through the *Bethe free energy functional* $f_{\text{Bethe}}(\{\hat{\nu}_{r\to i}\})$ defined as

$$-\beta f_{\text{Bethe}}(\{\hat{\nu}_{r\to i}\}) = \frac{1}{N} \sum_{r\in F} \ln \Big( \sum_{x_r} \psi(x_r) \prod_{j\in r} \prod_{s\in j\setminus r} \hat{\nu}_{s\to j}(x_j) \Big) \\ + \frac{1}{N} \sum_{i\in V} (1 - |\partial i|) \ln \Big( \sum_{x_i} \prod_{r\in i} \hat{\nu}_{r\to i}(x_i) \Big). \quad (2.45)$$

---

[1]That is true only for infinite graphs, for finite graphs we speak of quasi-solutions [15] or approximate solutions but this is unimportant in the present context.

Free entropy $\phi$ gives the correct Helmotz free energy ($-\beta f_H = \phi$) on tree models and is the RS approximation to the free energy on non tree models, known to be correct on some simple case (e.g. Ising model on random graphs, two index assignment and many others). It is also known as the Bethe free entropy. It can be shown that BP fixed point messages can be recovered through a variational principle applied to Bethe free entropy. More precisely each BP fixed point is also a local maximum of $\phi$ (considered as a function of $|E|$ messages) and vice versa [15].

On a factor graph ensemble we can use the (supposed to be unique) fixed point of Density Evolution $P_*$ and express the average Bethe free energy as

$$-\beta f_{\text{Bethe}} = \frac{d}{c} \overline{\ln \Big( \sum_{\{x_j\}} \psi_R(\{x_j\}) \prod_{j=1}^{C_0} \prod_{s=1}^{D_j} \hat{\nu}_{sj}(x_j) \Big)} + \overline{(1 - D_0) \ln \Big( \sum_x \prod_{s=1}^{D_0} \hat{\nu}_j(x) \Big)}. \quad (2.46)$$

Expectation are taken over random variables the random variable node and factor node degrees $D_0$ and $C_0$, the random variable node residual degrees $\{D_j\}$, the compatibility function node label $R$ and messages $\{\hat{\nu}_{sj}\}$ and $\{\hat{\nu}_j\}$ independently and identically distributed as $P_*$. $d$ and $c$ are the mean variable and factor node degrees respectively.

On locally tree-like factor graphs, where loops' length is $O(\ln N)$, Replica Symmetry Cavity Method achieves exact values for marginals and thermodynamic functions as long as correlations between variables remain short-ranged, so that independence of incoming messages assumption holds true. Obviously Cavity Method scores poorly on those graphs where short loops are present and near neighbours of a variable $i$ are strongly correlated even after $i$ removal, as in the case of most models on finite dimensional lattices. To cope with that generalizations of Belief Propagation algorithm have been proposed [82] and cluster-based free energy approximation (of which the Bethe free energy can be considered the simplest level) have been studied for a long time [83]. Long range correlations are responsible for the failure of Replica Symmetric assumptions on locally tree-like models and Replica Symmetry Breaking ansatz have been developed to dispatch a cure extending the Cavity Method formalism.

## 2.5 Combinatorial Optimization

### 2.5.1 Formulation and link with statistical physics

Optimization has a crucial role in every human activity, because whatever our purposes are, being them determined by chance, feelings, instinct and reason, it's our nature to speculate rationally about the best way to achieve them quickly and thoroughly. In mathematics and theoretical computer science a generic optimization problem is characterized by a feasible solution space $S$ and an objective function $f : S \to \mathbb{R}$ called the cost function. One is interested in finding the set of optimal solutions

$$S^* = \operatorname*{Arg\,min}_{s \in S} f(s) \tag{2.47}$$

and their cost, the optimal cost

$$E^* = \min_{s \in S} f(s). \tag{2.48}$$

Combinatorial optimization problems are a special class of optimization problems, the requisite being a finite feasible solution space $S$. The Matching problem and the Assignment problem we are going to cover in the next chapters are well known examples of optimization combinatorial problems, among many others as the Travelling Salesman Problem, the Minimum Weight Spanning Tree, graph coloring, boolean satisfiability problems and so on (see [84]) for an introduction to the theme). All the combinatorial optimizations problems we just mentioned have a natural representation in terms of factor graphs.

Link between combinatorial optimization and statistical physics is made through the partition function

$$Z = \sum_{s \in S} e^{-\beta f(s)} \tag{2.49}$$

and the solution to the problem is recovered in the limit $\beta \uparrow \infty$. cost function $f$ is thus associated to the total energy of a given configuration. To obtain a sensible statistical physic model the cost function has to be properly scaled (an operation that leaves $S^*$ unaltered) in the size of the problem $N$ (often the number of variables involved) so that thermodynamic quantities are linear in $N$. Moreover the large $N$ limit has to be taken, so that model (2.49) is apt only to study the asymptotic behaviour of an optimization problem. That's not a hindering limitation however, because threshold characterizing interesting transitions (e.g. the SAT-UNSAT transition in the random K-SAT problem) can be made sharp only for $N \uparrow \infty$ and many asymptotic properties tends to be acquired very fast as $N$ grows.

Very often, at least from a theoretical point of view, one is interested in elucidating the average behaviour of random instances of some optimization problems.

It turns out that techniques developed by physicists in the contest of spin glass theory, and the Cavity Method in particular, are well-suited to cope with these problems. The seminal paper in the field was published in 1985 by Parisi and Mézard [18]. They used the Replica Trick in the Replica Symmetric approximation to compute the asymptotic optimal total cost of the random link assignment problem. After their work many other appeared in the successive years, addressing completely connected or dense graphical model through the Replica Trick [19,85], which could be used in its original form and allowed for Replica Symmetry Breaking. At that time, on sparse graphical models only the RS ansatz could be applied using the Replica Trick [86]. The next leap forward was made in 2001 thanks again to Parisi and Mézard, who showed how to implement the first step of replica symmetry breaking on sparse graphical models through the Cavity Method [14]. The algorithmic version of the 1RSB cavity method was then proposed in Ref. [20].

Development in this field through the 2000s have been impressive. The connection between the Bethe approximation in physics and the Belief Propagation algorithm from computer science was recognized [87], and a class of heuristic algorithms based on Kikuchi's cluster-variation method [83] was proposed [82]. Random constraint satisfaction problems, such as K-SAT, where characterized in terms of the geometric structure of the space of solutions, a collective effort which involved many scholars for two decades, summarized in Ref. [22]. For example, in large instances of the random 4-SAT problem, calling $\alpha$ the parameter controlling the density of clauses that have to be satisfied, there exist three values $\alpha_d < \alpha_c < \alpha_s$ such that : for $\alpha < \alpha_d$ almost all solutions belong to unique cluster, that is any two of them are connected by a path of solution involving a small rearrangement of the configuration at each step; at $\alpha = \alpha_d$ there is the clustering (also said dynamical) transition, and for $\alpha_d < \alpha < \alpha_c$ solutions are organized in an exponential (in $N$) number of clusters; at $\alpha = \alpha_c$ the system undergoes a thermodynamic transition, called the static or the condensation transition; for $\alpha_c < \alpha < \alpha_s$ a finite number of clusters carries the whole Gibbs measure; finally, above $\alpha_s$, the satisfiability threshold, the problem admits no solutions at all [15].

In this thesis we will focus on a problem with a simple, replica symmetric structure, the Matching problem, that we present in the next paragraph.

### 2.5.2 The Matching Problem

The *matching problem* is a combinatorial optimization problem that has drawn the attention of both the computer science [84, 88–92] and the statistical physics [18, 19, 59, 62, 93–97] communities for many decades. Among its plethora of applications, we mention computer vision [98], control theory [99, 100] and pattern matching [101]. Variants of the matching problem are the multi-index matching problem [60] and the set packing problem [6]

Even in its most general formulation, the problem belongs to the $P$ computational complexity class, and many famous algorithms have been developed to solve it efficiently [88, 89, 91].

For a given $N \times N$ cost matrix $w_{ij}$, $i, j = 1, \ldots, N$, $w_{ij} = w_{ji}$, the minimum matching problem is formulated as the following integer programming problem:

$$
\begin{aligned}
\text{MINIMIZE} \quad & E[n;w] \equiv \sum_{i<j} w_{ij} n_{ij} \\
\text{SUBJECT TO} \quad & n_{ij} \in \{0,1\} && \forall i,j \\
& n_{ii} = 0 && \forall i \\
& n_{ij} = n_{ji} && \forall i,j \\
& \sum_{j} n_{ij} = 1 && \forall i
\end{aligned}
\qquad (2.50)
$$

More generally and from a graph theoretic perspective, for an arbitrary graph $G = (V,E)$ a *matching* $M$ is a subset of the nodes, $M \subseteq E$, such that no node in $V$ has two or more edges in $M$. Two nodes $i$ and $j$ are *matched* if $(i,j) \in M$, and in that case we call $(i,j)$ a *match*. A *perfect matching* is a matching where each node is matched. The *size* of a matching is its cardinality. A *maximum matching* is matching such that no other matching has greater size. Notice that maximum matchings generally are not unique. If we associate a real number $w_{ij}$, that we call *cost* or *weight*, we are considering the *weighted matchings*. Since in this thesis we address only weighted matchings we call them matchings as well. A *minimum weight maximum matching* is a matching that solves the following problem, that we will generally call *the matching problem*:

$$
\begin{aligned}
\text{MINIMIZE} \quad & E[n;w] \equiv \sum_{(i,j) \in E} w_{ij} n_{ij} \\
\text{SUBJECT TO} \quad & \sum_{(i,j) \in E} n_{ij} && \text{MAXIMIZED} \\
& n_{ij} \in \{0,1\} && \forall (i,j) \in E \\
& \sum_{j \in \partial i} n_{ij} \leq 1 && \forall i \in V
\end{aligned}
\qquad (2.51)
$$

Matching that satisfy this property are also called *optimal matchings*. The function $E[n;w]$ is the *cost function* and its mimum value $E^*[w]$ is the *optimal (total) cost*. We see that this problem reduces to the formulation 2.50 when the underlying graph is the complete graph. When $G$ is taken as a complete bipartite graph, we will call problem 2.51 an *assignement problem* or a *bipartite matching problem*.

In this thesis we deal with random instances of the matching problem. One kind of randomness that we do not examine is the one in the underlying graph $G$. Matching on random graphs, in fact, both in its weighted and unweighted versions, has been investigated in numerous papers [61, 90, 102, 103].

Here we investigate fully-connected models where the randomness is encoded in the cost matrix $\{w_{ij}\}$. The archetypal example of such problem is the *random link*, where the matrix elemens $w_{ij}$ are independent and identically distributed. The random link problem was investigated, using the replica method, by Parisi and Mézard [18]. They obtained, for exponentially distributed costs, the celebrated result

$$
\lim_{N \to +\infty} \mathbb{E}_w E^*[w] = \frac{\pi^2}{12}, \qquad (2.52)
$$

later proved by Aldous [104] and successively, using different techniques, by Linusson and Wästlund [105].

A much more difficult problem is the Euclidean matching problem. The cost matrix here is induce by the distances among $N$ points distributed uniformly in a box in $\mathbb{R}^d$, $w_{ij} = \|\mathbf{x}_i - \mathbf{x}_j\|$. Therefore the matrix element are highly correlated and analytical treatment is very difficult. The problem was introduced in Ref. [58] and studied with the replica method, although a crude approximation was made, considering only cost correlations among triple of points (triangles) and factorizing all the others. In Chapter 7 we show how to improve upon their approximations taking into account correlations in polygons of arbitrary lengths. In Chapter 6 instead we use a completely different approach, based on continuum approximation, to analyse the Eulidean bipartite matching problem, where density fluctuations among the two point sets lead to non-trivial scaling behaviour for the average optimal cost.

Noticeably, on complete graphs, although the BP algorithm does not converge, it gives the exact optimal configuration and total cost after a certain problem-dependent iteration step [81, 106]. Also the Cavity method predictions for diluted models have rigorously proved on locally tree-like structures [103]. .

For numerical analysis we make use of the solver implemented in the LEMON GRAPH LIBRARY [107]. It is based on Edmond's blossom algorithm [91], which has $\Theta(|V||E|\ln|V|)$ computational time complexity. This would make for a $\Theta(N^3 \ln N)$ complexity for the fully connected graphs we deal in this thesis. Since edges participating in optimal matchings belong to the $O(1/N)$ fraction with the lowest weights [108], in numerical experiments we prune from the graphs the edges with weights above a certain cut-off value $\gamma$. Obviously $\gamma$ has to be chosen high enough to not interfere with the optimal matching. The pruning procedure transforms the graph in a diluted random graph, therefore the algorithmic complexity is reduced to $\Theta(N^2 \ln N)$.

### 2.5.3 The Minimum Cut Algorithm

As an example of application of techniques from combinatorial optimization to disordered systems, in this paragraph we will show the equivalence between the problem of finding the ground state configuration in a RFIM and the minimum cut problem, an combinatorial optimization problem belonging to the $P$ computational complexity class, also known as min-cut problem). To our knowledge this equivalence is stated for the first time in Ref. [109]. Consider an instance of the RFIM on graph $G = (V, E)$. The Hamiltonian is given by

$$\mathcal{H}[\sigma] = - \sum_{(i,j) \in E} J_{ij} \sigma_i \sigma_j - \sum_{i \in V} h_i \, \sigma_i, \qquad (2.53)$$

with $J_ij > 0$. We want to find the spins configuration $\sigma^*$ that minimizes the energy of the system. We notice that a spin configuration $\sigma$ is specified by a partitioning of $V$ in two sets, $V^+$ and $V^-$, $V^+ \cup V^- = V$, such that $s_i = 1$ for $i \in V^+$ and $s_i = -1$ for $i \in V^-$. The energy of a configuration is determined by the "unsatisfied" bonds and fields. In fact, if we define the two sets $H^+ = \{i \in V : h_i > 0\}$ and

$H^- = \{i \in V : h_i < 0\}$, we have

$$\mathcal{H}[V^+, V^-] = E_{ref} + \sum_{\substack{i \in \tilde{V}^+ \\ j \in \tilde{V}^-}} 2J_{ij} + \sum_{i \in V^+ \cap H^-} 2|h_i| + \sum_{i \in V^- \cap H^+} 2|h_i|, \qquad (2.54)$$

where for convenience we set $J_{ij} = 0$ if $(i,j) \notin E$. The reference energy $E_{ref}$ is defined by

$$E_{ref} = -\sum_{(i,j) \in E} J_{ij} - \sum_{i \in V} |h_i| \qquad (2.55)$$

In order to transpose the energy minimization problem in a minimum cut problem, we construct an auxiliary *oriented* graph $\tilde{G} = (\tilde{V}, \tilde{A})$. The node set of the new graph is the old one plus two additional nodes representing the positive and negative field respectively, $\tilde{V} = V \cup \{+, -\}$. The arc set $A$ is the union of three sets, $\tilde{A} = \tilde{E} \cup E^+ \cup E^-$. $\tilde{E}$ contains two arcs for each edge in $E$, $\tilde{E} = \{(i \to j) : (i,j) \in E\}$. The other two sets are defined as $E^+ = \{(+ \to i) : i \in H^+\}$ and $E^- = \{(i \to -) : i \in H^-\}$. We associate a weight $\tilde{J}_{ij} > 0$ to each arc $(i \to j) \in \tilde{A}$ according to

$$\tilde{J}_{ij} = \begin{cases} J_{ij} & (i \to j) \in \tilde{E}, \\ h_j & (i \to j) \in E^+, \\ -h_i & (i \to j) \in E^-. \end{cases} \qquad (2.56)$$

For convenience we also set $\tilde{J}_{ij} = 0$ if $(i \to j) \notin A$. It is now easy to see that we can rewrite each configuration's energy in terms of the non-asymmetric matrix $\tilde{J}$, and that minimization of the energy given in Eqs. (2.53) and (2.54) is equivalent to finding

$$\min_{\tilde{V}^+, \tilde{V}^-} \sum_{\substack{i \in \tilde{V}^+ \\ j \in \tilde{V}^-}} \tilde{J}_{ij} \qquad (2.57)$$

Minimization is over all the partition of $\tilde{V}$ in two sets, $\tilde{V}^+$ and $\tilde{V}^-$, such that $+ \in \tilde{V}^+$ and $- \in \tilde{V}^-$. This the exact definition of the minimum cut problem for $\tilde{G}$ with *source* $+$ and *sink* $-$. Its connection to the ground state of the RFIM is thus established.

Therefore, in order to find the ground state configuration of the RFIM, we can go trough the construction of the auxiliary graph depicted above and then use one of the many algorithms available to solve the corresponding minimum cut problem. In particular *the max-flow min-cut theorem*, establishes the equivalence through duality of the minimum cut problem with another celebrated optimization problem on graphs, the maximum flow problem [84]. Therefore many algorithm used to solve for the min-cut exploit this connection with the max-flow. In this thesis we make use of the solver implemented in the LEMON GRAPH LIBRARY [107] based on a push-relabel maximum flow algorithm [110]. The computational complexity of this algorithm is $\Theta(|V|^2\sqrt{|E|})$. For finite connectivity systems the computational complexity is therefore $\Theta(N^{2.5})$. Average computational complexity is significantly inferior, as discussed in Figure 2.2.

Unfortunately minimum cut algorithms work only for positive weights $\tilde{J}_{ij}$, therefore they cannot be us find the ground state of Ising spin glasses. Energy minimization in spin glass can be stated as a maximum cut problem, and it belongs to the NP computational complexity class [111]. Best exact solver for spin glasses are usually based on brunch-and-bound approaches [111].

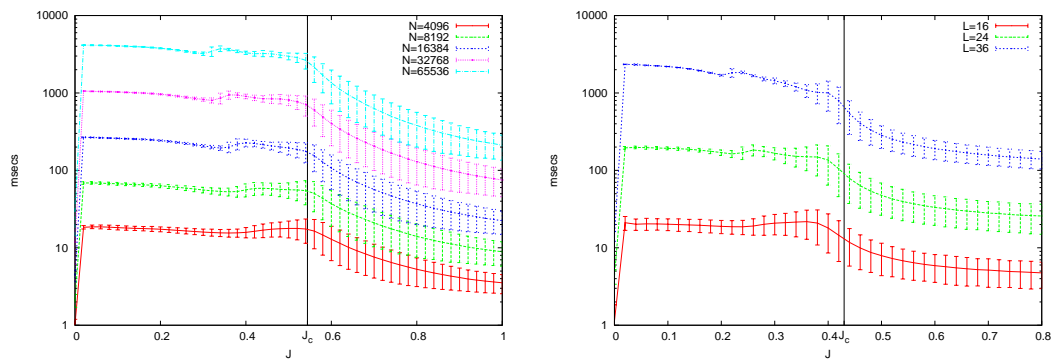

**Fig. 2.2.** Timing of the push-relabel minimum cut / maximum flow algorithm implemented in the Lemon graph library. Simulations were performed on a Core i5 2500k processor. (*Left*) RFIM on random regular graphs with conectivity $z = 4$. (*Right*) RFIM on a cubic lattice in dimension $d = 3$. Averages are taken over 10000 samples. The error bar is the standard deviation of the population. Quite interestingly the computational time remains constant in the whole paramagnetic phase up to the transition point and decreases only in the ferromagnetic phase. Average computational complexity varies approximately between $\Theta(N^{1.5})$ and $\Theta(N^2)$ in the range considered.

On the other hand many heuristic solver have been proposed to find spin glass ground states [111, 112]. Noticeably for the present discussion, one of these heuristics, the Cluster-Exact Approximation (CEA) , is a zero temperature Markov Chain Monte Carlo algorithm where the standard single spin flip is replaced with the exact minimization of the energy on a unfrustrated subgraph of the original graph [113]. In fact any unfrustrated Ising system can be gauge transformed in a ferromagnetic system, therefore the min-cut algorithm can be (locally) applied. Also combination of genetic algorithms [114] and CEA have been recently proposed and showed to be highly effective [115].

## 2.6 Finite size corrections

In finite dimensional systems with periodic boundary conditions one would expect exponentially decaying finite size corrections, due to arguments somewhat related to the Euler-McLaurin formula. The system in fact feels its finiteness, as far as observables are concerned, only through diagrams involving rewiring around the torus. This implies, for non-critical systems, an additive correction which is exponentially decaying in the length $L$ of the lattice, or some power of it. Divergence from this standard behaviour is an interesting phenomena and points towards criticality. The Euclidean monopartite and bipartite matching problems belong to this class, as we will investigate in Chapters 6 and 7.

Finite size corrections have been computed in the SK model, on the Almeda-Thouless line, by Parisi, Ritort and Slanina [116, 117]. Their approach is based on a replicated field theory. Due to the divergence of all the terms in the diagrammatic expansion around the saddle point, at the critical point and in the whole spin glass phase, they have to resum an infinite class of diagrams. They found a $O(\frac{\ln N}{N})$ correction to the free energy and a $O(N^{-\frac{2}{3}})$ to the energy. There is some evidence from numerical simulation of a $O(N^{-\frac{2}{3}})$ correction for both the energy and the free energy in the spin glass phase [118]. Understanding finite size corrections in finite dimensional spin glasses, a much difficult analytical and numerical (due to long equilibrating times) task, is of fundamental importance to asses the validity of any theory competing for the explanation of the glass transition [75, 119].

Regarding the deep spin glass phase, the zero temperature point has a special place. The search of ground state of spin glasses, a minimum cut problem with positive and negatives capacities, is a computationally hard problem. Nonetheless many successful heuristic algorithms, such as Cluster Exact Approximation (CEA) [113], Hierarchical Bayesan Optimization Algorithm (hBOA) [120, 121] and Extremal Optimization (EO) [122], have been developed to deal with large instances of the problem. In particular, using EO, the finite scaling behaviour of zero temperature mean field and finite dimensional spin glasses was deeply investigated in Refs. [123–125].

Finite size corrections in the glassy phase are outside the scope of this thesis though. In Chapter 4 we will discuss topology related corrections, since in diluted random graphs simple combinatorial arguments give the scaling of the average number of loopy subgraphs as powers of $1/N$. In Chapter 5 instead, $O(1/\sqrt{N})$ finite size correction in the RFIM are induced by fluctuations of the total random external field. Lastly, in the Euclidean Assignment problem investigate in Chapter 6, an anomalous subleading scaling for the average cost is also due to fluctuations in the disorder, namely to the differences in local densities among the two sets of points .

# Chapter 3

# The Replicated Transfer Matrix

## 3.1 Introduction

The study of one-dimensional Ising chain with random bonds and/or fields has a long tradition in the context of disordered systems. Over the years this field has experienced an interesting change in perspective. Earlier studies were essentially motivated by the need to obtain solvable version of three-dimensional models [126–129] and this line of research culminated with the introduction of specific random Hamiltonians that are actually solvable analytically [130–133]. In the last twenty years dynamical approaches have also been considered as alternatives to equilibrium approaches [134–137] while developments in the context of static studies [138, 139] have been mainly motivated by the connection between one dimensional systems and models defined on sparse random graphs. Random graphs in turn have many important applications in the context of computer science, artificial intelligence and information theory [13, 15]. In this broader context one is more interested in having a general formalism that can be applied to any given distribution of the quenched Hamiltonians at the price of obtaining the result through numerical solution of implicit equations.

In the general case one would like to study an Ising chain, either open or closed, of arbitrary length $L$ were the fields and couplings are i.i.d. random variables. Quantities of interest include the free energy but also all sort of averaged correlation functions. Indeed, at variance with pure systems, correlations can be averaged in two different ways: over thermal noise (conventionally referred as *connected correlations*) and over the quenched Hamiltonians (*disconnected correlations*). This difference is important both at the theoretical and the practical level. Indeed disconnected correlations happen to be much larger in random field systems (but not in Spin-Glasses) and lead to a very complex phenomenology, *e.g.* the increase in the critical dimension from $D = 4$ of the pure ferromagnet to $D = 6$ [80]. In this chapter we show how to complete this program by means of the replica method, more precisely by means of the replicated transfer matrix (RTM) approach. As long as the sources of disorder are independently distributed, one can express the integer moments of the partition function through traces of powers of the $2^n \times 2^n$ transfer matrix of a system of $n$ replicated spins. Then, as usual with replica calculations, the analytic continuation to $n = 0$ is performed. We will derive expressions for the aforementioned quantities in terms of the solutions of integral equations that can be solved for instance through population dynamics algorithms. In order to do so we build on the crucial contribution of Monasson and Weigt [138], who first characterized the spectral properties of the RTM. The motivation is not only to have a compilation of useful formulas but also to present some non-trivial features of their derivation. The most important is connected with the fact that in the limit $n \to 0$ two families of eigenvalues (corresponding to the Longitudinal and Anomalous sectors in the Spin-Glass jargon) become degenerate. From the theoretical perspective, this determines an anomalous behaviour of the disconnected correlation functions and of corrections to the free energy of closed chains. On a practical side this implies that one has to determine not only the eigenvalues and eigenvectors of the integral equations at $n = 0$ limit but also their first $O(n)$ correction.

While the replica method is at present the only way to derive expressions for all quantities of interest in a compact form, its well-known drawback is the assumption that one can make the analytical continuation $n \to 0$ of expressions whose derivation makes sense only for positive integer $n$. One is therefore interested in deriving the same expressions in a more direct way. Unfortunately there are no general results or strategies on how to do this and one has to proceed case by case. We will present a direct probabilistic derivation of many of the expressions obtained through the replica method. A particularly non-trivial result is the derivation of the formula for disconnected correlation functions that has been long sought for. Such a derivation is based on the fact that a direct physical meaning can be attributed to the continuation of replica expressions to real $n$, at variance with other classic analytical continuation tricks ( *e.g.* dimensional continuation in field theory). Therefore one can first derive rigorously an expression at any real $n$ and then safely take the limit $n \to 0$. The only replica expression whose derivation is left as an open problem is the free energy of closed chains. We recall that closed chains are rather important objects that appears in perturbative computations developed around the tree approximation [2].

We conclude this introduction by briefly discussing the connection between our results and the extensive literature on disordered Ising chains. As we said already earlier studies, appeared in the context of the random field Ising model (RFIM), were motivated essentially by the possibility of obtaining exact solutions when dealing with one-dimensional models. It was immediately recognized [126–128] that the free energy of an infinite chain can be expressed in terms of an iterative equation which corresponds to the Longitudinal sector in the terminology of RTM. Exact results can be obtained at zero temperature, where the equation can be solved explicitly [127] or by enumeration in the special case in which the random fields are either zero or infinity [129]. Finally it was discovered that some models with specific distributions of the disorder can be solved analytically [130–133]. Much effort has been put in the study of the iterative equation relevant to the free energy of the infinite chain, starting from the observation that when the random fields and couplings take a discrete number of values ( *e.g.* $H = \pm 1$) the solutions of the equations may display a multi-fractal structure [140–143]. The approaches taken to characterize the correlation functions have been less successful. Connected correlations where computed exactly for the aforementioned solvable models but the disconnected correlations resisted all efforts [131] to capture their expected peculiar features (the double pole) [80] up to this work. Later on correlations functions were also studied in a more general framework at zero temperature [144], but again non considered the disconnected correlations. The results of [131] are system-specific and not based on iterative methods, only recently [145–147] it has been recognized that general iterative expressions for connected correlations can be obtained by means of cavity arguments like those we will present in the following.

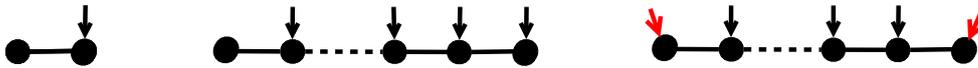

**Fig. 3.1.** Pictorial representation of the matrix $T_n$ (*left*), its powers $T_n^\ell$ (*center*) and the matrix $\tilde{T}_n^{(\ell)}$ (*right*).

The chapter is organized as follows: in Section 3.2 we define the model we are considering and expose the main results of this chapter; in Section 3.3 we develop all the spectral formalism of the RTM and we apply it in Section 3.4 to the computation of free energies and correlation functions. Most of the results obtained with the RTM are then rederived with a purely probabilistic approach in Section 3.5.

## 3.2 Definitions and main results

In this chapter we consider one-dimensional Ising spin system with i.i.d. random fields and couplings , e.g. an isolated chain or a chain embedded in a locally tree-like graph, therefore described by the product of uncorrelated $2 \times 2$ random transfer matrices $M_i$ defined by

$$M_i(\sigma_{i+1}, \sigma_i) = e^{\beta J_i \sigma_{i+1} \sigma_i + h_i \sigma_i}. \tag{3.1}$$

The partition function of a closed chain of length $\ell$ is then a random variable given by

$$Z_{\ell,c} = \text{Tr} \prod_{i=0}^{\ell-1} M_i. \tag{3.2}$$

A powerful technique to compute the statistical properties of this kind of objects is the well known replica method [13]. As we shall see, as long as the system stays in a replica symmetric phase, its statistical properties are encoded in the (replica symmetric) replicated transfer matrix $T_n$, the $2^n \times 2^n$ matrix defined by

$$T_n(\sigma, \tau) = \mathbb{E}_{J,h}\, e^{\beta J \sum_{a=1}^n \sigma^a \tau^a + \beta h \sum_{a=1}^n \tau^a}. \tag{3.3}$$

Here and in the following we denote with $\sigma$ the vector $(\sigma^1, \ldots, \sigma^n)$, with the $n$ replicated spins $\sigma^a$ taking values in $Z_2 = \{-1, 1\}$. A similar definition holds for $\tau$. As usual in the replica method [13] we shall work at integer value of $n$ and perform the analytic continuation for $n \downarrow 0$ at the end of the computations. We shall assume in the following that the field $h$ is an arbitrary distributed external random field, if we are considering an isolated chain, or, if we are considering a chain embedded in a locally tree-like graph, to be a random cavity field conditioned to act on a spin that is already connected to two other spins (its neighbours on the chain). See Figure 3.1 for a representation of $T_n$ and its powers $T_n^\ell$.

A first spectral analysis of $T_n$ was conducted by Weigt and Monasson [138]. Following their lead we take advantage of the replica index permutation symmetry of $T_n$ to choose an appropriate bases to express its right eigenvectors. There are $n+1$ non-equivalent irreducible representation of the permutation group, which can be glued together to form the *sectors* $D^{(q)}$, $q = 0, 1, \ldots, \lfloor \frac{n}{2} \rfloor$, partitioning $Z_2^{\otimes n}$. In the following, with some abuse of notation, we will denote with $D^{(q)}$ the set of eigenvalues of $T_n$ with eigenvector in that sector. The eigenvectors of $T_n$ in the sector $D^{(q)}$ can be parametrized by functions $g_q^\lambda(u)$ that, in the limit $n \downarrow 0$, satisfy the eigenvalue equation

$$\lambda \, g_q^\lambda(u) = \mathbb{E}_{J,h} \int dv \, g_q^\lambda(v) \, \delta\left(u - \hat{u}(J, v+h)\right) \, \left(\frac{\partial \hat{u}}{\partial v}\right)^q, \qquad (3.4)$$

where $\hat{u}(J,h) = \frac{1}{\beta} \operatorname{atanh}\left(\tanh(\beta J)\tanh(\beta h)\right)$ is the cavity iteration rule.

In this chapter we extend the analysis of the spectral properties of $T_n$ to achieve a complete description of the $n \downarrow 0$ limit, derive exact expressions for correlation functions and free energies of chains. Since $T_n$ is the product of two non-singular symmetric matrices, it possess a complete orthonormal (in the left-right sense) basis of left and right eigenvectors with real eigenvalue. The left eigenvector corresponding to a certain right $\psi_R$ is simply $\mathbb{E}_h e^{\beta h \sum_a \sigma^a} \psi_R(\sigma) \equiv \rho_h(\sigma)\psi_R(\sigma)$. Therefore the spectral decomposition of $T_n$ into the subspaces $D^{(q)}$ is given by

$$T_n(\sigma,\tau) = \sum_{q=0}^{\lfloor \frac{n}{2} \rfloor} \sum_{\lambda \in D^{(q)}} \lambda \, \rho_q^\lambda(\sigma)\rho_h(\tau)\rho_q^\lambda(\tau) \sum_{\substack{a_1 < \cdots < a_q \\ b_1 < \cdots < b_q}} Q_{a_1 \ldots a_q; b_1 \ldots b_q} \, \sigma^{a_1} \ldots \sigma^{a_q} \tau^{b_1} \ldots \tau^{b_q} \, . \qquad (3.5)$$

Here we have denoted with $\rho_q^\lambda(\tau)$ the replica symmetric part of the eigenvector in the sector $D^{(q)}$ with eigenvalue $\lambda$. The second sum is over all the eigenvalues of $T_n$ in the sector $D^{(q)}$, given in the $n \downarrow 0$ limit by the solutions of Eq. (3.4). The coefficients $Q_{a_1 \ldots a_q; b_1 \ldots b_q}$ have simple algebraic expressions in each sector (see Eqs. (3.27) and (3.28)) and are invariant under permutations of any of their two sets of indices. While a different left-right decomposition of $T_n$ has already been attempted [139], an unfortunate choice in the parametrization of the eigenvectors in terms of function of two variables led to an unmanageable formalism. Thanks to the spectral representation (3.5) we can easily take the powers of $T_n$ and contract the matrix with the quantities we want to average. In Section 3.3 we derive Eqs. (3.4) and (3.5), and discuss the non-trivial aspects of the small $n$ limit.

One of the applications of the spectral formalism is the computation of the average free energy of open and closed chains, as exposed in Section 3.4.1. Recently it has been shown [2] that the first finite size correction to thermodynamic free energy of systems on diluted graphs can be expressed as a linear combination of the free energies of closed and open chains. It has also been argued [148] [2] that a perturbative expansion around the Bethe approximation towards finite dimensional lattices, shall account for the presence of loops (closed chains) and will contain the free energies and the correlation function of one-dimensional objects, motivating the importance of exact and easily approximable expressions for their free energies.

Taking the trace of $T_n^\ell$ and performing the $n \downarrow 0$ limit one obtains the average free energy of a closed chain of size $\ell$:

$$-\beta f^c_\ell = -\beta \ell f_0 + \sum_{\lambda \in D^{(1)}} \Delta_\lambda \, \ell \, \lambda^{\ell-1} + \sum_{q=1}^\infty \hat{d}_q \sum_{\lambda \in D^{(q)}} \lambda^\ell \,. \tag{3.6}$$

The non-trivial features of this expression is the presence of a term $O(\ell \, \lambda^{\ell-1})$. This is typically not present in a ordinary eigenvalue decomposition that contains only $O(\lambda^\ell)$. Its presence is due to the $n \to 0$ limit combined with the fact that the longitudinal and anomalous eigenvalues become degenerate. As we said in the introduction this is a phenomenon that has dramatic physical consequences in the RFIM context [80].

The terms $\Delta_\lambda$, due to the degeneracy between the eigenvalues of the sectors $D^{(0)}$ and $D^{(1)}$ at $n=0$, are expressed in Eq. (3.48). The coefficients $\hat{d}_q$ are the analytic continuation of the degeneracies of the eigenvalues, and are given in Eq. (3.41). We note that the correction to the intensive free energy $f_0$ (expressed in Eq. (3.43)) is given by a linear combination of exponential and $\ell$ times exponential terms. The decaying part of $f^c_\ell$ is dominated by the largest eigenvalue among the various sectors.

In the computation of the free energy of an open chain, we allow for the incoming fields at the extremities of the chain to be distributed differently from the fields $h$ acting on the internal spins, and denote them by $\tilde{h}$. This is in fact what happens in general when considering an open chain embedded in a sparse graph. The expression we derived for the average free energy of an open chain of length $\ell$ is

$$\begin{aligned}-\beta f^o_\ell = &- \ell \beta f_0 + \mathbb{E}_{\tilde{h}} \int du \, P(u) \, 2 \log \cosh\left(\beta(u+\tilde{h})\right) \\ &- \mathbb{E}_h \int du \, dv \, P(u)P(v) \log \cosh(\beta(u+v+h)) \\ &+ \log 2 + \sum_{\lambda \in D^{(1)}} a^2_{\lambda,0} \, \lambda^\ell \,,\end{aligned} \tag{3.7}$$

where $P(u)$ is the distribution of cavity messages along the chain and the coefficients $a_{\lambda,0}$ are related to the left eigenvectors of the sector $D^{(0)}$ and given in Eq. (3.59).

Another result we will present is the expression of the connected correlation functions of two spin at distance $\ell$, in a form that is both analytically exact and easy to approximate numerically with high precision. In Section 3.4.2 we derive the formula

$$\overline{\langle \sigma_0 \sigma_\ell \rangle^q_c} = \sum_{\lambda \in D^{(q)}} a^2_{\lambda,q} \lambda^\ell \,, \tag{3.8}$$

where $a_{\lambda,k}$ can be computed through the eigenfunction $g^\lambda_q$ using Eq. (3.64). We indicate with $\overline{\bullet}$ the average over all kinds of disorder in the model considered. For Ising model on sparse random graphs with mean residual degree $z$, the susceptibility $\chi_q = \sum_{i<j} \frac{1}{N} \mathbb{E}\langle \sigma_i \sigma_j \rangle^q_c$ diverges when the greatest eigenvalue of $D^{(q)}$ reaches the value $1/z$. Therefore the sectors $D^{(1)}$ and $D^{(2)}$ are the relevant ones to the ferromagnetic and the spin-glass transitions respectively (see Figures 3.2 and 3.3).

The computation of the thermally disconnected correlation function, $\overline{\langle \sigma_0 \rangle \langle \sigma_\ell \rangle} - \overline{\langle \sigma_0 \rangle} \, \overline{\langle \sigma_\ell \rangle}$, particularly relevant to the RFIM [73], requires a careful treatment of the analytic continuation to $n=0$. The final expression we obtained, Eq. (3.71), is not a linear combination of terms involving only one eigenvalue, as in the previous formulas. The leading term for large $\ell$ is easily extracted though: let $\lambda_1$ be the greatest eigenvalue of the sector $D^{(1)}$, then

$$\overline{\langle\sigma_0\rangle\langle\sigma_\ell\rangle} - \overline{\langle\sigma_0\rangle}\,\overline{\langle\sigma_\ell\rangle} = \Delta_{\lambda_1}\,a_{\lambda_1,1}^2\,\ell\,\lambda_1^{\ell-1} + O(\lambda_1^\ell) \qquad \text{for } \ell \to +\infty, \qquad (3.9)$$

with $\Delta_\lambda$ and $a_{\lambda,1}$ given in Eq. (3.48) and Eq. (3.64) respectively. Therefore, on one-dimensional chains and sparse graphs, the susceptibility corresponding to the thermally disconnected correlation function present the characteristic double pole behaviour near the transition point, whose prefactor can also be computed by Eq. (3.9).

The expressions we found for free energies of chains, Eqs. (3.6) and (3.7), and the correlation function Eqs. (3.8) and (3.9), are exact for every value of the length $\ell$ of the chain but involve the computation of infinitely many terms. Fortunately it turns out from our numerical simulations that the spectrum of the integral operator in Eq. (3.4) is discrete and the eigenvalues are well spaced. Therefore considering only the first few highest eigenvalues one obtains very good approximations already at small values of $\ell$. They can be computed numerically, discretizing the kernel of the integral operator of Eq. (3.4) and directly computing the eigenvalues of the associated matrix. Moreover the leading eigenvector and eigenvalue of each sector can be efficiently selected with multiple applications of the discretized operator on an arbitrarily chosen vector (as it was done to obtain Figures 3.2 and 3.3).

All the results we obtained using the replicated partition function formalism, with the noticeable exception of the formula for the average free energies of closed chains Eq. (3.6), can be recovered using a purely probabilistic approach in the same spirit of the usual cavity method [13] [15].

In Section 3.5.1 we devise two alternative probabilistic derivation for the average free energies of open chains . The first is based on a recursive equation involving the moments of the partition function, which leads to an expression for the moment of the random partition function $Z_\ell^n$ of an asymmetric open chain in terms of the left and right eigenvector of an integral operator we also encountered in the RTM formalism:

$$\overline{Z_\ell^n(u;x)} = \sum_{\lambda\in D^{(0)}} \lambda^\ell(n)\,g_0^\lambda(u;n)\,S_0^\lambda(x;n)\,[2\cosh(\beta x)]^n. \qquad (3.10)$$

Here $n$ is not related to the number of replicas, since replicas are not present in this approach, but is an arbitrarily chosen real positive number. The other method presented in Section 3.5.1 the iteration of the average free energy itself during the construction of the chain, which requires to keep track of the message of $u_\ell$ the cavity message propagating through a chain at distance $\ell$ from one of the extremities, at each iteration. The two approaches are deeply related and obviously lead to the same result.

Crucial to the probabilistic computation of the connected correlation functions, ah has been noted recently [146] is the random variable $X_\ell$ defined by $X_\ell \equiv \frac{\partial u_\ell}{\partial H_0}$, where $H_0$ field acting on the same extremity. It turns out that the connected correlation function of Eq. (3.8) is encoded in the $q$-the moments of the joint law of $u_\ell$ and $X_\ell$ at fixed $u_\ell$. This object, the function

$$G_q^{(\ell)}(u) = \int \mathrm{d}X\,P_\ell(u,X)\,X^q, \qquad (3.11)$$

obeys a recursion rule, Eq. (3.87), containing the integral operator of Eq. (3.4). Expressing $G_q^{(\ell)}(u)$ in the basis of the eigenvalues of $D^{(q)}$ leads then straightforwardly to the expression (3.8) we obtained using replicas. Moreover in Section 3.5.2 a more general result is presented in Eq. (3.90).

The thermally disconnected correlation function $\overline{\langle \sigma_0 \sigma_\ell \rangle_c}$ is computed in Section 3.5.3, using some results we obtained for the connected correlation function and for the moments of the partition function of an open chain, thanks to the relation

$$\frac{\partial}{\partial H_0}\frac{\partial}{\partial H_\ell}Z_{\ell,o}^n = n \left\langle \sigma_0 \sigma_\ell \right\rangle_c Z_{\ell,o}^n + n^2 \left\langle \sigma_0 \right\rangle \left\langle \sigma_\ell \right\rangle Z_{\ell,o}^n. \qquad (3.12)$$

An alternative approach, technically more difficult, outlined in Section 3.5.3 involves the resolution of an iterative equation for the function $R^{(\ell)}(u) = \overline{\delta(u - u_\ell)\langle\sigma_0\rangle^{(\ell)}}$, which takes into account the shift in the magnetization of the (initial) spin at the other side of the chain with respect to the spin where a new spin is attached to increase the length of the chain.

In the following Sections we fill-in all the technical details associated to the previous claims.

## 3.3 Spectral decomposition

We present an in-depth treatment of the spectral theory of the replica symmetric RTM. In Section 3.3.1 we discuss the spectral decomposition of the matrix for integer values of the number of replicas $n$. We introduce in Section 3.3.2 an integral representations of the eigenvectors, in order to discuss the main features of the analytic continuation to small values of $n$ in Section 3.3.3. In Section 3.3.4 we discuss some technicalities related to a peculiar aspect of the $n \downarrow 0$ limit, the degeneracy between the Longitudinal and the Anomalous sectors.

### 3.3.1 The Permutation Group

The $2^n \times 2^n$ matrix $T_n$ defined by Eq. (3.3) is invariant under the action of the group of permutations among the replicated spins: for each permutation $\pi$ acting on the $n$ spin, we have the equivalence $T_n(\pi(\sigma), \pi(\tau)) = T_n(\sigma, \tau)$. This symmetry allows us to block-diagonalize $T_n$ according to the irreducible representations of the permutation group. This idea has been first introduced by Weigt and Monasson [138] in order to compute the eigenvalue spectrum of $T_n$.

For the sake of completeness we now review Weigt and Monasson's method, then we extend it further, in order to achieve the decomposition of the transfer matrix in terms of left and right eigenvectors. The replicated space is $Z_2^{\otimes n}$. Let's call $\Delta_m$, with $m = 0, ..., n$, the subspace of configurations having exactly $m$ spins up. These subspaces are clearly invariant under any permutation of the replicas, therefore we can consider the representation of the permutation group in the $n + 1$ subspaces $\Delta_m$ and look for the irreducible ones. The complete decomposition of $\Delta_m$ into irreducible subspaces $D^{(m,q)}$ has been done by Wigner [149]. It reads:

$$\begin{aligned}
\Delta_0 &= D^{(0,0)} \,, \\
\Delta_1 &= D^{(1,0)} \oplus D^{(1,1)} \,, \\
&\ldots \\
\Delta_m &= D^{(m,0)} \oplus \ldots \oplus D^{(m,\,\min(m,n-m))} \,, \\
&\ldots \\
\Delta_{n-1} &= D^{(n-1,0)} \oplus D^{(n-1,1)} \,, \\
\Delta_n &= D^{(n,0)} \,.
\end{aligned} \quad (3.13)$$

Representations $D^{(m,q)}$, at fixed $q$, are isomorphic and have dimension

$$d_q \equiv \dim\left(D^{(m,q)}\right) = \binom{n}{q} - \binom{n}{q-1} \qquad q = 0, ..., \lfloor n/2 \rfloor \,, \quad (3.14)$$

where $\lfloor x \rfloor$ is the smallest integer part of $x$. Notice that by definition $d_0 = 0$. As we have $(n+1-2q)$ subspaces $D^{(m,q)}$, the $q$-sector of our matrix $T_n$ will contain $(n+1-2q)$ eigenvalues with degeneracy $d_q$. One can check that $\sum_{q=0}^{\lfloor n/2 \rfloor} d_q \, (n+1-2q) = 2^n$.

A vector of the space $D^{(m,q)}$ can be constructed using Young tableaux [145], and has the form

$$|m,q\rangle = (|+\rangle|-\rangle - |-\rangle|+\rangle)^q \operatorname{SYM} \left(|+\rangle^{m-q}|-\rangle^{n-m-q}\right) \,, \quad (3.15)$$

where the operation SYM means a complete symmetrization with respect to the $n - 2q$ last entries (the first $2q$ entries are, instead, anti-symmetrized). A basis of the subspace $D^{(m,q)}$ can be constructed by applying all the transformations of the permutation group to the vector $|m,q\rangle$ in Eq. (3.15) and choosing a maximal linearly independent subset.

We look for the eigenvectors of $T_n$ in the subspaces

$$D^{(q)} = \bigoplus_{m=q}^{n-q} D^{(m,q)} \qquad q = 0, \ldots, \left\lfloor \frac{n}{2} \right\rfloor \quad (3.16)$$

of dimension $d_q(n + 1 - 2q)$. Since $T_n$ has no symmetries beside the replicas permutation one, it has $n+1-2q$ different eigenvalues in $D^{(q)}$, each with multiplicity $d_q$. In the following we will refer to the subspaces $D^{(q)}$ as to *sectors*. Moreover, with some abuse of notation, we shall use the symbol $D^{(q)}$ for the set of eigenvalues corresponding to eigenvectors in that sector.

Of particular relevance are the sectors $D^{(0)}$, $D^{(1)}$ and $D_0^{(2)}$ since they are associated to the Longitudinal, Anomalous and Replicon modes respectively from mean-field spin-glass theory [150], as we will later show when discussing correlation functions in Section 3.4.2.

By Eqs. (3.15) and (3.16) it is possible to factorize the replica symmetric part in the eigenvectors $\psi_q^\lambda(\sigma)$ of the transfer matrix in the sector $D^{(q)}$, that is we can write

$$\psi_q^\lambda(\sigma) = \rho_q^\lambda \left(\sum_a \sigma^a\right) \sum_{a_1 < \cdots < a_q} C_{a_1 \ldots a_q} \, \sigma^{a_1} \ldots \sigma^{a_q} \,, \quad (3.17)$$

where the replica symmetric part $\rho_q^\lambda$ of the eigenvectors is the one relevant to the computation of the eigenvalues. By last equation the eigenvectors of the sector $D^{(0)}$ are completely replica symmetric. The coefficients $C_{a_1\ldots a_q}$ are invariant for any permutation of the indices and are equal to zero if any two of the indices are equal. Moreover they have to satisfy the constraint

$$\sum_{a_1=1}^{n} C_{a_1\ldots a_q} = 0, \tag{3.18}$$

which is a necessary and sufficient condition for any vector of the form of Eq. (3.17) to belong to the subspace $D^{(q)}$. Any set of $d_q$ linearly independent coefficient vectors $C$ can be chosen as an appropriate basis for the subspace. It is easy to prove that the product of two non-singular symmetric matrices possess a complete orthonormal (in the left-right sense) basis of left and right eigenvectors with real eigenvalues, and this is indeed case for $T_n$. In fact if we define, with a little abuse of notation, the vector

$$\rho_h(\sigma) \equiv \mathbb{E}_h e^{\beta h \sum_a \sigma^a}, \tag{3.19}$$

than $T_n(\sigma,\tau) = \sum_{\sigma'} \mathbb{E}_J e^{\beta J \sigma \sigma'} \times \delta_{\sigma'\tau} \rho_h(\tau)$. Moreover the left eigenvector $\psi_L$ corresponding to a certain right $\psi_R(\sigma;\lambda,k)$ is simply given by

$$\psi_L(\sigma;\lambda,k) = \rho_h(\sigma)\psi_R(\sigma;\lambda,k), \tag{3.20}$$

where $k$ denotes one choice of the coefficients $C_{a_1\ldots a_q}$ among the $d_q$ possible. Imposing the orthonormality condition

$$\sum_\sigma \psi_L(\sigma;\lambda,k)\,\psi_R(\sigma;\lambda',k') = \delta_{\lambda\lambda'}\,\delta_{kk'}, \tag{3.21}$$

with the sum ranging over all the $2^n$ configuration of the replicated spin, and after successive application of Eq. (3.18), we obtain

$$\sum_\sigma \rho_q^\lambda(\sigma)\rho_h(\sigma)\rho_q^{\lambda'}(\sigma) \prod_{a=1}^{q}(1 - \sigma^{2a-1}\sigma^{2a}) = \delta_{\lambda\lambda'} \tag{3.22}$$

along with

$$\sum_{a_1<\cdots<a_q} C_{a_1\ldots a_q}^k C_{a_1\ldots a_q}^{k'} = \delta_{kk'}. \tag{3.23}$$

We are now able to write down the transfer matrix in the spectral form

$$T_n(\sigma,\tau) = \sum_{q=0}^{\lfloor \frac{n}{2} \rfloor} T_{n,q}(\sigma,\tau) \tag{3.24}$$

where $T_{n,q}$ is the restriction of $T_n$ to the subspace $D^{(q)}$, defined by

$$T_{n,q}(\sigma,\tau) = \sum_{\lambda \in D^{(q)}} \lambda\, \rho_q^\lambda(\sigma)\rho_h(\tau)\rho_q^\lambda(\tau) \sum_{\substack{a_1<\cdots<a_q \\ b_1<\cdots<b_q}} Q_{a_1\ldots a_q;b_1\ldots b_q}\, \sigma^{a_1}\ldots\sigma^{a_q}\tau^{b_1}\ldots\tau^{b_q}. \tag{3.25}$$

The coefficients $Q$ appearing in last expression are invariant for any permutation of the set of indices $a$ or $b$, therefore they depended only on the number of equal indexes in the sets $\{a_1, \ldots, a_q\}$ and $\{b_1, \ldots, b_q\}$. They are defined by

$$Q_{a_1 \ldots a_q; b_1 \ldots b_q} = \sum_{k=1}^{d_q} C^k_{a_1 \ldots a_q} C^k_{b_1 \ldots b_q}, \qquad (3.26)$$

and their $(q+1)$ different values can be computed applying recursively Eqs. 3.18 and 3.23. If we denote $Q_p^{(q)}$ the coefficient in the sector $D^{(q)}$ with $p$ pairs of different indexes, for the first sectors we have

$$Q_0^{(1)} = \frac{n-1}{n} \qquad\qquad Q_1^{(1)} = -\frac{1}{n} \qquad (3.27)$$

$$Q_0^{(2)} = \frac{n-3}{2(n-1)} \qquad\qquad Q_1^{(2)} = -\frac{Q_0^{(2)}}{n-2} \qquad\qquad Q_2^{(2)} = -\frac{2Q_1^{(2)}}{n-3} \qquad (3.28)$$

### 3.3.2 Integral representations

In order to perform the limit $n \downarrow 0$ it is convenient to find a suitable parametrization for the eigenvectors of the form (3.17). For the replica symmetric part of the eigenvectors $\psi_q^\lambda$, see Eq. (3.17), we employ the standard parametrization

$$\rho_q^\lambda(\sigma) = \int du \; g_q^\lambda(u; n) \frac{e^{\beta u \sum_a \sigma^a}}{[2\cosh(\beta u)]^n}, \qquad (3.29)$$

in terms of the functions $g_q^\lambda(u; n)$. Turns out that all the functions $g_q^\lambda$ parameterizing the eigenvectors of the sector $D^{(0)}$, are by themselves the eigenfunctions of an integral operator associated to that sector. In fact, expressing the linear terms in Eq. (3.17) through the identity

$$\sigma_{a_1} \ldots \sigma_{a_q} = \left. \frac{\partial}{\partial \epsilon_{a_1}} \ldots \frac{\partial}{\partial \epsilon_{a_q}} \right|_{\epsilon=0} e^{\sum_a \epsilon_a \sigma^a} \qquad (3.30)$$

and plugging Eq. (3.29) into the eigenvalue equation $T_n \psi_q = \lambda \psi_q$, we obtain, after some manipulations, the new eigenvalue equation

$$\lambda \, g_q^\lambda(u; n) = \mathbb{E}_{J,h} \int dv \; \delta(u - \hat{u}(J, h+v)) \left( \frac{\partial \hat{u}}{\partial v} \right)^q Z^n(J, h, v) \, g_q^\lambda(v; n). \qquad (3.31)$$

The function $\hat{u}(J, x)$, defined by

$$\hat{u}(J, x) = \frac{1}{\beta} \operatorname{atanh}\left( \tanh(\beta J) \tanh(\beta x) \right), \qquad (3.32)$$

will be recognized by the learned reader as the update rule for cavity messages. As we shall see, the function

$$Z(J, h, v) = \frac{2\cosh(\beta J) \cosh(\beta(v+h))}{\cosh(\beta v)} \qquad (3.33)$$

is related to the intensive free energy of an chain. Notice that in writing down Eq. (3.31) we have shifted the problem of finding a complete bases of eigenvectors for the matrix $T_n$ to the equivalent problem of the spectral decomposition of the integral operators of Eq. (3.31), for $q = 0, \ldots, \lfloor \frac{n}{2} \rfloor$. Turns out that, for a given sector $D^{(q)}$, the integral operator has a set of left eigenfunctions in the form

$$S_q^\lambda(v; n) = \mathbb{E}_h \int du\ g_q^\lambda(u; n) \left[ \frac{\cosh(\beta(u+v+h))}{2\cosh(\beta u)\cosh(\beta v)} \right]^n \left[ 1 - \tanh^2(\beta(u+v+h)) \right]^q, \quad (3.34)$$

as can be inferred from Eq. (3.20) and can be directly verified. In the rest of the chapter we will assume that the left and right eigenfunctions of the sector $D^{(q)}$ satisfy the normalization condition

$$\int du\ S_q^\lambda(u; n)\, g_q^{\lambda'}(u; n) = \delta_{\lambda\lambda'} \quad (3.35)$$

derived from Eq. (3.22). We are now ready to take the $n \downarrow 0$ limit and discuss its non trivial aspects.

### 3.3.3 The small $n$ limit

In the limit $n \downarrow 0$ we obtain an infinite number of sectors $D^{(q)}$, $q = 0, 1, \ldots$, in a fashion that is characteristic to replicas computations. Setting $n = 0$ in Eq. (3.31) we obtain Eq. (3.4), which we rewrite for convenience:

$$\lambda\, g_q^\lambda(u) = \mathbb{E}_{J,h} \int dv\ \delta\left(u - \hat{u}(J, h+v)\right) \left( \frac{\partial \hat{u}}{\partial v} \right)^q g_q^\lambda(v)\ . \quad (3.36)$$

From now on we shall refer to $g_q^\lambda(v)$ as a solution of last equation and shall explicitly express the $n$ dependence for the solutions of (3.31) at finite $n$. In Figure 3.2 and Figure 3.3 we show two examples of eigenvalues and eigenfunctions in the sector $D^{(1)}$ and $D^{(2)}$ respectively.

For $q = 0$, i.e. in the sector $D^{(0)}$, Eq. (3.36) admits a unique maximum eigenvalue $\lambda = 1$ by Perron-Frobenius theorem. The corresponding eigenfunction is the probability distribution of cavity biases, which we call $P(u)$ [15]. We have thus established a first connection between the cavity method and the RTM formalism, and we shall enforce this connection in Section 3.5. The other eigenfunctions of $D^{(0)}$ are characterized by $\int du\ g_0^\lambda(u) = 0$ at $n = 0$. It is convenient, to held compatibility with the normalization condition Eq. (3.35) as we will see, to impose a diverging scaling for all the eigenfunctions of $D^{(0)}$ except for the first one:

$$g_0^\lambda(u; n) \sim \frac{1}{\sqrt{n}} \left( g_0^\lambda(u) + n\, \tilde{g}_0^\lambda(u) \right). \quad (3.37)$$

The symbol $\sim$ denotes equivalence between the r.h.s. ad l.h.s. up to higher order correction in $n$, and $\tilde{g}_0^\lambda$ is the first correction to the leading order of the eigenfunction in $D^{(0)}$. Using Eq. (3.37) for the right eigenfunctions and considering also the correction in $n$ to the eigenvalues, we can compute the left eigenfunctions of $D^{(0)}$ from Eq. (3.34). In fact we obtain at the leading order

$$S_0^\lambda(v; n) \sim \sqrt{n}\, S_0^\lambda(v) = \sqrt{n} \left[ c_\lambda + \mathbb{E}_h \int du\ g_0^\lambda(u) \log\left( \frac{\cosh(\beta(u+v+h))}{\cosh(\beta u)} \right) \right], \quad (3.38)$$

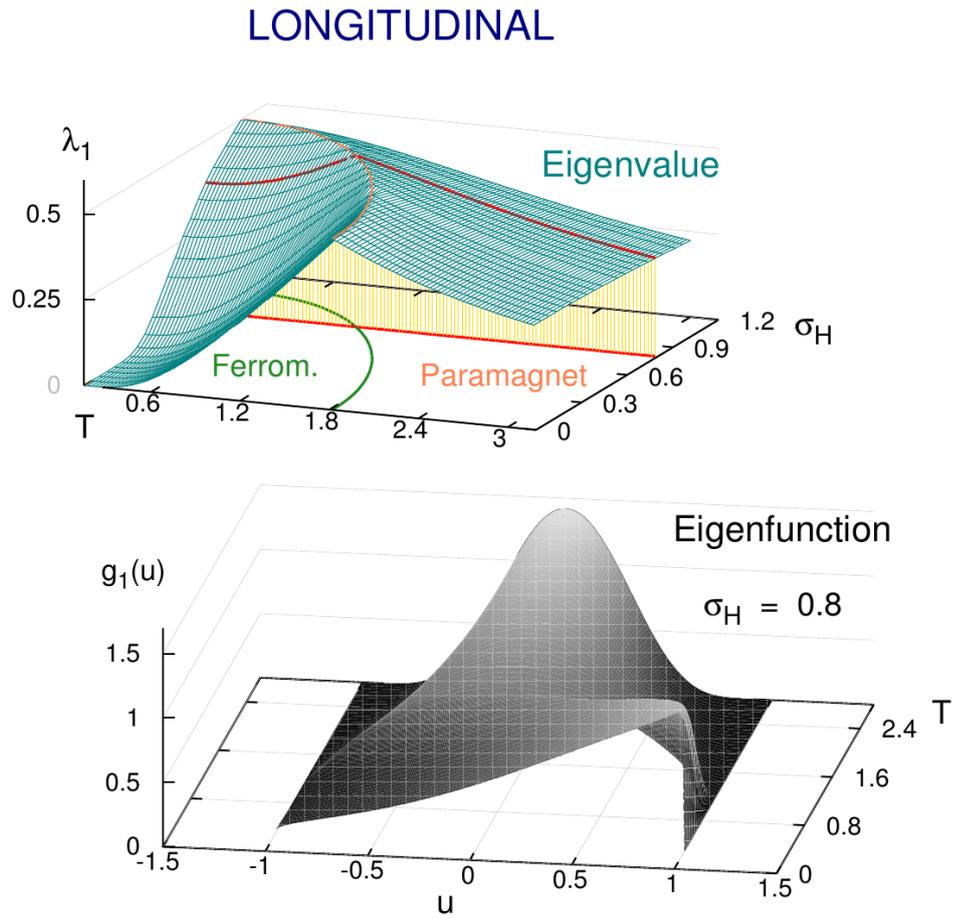

**Fig. 3.2.** (*Top*) The leading eigenvalue $\lambda_1$ of the sector $D^{(1)}$ in the RFIM, as a function of the temperature and of the gaussian external field with variance $\sigma_H^2$. (*Bottom*) The corresponding right eigenfunction $g_1(u)$ at $\sigma_H = 0.8$. The random fields $h$ and $\tilde{h}$ are distributed as the cavity fields arriving on a chain embedded in a RRG with connectivity $z = 3$, therefore the transition point is localized at $\lambda_1 = \frac{1}{2}$.

where $c_\lambda$ is the normalization of the first order correction to the eigenfunction $g_0^\lambda$, that is

$$c_\lambda \equiv \int du \, \tilde{g}_0^\lambda(u) = \frac{1}{\lambda - 1} \, \mathbb{E}_h \int du \, g_0^\lambda(u) \log\left(\frac{\cosh(\beta(u+h))}{\cosh(\beta u)}\right) . \qquad (3.39)$$

In all calculations involving the sector $D^{(0)}$ we will express the eigenvectors using Eqs. (3.37) and (3.38), then proceed carefully to take the $n \downarrow 0$ limit.

To find an expression for the left eigenfunctions in the other sectors no such care is needed to take the $n \downarrow 0$ limit in Eq. (3.34), therefore we straightly obtain

$$S_q^\lambda(v) = \mathbb{E}_h \int du \, g_q^\lambda(u) \left[1 - \tanh^2(\beta(u+v+h))\right]^q \qquad \text{for } q \geq 1 . \qquad (3.40)$$

The degeneracy between $D^{(0)}$ and $D^{(1)}$ corresponds to the degeneracy between the Longitudinal and Anomalous eigenvalues in the Hessian of the Sherrington-Kirkpatrick model [150, 151]. The multiplicity of the eigenvalues in the two sectors, $d_0 = 1$ and $d_1 = n - 1$, sum up to give an $O(n)$ contribution as should be expected, while from Eq. (3.14) the other sectors have degeneracies of order $O(n)$ without the need of further elisions. Therefore it is convenient to define

$$\hat{d}_q = \begin{cases} 1 & \text{for } q = 1 , \\ \lim_{n \to 0} \frac{d_q}{n} = (-1)^{q+1} \frac{2q-1}{q(q-1)} & \text{for } q \geq 2 . \end{cases} \qquad (3.41)$$

The first eigenvalue of $D^{(0)}$ requires separate considerations. We define the coefficient $f_0$ from its $n$ expansion:

$$\lambda(n) \sim 1 - \beta f_0 n. \qquad (3.42)$$

As we already noted, the cavity messages distribution $P(u)$ is the eigenvector associated to the largest eigenvalue of the sector $D^{(0)}$ for $n = 0$. The corresponding left eigenvalue is $S(u) \equiv 1$. In Section 3.4.1 we shall see that $f_0$ is the intensive free energy of a chain. From Eq. (3.47) we obtain

$$-\beta f_0 = \mathbb{E}_{J,h} \int dv \, \log[Z(J,h,v)] P(v) . \qquad (3.43)$$

### 3.3.4 The degeneracy between $D^{(0)}$ and $D^{(1)}$

A close inspection of the eigenvalue equation (3.36) reveals a surprising relation between the sectors $D^{(0)}$ and $D^{(1)}$ at $n = 0$. It can be shown, respectively deriving or integrating both members of Eq. (3.36) for $q = 1$ and $q = 0$, that all the eigenfunctions of $D^{(1)}$ have a corresponding eigenfunction in $D^{(0)}$ with the same eigenvalue. On the other hand, all the eigenfunctions of $D^{(0)}$, except for the first one, i.e. the ones having zero sum, have a corresponding eigenfunction in $D^{(1)}$ with the same eigenvalue. We have thus established a degeneracy between the Longitudinal and the Anomalous sectors. The following relations hold:

$$g_0^\lambda(u) = \frac{1}{\beta} \partial_u \, g_1^\lambda(u); \qquad \frac{1}{\beta} \partial_u \, S_0^\lambda(u) = -S_1^\lambda(u). \qquad (3.44)$$

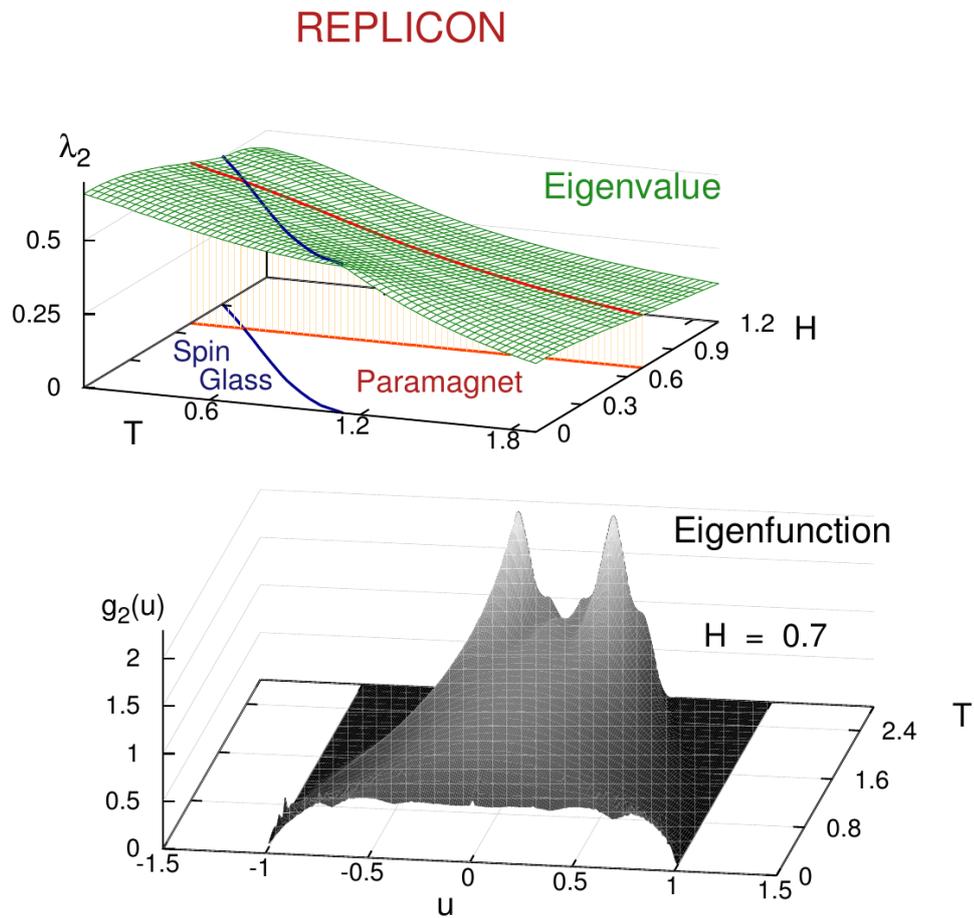

**Fig. 3.3.** (*Top*) The leading eigenvalue $\lambda_2$ of the sector $D^{(2)}$ in a $J = \pm 1$ spin-glass, as a function of the temperature and of the uniform external field $H$. The phase diagram is also shown in the $H - T$ plane. (*Bottom*) The corresponding right eigenfunctions $g_2(u)$ along the orange line of the top picture. The random fields $h$ and $\tilde{h}$ are distributed as the cavity fields arriving on a chain embedded in a RRG with connectivity $z = 3$, therefore the transition point is localized at $\lambda_2 = \frac{1}{2}$.

Particular attention has to be taken in the limits involving these two sectors, keeping track of the $O(n)$ corrections both to eigenvalues and eigenvectors. A double pole contribution to some observables, as we shall later see, stems from the first correction in $n$ to the paired eigenvalues in $D^{(0)}$ and $D^{(1)}$. In fact if we define the eigenvalue shifts $\delta\lambda_0$ and $\delta\lambda_1$ by

$$\lambda_0(n) \sim \lambda + n\,\delta\lambda_0, \tag{3.45}$$

$$\lambda_1(n) \sim \lambda + n\,\delta\lambda_1, \tag{3.46}$$

and consider the expansion to the first order in $n$ of the eigenvalue equation (3.31) for $q =$, from standard perturbation theory we have

$$\delta\lambda_q = \mathbb{E}_{J,h} \int \mathrm{d}u \mathrm{d}v \; S_q^\lambda(u) \log\left[Z(J,h,v)\right] \delta\left(u - \hat{u}(J, h+v)\right) \left(\frac{\partial \hat{u}}{\partial v}\right)^q g_q^\lambda(v). \tag{3.47}$$

The shift difference $\Delta_\lambda = \delta\lambda_0 - \delta\lambda_1$ is the relevant quantity we are looking after, since it arises in the calculation of the free energies of closed chains and of the thermally disconnecter correlation function, see Section 3.4. Using Eq. (3.47) and the relation (3.44) between the eigenfunctions in the two sectors, we obtain the expression

$$\Delta_\lambda = -\mathbb{E}_{J,h} \int \mathrm{d}u \mathrm{d}v \; S_0^\lambda(u) \delta\left(u - \hat{u}(J, v+h)\right) \left[\tanh\left(\beta(v+h)\right) - \tanh(\beta v)\right] g_1^\lambda(v) \tag{3.48}$$

If we call $(\bullet, \bullet)$ the scalar product in $L^2$ and define the kernel

$$Q(u,v) = \mathbb{E}_{J,h} \, \delta\left(u - \hat{u}(J, v+h)\right) \left[\tanh\left(\beta(v+h)\right) - \tanh(\beta v)\right], \tag{3.49}$$

then Eq. (3.48) can be rewritten as

$$\Delta_\lambda = -(S_0^\lambda, Q\, g_1^\lambda). \tag{3.50}$$

In the next Section we shall apply the formalism we have developed to the computation of some physically relevant quantities.

## 3.4 Some Applications of the formalism

### 3.4.1 Free energy of chains

Let us first consider the average free energy of a closed chain of length $\ell$, each node receiving i.i.d. random fields $h$, and call it $f_\ell^c$. If the chain considered is embedded in a locally tree-like graph, the random fields $h$ are distributed according to the cavity messages distribution on that graph ensemble. Since $\mathrm{Tr}\, T_n^\ell$ is the replicated partition function of this system, the free energy is given by

$$-\beta f_\ell^c = \lim_{n \to 0} \partial_n \mathrm{Tr}\, T_n^\ell, \tag{3.51}$$

where, thanks the orthonormal decomposition of $T_n$, the trace can be written in the form

$$\mathrm{Tr}\, T_n^\ell = \sum_{q=0}^{\frac{n}{2}} d_q \sum_{\lambda \in D^{(q)}} \lambda^\ell. \tag{3.52}$$

In last equation the eigenvalue degeneracies $d_q$ are given in Eq. (3.14), and the eigenvalues $\lambda$ depends implicitly on $n$. In the small $n$ limit the sum over $q$ can be extended to infinity. The considerations over the eigenvalues' shifts and degeneracies of last Section lead to the final expression

$$-\beta f_\ell^c = -\beta \ell f_0 + \sum_{\lambda \in D^{(1)}} \Delta_\lambda \, \ell \, \lambda^{\ell-1} + \sum_{q=1}^{\infty} \hat{d}_q \sum_{\lambda \in D^{(q)}} \lambda^\ell \ . \qquad (3.53)$$

The coefficients $\hat{d}_q$ are given by Eq. (3.41), the shift differences $\Delta_\lambda$ given by Eq. (3.48) and an expression for the intensive free energy $f_0$ is found in Eq. (3.43). We notice that all the quantities entering Eq. (3.53) can be expressed in terms of the eigenvalues and eigenfunctions of Eq. (3.36).

The computation of the average free energy of open chains is a little more involved. In the definition of open chains, we allow the spins at the extremities to receive a random field $\tilde{h}$ that could have a distribution different from the one of the fields acting on the internal spins of the chain. We introduce this relaxation of the model in order to apply our formalism to the case of open chains embedded in a generic tree-like random graph.

It is convenient to define the replicated partition function of an open chain of length $\ell$, conditioned on the configuration of the replicated spins at its extrema in the following way: starting from $T_n^\ell$, we remove the field $h$ on the right and substitute it with a field $\tilde{h}$, then we add the other field $\tilde{h}$ on the left (see Figure 3.1). Therefore we define

$$\tilde{T}_n^{(\ell)}(\sigma, \tau) \equiv \rho_{\tilde{h}}(\sigma) \, T_n^\ell(\sigma, \tau) \, \rho_h^{-1}(\tau) \rho_{\tilde{h}}(\tau), \qquad (3.54)$$

where, with a little abuse of notation, the vector $\rho_{\tilde{h}}$ is defined by

$$\rho_{\tilde{h}}(\sigma) \equiv \mathbb{E}_{\tilde{h}} \, e^{\beta \tilde{h} \sum_a \sigma^a}. \qquad (3.55)$$

By definition the matrix $\tilde{T}_n^{(\ell)}$ is symmetric (see Figure 3.1 for a pictorial representation). From Eq. (3.54) and Eq. (3.24) we obtain the spectral decomposition

$$\tilde{T}_n^{(\ell)}(\sigma, \tau) = \sum_{q=0}^{\lfloor \frac{n}{2} \rfloor} \sum_{\lambda \in D^{(q)}} \lambda^\ell \, \rho_{\tilde{h}}(\sigma) \rho_q^\lambda(\sigma) \rho_{\tilde{h}}(\tau) \rho_q^\lambda(\tau) \sum_{\substack{a_1<\cdots<a_q \\ b_1<\cdots<b_q}} Q_{a_1\ldots a_q;b_1\ldots b_q} \, \sigma^{a_1}\ldots\sigma^{a_q}\tau^{b_1}\ldots\tau^{b_q}.$$

$$(3.56)$$

The average free energy of an open chain of length $\ell$ is then given by

$$-\beta f_\ell^o = \lim_{n \to 0} \partial_n \sum_{\sigma, \tau} \tilde{T}_n^{(\ell)}(\sigma, \tau) \ . \qquad (3.57)$$

From Eq. (3.56) it easy to see that only the $D^{(0)}$ sector of $\tilde{T}_n^{(\ell)}$ contributes to last equation.

A different behaviour characterize the terms corresponding to the leading eigenvalue at $n = 0$ (the cavity one) from the others. As in the case of the closed chain, the extensive contribution to the free energy comes from the leading eigenvalue of $D^{(0)}$, $\lambda \sim 1 - n\,\beta f_0$. An $O(1)$ contribution comes from the leading eigenfunction $g_0^\lambda(u; n) = P(u) + O(n)$, while each other eigenvalue, the ones degenerate with $D^{(0)}$, gives an exponential term. Therefore, after a careful treatment of the small $n$ limit, we arrive to the expression

$$-\beta f_\ell^o = -\ell\beta f_0 + \mathbb{E}_{\tilde{h}} \int \mathrm{d}u\ P(u)\, 2\log\cosh\left(\beta(u+\tilde{h})\right)$$
$$- \mathbb{E}_h \int \mathrm{d}u\mathrm{d}v\ P(u)P(v) \log\cosh\left(\beta(u+v+h)\right) \quad (3.58)$$
$$+ \log 2 + \sum_{\lambda \in D^{(1)}} a_{\lambda,0}^2\, \lambda^\ell ,$$

with

$$a_{\lambda,0} = \frac{1}{\lambda - 1}\, \mathbb{E}_h \int \mathrm{d}u\ g_0^\lambda(u) \log\left[\frac{\cosh\left(\beta(u+h)\right)}{\cosh(\beta u)}\right]$$
$$+ \mathbb{E}_{\tilde{h}} \int \mathrm{d}u\ g_0^\lambda(u) \log\left[\frac{\cosh(\beta(u+\tilde{h}))}{\cosh(\beta u)}\right] \quad (3.59)$$

In Eq. (3.58) it is clearly expressed at the order $O(1)$ in $\ell$ the free energy shift, with respect to the free energy of a closed chain, due to the addition of two extremal spins and the removal of an internal one.

The coefficients $a_{\lambda,0}$ are strictly related to the left eigenfunctions $S_0^\lambda$ defined in Eq. (3.38). In fact if the random field at the extremities of the chain are distributed as the one on the internal spins, i.e. $\tilde{h} \stackrel{d}{=} h$ as in the case of a chain embedded in a Poissonian random graph, then $a_{\lambda,0} = S_0^\lambda(0)$. More generally if a probability distribution $G^{(0)}(u)$ exists such that

$$\tilde{P}(\tilde{h}) = \mathbb{E}_h \int \mathrm{d}u\ G^{(0)}(u)\ \delta(\tilde{h} - (u+h)) \quad (3.60)$$

holds, then Eq. (3.59) can be written in the compact form $a_{\lambda,0} = (S_0^\lambda, G^{(0)})$. Obviously if $\tilde{h} \stackrel{d}{=} h$ we have $G^{(0)}(u) = \delta(u)$. For a chain embedded in a random regular graphs ensemble instead, $G^{(0)}(u)$ is given by the distribution of cavity biases $P_{cav}(u)$ [15], which corresponds to the first eigenvector of the Longitudinal Sector. Therefore in the random regular graph ensemble $(S_0^\lambda, G^{(0)}) = 0$ and no exponential decays are present in the expression (3.58) for the free energy of open chains.

### 3.4.2 Correlation functions

We take advantage of the spectral representation of the RTM to find some analytical expressions for the two-point correlation functions. We consider two spins, $\sigma_0$ and $\sigma_\ell$, at distance $\ell$ along a chain. As in the previous paragraph, we admit the possibility for the chain to be embedded in a locally tree-like graph, therefore the random fields $\tilde{h}$ acting on $\sigma_0$ and $\sigma_\ell$ can be distributed differently from the fields $h$ on the internal spin of the chain. The decomposition of $\tilde{T}_n^{(\ell)}(\sigma,\tau)$ in Eq. (3.56) can be exploited to obtain the correlation functions. In fact contracting $\tilde{T}_n^\ell(\sigma,\tau)$ with two spins having the same replica index constrains them to be in the same thermal state, as in $\overline{\langle \sigma_0 \sigma_\ell \rangle} = \lim_{n \to 0} \sum_{\sigma,\tau} \sigma^1\, \tilde{T}_n^{(\ell)}(\sigma,\tau)\, \tau^1$. Choosing different replica indexes instead corresponds to choosing different thermal states, e.g.

$\overline{\langle\sigma_0\rangle\langle\sigma_\ell\rangle} = \lim_{n\to 0}\sum_{\sigma,\tau}\sigma^1\,\tilde{T}_n^{(\ell)}(\sigma,\tau)\,\tau^2$. Generalizing this considerations is easy to obtain

$$\overline{\langle\sigma_0\sigma_\ell\rangle^k} = \lim_{n\to 0}\sum_{\sigma,\tau}\sigma^1\ldots\sigma^k\,\tilde{T}_n^{(\ell)}(\sigma,\tau)\,\tau^1\ldots\tau^k. \tag{3.61}$$

Since vectors of the form $\sigma^1\ldots\sigma^k$ have non-zero projections in $D^{(q)}$ only for $q\leq k$, only these sectors of the spectral representation of $\tilde{T}_n^{(\ell)}$ contribute to Eq. (3.61). The expression for $\overline{\langle\sigma_0\sigma_\ell\rangle^k}$ is quite complicated and it involves also the correction for small $n$ to the eigenfunction of $D^{(0)}$ and $D^{(1)}$, as in the case of the thermally disconnected correlation function we shall later see. Therefore, since this kind of correlation function has little physical relevance, we won't report its expression in terms of the transfer matrix eigenvalues and eigenfunctions.

Far more interesting from the physical viewpoint are the connected correlation functions. The ferromagnetic connected correlation functions can be expressed as $\overline{\langle\sigma_0\sigma_\ell\rangle_c} = \lim_{n\to 0}\frac{1}{2}\sum_{\sigma,\tau}(\sigma^1-\sigma^2)\,\tilde{T}_n^{(\ell)}(\sigma,\tau)\,(\tau^1-\tau^2)$, as one can rapidly check, and this expression can be easily generalized to

$$\overline{\langle\sigma_0\sigma_\ell\rangle_c^k} = \lim_{n\to 0}\frac{1}{2^k}\sum_{\sigma,\tau}\left(\sigma^1-\sigma^2\right)\ldots\left(\sigma^{2k-1}-\sigma^{2k}\right)\tilde{T}_n^{(\ell)}(\sigma,\tau)\left(\tau^1-\tau^2\right)\ldots\left(\tau^{2k-1}-\tau^{2k}\right). \tag{3.62}$$

It is worth noticing that the vector $v = \left(\sigma^1-\sigma^2\right)\ldots\left(\sigma^{2k-1}-\sigma^{2k}\right)$ belongs to the subspace $D^{(k)}$, therefore we can choose a basis for the spectral representation of $\tilde{T}_n^{(\ell)}$ such that all but one vectors are orthogonal to $v$. This leads to the following compact expression for the connected correlation functions:

$$\overline{\langle\sigma_0\sigma_\ell\rangle_c^k} = \sum_{\lambda\in D^{(k)}} a_{\lambda,k}^2\,\lambda^\ell, \tag{3.63}$$

with the coefficients $a_{\lambda,k}$ given by

$$a_{\lambda,k} = \mathbb{E}_{\tilde{h}}\int du\,g_k^\lambda(u)\left[1-\tanh^2\left(\beta(u+\tilde{h})\right)\right]^k. \tag{3.64}$$

As in the case of the coefficient $a_{\lambda,0}$ defined in Eq. (3.59), if a solution $G^{(0)}$ of (3.60) exist then $a_{\lambda,k}$ is simply given by the projection of $G^{(0)}$ on $S_k^\lambda$, that is $a_{\lambda,k} = (S_k^\lambda, G^{(0)})$.

The susceptibilities $\chi_k = \lim_{N\to\infty}\frac{1}{N}\sum_{i,j}\overline{\langle\sigma_i\sigma_j\rangle_c^k}$ can be easily computed through equation (3.63) in a random graph with mean degree and mean residual degree $z_0$ and $z$ respectively. In fact in thermodynamic limit we have

$$\begin{aligned}\chi_k &= \overline{(1-m^2)^k} + \sum_{\ell=1}^\infty z_0 z^{\ell-1}\,\overline{\langle\sigma_0\sigma_\ell\rangle_c^k} \\ &= \overline{(1-m^2)^k} + z_0\sum_{\lambda\in D^{(k)}} a_{\lambda,k}^2\,\frac{\lambda}{1-z\lambda}\end{aligned} \tag{3.65}$$

At a transition point the largest eigenvalue of one of the sectors $D^{(q)}$ reaches the value $\frac{1}{z}$ and the corresponding susceptibility diverges. Assuming a smooth behavior for the eigenvalue in the high temperature region before the transition, $\lambda(T) = \frac{1}{z} + O(T - T_c)$ for $T \to T_c^+$, we obtain the mean-field critical exponent $\gamma = 1$.

The computation of the thermal disconnected correlation function $\overline{\langle\sigma_0\rangle\langle\sigma_\ell\rangle}$, relevant to the RFIM transition, is more complicated, since it involves the subleading corrections in $n$ to the eigenvectors of $T_n$. Great care has to be taken in the limit $\lim_{n\to 0}\sum_{\sigma,\tau}\sigma^1\,\tilde{T}_n^{(\ell)}(\sigma,\tau)\,\tau^2 = \overline{\langle\sigma_0\rangle\langle\sigma_\ell\rangle}$. As in Eq. (3.37), let us call $\tilde{g}_0^\lambda(u)$ the correction to the eigenfunction $g_0^\lambda(u)$. We denote with $\langle\bullet\rangle_q$ the expectation over $\tilde{T}_n^{(\ell)}(\sigma,\tau)$ restricted to the sector $D^{(q)}$. Than in $D^{(0)}$ we obtain

$$\langle\sigma^1\tau^2\rangle_0 \sim \overline{\langle\sigma_\infty\rangle}^2 + \sum_{\lambda\in D^{(1)}} \frac{1}{n}a_{\lambda,1}^2\,\lambda^\ell + a_{\lambda,1}^2\,\ell\,\delta\lambda_0\,\lambda^{\ell-1}$$

$$- 2a_{\lambda,1}\,\lambda^\ell\bigg[\int du\,\tilde{g}_0^\lambda(u)\tanh(\beta(u+\tilde{h})) \qquad (3.66)$$

$$+ \int du\,g_0^\lambda(u)\tanh(\beta(u+\tilde{h}))\log\frac{\cosh(\beta(u+\tilde{h}))}{\cosh(\beta u)}\bigg],$$

where the contribution $\overline{\langle\sigma_\infty\rangle}$ comes from the cavity eigenvector and is the average magnetization of a spin at the end of an infinite chain.

Similarly, if we define $\tilde{g}_1^\lambda(u)$ by $g_1^\lambda(u;n) \sim g_1^\lambda(u) + n\,\tilde{g}_1^\lambda(u)$, in the sector $D^{(1)}$ we have

$$\langle\sigma^1\tau^2\rangle_1 \sim \sum_{\lambda\in D^{(1)}} \frac{-1}{n}a_{\lambda,1}^2\,\lambda^\ell - a_{\lambda,1}^2\,\ell\,\delta\lambda_1\,\lambda^{\ell-1}$$

$$- 2a_{\lambda,1}\,\lambda^\ell\bigg[\int du\,\tilde{g}_1^\lambda(u)\left(1 - \tanh^2(\beta(u+\tilde{h}))\right) \qquad (3.67)$$

$$+ \int du\,g_1^\lambda(u)\left(1 - \tanh^2(\beta(u+\tilde{h}))\right)\log\frac{\cosh(\beta(u+\tilde{h}))}{\cosh(\beta u)}\bigg]$$

Summing the two contributions, the final result for the disconnected correlation function is

$$\overline{\langle\sigma_0\rangle\langle\sigma_\ell\rangle} = \overline{\langle\sigma_\infty\rangle}^2 + \sum_{\lambda\in D^{(1)}} \Delta_\lambda\,a_{\lambda,1}^2\,\ell\,\lambda^{\ell-1} + \alpha_\lambda\,\lambda^\ell. \qquad (3.68)$$

Therefore each eigenvalue of the Anomalous sector contributes to $\overline{\langle\sigma_0\rangle\langle\sigma_\ell\rangle}$ with an simple exponential term and with a term that leads to a double pole behaviour in the associated susceptibility, with coefficients $\Delta_\lambda$ given in Eq. (3.48). The coefficients $\alpha_\lambda$ of the exponential decays instead are given by

$$\alpha_\lambda = 2\,a_{\lambda,1}\bigg[\int du\,g_1^\lambda(u)\tanh(\beta(u+\tilde{h}))\left(\tanh(\beta(u+\tilde{h})) - \tanh(\beta u)\right)$$

$$- \int du\,\tilde{g}_1^\lambda(u)\left(1 - \tanh^2(\beta(u+\tilde{h}))\right) - \int du\,\tilde{g}_0^\lambda(u)\tanh(\beta(u+\tilde{h}))\bigg] \qquad (3.69)$$

Since the magnetization of a spin conditioned to be to be the extremity of a chain of size $\ell$ is given by

$$\overline{\langle\sigma_\ell\rangle} = \lim_{n\to 0}\sum_{\sigma,\tau}\sigma^1\,\tilde{T}_n^{(\ell)}(\sigma,\tau) = \overline{\langle\sigma_\infty\rangle} - \sum_{\lambda\in D^{(1)}} a_{\lambda,1}\,a_{\lambda,0}\,\lambda^\ell, \qquad (3.70)$$

if we call $\Lambda$ the highest eigenvalue of the sector $D^{(1)}$, the most relevant contributions to the thermally-disconnected disorder-connected correlation function is given, for $\ell \to +\infty$, by

$$\overline{\langle\sigma_0\rangle\langle\sigma_\ell\rangle} - \overline{\langle\sigma_0\rangle}\,\overline{\langle\sigma_\ell\rangle} \approx \Delta_\Lambda\, a_{\Lambda,1}^2\, \ell\, \Lambda^{\ell-1} + \left(\alpha_\Lambda - 2\overline{\langle\sigma_\infty\rangle}\, a_{\Lambda,1} a_{\Lambda,o}\right) \Lambda^\ell. \qquad (3.71)$$

We notice that, while the coefficient of the exponential term is quite hard to compute, the coefficient $a_{\Lambda,1}^2 \Delta_\Lambda$, which regulates the leading behaviour, has a much simpler expression given in Eq. (3.48) and Eq. (3.64). From Eq. 3.71 turns out that near a ferromagnetic transition point, i.e. $\Lambda = \frac{1}{z}$, as long as $\Delta_\Lambda$ is not zero, the leading behavior of the disconnected susceptibility $\chi_{disc} = \sum_{i,j} \overline{\langle\sigma_0\rangle\langle\sigma_\ell\rangle} - \overline{\langle\sigma_0\rangle}\,\overline{\langle\sigma_\ell\rangle}$ reads

$$\chi_{disc} \simeq z_0\, \Delta_\Lambda\, a_{\Lambda,1}^2 \frac{1}{(1-z\Lambda)^2}. \qquad (3.72)$$

The expected double-pole behavior of the disconnected susceptibility is thus recovered.

## 3.5 Cavity derivation

In this section we present the derivation of several of the results of last Section using a probabilistic approach, in the same spirit of the usual cavity method calculations [13, 15]. While this approach is more physically intuitive than the RTM formalism, it requires the set up of an ad-hoc recursion rule for each observable. Noticeably we could not recover Eq. (3.53) for the free energy of closed chains.

### 3.5.1 Open chains

We want to study the statistical properties of a random Ising open chain without the use of replicas. We start with an asymmetric chain of length $\ell$, whose random partition function we denote with $Z_\ell$, constructed iteratively according to the following procedure: $Z_0$ is the partition function of a single spin receiving a random field $u_0$, i.e. $Z_0 = 2\cosh(\beta u_0)$; at the $i$-th step of the construction we add a spin $\sigma_i$, a random coupling $J_i$ between $\sigma_i$ and $\sigma_{i-1}$ and a random field $h_{i-1}$ on $\sigma_{i-1}$; the random variable $Z_\ell$ is the partition function of the system obtained after the $\ell$-th step of the procedure. Note that the last spin added to the chain has no external fields acting on it. The following distributional identity can be easily derived:

$$Z_{\ell+1} = \frac{2\cosh(\beta J_\ell)\,\cosh(\beta(u_\ell + h_\ell))}{\cosh(\beta u_\ell)} \times Z_\ell \equiv Z(J_\ell, h_\ell, u_\ell) \times Z_\ell. \qquad (3.73)$$

It is convenient to introduce the quantity $\overline{Z_\ell^n(u)} \equiv \overline{\delta(u - u_\ell)\, Z_\ell^n}$, which corresponds to the expectation of $Z_\ell^n$ along with the indicator function of the event $u_\ell = u$. Here $n$ is an arbitrary chosen positive real number, the symbol being chosen to stress the analogy with the replica formalism where the quantity $n$ (integer in this case) is the number of replicated systems. Using this definition from Eq. (3.73) follows readily

$$\overline{Z_{\ell+1}^n(u)} = \mathbb{E}_{J,h} \int dv\, \delta\left(u - \hat{u}(J, v+h)\right)\, Z^n(J,h,v)\, \overline{Z_\ell^n(v)}, \qquad (3.74)$$

where $\tilde{u}(J, x) = \frac{1}{\beta}\operatorname{atanh}(\tanh(\beta J)\tanh(\beta x))$ is the usual message passing rule. The integral operator of Eq. (3.74) is the same we found in the RTM formalism in Eq. (3.31) for the sector $D^{(0)}$, therefore we can make use of the spectral analysis result from those paragraphs, in particular of the completeness relation

$$\mathbb{E}_{J,h}\ \delta\left(u - \hat{u}(J, v+h)\right) Z^n(J, h, v) = \sum_{\lambda \in D^{(0)}} \lambda(n)\, g_0^\lambda(u; n)\, S_0^\lambda(v; n), \qquad (3.75)$$

between left and right eigenvectors. The definition of the left eigenfunctions of $D^{(0)}$ was already given in Eq. (3.34), but we rewrite it for convenience:

$$S_0^\lambda(v; n) = \mathbb{E}_h \int du\ g_0^\lambda(u; n) \left[\frac{\cosh\left(\beta(u+v+h)\right)}{2\cosh(\beta u)\cosh(\beta v)}\right]^n. \qquad (3.76)$$

Let us define another random partition function, $Z_\ell(u; x)$, obtained from $Z_\ell(u)$ conditioning on the value of the message $u_0$ on the first spin, that is $Z_\ell(u; x) = Z_\ell(u)|(u_0 = x)$. Since also $\overline{Z_\ell^n(u; x)}$ as a function of $u$ obeys equation (3.74), using the decomposition Eq. (3.75) and the initial condition $Z_0 = 2\cosh(\beta u_0)$ we arrive to the important result

$$\overline{Z_\ell^n(u; x)} = \sum_{\lambda \in D^{(0)}} \lambda^\ell(n)\, g_0^\lambda(u; n)\, S_0^\lambda(x; n)\, [2\cosh(\beta x)]^n. \qquad (3.77)$$

Using last equation it is easy to compute any moment $\overline{Z_\ell^n}$, $n$ not necessarily integer, of the partition function of a random asymmetric Ising chain of length $\ell$. More interesting is the computation of the properties of a symmetric Ising open chain, the one considered in Section 3.4.1, which receives on each extremity an external field distributed according to a certain probability distribution $\tilde{P}(\tilde{h})$. As already stated, this is definition stems from the need to cover the important case of a chain embedded in a locally tree-like graph. Let us call $Z_{\ell,o}$ the random partition function of this open chain. It is related to the random partition function $Z_\ell$ of the asymmetric open chain by

$$Z_{\ell,o} = \frac{\cosh(\beta(u_\ell + \tilde{h}_\ell))}{\cosh(\beta u_\ell)} \times Z_{\ell-1}(u_\ell; u_1) \times \frac{2\cosh(\beta J_0)\cosh(\beta \tilde{h}_0)}{\cosh(\beta u_1)} \qquad (3.78)$$

where $u_1$ is distributed as $\tilde{u}(J_0, \tilde{h}_0)$. From Eq. (3.78) along with Eq. (3.77) and Eq. (3.76), we derive the main result of this paragraph:

$$\overline{Z_{\ell,o}^n} = \sum_{\lambda \in D^{(0)}} \lambda^\ell(n)\, a_{\lambda,0}^2(n). \qquad (3.79)$$

where $a_{\lambda,0}(n)$ is defined by

$$a_{\lambda,0}(n) \equiv \mathbb{E}_{\tilde{h}} \int du\ \left[\frac{\cosh(\beta(u+\tilde{h}))}{\cosh(\beta u)}\right]^n g_0^\lambda(u; n). \qquad (3.80)$$

In the RTM formalism of Section 3.3 and 3.4, last expression could be derived from $\tilde{T}_n^{(\ell)}$ defined in Eq. (3.56) by analytic continuation of $\overline{Z_{\ell,o}^n} = \sum_{\sigma,\tau} \tilde{T}_n^{(\ell)}(\sigma, \tau)$ to non-integer $n$.

The average free energy of an open chain of length $\ell$ can then be obtained by

$$-\beta f_\ell^o = \lim_{n \to 0} \partial_n \overline{Z_{\ell,o}^n}. \tag{3.81}$$

The computation involves computing the order $n$ of all the quantities present in Eq. (3.79), as it was done in Section (3.4.1). In this paragraph however, without any use of replicas, we gave a purely probabilistic argument valid for any real value of $n$. We refer therefore to Section (3.4.1) for the successive step of the computation of $f_\ell^o$, leading to the final result Eq. (3.58). Notice that in the notation of that paragraph $a_{\lambda,0}$ is related to $a_{\lambda,0}(n)$ defined in Eq. (3.80) by $a_{\lambda,0}(n) \sim \sqrt{n}\, a_{\lambda,0}$.

The expression (3.58) for $f_\ell^o$ could also be obtained by a different approach that does not involve any limit $n \downarrow 0$ but is technically more difficult. We define the function $\varphi^{(\ell)}(u)$ by

$$\varphi^{(\ell)}(u) \equiv \overline{\delta(u - u_\ell) \log Z_\ell}, \tag{3.82}$$

and observe that given the distribution of the cavity message at distance $\ell$ along the chain, $u_\ell$, which we call $G_0^{(\ell)}(u)$, it obeys the iterative rule

$$\begin{aligned}\varphi^{(\ell+1)}(u) &= \mathbb{E}_{J,h} \int dv\, \delta\left(u - \hat{u}(J, h + v)\right) \varphi^{(\ell)}(v) \\ &+ \mathbb{E}_{J,h} \int dv\, \delta\left(u - \hat{u}(J, h + v)\right) \log\left[\frac{2 \cosh(\beta J) \cosh(\beta(v + h))}{\cosh(\beta v)}\right] G_0^{(\ell)}(v)\end{aligned} \tag{3.83}$$

Last equation can be solved decomposing $\varphi^{(\ell)}(u)$ and $G_0^{(\ell)}(v)$ along the eigenfunctions of $D^{(0)}$ at $n = 0$, then $\varphi^{(\ell)}(u)$ can be used to obtain $f_\ell^o$.

### 3.5.2 Connected correlation functions

Let us derive the eigenvalue equation (3.4) and the expression for the connected correlation functions Eq. (3.8), without making any use of replicas. Here we consider straightly the random open chain with partition function $Z_{\ell,o}$, characterized by independent random external field distributes a $h$ on the internal spins and as $\tilde{h}$ on the extremities. The connected correlation function $\langle \sigma_0 \sigma_\ell \rangle_c = \frac{1}{\beta} \frac{\partial \langle \sigma_\ell \rangle}{\partial H_0}$, where $H_0$ is an auxiliary field acting on $\sigma_0$, can be expressed as a function of the message $u_\ell$, coming through the chain to the spin $\sigma_\ell$, and its derivative with respect to $H_0$. In fact we have

$$\langle \sigma_0 \sigma_\ell \rangle_c = \left(1 - \tanh^2(\beta(\tilde{h}_\ell + u_\ell))\right) \frac{\partial u_\ell}{\partial H_0}, \tag{3.84}$$

where $\tilde{h}_\ell$, as usual, is the random effective field acting on $\sigma_\ell$ and coming eventually from the rest of the graph. Let us define the random variable $X_\ell$ by $X_\ell \equiv \frac{\partial u_\ell}{\partial H_0}$. The average over disorder of Eq. (3.84) and its moments $\overline{\langle \sigma_0 \sigma_\ell \rangle_c^q}$ can then be computed once we know the joint law of the random variables $u_\ell$ and $X_\ell$, which we call $P_\ell(u, X)$. Since $X_\ell$ obeys the chain rule $X_{\ell+1} = \frac{\partial u_{\ell+1}}{\partial u_\ell} X_\ell$ the recursion rule for $P_\ell$ reads

$$P_{\ell+1}(u, X) = \mathbb{E}_{J,h} \int dv\, dY\, \delta\left(X - \frac{\partial \hat{u}}{\partial v} Y\right) \delta\left(u - \hat{u}(J, h + v)\right) P_\ell(v, Y), \tag{3.85}$$

where $\hat{u}$ is the message passing rule defined in Eq. (3.32). From last expression it turns out we can write an iteration rule for the momenta of $X_\ell$ at fixed $u_\ell$,

$$G_q^{(\ell)}(u) = \int \mathrm{d}X \; P_\ell(u, X) \; X^q \;, \tag{3.86}$$

which reads

$$G_q^{(\ell+1)}(u) \;=\; \mathbb{E}_{J,h} \int \mathrm{d}v \; \delta\left(u - \hat{u}(J, h + v)\right) \left(\frac{\partial \hat{u}}{\partial v}\right)^q G_q^{(\ell)}(v). \tag{3.87}$$

Equations (3.85) and (3.87) with $q = 2$ have been recently introduced in literature [146] in order to derive an analytical expression for the spin-glass susceptibility.

We note that the knowledge of the maximum eigenvalue of the integral operator of Eq. (3.87) for a generic $q$ allows one to reconstruct the full distribution of the connected correlation function at large distance [152].

From last equation it is clear the relation of $G_q^{(\ell)}$ with the eigenfunctions $g_q^\lambda$ of Eq. (3.4). In fact, decomposing $G_q^{(\ell)}(u)$ along the eigenfunctions of $D^{(0)}$, projecting Eq. (3.87) on the left eigenvectors $S_q^\lambda(u)$ and with some computations analogue to the ones leading from Eq. (3.79) to Eq. (3.80), we arrive to

$$G_q^{(\ell)}(u) \;=\; \sum_{\lambda \in D^{(q)}} a_{\lambda, q} \, \lambda^\ell \, g_q^\lambda(u), \tag{3.88}$$

where $a_{\lambda, q}$ is defined in Eq. (3.64). Equation (3.88), along with Eq. (3.84), gives the expression (3.63) obtained with the RTM formalism for the connected correlation functions.

Following the lead of the previous paragraph, we can extend the above derivation to compute the disorder averages $\langle \sigma_0 \sigma_\ell \rangle_c^q$ along with an arbitrary power of $Z_{\ell,o}$. The generalization of Eq. (3.87) in fact becomes

$$G_q^{(\ell+1)}(u; n) \;=\; \mathbb{E}_{J,h} \int \mathrm{d}v \; \delta\left(u - \hat{u}(J, h + v)\right) \left(\frac{\partial \hat{u}}{\partial v}\right)^q Z(J, h, v) \, G_q^{(\ell)}(v; n), \tag{3.89}$$

and Eq. (3.88) generalizes trivially as well. The final result is

$$\overline{\langle \sigma_0 \sigma_\ell \rangle_c^q \, Z_{\ell,o}^n} = \sum_{\lambda \in D^{(q)}} \lambda^\ell(n) \, a_{\lambda,q}^2(n), \tag{3.90}$$

which extrapolates smoothly to the result we obtained for $n = 0$, i.e. Eq. (3.63). In last equation the coefficients $a_{\lambda,q}(n)$ are defined by

$$a_{\lambda,q}(n) \equiv \mathbb{E}_{\tilde{h}} \int \mathrm{d}u \; \left[\frac{\cosh(\beta(u + \tilde{h}))}{\cosh(\beta u)}\right]^n \left[1 - \tanh^2(\beta(u + \tilde{h}))\right]^q g_q^\lambda(u; n), \tag{3.91}$$

such that $a_{\lambda,q}(0) = a_{\lambda,q}$.

### 3.5.3 The disconnected correlation function

The computations of the thermally disconnected correlation function $\overline{\langle\sigma_0\rangle\langle\sigma_\ell\rangle}$ is straightforward once we use the results we obtained in the two preceding paragraphs. In fact, calling $H_0$ and $H_\ell$ two auxiliary fields we add to the first and the last spin of the chain respectively and set to zero after the computation, the following relation holds:

$$\frac{\partial}{\partial H_0}\frac{\partial}{\partial H_\ell}\overline{Z_{\ell,o}^n} = n\,\overline{\langle\sigma_0\sigma_\ell\rangle_c\,Z_{\ell,o}^n} + n^2\,\overline{\langle\sigma_0\rangle\langle\sigma_\ell\rangle\,Z_{\ell,o}^n}. \qquad (3.92)$$

Using Eqs. (3.79) and (3.90) last expression leads to the main result of this paragraph, that is

$$\overline{\langle\sigma_0\rangle\langle\sigma_\ell\rangle\,Z_{\ell,o}^n} = \frac{1}{n}\left[\sum_{\lambda\in D^{(0)}}\lambda^\ell(n)\,b_{\lambda,0}^2(n) - \sum_{\lambda\in D^{(1)}}\lambda^\ell(n)\,a_{\lambda,1}^2(n),\right] \qquad (3.93)$$

where the coefficient $b_{\lambda,0}(n) \equiv \frac{\partial\,a_{\lambda,0}(n)}{\partial H_{0/\ell}}$ reads

$$b_{\lambda,0}(n) = \mathbb{E}_{\tilde{h}}\int\mathrm{d}u\,\left[\frac{\cosh(\beta(u+\tilde{h}))}{\cosh(\beta u)}\right]^n\tanh(\beta(u+\tilde{h}))\,\sqrt{n}\,g_0^\lambda(u;n). \qquad (3.94)$$

We included a factor $\sqrt{n}$ in the definition of $b_{\lambda,0}(n)$ to facilitate the extrapolation of Eq. (3.93) to small $n$. In fact for all but the first eigenfunctions of $D^{(0)}$ the normalization condition imposes the scaling $g_0^\lambda(u;n)\sim\frac{1}{\sqrt{n}}[g_0^\lambda(u) + n\,\tilde{g}_0^\lambda(u)]$.

We could derive Eq. (3.93) also in the RTM formalism for integer values of $n$ and then perform an analytic continuation to arbitrary real $n$. In the limit $n\downarrow 0$ it is easy to see that the contribution to $\overline{\langle\sigma_0\rangle\langle\sigma_\ell\rangle}$ from the first and the second sums of Eq. (3.93) are given in Eqs. (3.66) and (3.67) of Section 3.4.2 respectively.

An alternative probabilistic derivation of the formula (3.68) for $\overline{\langle\sigma_0\rangle\langle\sigma_\ell\rangle}$, which does not require the knowledge of the moments of the partition function and of $\overline{\langle\sigma_0\sigma_\ell\rangle_c\,Z_{\ell,o}^n}$, goes through the definition of

$$R^{(\ell)}(u) \equiv \overline{\delta(u-u_\ell)\langle\sigma_0\rangle^{(\ell)}}. \qquad (3.95)$$

We used the symbol $\langle\sigma_0\rangle^{(\ell)}$ to denote the magnetization of the first spin at the $\ell$-th iteration of the construction of the asymmetric chain described in Section 3.5.1. It can be easily shown that the knowledge of $R^{(\ell)}(u)$ allows the computation of $\overline{\langle\sigma_0\rangle\langle\sigma_\ell\rangle}$. Since $\langle\sigma_0\rangle^{(\ell)}$ is given by the derivative of the free energy of the chain at the step $\ell$ with respect to a field on the the first spin, considering the free energy difference after an iteration it is easy to arrive to the relation

$$\langle\sigma_0\rangle^{(\ell+1)} = \langle\sigma_0\rangle^{(\ell)} + [\tanh(\beta(u_\ell+h)) - \tanh(\beta u_\ell)]\frac{\partial u_\ell}{\partial h_0}. \qquad (3.96)$$

Therefore the recursion rule for $R^{(\ell)}(u)$ is given by

$$\begin{aligned}R^{(\ell+1)}(u) &= \mathbb{E}_{J,h}\int\mathrm{d}v\,R^{(\ell)}(v)\,\delta\left(u-\hat{u}(J,h+v)\right)\\&\quad + \mathbb{E}_{J,h}\int\mathrm{d}v\,G_1^{(\ell)}(v)\,\delta\left(u-\hat{u}(J,h+v)\right)[\tanh(\beta(v+h))-\tanh(\beta v)],\end{aligned}$$
$$(3.97)$$

where $G_1^{(\ell)}(v)$ was defined in Eq. (3.88) in last paragraph. Last equation can be solved decomposing $R^{(\ell)}(u)$ along the eigenfunctions of $D^{(0)}$ at $n = 0$, and using Eq. (3.88) for $G_1^{(\ell)}(v)$. The computation is lengthy and not trivial, since it involves expressing $\tilde{g}_0^\lambda$ and $\tilde{g}_1^\lambda$ (defined in Section 3.4.2) respectively in terms of the basis of $D^{(0)}$ and $D^{(1)}$ at $n = 0$. In the end though one arrives at the expression (3.68) for the disconnected correlation function.

## 3.6 Conclusions

In the present chapter we presented a thorough analysis of the spectral properties of the RTM. We have developed a formalism that is suitable to compute many different types of connected and disconnected correlation functions and can be applied both to one-dimensional systems and to locally tree-like graphs. The expressions we found are exact for any value $\ell$ of the spin distance and can be approximated numerically considering only the top eigenvalues of certain integral operators. Also the formalism can be trivially adapted to perform the same computations in diluted p-spin models.

We also managed to obtain exact formulas for the moments of the partition function and of the average free energies of open and closed chains of finite length. It has been recently found that short chains have an important role in the finite size corrections to disordered models on diluted graphs [2] and in perturbative expansions around the Bethe approximation on Euclidean systems [148]. Therefore the analytical tools we have developed also apply to these contexts.

Most of the results have also been derived using rigorous probabilistic arguments. This approach has the merits of avoiding the complication of the decomposition of the replicated space $Z_2^{\otimes n}$ and of being more physically intuitive than the replica one. The advantage of the replica method instead is that once the spectral representation of the RTM is obtained all the observables can be computed just with opportune contraction. In the cavity analysis an ad-hoc iterative function or a computation strategy has to be devised for each observable.

Noticeably we did not manage to derive Eq. (3.53) for the free energy of closed chains using a cavity argument. This is the only point withstanding the proof of the complete equivalence between the two methods.

A limitation of both the RTM formalism and of its cavity counterpart, is the fact that it is applicable to the analysis of disordered Ising models only in their replica symmetric phase. This includes all isolated one-dimensional systems but not diluted models in the spin glass phase. Therefore an investigation of the spectral properties of the 1RSB replicated transfer matrix, extending Wigner's decomposition [149] to the 1RSB symmetry group, is desirable. Another direction for the extension of our results, which should not require too much analytical effort [153], is toward the investigation of Potts models.

# Part II

# Perturbative Finite Size Corrections

# Chapter 4

# Finite Size Corrections On Random Graphs

## 4.1 Introduction

Finite size corrections to the thermodinamic free energy have been investigated in fully connected systems [116, 117, 154], mean field optimization problems [93, 95] and some simple disorder system [155], sometimes as a byproduct of the Hessian diagonalization [156]. However, to our knowledge, only a solution in zero external field has been derived for sparse random graphs [157, 158] in the replica symmetric phase.

In the domain of physical spin systems, diluted models represent a class of mean-field like systems sharing an essential feature of the finite-dimensional ones, that is the finite coordination number. By consequence diluted models should mimic the physics of real systems better than the fully-connected ones (we have already remarked that this is what happens for zero temperature ferromagnets in random magnetic fields). Moreover when dealing with finite systems, the peculiar structure of diluted networks should give a first insight on how the topology can modify thermodynamic quantities. Indeed diluted models are defined on random graphs which are locally tree-like and have typical loops of size $O(\log N)$. However for finite (and small) sizes these loops become short and much more similar to the short loops which are abundant in any finite-dimensional network (think e.g. to lattice models). In this sense we can interpret the $1/N$ corrections in diluted models as a way to expand towards finite dimensional models.

In the thermodynamic limit the free energy of diluted systems is exactly the Bethe free energy. When the number of vertices $N$ in the graph is finite though, the average free energy density $f(N)$ resents the presence of loops. If $f(N)$ has a regular expansion around $N = \infty$, each term of the $1/N$ expansion $f(N) = f_0 + f_1/N + o(1/N)$ would account for the contribution of a certain class of loopy structures. We see that in the context of diluted systems, finite size corrections and loop expansions are strictly related concepts. In this Cshapter we set up a formalism, based on a replicated action, apt to the systematic computation of the $f(N)$ expansion for diluted disordered Ising systems in the replica symmetric phase. We calculate explicitly the first correction $f_1$ to the thermodynamic free energy. It is simple combinatorics to show that only simple (i.e. non-intersecting) loops can participate to the $O(1/N)$ correction $f_1$. In fact more complicated loopy subgraphs, with no dangling edges, typically involve only a fraction $O(1/N^2)$ of the total number of nodes, therefore can only contribute to higher order terms in the free energy expansion. Obviously this fact naturally emerges from the analytic computation as well.

In the following we will use the replica method in order to compute disorder-averaged corrections to the free energy. Then we will rederive the same results using a probabilistic approach.

We will deal with random regular graphs (RRG) and Erdős-Rényi (ER) graphs topologies. The replica procedure used is very similar, but is more involved in the RRG case, so we will describe in greater detail in the ER case, and leave most details of the RRG computations to Appendix B. On the other hand the combinatorial/probabilistic arguments are very different instead, the one for RRGs much shorter since it relies on the expression for the free energies of open and closed chains obtained in Chapter 3. In both ensembles we corroborate the analytical results with numerical simulations, conducted on RFIMs and spin-glass systems.

The results obtained and the method used in this Chapter, in particular for the RRG ensemble, are the foundations for the large $M$ expansion for finite dimensional systems presented in Chapter 8.

## 4.2 Erdős-Rényi

### 4.2.1 The model

We consider a model of $N$ interacting Ising spins $\{\sigma_i = \pm 1\}_{i=1}^N$ defined by the following Hamiltonian:

$$\mathcal{H} = -\sum_{i<j} C_{ij}\, J_{ij} \sigma_i \sigma_j - \sum_i h_i \sigma_i \,, \tag{4.1}$$

where we have decoupled the topology of the underlying graph, encoded in the symmetric adjacency matrix $\{C_{ij}\}$, from the exchange interactions $\{J_{ij}\}$. The numbers $C_{ij}$ specify the particular graph considered and take values $C_{ij} = 1$ or 0 whether the sites $i$ and $j$ are connected or not. In the case of Erdös-Rényi random graphs [159], the matrix $C$ has the following distribution [160]:

$$\mathcal{P}(\{C_{ij}\}) = \prod_{i<j} \left[ \frac{z}{N} \delta(C_{ij} - 1) + \left(1 - \frac{z}{N}\right) \delta(C_{ij}) \right] \,. \tag{4.2}$$

The spins interact among each other via quenched random couplings $J_{ij}$, which are assumed to be identically independently distributed (or fixed to a single value $J$). Moreover we allow the spins to interact with a local magnetic field (random or non-random). The disorder averaged free energy density of the system, at the temperature $T = \beta^{-1}$, is defined as

$$f(\beta, N) = -(\beta N)^{-1} \left[\log Z_N(\beta)\right]_{\text{av}} = f_0(\beta) + \frac{1}{N} f_1(\beta) + o\left(\frac{1}{N}\right), \tag{4.3}$$

where the average has to be performed over the topological disorder and the quenched randomness. The main part of this work is devoted to the analytical computation of the $f_1(\beta)$ term, the finite size correction to the free energy. The calculation can be performed in two different ways, known as the replica method and the cavity method. The latter derivation is particularly useful in order to better understand the physical meaning of the results, which is less clear in the replica picture.

### 4.2.2 Computing the free energy density with replicas

The replica calculation of the free energy density starts from the well known identity:

$$[\log Z_N(\beta)]_{\text{av}} = \lim_{n \to 0} \frac{\partial}{\partial n} \log \left[(Z_N(\beta))^n\right]_{\text{av}} . \tag{4.4}$$

The moments of the partition function $[(Z_N(\beta))^n]_{\text{av}}$ are then evaluated for integer values of number of replicas $n$. At the end of the calculation, the analytical continuation to real values of $n$ allows us to take the limit $n \to 0$. The replicated averaged partition function reads (from now on we drop the dependence of $Z_N$ on $\beta$):

$$[(Z_N)^n]_{\text{av}} = \left[ \text{Tr} \left( \prod_{i<j} \exp\left(\beta J_{ij} C_{ij} \sum_a^n \sigma_i^a \sigma_j^a\right) \prod_i \exp\left(\beta h_i \sum_a^n \sigma_i^a\right) \right) \right]_{\text{av}} . \tag{4.5}$$

Performing the average over the topological disorder using the distribution (4.2), and setting

$$V(\sigma,\tau) \equiv N \log \left[ 1 + \frac{z}{N} \left( \overline{\exp\left(\beta J \sum_a \sigma^a \tau^a\right)}^J - 1 \right) \right],$$

$$B(\sigma) \equiv \log \left[ \overline{\exp\left(\beta h \sum_a \sigma^a\right)}^h \right] - \frac{1}{2N} V(\sigma,\sigma), \tag{4.6}$$

eq. (4.5) takes the following form:

$$[(Z_N)^n]_{\text{av}} = \text{Tr} \left[ \exp\left\{ \frac{1}{2N} \sum_{i,j} V(\sigma_i, \sigma_j) + \sum_i B(\sigma_i) \right\} \right] . \tag{4.7}$$

We can achieve the site factorization of eq. (4.7) by means of the order parameter

$$\tilde{\rho}(\sigma) = N^{-1} \sum_i \prod_a \delta(\sigma^a - \sigma_i^a) , \tag{4.8}$$

Enforcing Eq. (4.8) with a $\delta$ functional in Eq. (4.7), we trace over the decoupled sites and then integrate out $\tilde{\rho}(\sigma)$ which appears in Gaussian form. We arrive at an expression for the auxiliary fields $\rho(\sigma)$ suitable for saddle-point evaluation: [1]

$$[(Z_N)^n]_{\text{av}} = \sqrt{\det(V)} \int [D\rho] e^{-NS[\rho]} . \tag{4.9}$$

---

[1]The functional measure is $[D\rho] = \prod_\sigma \sqrt{\frac{N}{2\pi}} d\rho(\sigma)$

The replicated action $S[\rho]$ is given by

$$S[\rho] = \frac{1}{2}\int d\sigma d\tau\, \rho(\sigma)\, V(\sigma,\tau)\rho(\tau) - \log\int d\sigma \exp\left[\int d\tau\, V(\sigma,\tau)\rho(\tau) + B(\sigma)\right], \quad (4.10)$$

where the symbol "$\int d\sigma$" is a proxy for the more cumbersome notation $\int d\sigma \equiv \prod_{a=1}^{n}\sum_{\sigma^a=\pm 1}$. Let us now extract the leading order contribution in the replicated action $S[\rho]$. We define the matrix $U(\sigma,\tau)$ and the vector $H(\sigma)$ from the first order expansion in $N$ of eq. (4.6) to be

$$U(\sigma,\tau) \equiv \overline{\exp\left(\beta J \sum_{a}\sigma^a\tau^a\right)}^{J},$$

$$H(\sigma) \equiv \log\left[\overline{\exp\left(\beta h \sum_{a}\sigma^a\right)}^{h}\right], \quad (4.11)$$

and write the thermodynamically relevant part of the action (4.10) as $S[\rho] = S_0[\rho] + o(1)$, where

$$S_0[\rho] = \frac{z}{2}\int d\sigma d\tau\, \rho(\sigma)\,(U(\sigma,\tau)-1)\rho(\tau) - \log\int d\sigma \exp\left[z\int d\tau\,(U(\sigma,\tau)-1)\rho(\tau) + H(\sigma)\right]. \quad (4.12)$$

The leading order free energy $f_0$ comes from the saddle point of eq. (4.12), followed by the limit $n\to 0$, as we will see in the next section. A first $O\left(\frac{1}{N}\right)$ correction to the free energy comes from the $O(1/N)$ term in eq. (4.10) evaluated at the saddle point.

**Leading free energy**

We now evaluate the functional integral (4.9) by the steepest descent method:

$$\lim_{N\to+\infty} -\frac{1}{N}\log\left[(Z_N)^n\right]_{\text{av}} = S_0[\rho_*], \quad (4.13)$$

where $\rho_*(\sigma)$ is the solution of the the saddle-point equation:

$$\frac{\delta S_0[\rho]}{\delta\rho(\sigma)} = 0 \quad\longrightarrow\quad \rho_*(\sigma) = \frac{\exp\left[z\int d\sigma' U(\sigma,\sigma')\rho_*(\sigma') + H(\sigma)\right]}{\int d\sigma \exp\left[z\int d\sigma' U(\sigma,\sigma')\rho_*(\sigma') + H(\sigma)\right]}. \quad (4.14)$$

In order to take the small $n$ limit we have to use an appropriate parametrization for the order parameter $\rho_*(\sigma)$. If we assume a Replica Symmetric (RS) ansatz, a convenient parametrization for $\rho_*(\sigma)$ is given by

$$\rho_*(\sigma) = \int dh P(h)\left[\frac{\exp\left(\beta h\sum_a\sigma^a\right)}{(2\text{ch}(\beta h))^n}\right]. \quad (4.15)$$

Inserting this parametrization in eq. (4.14) and taking the limit $n\to 0$ we obtain the usual self-consistent Cavity equations for the distribution $P(h)$ and $Q(u)$ of cavity fields and bias respectively:

$$P(h) = \sum_{k=0}^{\infty}\frac{z^k}{k!}e^{-z}\overline{\int\left[\prod_{i=1}^{k}dQ(u_i)\right]\delta\left(h - h_R - \sum_{i=1}^{k}u_i\right)}^{h_R},$$

$$Q(u) = \overline{\int dP(h)\,\delta\left[u - \frac{1}{\beta}\text{th}^{-1}\left[\text{th}(\beta J)\text{th}(\beta h)\right]\right]}^{J}. \quad (4.16)$$

The RS free energy density can then be estimated as

$$f_0(\beta) = \beta^{-1} \lim_{n \to 0} \frac{\partial}{\partial n} S_0[\rho_*] \qquad (4.17)$$

and can be explicitly written in term of the distributions $P(h)$ and $Q(u)$ [15].

**Fluctuations around the RS saddle point**

The Gaussian integral obtained by expanding eq.(4.10) around the saddle point generates the order $1/N$ corrections. We set

$$\rho(\sigma) = \rho_*(\sigma) + \frac{\chi(\sigma)}{\sqrt{N}},$$
$$S^{(2)}(\sigma, \sigma'; \rho) = \frac{\delta^2 S_0[\rho]}{\delta\rho(\sigma)\delta\rho(\sigma')}. \qquad (4.18)$$

Expanding the action in powers of $1/N$ we find

$$S[\rho] = S_0[\rho_*] + \frac{1}{N} S_1[\rho_*] + \frac{1}{2N} \int d\sigma d\sigma' \chi(\sigma) S^{(2)}(\sigma, \sigma'; \rho_*) \chi(\sigma') + o(N^{-1}), \qquad (4.19)$$

where $S_1[\rho_*]$ is given by the following expression:

$$S_1[\rho_*] = \frac{z}{2} \int d\sigma \left[(U(\sigma,\sigma) - 1\right] \rho_*(\sigma) + \frac{z^2}{4} \int d\sigma d\sigma' \rho_*(\sigma) \left[U(\sigma,\sigma') - 1\right]^2 \rho_*(\sigma'). \qquad (4.20)$$

The functional integral (4.9) at this order evaluates:

$$-\frac{1}{N} \log \left[(Z_N)^n\right]_{\mathrm{av}} = S_0[\rho_*] + \frac{1}{N} S_1[\rho_*] + \frac{1}{2N} \log \det (1 - T) + o(N^{-1})$$
$$= S_0[\rho_*] + \frac{1}{N} S_1[\rho_*] - \frac{1}{2N} \sum_{L=1}^{\infty} \frac{\mathrm{Tr}[T^L]}{L} + o(N^{-1}), \qquad (4.21)$$

where the matrix $T(\sigma, \sigma')$ reads

$$T(\sigma, \sigma') = z \left[ U(\sigma, \sigma') \rho_*(\sigma') - \left( \int d\tau U(\sigma,\tau) \rho_*(\tau) \right) \rho_*(\sigma') \right]. \qquad (4.22)$$

Using the RS parametrization (4.15), it turns out that in the limit $n \downarrow 0$ the trace $\mathrm{Tr}\left(T^L\right)$ can be arranged in a linear combination of free energies of closed and open chains. It all comes down to the fact that the term $U(\sigma, \sigma')\rho_*(\sigma')$, present in $T(\sigma, \sigma')$, can be linked to the replicated transfer matrix of an edge receiving a cavity field at one of its extremities. In Appendix A.2 we prove the following formula:

$$\frac{\partial}{\partial n} \mathrm{Tr}(T^L) = -\beta z^L \left[ \phi_L^c - L \left( \phi_L^a - \phi_{L-1}^a \right) \right] + O(n), \qquad (4.23)$$

where $\phi_L^{c/a}$ are **free energies** of closed and open spin chains in the graph of length $L \geq 1$, with $\phi_0^a$ defined as $\phi_0^a \equiv -\beta^{-1} \mathbb{E}_h \log 2 \cosh(\beta h)$. Writing the RS free energy density as:

$$f_{\mathrm{RS}} = f_{\mathrm{RS}}^{(0)} + \frac{1}{N} f_{\mathrm{RS}}^{(1)} + o\left(N^{-1}\right), \qquad (4.24)$$

and observing that the term $S_1[\rho_*]/N$ [viz. eq. (4.20)] cancels out with part of the first two terms in the sum $\sum_{L=1}^{\infty} \text{Tr}\left(T^L\right)/(2NL)$, the finite size correction of the RS free energy density $f_{\text{RS}}^{(1)}$ can be evaluated as:

$$f_{\text{RS}}^{(1)} = \left(z - \frac{z^2}{2}\right)\phi_0^o - \frac{z}{2}\phi_1^o - \frac{z^2}{2}(\phi_2^o - 2\phi_1^o) + \frac{1}{2}\sum_{L=3}^{\infty} \frac{z^L}{L}\left[\phi_L^c - L(\phi_L^o - \phi_{L-1}^o)\right]. \tag{4.25}$$

The sum entering the previous formula can be considered as a sum over independent loops weighted with the factor $\left[\phi_L^c - L(\phi_L^o - \phi_{L-1}^o)\right]$, by noticing that, in the thermodynamic limit, $z^L/(2L)$ is exactly the average number of loops of length $L$ in a Erdös-Rényi random graph of mean connectivity $z$. The same formula holds true also on the Erdös-Rényi ensemble $\mathbb{G}(N,M)$, where $M = zN/2$ is the fixed number of edges, since the distribution of topological structures such as the number of finite loops remains the same at the $1/N$ order.

In the limit of vanishing external field, eq.(4.25), evaluated in the paramagnetic phase, takes the following simpler form:

$$f_{\text{RS}}^{(1)} = \frac{z}{2\beta}\mathbb{E}_J \log \text{ch}(\beta J) - \frac{1}{2\beta}\sum_{L=3}^{\infty} \frac{z^L}{L}\mathbb{E}_{\{J_i\}} \log\left[1 + \prod_{i=1}^{L} \text{th}(\beta J_i)\right], \tag{4.26}$$

where the first term takes into account the fact that the average number of links is $z(N-1)/2$ and the second one is the contribution of all the loops of length $L \geq 3$. The loops we are talking about are topologically defined as non-self-intersecting closed paths. Self-intersecting closed paths would give contributions proportional to $N^{-2}$, since the self-intersection is observed, on average, in a fraction $N^{-2}$ of the total number of vertices. While eq. (4.25) is an original contribution to the literature, its zero field counterpart eq. (4.26) has been already presented [157]. Moreover the full distribution of $f^{(1)}$ in the absence of external field and in the RS phase has been rigorously computed [158] an it is consistent with the mean value given by eq. (4.26).

### 4.2.3 Computing the free energy density with cavity method

We now show how to compute the finite size corrections to the free energy density using the cavity method. The reason to be interested in such a kind of calculation is twofold. Firstly we have to corroborate the physical insight gained from replicas; secondly we want to establish the equivalence of the two methods beyond the leading order, showing how both procedures give the same result also at order $1/N$.

The cavity method is well defined only in thermodynamic limit. In order to study $1/N$ corrections to the free energy density of a model defined on a Erdös-Rényi random graph (ERRG), we need to define a new ensemble of random graphs of $\mathcal{N}$ vertices, such that in the limit $\mathcal{N} \to \infty$ any topological structure appears with the same density it has in the ERRG of $N$ vertices. Here we are assuming that the free energy of a model of $N$ variables can be written as $F_N = Nf(\{d_i\})$, where $f(\{d_i\})$ is the free energy density computed in the thermodynamic limit on a model having the same *densities* $d_i$ of topological structures appearing in the finite $N$ model. The new ensemble we are going to define is required to compute such a free energy density.

The topological structures we are interested in are the only ones that give contributions up to order $O(1/N)$, i.e linear chains of length $L$ (i.e. with $L$ edges and $L+1$ vertices) and loops of length $L$. Let us start by computing their densities in a ERRG of $N$ sites, where each link is present with probability $z/N$. The density of linear chains of size $L$ (i.e. the number of linear chains per node) is

$$d_L^{chain} = \frac{1}{N}\left(\frac{z}{N}\right)^L \frac{1}{2} N(N-1)\ldots(N-L) \simeq \frac{z^L}{2}\left(1 - \frac{L(L+1)}{2N}\right), \qquad (4.27)$$

and the density of loops of length $L$ is

$$d_L^{loop} = \frac{1}{N}\left(\frac{z}{N}\right)^L \frac{1}{2L} N(N-1)\ldots(N-L+1) \simeq \frac{1}{N}\frac{z^L}{2L}. \qquad (4.28)$$

In the new ensemble a random graph of $\mathcal{N}$ nodes can be viewed as the union of basic topological structures (BTS), that, for the present purposes, are chains and loops. The graph can be build in the following way. For each $L \geq 1$, consider all sequences of $L+1$ different indices $(i_0, i_1, \ldots, i_L)$ with the condition $i_0 < i_L$, that avoids double counting of a chain; for each sequence of indices draw the edges $(i_0, i_1), (i_1, i_2), \ldots, (i_{L-1}, i_L)$ with probability $a_L/\mathcal{N}^L$. Then, for each $L \geq 3$, consider all sequences of $L$ different indices $(i_1, i_2, \ldots, i_L)$ with the conditions that $i_1$ is the smallest among the $L$ indices and $i_2 < i_L$ (these two conditions ensure that each loop is counted only once); for each sequence of indices draw the edges $(i_1, i_2), (i_2, i_3), \ldots, (i_{L-1}, i_L), (i_L, i_1)$ with probability $c_L/\mathcal{N}^{L-1}$.

A useful representation of this graph is in terms of a factor graph, where the variable nodes are the graph nodes and the factor nodes are the BTS. Thanks to the scaling of the probabilities used in the building of the graph, the corresponding factor graph is sparse, since the total number of BTS (i.e., of factor nodes) is given by

$$\sum_{L=1}^{\infty} \frac{\mathcal{N}(\mathcal{N}-1)\ldots(\mathcal{N}-L)}{2}\frac{a_L}{\mathcal{N}^L} + \sum_{L=3}^{\infty} \frac{\mathcal{N}(\mathcal{N}-1)\ldots(\mathcal{N}-L+1)}{2L}\frac{c_L}{\mathcal{N}^{L-1}} \simeq \mathcal{N}\left(\sum_{L=1}^{\infty}\frac{a_L}{2} + \sum_{L=3}^{\infty}\frac{c_L}{2L}\right) \qquad (4.29)$$

and the coefficients $a_L$ and $c_L$ are constants.

The sparsity of the factor graph ensures that the whole construction is consistent in the $\mathcal{N} \to \infty$ limit. Indeed the probability that any pair of graph nodes enters in more than one BTS is $O(1/\mathcal{N})$. Since in the new ensemble we are interested in computing the free energy density to leading order, we can safely assume that any two graph nodes interact through at most one BTS; and this BTS uniquely determines whether the edge between the two graph nodes is present or not.

The factor graph representation also allows us to write down the free energy density in a standard way by summing factor nodes and variable nodes contributions

$$f = \frac{1}{2}\sum_{k=1}^{\infty} a_k \phi_k^a + \frac{1}{2}\sum_{k=3}^{\infty} \frac{c_k}{k}\phi_k^c + \phi_{site}, \qquad (4.30)$$

where $\phi_k^o$ and $\phi_k^c$ are respectively the free energies of chains and loops of length $k$ and

$$\phi_{site} = \frac{T}{\mathcal{N}}\sum_i (1-n_i)\sum_{\sigma_i} \mu_i(\sigma_i)\log\mu_i(\sigma_i), \qquad (4.31)$$

with $\mu_i(\sigma_i)$ being the single spin marginal and $n_i$ the number of BTS where the variable $i$ enters.

We should now determine the values of the coefficients $a_k$ and $c_k$ such that the densities of chains and loops in a typical graph of the new ensemble match those in eqs. (4.27) and (4.28) in the large $\mathcal{N}$ limit. When computing the actual density of a given topological structure (e.g. a chain or a loop) one should consider that such a topological structure can coincide with a BTS, or be part of a BST or involve more than one BST.

As a warm-up, let us compute the density of links (chains of length $L = 1$) in the limit $\mathcal{N} \to \infty$:

$$d_1^{chain} = \lim_{\mathcal{N}\to\infty} \frac{1}{\mathcal{N}} \frac{\mathcal{N}^2}{2} \left[ \sum_{k=1}^{\infty} k \mathcal{N}^{k-1} \frac{a_k}{\mathcal{N}^k} + \sum_{k=3}^{\infty} \mathcal{N}^{k-2} \frac{c_k}{\mathcal{N}^{k-1}} \right] = \frac{1}{2} \left[ \sum_{k=1}^{\infty} k a_k + \sum_{k=3}^{\infty} c_k \right], \tag{4.32}$$

where $k\mathcal{N}^{k-1}$ in the first sum and $\mathcal{N}^{k-2}$ in the second sum are respectively the number of chains and loops of length $k$ passing trough a given link, i.e., the number of possible BTS containing the two variables connected by a given link.

When computing the density of topological structures made of more than one link, we need to consider that such structures can overlap with more than one BTS. In order to be concrete let us consider the density of chains of length $L = 2$:

$$d_2^{chain} = \lim_{\mathcal{N}\to\infty} \frac{1}{\mathcal{N}} \frac{\mathcal{N}^3}{2} \left[ \left( \frac{2d_1^{chain}}{\mathcal{N}} \right)^2 + \sum_{k=2}^{\infty} (k-1)\mathcal{N}^{k-2} \frac{a_k}{\mathcal{N}^k} + \sum_{k=3}^{\infty} \mathcal{N}^{k-3} \frac{c_k}{\mathcal{N}^{k-1}} \right] \tag{4.33}$$

where $2d_1^{chain}/\mathcal{N} \equiv p_1$ is the probability of having a link[2]. The general expression for densities of linear chains of length $L \geq 3$ is the following

$$d_L^{chain} = \lim_{\mathcal{N}\to\infty} \frac{1}{\mathcal{N}} \frac{\mathcal{N}^{L+1}}{2} \left[ S_L\left(\frac{2d_1^{chain}}{\mathcal{N}}, \ldots, \frac{2d_{L-1}^{chain}}{\mathcal{N}^{L-1}}\right) + \sum_{k=L}^{\infty} (k-L+1)\frac{a_k}{\mathcal{N}^L} + \sum_{k=L+1}^{\infty} \frac{c_k}{\mathcal{N}^L} \right], \tag{4.34}$$

where the function $S_L$ gives the probability that the $L$ consecutive links comes from more than one BTS and can be written (see Appendix A) in terms of the probabilities of having $k(<L)$ consecutive links: $p_k \equiv 2d_k^{chain}/\mathcal{N}^k$. Since each term in function $S_L$ is of order $\mathcal{N}^{-L}$, in the limit $\mathcal{N} \to \infty$ we have

$$2d_L^{chain} = \mathcal{N}^L S_L(p_1,\ldots,p_{L-1}) + \sum_{k=L}^{\infty}(k-L+1)a_k + \sum_{k=L+1}^{\infty} c_k =$$

$$S_L(2d_1^{chain},\ldots,2d_{L-1}^{chain}) + \sum_{k=L}^{\infty}(k-L+1)a_k + \sum_{k=L+1}^{\infty} c_k = z^L\left(1 - \frac{L(L+1)}{2N}\right). \tag{4.35}$$

Note that eq.(4.35) is valid for any $L \geq 1$ since $S_1 \equiv 0$ and $c_2 \equiv 0$.

A similar expression can be written for the densities of loops of length $L$:

$$d_L^{loop} = \lim_{\mathcal{N}\to\infty} \frac{1}{\mathcal{N}} \frac{\mathcal{N}^L}{2L} \left[ R_L(p_1,\ldots,p_{L-1}) + \frac{c_L}{\mathcal{N}^{L-1}} \right], \tag{4.36}$$

---

[2]In principle in $p_1$ there are already some contributions entering the sums, but this over-counting effect is irrelevant in the $\mathcal{N} \to \infty$ limit.

where again the function $R_L$ represents the probability of generating a loop of size $L$ by more than one BTS. Since the probability of having $k$ consecutive links is $O(\mathcal{N}^{-k})$ the function $R_L$ is $O(\mathcal{N}^{-L})$ and then

$$d_L^{loop} = \frac{c_L}{2L} = \frac{1}{N}\frac{z^L}{2L} \qquad \Longrightarrow \qquad c_L = \frac{z^L}{N} \quad \text{for } L \geq 3\,. \qquad (4.37)$$

In other words, making a loop by randomly choosing smaller structures is more improbable than randomly generating directly such a loop.

The detailed computation of the coefficients $a_k$ from Eq. (4.35) is made in Appendix A. Here we just quote the result

$$a_1 = z + \frac{1}{N}(2z^2 - z)\,, \qquad (4.38)$$

$$a_L = \frac{1}{N}(z^{L+1} - z^L) \ \text{ for } L \geq 2\,. \qquad (4.39)$$

Plugging these coefficients in Eq. (4.30) we finally get

$$f = \frac{z}{2}\left(1 - \frac{1}{N}\right)\phi_1^o + \frac{z^2}{2N}(2\phi_1^o - \phi_2^o) + \frac{1}{N}\sum_{L=3}^{\infty}\frac{z^L}{2L}\left[\phi_L^c - L(\phi_L^o - \phi_{L-1}^o)\right] + \phi_{site}\,. \qquad (4.40)$$

We observe that the sum on the r.h.s matches the sum over loops entering eq.(4.25). Moreover, if the term $\phi_{site}$ is expressed by means of cavity fields, one finds exactly eq.(4.25). This can be immediately seen in the case of zero external field in all the paramagnetic phase, where variables are unbiased and we have $\phi_{site} = -T(1-\ell)\log 2$ with $\ell = z + (z^2/2 - z)/N$ being the density of edges in the factor graph (i.e., the number of edges in the factor graph per variable node). Substituting this expression for $\phi_{site}$ in Eq. (4.40), simply gives:

$$f = -T\left(\log 2 - \frac{z}{2}\mathbb{E}_J \log \text{ch}(\beta J)\right) + \frac{z}{2N}T\,\mathbb{E}_J \log \text{ch}(\beta J) - \frac{T}{2N}\sum_{L=3}^{\infty}\frac{z^L}{L}\mathbb{E}_{\{J_i\}}\log\left[1 + \prod_{i=1}^{L}\text{th}(\beta J_i)\right], \qquad (4.41)$$

thus recovering the simplified replica result of Eq. (4.26).

We can conclude that the replica calculation reproduces correctly all the topological structures involved in the $1/N$ corrections to the free energy density. Incidentally we note that self intersecting loops occur only with probability $N^{-2}$ and they do not contribute to $1/N$ corrections.

Let us finish this section by giving a different interpretation to the present results. We have seen that under the assumption that finite size corrections can be computed by the cavity method in a graph with finite densities of certain topological structures, we have been able to reproduce the replica result (and give to it a more physical intuition). However, we could assume that replica and cavity methods should provide the correct free-energy for a very large, but finite, system, and then conclude that the free-energy of a model only depends on the densities of certain topological structures. This alternative view can be useful if one aims at computing the free-energy of a model defined on a finite dimensional lattice, by considering a lattice as a random graph with strong topological correlations, and making an expansion in these topological correlations (e.g., number of loops, but not only that).

### 4.2.4 Numerical Analysis

In this Section we check the validity of our analytical expressions for the free energy corrections, eq. (4.25), against numerical simulations. Since from Monte Carlo simulations one obtains the energy of the systems, in order to avoid an integration in temperature we decided to work perform the simulations at zero temperature, where energy and free energy coincides. Moreover since eq. (4.25) holds for arbitrary disorder in the interaction and in the external field, we choose to keep the former deterministic and the latter randomly distributed. In this case in fact an exact polynomial algorithm is available to calculate the ground state. Therefore we apply eq. (4.25) to compute the finite size corrections to the energy density of the zero-temperature Random Field Ising Model (zt-RFIM) and compare with numerical simulations. The model is defined by the following Hamiltonian:

$$\mathcal{H} = -J \sum_{i,j} C_{ij}\sigma_i\sigma_j - \sum_i h_i\sigma_i \; , \qquad (4.42)$$

where the random magnetic fields are Gaussian random variables of zero mean and variance $\overline{h_i^2} = 1$ and the ferromagnetic exchange coupling $J$ take values in the interval $[0, \infty)$. The underlying graph topology is that of a Erdös-Rényi random graph. Due to the FKG [161] inequality the model does not undergo replica-symmetry-breaking [78] at any value of the ferromagnetic interaction strength $J$, so that our formulae for the finite size (free) energy density corrections remain valid also below the critical point, provided that a single pure state is selected. In the ferromagnetic phase the existence of two energy minima generates additional finite size fluctuations, which are proportional to $N^{-1/2}$. These kind of *interstate* fluctuations overcomes the $1/N$ *intrastate* contribution, which becomes practically invisible in numerical experiments. In this work we compare analytical predictions and numerical results only in the paramagnetic phase $J < J_c$.

The uniqueness of the ground state of the model allows to translate the formula (4.25) for the free energy density corrections into the corresponding expression for the ground state energy density corrections. We write the ground state energy density as the leading term plus the $O(1/N)$ correction:

$$e^{\mathrm{GS}}(N) = e_0^{\mathrm{GS}} + \frac{1}{N}e^{(1)} + o\left(\frac{1}{N}\right) \; , \qquad (4.43)$$

where $e^{(1)}$ reads:

$$e^{(1)} = -\left(z - \frac{z^2}{2}\right)\overline{|h^c|} - \frac{z}{2}e_1^o - \frac{z^2}{2}(e_2^o - 2e_1^o) + \frac{1}{2}\sum_{L=3}^{\infty}\frac{z^L}{L}\left[e_L^c - L(e_L^o - e_{L-1}^o)\right] \; . \qquad (4.44)$$

The random variable $h^c$ is the cavity field, distributed according to the zero temperature solution of eq. (4.16), while $e_L^{a/c}$ are the energies of open and closed chains in the graph. The computational time cost of computing the energy density of a chain of size $L$ by enumeration is exponentially increasing in $N$, therefore only partial sums up to $L = 7$ in eq. (4.44) have been considered in Figure 4.1. To accurately compute the whole $L$ series, especially near the critical point, some assumptions has to be made about the large $L$ behaviour of its term. Some of the authors have been developing a formalism through which a spectral representation of the replicated transfer matrix [1] [138, 145] can be obtained. Using this result the leading behaviour

$$e_L^c - L(e_L^o - e_{L-1}^o) \sim AL\lambda^L \tag{4.45}$$

has been established for the zero temperature RFIM, which allows to analytically sum the remaining terms of the series (from $L = 8$ to infinity). The coefficient $\lambda$ is given by the first eigenvalue of the replicated transfer matrix and gives the decay rate of ferromagnetic correlation functions. It can be computed to high precision with population dynamics techniques or as the first eigenvalue of an integral operator. The coefficient $A$ instead has been obtain from a fit of the first five point of the series. As an alternative approach assuming the validity of the $AL\lambda^L$ behaviour (which fares much better then a simple exponential decay assumptions) both $A$ and $\lambda$ could be inferred from a fit of the first terms of the sum. The finite size corrections of the energy in the RFIM at zero temperature diverges as $e^{(1)} \propto \frac{1}{1-z\lambda}$, at odds with the double pole divergence $e^{(1)} \propto \frac{1}{(1-z\lambda)^2}$ which can be found at finite temperature. This matter has been elucidated in Chapter 3.

At the critical point a scaling analysis of the correction $e^{(1)}$ can be performed. Calling $\tau = |J - J_c|$ the distance from the critical point, mean field theory [80] predicts the following finite size scaling for $\tau$ and $e^{(1)}$ in the critical region:

$$\tau = \frac{\tilde{\tau}}{N^{1/3}}, \tag{4.46}$$

$$e^{(1)} = \tilde{e}^{(1)} N^{1/3}, \tag{4.47}$$

The leading correction to the thermodynamic ground state energy density is of order $O(N^{-2/3})$ in the whole critical region:

$$e^{\text{GS}}(N) = e_0^{\text{GS}} + \frac{1}{N^{2/3}} \tilde{e}^{(1)} + o\left(\frac{1}{N^{2/3}}\right) \quad \text{for} \quad J \to J_c. \tag{4.48}$$

Furthermore eq. (4.43) is not valid in the ferromagnetic phase (for reasons mentioned in the beginning of this Section), where the leading correction happens to be of order $O(N^{-1/2})$:

$$e^{\text{GS}}(N) = e_0^{\text{GS}} + \frac{1}{N^{1/2}} e'^{(1)} + o\left(\frac{1}{N^{1/2}}\right) \quad \text{for} \quad J > J_c. \tag{4.49}$$

The numerical experiment is performed on a Erdös-Rényi random graph with average connectivity $z = 4$. We compute the ground state energy with the Minimum-cut algorithm [162, 163], using the Lemon Library [107]. To draw the profiles of the energy density corrections in Figure 4.1 we took the average over $10^8$ samples for each system size. In the same figure we compare the numerical data with the analytical prediction given by eq. (4.44) and check the finite size scaling relation given by equation (4.47). In Figure 4.2a we report the Binder cumulant:

$$\text{Bi} = \frac{3}{2}\left[1 - \frac{\overline{m^4}}{3\left(\overline{m^2}\right)^2}\right], \tag{4.50}$$

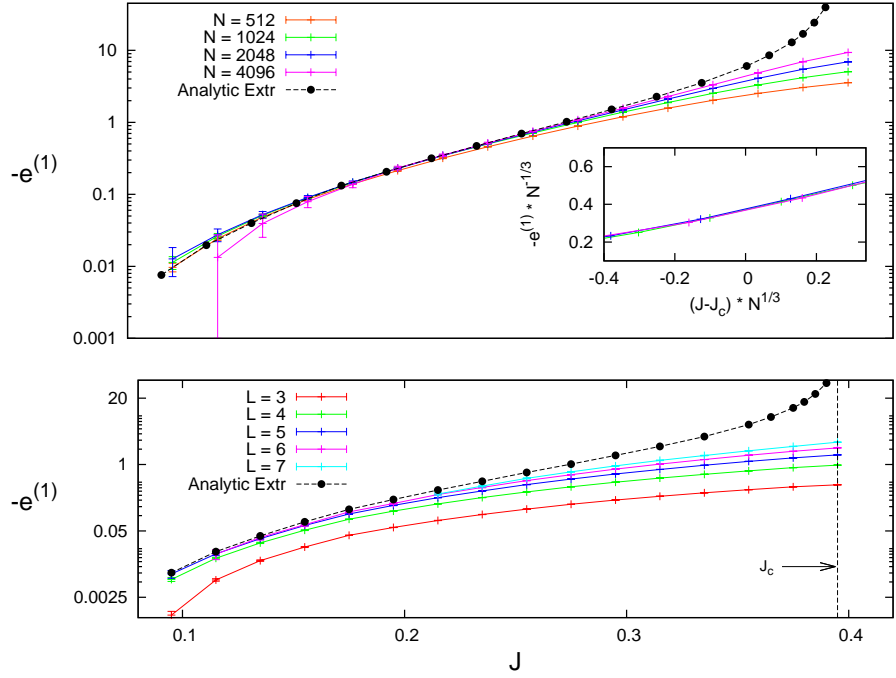

**Fig. 4.1.** Finite size corrections of the ground state energy density in the $T = 0$ RFIM on Erdös-Rényi random graphs with mean connectivity $z = 4$. In the upper panel we plot numerical data for different system sizes and the analytical formula given by eq. (4.44). Close to the critical point (which is $J_c \approx 0.395$) the scaling of the energy corrections is given by the mean field prediction (4.47) as confirmed by the data collapse shown in the inset. In the lower panel we show the estimates of the formula (4.44), truncating the sum over loops with a cutoff $L = 3, 4, 5, 6, 7$ and extrapolating the whole series as explained in the main text.

for system sizes ranging from $N = 256$ to $N = 2048$. From the intersection of the curves we identify the critical point, obtaining $J_c \sim 0.395(1)$.

Figure 4.2b shows the behaviour of the averaged squared magnetization $\overline{m^2}$. The finite size scaling of $\overline{m^2}$ in the critical region is given by the following scaling relation:

$$\overline{m^2} = O(N^{-1/3}) = O(\tau) \quad \text{for} \quad \tau \to 0. \tag{4.51}$$

This scaling form is confirmed by the data collapse shown in the inset of Figure 4.2b.

## 4.3  Random Regular Graphs

### 4.3.1  Replica formalism for random regular graphs

Let us turn now to the computation of the first finite size correction to the free energy in RRGs. As usual the Hamiltonian is given by

$$\mathcal{H} = -\sum_{i<j} C_{ij}\sigma_i J_{ij}\sigma_j - \sum_i H_i \sigma_i, \tag{4.52}$$

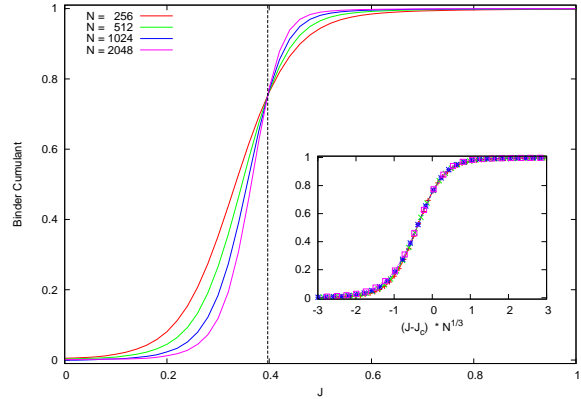

**(a)** Binder cumulant for the $T = 0$ RFIM on Erdös-Rényi random graphs with mean connectivity $z = 4$ for different system sizes as a function of the exchange interaction $J$. A vertical dashed line is drawn in correspondence of the critical point $J_c \sim 0.395$. In the inset it is shown the data collapse in the critical region using the scaling variable $(J - J_c)N^{1/3}$ for the reduced interaction.

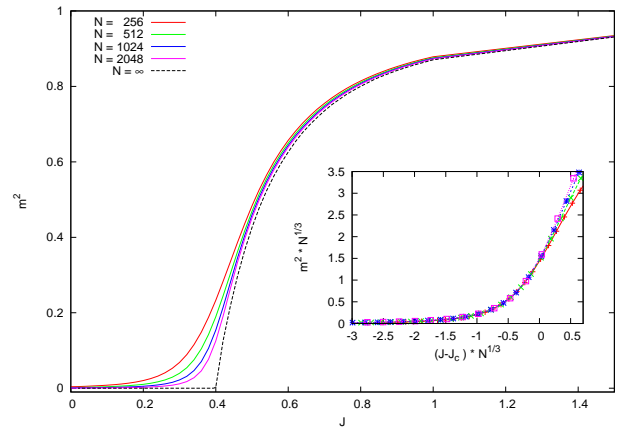

**(b)** Average squared magnetization for the $T = 0$ RFIM on Erdös-Rényi random graphs with mean connectivity $z = 4$ for different system sizes as a function of the exchange interaction $J$. In the inset it is shown the data collapse in the critical region using the scaling $N^{\frac{1}{3}}$ both for the reduced interaction and for the squared magnetization.

where the exchange couplings $J_{ij}$ and/or the local magnetic fields $H_i$ are quenched independent random variables. The matrix $C_{ij}$ represent the entries of the adjacency matrix of a graph $G$ extracted from the RRG ensemble. They take the values $C_{ij} = 1, 0$ depending on whether or not the vertices $i$ and $j$ are connected. Here we study $c$-RRGs, i.e., random graphs with vertices having uniform degree $c$. The probability measure of the $c$-RRG ensemble is uniform over all the regular graphs of degree $c$.

To solve the model in the thermodynamic limit (i.e. $N \to \infty$) and to obtain the finite size correction to this limit, we have to cast the averaged replicated partition function $[Z^n]_{\text{av}}$ into an integral form suited to steepest descent evaluation, as we did for ER ensemble. This procedure uses standard techniques [160, 164–166] and is reported in detail in the Appendix B.1. Here we quote only the final result, which reads

$$[Z^n]_{\text{av}} = [\det(cU)]^{1/2} e^{\mathcal{A}(N,c)} \int \mathcal{D}\rho \ e^{-N\mathcal{S}[\rho,N]}. \tag{4.53}$$

In the last equation the integral is performed over the space of all possible functions $\rho(\sigma) \equiv \rho(\sigma_1, \ldots, \sigma_n)$ of a $n$-replicated spin, taking $2^n$ different values. The action $\mathcal{S}[\rho, N]$ is a functional of $\rho(\sigma)$ and $N$, and at the leading order in $N$ can be written as

$$\mathcal{S}_0[\rho] = \frac{c}{2} \int \mathrm{d}\sigma \mathrm{d}\tau \ \rho(\sigma)U(\sigma,\tau)\rho(\tau) \\ - \log \int \mathrm{d}\sigma \ e^{B(\sigma)} \left[ \int \mathrm{d}\tau U(\sigma,\tau)\rho(\tau) \right]^c \tag{4.54}$$

The action $\mathcal{S}_0$ will be optimized through the steepest-descent method. After that, we will integrate the Gaussian fluctuations around the optimal saddle point, thus obtaining the desired finite size corrections. The quantities $U(\sigma, \tau)$ and $B(\sigma)$ appearing in Eq. (4.54) are defined by

$$U(\sigma, \tau) = \mathbb{E}_J \left[ \exp \left( \beta J \sum_{a=1}^n \sigma^a \tau^a \right) \right] \tag{4.55}$$

and

$$B(\sigma) = \log \mathbb{E}_H \left[ \exp \left( \beta H \sum_{a=1}^n \sigma^a \right) \right] \tag{4.56}$$

The explicit expression for the constant $\mathcal{A}(N, c)$ appearing in Eq. (4.53) can be found in the Appendix B.1.

Saddle point evaluation of $\mathcal{S}_0$ leads to the following self-consistence equation for the order parameter $\rho$:

$$\rho_*(\sigma) = \frac{e^{B(\sigma)} \left[ \int \mathrm{d}\sigma' U(\sigma, \sigma')\rho_*(\sigma') \right]^{c-1}}{\int \mathrm{d}\sigma'' e^{B(\sigma'')} \left[ \int \mathrm{d}\sigma' U(\sigma'', \sigma')\rho_*(\sigma') \right]^c}. \tag{4.57}$$

Once a solution of Eq. (4.57) has been found, using Eq. (4.54) one gets the thermodynamic free energy density $f_0 = \lim_{N\to\infty} f(N)$. The difficulties of the problem are all hidden in the solution $\rho_*(\sigma)$ of the saddle point equation. The function $\rho_*(\sigma)$ depends on the replicated spin $(\sigma_1, \ldots, \sigma_n)$, and hence, it is uniquely determined by the set of the $2^n$ possible values it can take. The most general solution should specify all these $2^n$ values. Here we limit ourselves to the simplest solution, i.e., the replica symmetric one. This solution has the property to be invariant under the group of permutations of the replica indexes, therefore $\rho_*(\sigma)$ can depend only on the sum $\sum_{a=1}^n \sigma_a$. The number of parameters necessary to fully specify a replica symmetric function is $n+1$ (and hence much smaller than $2^n$). The most general replica symmetric parametrization of $\rho_*(\sigma)$ can be written in the form:

$$\rho(\sigma) = \int dh\, P(h) \frac{e^{\beta h \sum_{a=1}^n \sigma_a}}{[2\cosh(\beta h)]^n}, \qquad (4.58)$$

where the function $P(h)$ depends implicitly on $n$ and is non-negative and normalized to one in the limit $n \to 0$.

Inserting the parametrization (4.58) into the saddle point equation (4.57), and performing the limit $n \to 0$, we obtain a self consistent equation for the density $P(h)$:

$$P(h) = \mathbb{E}_{J,H} \int \prod_{k=1}^{c-1} dh_k\, P(h_k)\, \delta\left[h - H - \sum_{k=1}^{c-1} \hat{u}(\beta, J, h_k)\right], \qquad (4.59)$$

where $\hat{u}(\beta, x, y) = \beta^{-1} \tanh^{-1}[\tanh(\beta x) \tanh(\beta y)]$. We recognize Eq. (4.59) as the self-consistent equation for the probability distribution $P(h)$ of the cavity field on a RRG of connectivity $c$. Solving the last equation for $P(h)$ one can eventually evaluate the $n = 0$ limit of Eq. (4.54) and recover the thermodynamic free energy $f_0 \equiv \lim_{N \to \infty} \frac{f(N)}{N}$, given by the Bethe free energy approximation [15].

### 4.3.2 Finite size corrections

In this section we present the analytical expression of the first finite size corrections to the free energy density of disordered Ising models on the RRG ensemble. As we anticipated in the introduction, we assume the leading correction to the thermodynamical free energy density to be proportional to $1/N$. Therefore, we split $f(N)$ into the sum of the leading term plus the $1/N$ correction, that is

$$f(N) = f_0 + \frac{f_1}{N} + o\left(\frac{1}{N}\right). \qquad (4.60)$$

The detailed calculation of the coefficient $f_1$, the main result of this Section, is given in the Appendices B.2 and B.3. The derivation is based on the expansion of the contributions of the Gaussian fluctuations of the replicated action around the saddle point, given by

$$-\frac{1}{2} \log \det \left( \left.\frac{\partial^2 S_0}{\partial \rho(\sigma) \partial \rho(\tau)}\right|_{\rho_*} \right), \qquad (4.61)$$

as a power series containing the replicated transfer matrix of the system [138] [1]. The final result reads

$$f_1 = \sum_{\ell=3}^{\infty} \frac{(c-1)^\ell}{2\ell} \Delta \phi_\ell. \qquad (4.62)$$

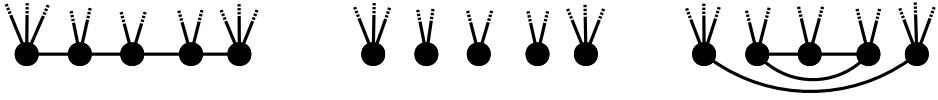

**Fig. 4.3.** Pictorial representation of the argument, given in Section 3.5, to compute the free energy shift due to the addition of a loop to a large tree graph. It is shown an open chain embedded in a tree graph (*left*), its removal from the tree (*center*) and the addition of a loop (*right*).

The terms appearing in last equation and computed in the replica formalism have a clear physical meaning, as we will readily explain. We call $\Delta\phi_\ell$ the quantity defined by

$$\Delta\phi_\ell = \phi_\ell^c - \ell\,\phi. \tag{4.63}$$

where $\phi_\ell^c$ is the average free energy of a closed chain (loop) of length $\ell$ embedded in the graph, that is

$$\phi_\ell^c \equiv -\frac{1}{\beta}\big[\log Z_\ell^c\big]_{\mathrm{av}} \tag{4.64}$$

with

$$Z_\ell^c \equiv \sum_{\sigma_1,\ldots,\sigma_\ell} e^{\beta(r_1\sigma_1 + J_1\sigma_1\sigma_2 + \cdots + r_\ell\sigma_\ell + J_\ell\sigma_\ell\sigma_1)}. \tag{4.65}$$

The cavity fields $r_i$ are i.i.d. random variables sampled from the distribution

$$R(r) = \mathbb{E}_{J,H} \int \prod_{k=1}^{c-2} \mathrm{d}h_k\, P(h_k)\, \delta\left[r - H - \sum_{k=1}^{c-2} \hat{u}(\beta, J, h_k)\right]. \tag{4.66}$$

In other words, the cavity fields $r_i$ represent the effective fields coming from the rest of the graph on the nodes in a loop. The quantity $\phi$ is the intensive average free energy of a closed chain with random couplings $J_i$, and random fields $r_i$, i.e. $\phi \equiv \lim_{\ell\to\infty} \frac{\phi_\ell^c}{\ell}$, and can be easily computed through cavity method [138] [1].

The fact that the fields $r_i$ are independently distributed and that they obey Eq. 4.66, containing the fixed point distribution $P(h)$, indicates that the contribution of each loop can be considered independently from the others. In fact the factor $(c-1)^\ell/2\ell$ in Eq. (4.62) is exactly the average number of loops of length $\ell$ in a RRG of connectivity $c$. Therefore, the coefficient $f_1$ of the $O(1/N)$ correction can be expressed as a sum over all the loops in a graph, each one contributing with the amount $\Delta\phi_\ell$ to the free energy. We call $\Delta\phi_\ell$ a free energy shift since it is the free energy difference observes in a infinite tree after the addition of a single loop of size $\ell$, as we will argue in the next Section.

It is yet to be investigated the relation between (4.62) for $f_1$ and an analogous result that one could derive using the loop calculus formalism [167, 168].

We notice that the loops considered here are defined as non-self intersecting closed paths. In fact, self-intersecting loops would give a contribution of order $O(1/N^2)$ to the average free energy for simple combinatorial arguments.

### 4.3.3 Probabilistic argument

The computation of the $O(1/N)$ correction to the free energy in the RRG ensemble can be easily done through simple probabilistic arguments, as one realizes a posteriori analysing the final result Eq. (4.62) obtained with the replica formalism. In fact, as already discussed at the end of the previous Section, at the $O(1/N)$ order loops are sparsely distributed in the graph and do not interact with each other. Therefore their contributions to the free energy can be summed up separately and each one of them can be considered as embedded in an infinite tree. In order to compute the free energy shift due to the presence of a loop of length $\ell$, we consider a very large random tree, with partition function $Z_T$, and remove the $\ell + 1$ edges of an open chain of length $\ell + 1$, as showed in Figure 4.3. We call $\sigma_0, \ldots, \sigma_\ell + 1$ the cavity spins of the new graph, that is the ones who lost one (this is the case of $\sigma_0$ and $\sigma_{\ell+1}$) or two $(\sigma_1, \ldots, \sigma_\ell)$ of their adjacent edges. We call $Z_{cav}(\sigma_0, \ldots, \sigma_{\ell+1})$ the partition function of this new system, conditioned on the values of the cavity spins. Since we assumed to start from a tree graph, the partition function $Z_{cav}$ takes the form

$$Z_{cav}(\sigma_0, \ldots, \sigma_{\ell+1}) = \tilde{Z} e^{h_0 \sigma_0 + r_1 \sigma_1 + \ldots + r_\ell \sigma_\ell + h_{\ell+1} \sigma_{\ell+1}}, \qquad (4.67)$$

where $\tilde{Z} \geq 0$ and the cavity fields $h_i$ and $r_{0/\ell+1}$ are independently distributed according to $P(h)$ from (4.59) and $R(r)$ from Eq. (4.66) respectively. We recover the partition function of the original tree adding back the missing links, therefore we establish the relation

$$Z_T = \tilde{Z} \times Z^o_{\ell+1}, \qquad (4.68)$$

where $Z^o_{\ell+1}$ is the partition function of an open chain of length $\ell + 1$ with incoming fields $h_0, r_1, \ldots, r_\ell, h_{\ell+1}$. On the other hand, starting from the cavity graph, we can create another graph $G$ containing exactly one loop. This can be achieved adding an edge between the spin $\sigma_0$ and $\sigma_{\ell+1}$, and adding other $\ell$ edges to form a loop among the $\ell$ internal cavity spins (see Figure 4.3). Notice that with this construction all the spins retain the same degree that they had in the original graph $T$. The partition function of the system defined on $G$ is then given by

$$Z_G = A \times Z^o_1 \times Z^c_\ell. \qquad (4.69)$$

We are interested in the difference of the average free energy between the system $G$ an $T$ in the large graph limit. Let us call N the number of nodes in $T$ and $G$. The free energy shift is then given by

$$\Delta \phi_\ell = -\frac{1}{\beta} \lim_{N \to \infty} [\log Z_G - \log Z_T]_{av}. \qquad (4.70)$$

For the average free energy $\phi^o_L$ of an open chain of length $L$ embedded in a RRG the following relation holds [1]:

$$\phi^o_L = L \phi + \phi_s, \qquad (4.71)$$

where $\phi_s$ is a site term that does not depend on $L$ [1]. It is therefore easy to derive the expected result:

$$\Delta \phi_\ell = \phi^c_\ell - \ell \phi. \qquad (4.72)$$

We have proven that the free energy difference $\Delta \phi_\ell$ as defined by Eq. (4.70) corresponds to the quantity $\phi^c_\ell - \ell \phi$, as it was defined in the last Section. Taking into account that the average number of loops of length $\ell$ in a graph of the RRG ensemble is $\frac{z^\ell}{2\ell}$ in the thermodynamic limit, we re-obtain Eq. (4.62) without making any resort to replicas.

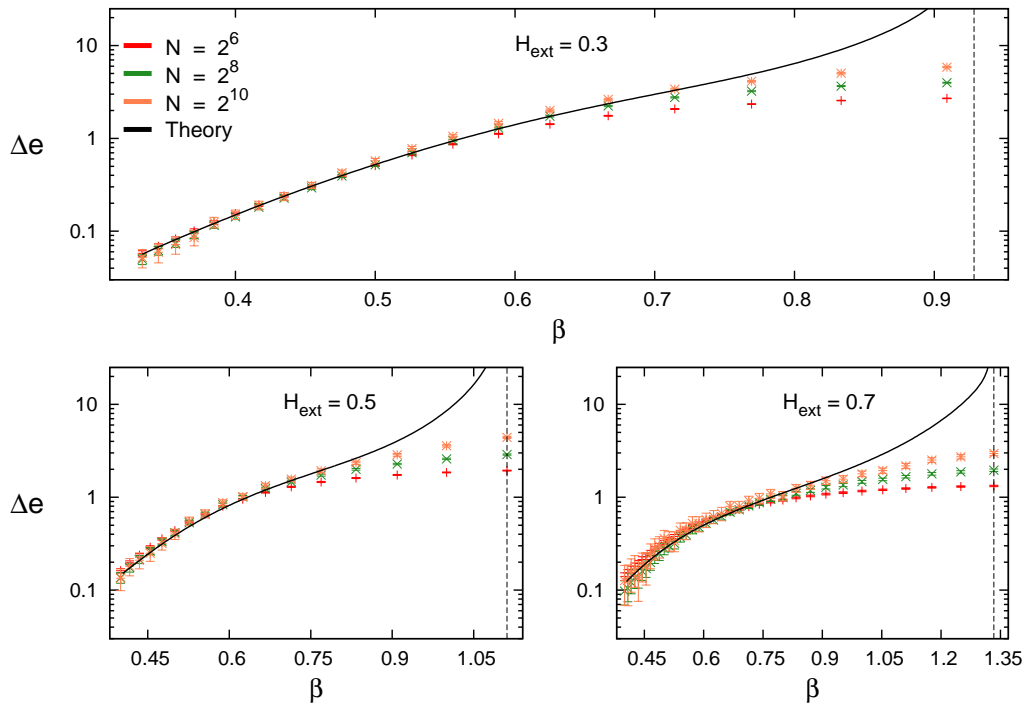

**Fig. 4.4.** Finite size corrections to the energy density of a spin glass model on a RRG with connectivity $c = 4$ and bimodal random couplings $J = \pm 1$. The various panels refer to different values of the external uniform magnetic field. The results obtained from Monte Carlo simulations are compared with the analytical values predicted from Eqs. (4.62) and (4.78). The vertical dashed lines mark the positions of the critical temperatures.

The argument we gave in this Section to compute the first finite size correction to the free energy is strictly limited to the RRG ensemble. In fact it relies heavily on the homogeneity of the graphs. On different graph ensembles more refined combinatorial arguments, as the one given in [2] for Erdös-Rényi random graphs, have to be used.

### 4.3.4 Numerical Experiment: Spin Glass in a Magnetic Field

In this Section we test our analytical prediction for the finite size correction to the free energy, Eq. (4.62), on the spin glass in a uniform magnetic field. The connectivity of the graph is $c = 4$. In the experiment the couplings $J_{ij}$ are bimodal random variables, taking value $J_{ij} = \pm 1$ with equal probability. We simulate the model using a parallel tempering Monte Carlo algorithm and three different values of the external field $H = 0.3, 0.5$ and $0.7$. For each value of the magnetic field $H$ we simulate systems of three different sizes: $N = 2^6, 2^8$ and $2^{10}$.

In order to compute the analytical estimate of $f_1$ we proceed in two steps: we explicitly calculate the first terms of the sum. We computed by transfer matrix multiplication the partition function and the free energy of closed chain of length $\ell$, for many realizations of the disorder and up to $\ell = 7$. We then resummed the remaining terms of the series using the criterion explained in Ref. [2], that we briefly recap. Using the formalism of the replicated transfer matrix developed in Ref. [1], one can show that, in a spin glass, the dominant contribution to $f_1$ comes from the replicon eigenvalue. Therefore, we use only the knowledge of this eigenvalue to analytically resum the remaining terms of the series (from $\ell = 8$ to $\infty$). The large $\ell$ behaviour of the shift $\Delta\phi_\ell$ is given by the expression

$$\Delta\phi_\ell \sim A\lambda^\ell \qquad \text{for} \quad \ell \gg 1, \tag{4.73}$$

where $\lambda$ is the replicon eigenvalue, the largest eigenvalue satisfying the following integral equation:

$$\lambda g_\lambda(u) = \mathbb{E}_{J,r} \int du' \, g_\lambda(u')\delta[u - \hat{u}(\beta J, r + u')] \left(\frac{\partial \hat{u}}{\partial u}\right)^2. \tag{4.74}$$

Here $r$ is distributed as $R(r)$ defined in Eq. (4.66). The maximum eigenvalue of the integral operator in last equation can be obtained numerically by population dynamics techniques. The coefficient $A$ instead can be computed analytically, as shown in Ref. [1], and takes value $A = 3/(2\beta)$. We can split the quantity $f_1$ in two pieces:

$$f_1 \sim S(L) - \frac{3}{4\beta}\mathcal{L}\text{og}_{L+1}\left[1 - (c-1)\lambda\right], \tag{4.75}$$

where $S(L)$ is the partial sum over the loops up to $\ell = L$, and the second term is the resummation of the remaining series from $\ell = L+1$ to $\ell = \infty$, which we represented via the function $\mathcal{L}\text{og}_p(1-x)$, defined as:

$$\mathcal{L}\text{og}_p(1-x) = -\sum_{\ell=p}^{\infty} \frac{x^\ell}{\ell}. \tag{4.76}$$

In our concrete case we can compute explicitly the first $L = 7$ terms of the series, and so, the approximated analytic form of $f_1$ is

$$f_1 \sim S(7) - \frac{3}{4\beta}\mathcal{L}\text{og}_8\left[1 - 3\lambda\right] \qquad \text{for} \quad c = 4. \tag{4.77}$$

In a numerical simulation, measuring the energy is, actually, much simpler than the free energy (since the last one involves an estimate of the entropy). As a consequence we preferred to compare analytical and numerical results for the finite size corrections to the energy density $e_1$. Analytically, the quantity $e_1$ is given by the usual formula relating energy and free energy:

$$e_1 = f_1 + \beta \frac{\partial f_1}{\partial \beta}. \tag{4.78}$$

In Figure 4.4 we show the comparison between the experiments and our theoretical result. The agreement is good at high temperatures, while it deteriorates close to the critical point. At the critical point in fact every order of the $O(1/N)$ expansion of the free energy diverges, therefore near the critical point subleading finite size corrections become increasingly important and extrapolation of $e_1$ obtained from numerical simulations to its large $N$ limit, that can be derived by our analytical expression (4.62), is difficult to achieve.

## 4.4 Conclusions

In this Chapter we performed a thorough analysis of the $O\left(1/N\right)$ correction to the free energy density in disordered Ising models defined on Erdös-Rényi and random regular graphs. We derived an analytical formula which can be easily used to quantify finite size effects, avoiding the subtleties associated with the diagonalization of the Hessian. We also checked the correctness of our results through a numerical study of the RFIM at zero temperature in the ER case and of a a spin glass at finite temperature in the RRG case. In both case we found excellent agreement with the analytic prediction. Using this formalism though we could not compute the finite size corrections in the ferromagnetic ordered phase of the RFIM. In this phase in fact the $O\left(1/N\right)$ correction is overshadowed by a $O\left(N^{-\frac{1}{2}}\right)$ correction. This anomalous correction will be addressed analytically in Chapter 5.

A much more serious problem arises in the glassy phase of thespin glass model, where exponentially many pure states are involved. The leading finite size correction are much bigger than $O\left(1/N\right)$, in fact numerical results for the ground state of spin glasses on RRGs suggest that they are order $O\left(N^{-\frac{2}{3}}\right)$ or $O\left(N^{-\frac{4}{5}}\right)$, depending (surprisingly) on the bond distribution [124]. The question if our formalism can be adapted to cover the fullRSB case remains open.

In the ER ensemble we showed how replica results for the $1/N$ corrections to the free energy density can be derived also in the cavity formalism, resorting to an auxiliary graph ensemble which in some sense lifts the $O\left(1/N\right)$ *contributions* to the leading order. It would be interesting to see if this combinatorial derivation could be transposed to other graph ensembles.

Our results seems to hold some degree of universality, in fact the free energy contribution from simple loops at the order $O(1/N)$, can expressed in both ensembles as

$$\sum_{\ell \geq 3} \mathcal{N}_\ell \, \Delta\phi_\ell, \tag{4.79}$$

where $\mathcal{N}_\ell$ is the mean number of simple loops of length $\ell$, and $\Delta\phi_\ell$ is the free energy shift due to the addition of a loop to an infinite tree, defined as

$$\Delta\phi_\ell = \phi_\ell^c - \ell(\phi_\ell^o - \phi_{\ell-1}^o). \tag{4.80}$$

Moreover using the replicated transfer matrix formalism of the previous Chapter we can approximate $\Delta\phi_\ell$, with the leading eigenvalue and corresponding coefficients of the Longitudinal or the Replicon sector, with good accuracy even for small $\ell$. We will show in Chapter 8 how to make use of these results to compute the first order term in the perturbation theory around the Bethe free energy for finite dimensional systems. It is given also in that case by Eq. (4.79), with only a simple redefinition of the coefficients $\mathcal{N}_\ell$.

# Part III

# Non-Perturbative Finite Size Corrections

In the last Chapter we computed the first finite size correction to the free energy in diluted random graphs. The computation was done in a (replicated) field theoretical frameworks, examining the Gaussian fluctuation around the saddle point. Eventually higher order terms in the $1/N$ can be computed in this framework by standard perturbation theory, and all the terms in this expansion would have a direct topological interpretation in terms of free energy shifts due to the presence of loopy structures in the graph.

In same cases though, when there are other competing sources of finite size corrections or multiplicity of saddle points, this approach fails to capture the most relevant contributions. Problem specific techniques have to be devised to deal with them. Here we examine two such cases, where leading finite size correction are ultimately due to non-perturbative effects.

The first case is the RFIM in the ferromagnetic phase, that, as we mentioned in last Chapter, presents $O(1/\sqrt{N})$ corrections to the free energy. This phenomena is present both mean field and finite dimensional systems. We set up a formalism based on replica theory in order to compute such corrections. While the approach is general to any random field model, only in mean field topologies the computations can be carried on exactly.

In the second part of the Chapter we investigate the scaling behaviour of the average cost in the Euclidean Assignment Problem. We show how the difference in the number of blue and red points in a small region of space is a source for the transport field. This is formalized in a Poisson-like equations that yields a surprisingly ample set of predictions for the leading and subleading behaviour and coefficients of the average cost for each dimension $d$.

# Chapter 5

# The Random Field Ising Model

## 5.1 Introduction

We consider for concreteness a system of $N$ Ising spins, $\sigma_i = \pm 1$, and external random fields. The following arguments though apply to a general class of models possessing $O(m)$ symmetry (once the average over disorder is taken). On a given graph $G$, the Hamiltonian of the RFIM is

$$\mathcal{H} = -J \sum_{(i,j)} \sigma_i \sigma_j - \sum_{i=1}^{N} h_i \, \sigma_i, \tag{5.1}$$

where the first sum is over adjacent spins, $J \geq 0$ is a ferromagnetic coupling and the the fields $h_i$ are quenched i.i.d. random variables with zero mean and unit variance, i.e. $\mathbb{E}[h_i] = 0$ and $\mathbb{E}[h_i \, h_j] = \delta_{ij}$.

It takes a simple argument to show that, at least at zero temperature and for $J$ large enough, the subleading term in $N$ to the average energy $E(N)$ is of order $O(\sqrt{N})$. In fact in this case, for a given realization of the external fields $\{h_i\}$, the Gibbs measure is concentrated on the configuration with minimum energy, the candidates being the one with all the spins up and the one with all the spins down. For a given graph with $M = O(N)$ edges, the energies of the two states, let us call them $E_+$ and are $E_-$ respectively, are given by

$$E_\pm = -MJ \mp \sum_{i=1}^{N} h_i \qquad \text{for } J \gg 1. \tag{5.2}$$

The sum in the r.h.s. is a random variable of variance $N$, converging to a Gaussian variable in the thermodynamic limit. Therefore it is easy to see that for the average energy $E(N) = \mathbb{E}[\min(E_+, E_-)]$ we have

$$E(N) = -MJ - \sqrt{\frac{2}{\pi}} \sqrt{N} + o(\sqrt{N}) \qquad \text{for } J \gg 1. \tag{5.3}$$

It turns out that the $\sqrt{N}$ subleading behaviour we found in this limit case is present in the whole ferromagnetic region in the $J - T$ (coupling-temperature) plane.

In fact in the ferromagnetic phase the statistical weight is concentrated on two disconnected regions of the configuration space, having positive and negative magnetization respectively, separated by a free energy barrier exponentially increasing in $N$. While the two regions are completely equivalent in the pure ferromagnetic system, once the disordered external field is turned on their free energies start to differ of a random quantity of order $O(\sqrt{N})$. The total free energy of the system is then simply given by the minimum among the two, except for some exponentially decaying terms. Let us first define, for a given realization of the disorder, the free-energies $F_+$ and $F_-$ as the ones corresponding to configurations having positive and negative magnetization respectively (for simplicity we assume $N$ to be odd). We assume, and verify a posteriori for the RFIM, that in the ferromagnetic phase the difference between the free energies of the two states, $F_+ - F_-$, is of order $O(\sqrt{N})$. Then the average free energy of the system is given by

$$F(N) = -\frac{1}{\beta} \mathbb{E}\left[\log\left(e^{-\beta F_+} + e^{-\beta F_-}\right)\right]$$
$$= \mathbb{E}\left[\min(F_+, F_-)\right] + \textit{exp. vanish. terms}. \tag{5.4}$$

From these premises it follows simply that the average free energy has an expansion in $N$ of the form

$$F(N) = f_0 N + f_1 \sqrt{N} + o(\sqrt{N}), \tag{5.5}$$

where the coefficients $f_0$ and $f_1$ of the expansion are defined by

$$f_0 = \lim_{N \to +\infty} \frac{1}{N} \mathbb{E}[F_+] = \lim_{N \to +\infty} \frac{1}{N} \mathbb{E}[F_-], \tag{5.6}$$

$$f_1 = \lim_{N \to +\infty} \frac{-1}{\sqrt{N}} \frac{\mathbb{E}[|F_+ - F_-|]}{2}. \tag{5.7}$$

To compute $f_1$ we extend a method developed in Refs. [169] and [170–172]. At fixed $N$ and for a given realization of the disorder, let us call $Z_+$ and $Z_-$, the partition function constrained to the configurations with positive and negative magnetization. Considering $n$ and $m$ replicated systems respectively, one can easily see that for small $n$ and $m$ we have

$$-\frac{1}{\beta} \log \mathbb{E}[Z_+^n Z_-^m] \sim n \mathbb{E}[F_+] + m \mathbb{E}[F_-] - \frac{n^2}{2} \beta \operatorname{Var}(F_+)$$
$$- \frac{m^2}{2} \beta \operatorname{Var}(F_-) - nm\beta \operatorname{Covar}(F_+, F_-). \tag{5.8}$$

Obviously, for symmetry reasons, $\mathbb{E}[F_+] = \mathbb{E}[F_-]$ and $\operatorname{Var}[F_+] = \operatorname{Var}[F_-]$. The free energies $F_+$ and $F_-$ are jointly distributed random variables. If the limit

$$\Delta^2 \equiv \lim_{N \to \infty} \frac{1}{2N} \left[\operatorname{Var}(F_+) - \operatorname{Covar}(F_+, F_-)\right]$$
$$= \lim_{N \to \infty} \frac{1}{4N} \mathbb{E}\left[(F_+ - F_-)^2\right] \tag{5.9}$$

is non-zero, the rescaled variables $f_+$ and $f_-$, defined by

$$f_\pm = \frac{F_\pm - f_0 N}{\sqrt{N}}, \tag{5.10}$$

become non-trivially jointly distributed Gaussian random variables for large $N$. In fact, while higher orders cumulants in the $n$ and $m$ expansion of Eq. (5.8) are of order $O(N)$, with this rescaling they give vanishing contributions.

Since asymptotically $\frac{f_+ - f_-}{2}$ is itself a Gaussian random variable of zero mean and variance $\Delta^2$, it follows that the coefficient $f_1$ of Eqs. (5.5) and (5.7) is given by

$$f_1 = -\int \frac{\mathrm{d}z}{\sqrt{2\pi\Delta^2}} \, |z| \, e^{-\frac{z^2}{2\Delta^2}} = -\Delta\sqrt{\frac{2}{\pi}}. \tag{5.11}$$

The computation of the $n$ and $m$ expansions in Eq. (5.8) can be performed using the standard replica techniques. Since it is impractical to work with the partial partition functions $Z_+$ and $Z_-$, we define the partition function $Z_H$ of the system having an additional deterministic external field $H$ acting on all the spins. Then we define the replicated free energy $\phi(n,m)$ as

$$\phi(n,m) = \lim_{H\to 0^+} \lim_{N\to\infty} -\frac{1}{\beta N} \log \mathbb{E}\big[\, Z_H^n \, Z_{-H}^m \,\big]. \tag{5.12}$$

This definition is completely consistent with the *l.h.s.* of Eq. (5.8) in the ferromagnetic region, and is even more physically sound in the paramagnetic region, allowing to treat with a unified formalism the whole phase space.

We compute $\phi(n,m)$ for integer values of $n$ and $m$ using the saddle point technique for $N\to\infty$. Then we make an analytic continuation of the solution to real value of $n$ and $m$. By definition (5.12) the $0-th$ order term of the expansion is zero. Since we are interested only in the quadratic terms of the $\phi(n,m)$ expansion, and since the $\phi(n,m)$ is variational in some order parameter, our calculations will involve only the order parameter computed at $n = m = 0$ and we do not have to consider its $n$ and $m$ dependence. As always happens with calculations involving replicas, the exchange of the $n$ and $N$ limits will be done without particular care.

In the following Sections we will explicitly compute $\phi(n,m)$ given by Eq. (5.12) in two mean field models.

## 5.2 Fully connected

The first case we consider is that of the RFIM on the fully connected graph. Here the first sum in the Hamiltonian (5.1) runs over $N(N-1)/2$ edges, and the coupling $J$ has to be rescaled by a factor $N^{-1}$ in order to obtain an extensive thermodynamic behaviour. It is then easy to compute the replicated free energy of the system, defined in Eq. (5.12), through standard Hubbard-Stratonovich transformation is given by

$$\begin{aligned}\phi_{FC}(n,m) = \min_{x\geq 0} \bigg\{ &\frac{1}{2} J(n+m)\, x^2 \\ &- \frac{1}{\beta} \log \mathbb{E}_h \left[ \big(2\cosh\beta(Jx+h)\big)^n \big(2\cosh\beta(-Jx+h)\big)^m \right] \bigg\}.\end{aligned} \tag{5.13}$$

Notice that the order parameter $x$, the magnetization of the up-state, enters in the last term of Eq. (5.13) both with a plus and a minus sign. As a technical note, in Eq.

(5.13) the minimization condition does not ever turn to a maximization when the number of replicas is small, as it happens in glassy models [13], since the dimension of the order parameter in the full replica space is $n + m$ and is always greater than zero in our calculations.

To obtain the $O(\sqrt{N})$ contribution to the free energy we have to compute the small and $n$ and $m$ expansion of $\phi_{FC}(n, m)$. As already discussed at the end of the previous section, only the $n = m = 0$ saddle point solution, given by the non-negative solution of

$$x = \mathbb{E}_h \tanh \beta(Jx + h), \quad (5.14)$$

appears in the second order expansion of the replicated action. Therefore the coefficient $\Delta$, defined in Eq. (5.9) and characterizing the $O(\sqrt{N})$ correction though (5.11), is given by

$$\Delta_{FC}^2 = \frac{1}{2\beta^2} \{ \mathbb{E}_h[(\log \cosh \beta(Jx + h))^2] \\ - \mathbb{E}_h[\log \cosh \beta(Jx + h) \log \cosh \beta(Jx - h)] \}. \quad (5.15)$$

Last expression can be easily computed for any value of $J$ and $\beta$. In order to compare the analytic prediction with exact results, it is easier to consider the zero temperature limit of (5.15). This is given by

$$\lim_{T \to 0} \Delta_{FC}^2 = \frac{1}{4} \mathbb{E}_h \big(|Jx + h| - |Jx - h|\big)^2 \quad (5.16)$$

with $x$ non-negative solution of

$$x = \mathbb{E}_h \operatorname{sign}(Jx + h). \quad (5.17)$$

We focus on the zero temperature limit for two reason: the zero temperature fixed point is the one controlling the flow of the renormalization group also starting from finite temperature [53]; the ground state of the RFIM can be computed efficiently using exact numerical algorithms.

We implemented an exact and very efficient algorithm to compute the ground state of a fully connected RFIM, that takes advantage of the topological equivalence of all the spins. In fact it is easy to realize that among all the configurations having a certain total magnetization $M = \sum_i \sigma_i$, the one with lowest energy is the one where only the first $\frac{N-M}{2}$ spins with the lowest external field are down.

Therefore, once the spins are sorted according to their external fields (an operation of time complexity $\Theta(N \log N)$), we have to look only to these $N + 1$ configurations characterized by $M = -N, -N + 2, \ldots, N$ to find the ground state (an $\Theta(N)$ operation). We have thus produced an algorithm of time complexity $\Theta(N \log N)$ and with $\Theta(N)$ memory requirements. This is a great improvement over the min-cut algorithm that we used on diluted graphs (as we shall explain in next Section), that has time and memory complexity $\Theta(N^3)$ for fully connected graphs. With this algorithm we where able to perform highly precise averages of systems up to $\sim 10^6$ spins. We then subtract the leading order (in $N$) term $f_0 N$, that can be computed exactly, to obtain the numeric estimate of the coefficient $f_1$ up to subleading finite size effects. In Fig. 5.1 we show the perfect agreement between the results of our exact algorithm and the analytic prediction $f_1 = -\Delta\sqrt{\frac{2}{\pi}}$, with $\Delta$ given by Eq. (5.16), for the RFIM at $T = 0$.

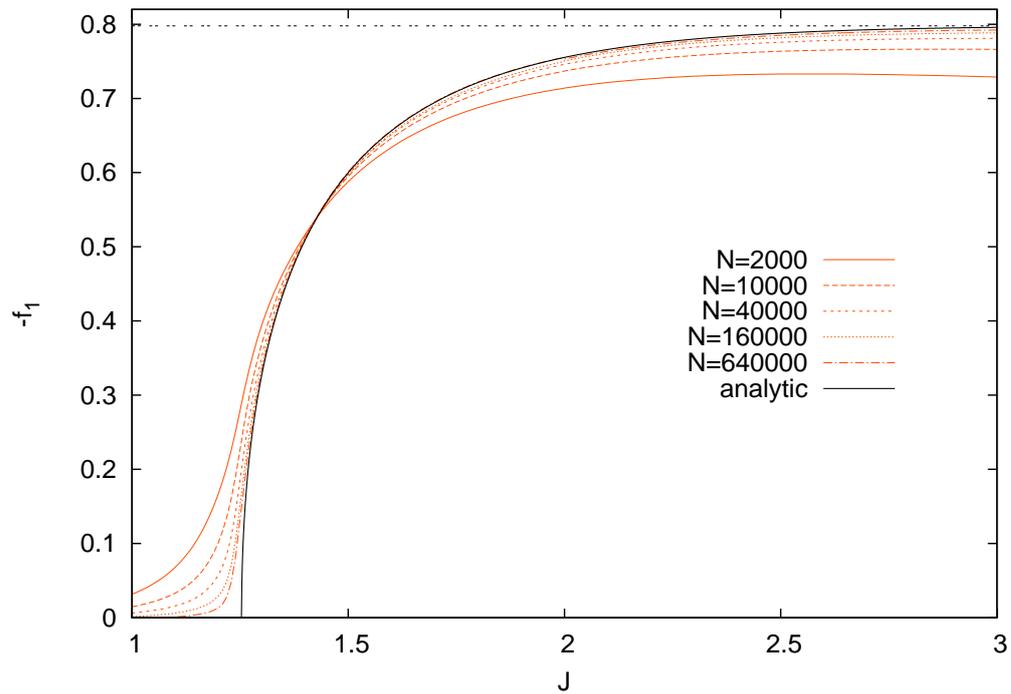

**Fig. 5.1.** The numerical estimates, obtained through the algorithm described in the text, of the coefficient of the $O(\sqrt{N})$ correction to the average free energy in the RFIM at zero temperature on the Fully Connected Graph, for many sizes of the system (*orange lines*). The numerical curves extrapolate to the analytical prediction of Eqs. (5.11) and (5.16) for large $N$ (*black line*). The dashed black line is the asymptotic value $\sqrt{\frac{2}{\pi}}$.

## 5.3 Random regular graphs

The analytic computation of the $O(\sqrt{N})$ correction to the average free energy in diluted models is slightly more involved. Here we focus on the random regular graph (RRG) ensemble. Each element of the ensemble is chosen uniformly at random among all the graphs where each node has fixed connectivity $z$. The RRG and other diluted graph ensembles have the property of being locally tree-like, in the sense that each finite neighborhood of a randomly chosen node is with high probability a tree in the large $N$ limit, and the density of finite loops goes to zero [15]. This property allows for the analytical tractation of such models, at list at the leading order in $N$.

Following the lead of Ref. [172], we apply the replica technique to Eq. (5.12), to obtain the replicated free energy

$$\phi_{RRG}(n,m) = \min_\rho \left\{ \frac{z}{2} \phi_{edge}[\rho] - (z-1) \phi_{site}[\rho] \right\}, \qquad (5.18)$$

where

$$\phi_{edge}[\rho] = \log \Big[ \mathbb{E}_{h_1,h_2} \sum_{\sigma_1,\sigma_2} \rho^{z-1}(\sigma_1) \\ \times e^{\beta(h_1 \sum_a \sigma_1^a + J \sum_a \sigma_1^a \sigma_2^a + h_2 \sum_a \sigma_2^a)} \rho^{z-1}(\sigma_2) \Big]. \qquad (5.19)$$

and

$$\phi_{site}[\rho] = \log \Big[ \mathbb{E}_h \sum_\sigma e^{\beta h \sum_{a=1}^{n+m} \sigma^a} \rho^z(\sigma) \Big] \qquad (5.20)$$

A similar expression, for the replicated free energy of spin glasses on RRG, was derived in Ref. [173] (see also Eq. (7) of Ref. [174] for a more general formulation) and relies on the hypothesis that the graph contains few short loops.

Eq. (5.18) is a straightforward generalization of these results, the difference being that the replicas are divided into two blocks of size $n$ and $m$ respectively. The minimization condition in Eq. (5.18) is imposed over all the functions $\rho(\sigma) = \rho(\sigma^1,\ldots,\sigma^{n+m})$ of a $n+m$ replicated Ising spin, taking particular attention to constrain the first $n$ and last $m$ replicas to be in the up and down state respectively. Here as in the fully connected case, the dimension of the order parameter, $2^{n+m}$, is always positive, and we do not incur in the usual peculiarity of replica calculations, the exchange between minimum and maximum conditions.

Under replica symmetry assumptions the order parameter $\rho(\sigma)$ can be parametrized in the form

$$\rho(\sigma) = \int \frac{\mathrm{d}P(u^+,u^-)}{(2\cosh\beta u^+)^n (2\cosh\beta u^-)^m} e^{\beta(u^+ \sum_{a=1}^n \sigma^a + u^- \sum_{a=n+1}^m \sigma^a)}. \qquad (5.21)$$

In the small $n$ and $m$ limit the minimum condition gives

$$P(u^+,u^-) = \mathbb{E}_h \int \prod_{k=1}^{z-1} \mathrm{d}P(u_k^+,u_k^-) \, \delta\Big(u^+ - g(h + \sum_k u_k^+)\Big) \delta\Big(u^- - g(h + \sum_k u_k^-)\Big), \qquad (5.22)$$

where the function $g(x)$ is the usual cavity iteration rule $g(x) = \frac{1}{\beta} \operatorname{atanh}(\tanh(\beta J) \tanh(\beta x))$. The vanishing auxiliary external field $H$ of Eq. (5.12) selects the solution of Eq. (5.22), supposed to be unique, such that $\overline{u^+} \geq 0 \geq \overline{u^-}$ (denoting with $\overline{\bullet}$ the expectation over $P(u^+, u^-)$).

Expanding the replicated free energy $\phi_{RRG}(n, m)$ to the second order in $n$ and $m$ we can than derive the $O(\sqrt{N})$ coefficient of the free energy according to Eqs. (5.8) and (5.11). We note that only the fixed point distribution $P(u^+, u^-)$ computed at $n = m = 0$, that is the solution of (5.22), contributes to the quadratic order of $\phi(n, m)$. The coefficient $\Delta$ appearing in $f_1 = -\Delta\sqrt{2/\pi}$ than reads

$$\Delta^2_{RRG} = \frac{1}{2\beta^2} \left\{ \frac{z}{2} (\mathbb{E}[A_+^2] - \mathbb{E}[A_+ A_-]) \right. \\ \left. - (z-1)(\mathbb{E}[B_+^2] - \mathbb{E}[B_+ B_-]) \right\}. \quad (5.23)$$

The terms $A_\pm$ and $B_\pm$ stem from the edge and site replicated free energies ($\phi_{site}$ and $\phi_{edge}$) respectively. The expectations $\mathbb{E}[\bullet]$ are over both the distribution of the external random field and of the cavity fields. The site terms $B_+$ and $B_-$ are defined by

$$B_\pm = \log \frac{2 \cosh \beta(h + \sum_{k=1}^z u_k^\pm)}{\prod_{k=1}^z 2 \cosh \beta u_k^\pm}. \quad (5.24)$$

Here the fields $u_k^+$ and $u_k^-$ are distributed according to $P(u_k^+, u_k^-)$, solution of Eq. (5.22), and $h$ is distributed as the external random fields. The edge terms $A_+$ and $A_-$, appearing in Eq. (5.23), are defined by

$$A_\pm = \log \sum_{\sigma_1, \sigma_2} \frac{\exp \beta(h_1^\pm \sigma_1 + J\sigma_1\sigma_2 + h_2^\pm \sigma_2)}{\prod_{k=1}^{z-1} 4 \cosh \beta u_{1k}^\pm \cosh \beta u_{2k}^\pm} \quad (5.25)$$

The random field $h_1^+$ is distributed as $h + \sum_{k=1}^{z-1} u_{1k}^+$, where $h$ is an external random fields, and analogous definitions follows for the other cavity fields.

We notice that only in the terms $\mathbb{E}[A_+ A_-]$ and $\mathbb{E}[B_+ B_-]$ of Eq. (5.23) the full joint distribution $P(u^+, u^-)$ is needed, not only its marginals.

The computation of $\Delta^2$, and therefore of the analytic finite size correction $f_1$, to a high level of precision through Eq. (5.23), is a computationally easy task. We solved numerically the fixed point condition Eq. (5.22) through a population dynamic algorithm. In this case each element of the population is a couple of cavity messages, $u^+$ and $u^-$, each of them encountering the same external random fields $h$ during the iterations of the algorithm. As initial condition, in each couple the message $u^+$ is set to a high positive value, while the message $u^-$ is set to a low negative value.

In the paramagnetic phase the stable solution of Eq. (5.22) takes the trivial form $P(u^+, u^-) = P(u^+)\delta(u^+ - u^-)$. The $O(\sqrt{N})$ finite size correction is thus zero. In the ferromagnetic phase instead, the messages $u^+$ and $u^-$ become non-trivially correlated.

We computed with the population dynamics algorithm the solution of the fixed point Eq. (5.22) at temperature $T = 0$, for many values of the coupling $J$ and for connectivity $z = 4$. The expectations we find in the expression of $\Delta^2$ given Eq. (5.23) are then computed sampling from the population.

As in the case of the fully connected model, it is easier to verify the analytic predictions working at zero temperature, since the free energy and the energy coincides and the ground state of the system can be obtained through exact polynomial algorithms.

Therefore the analytical result is compared with an exact numerical algorithm that exploit the equivalence between the problem of finding the ground state of the RFIM on an arbitrary graph and the minimum cut optimization problem [109]. We used the implementation of the Goldberg-Tarjan's preflow push-relabel algorithm provided by the open source Lemon Graph Library [107], whose worst case complexity is $\Theta(N^{\frac{5}{2}})$ for instances of the RRG ensemble. The numerical estimate of the $O(\sqrt{N})$ coefficient is obtained from a linear combinations of the free energy of systems of different sizes, and it is given by

$$\tilde{f}_1(N) = \frac{\mathbb{E}[F(2N) - 2F(N)]}{c\sqrt{N}} \qquad (5.26)$$

with $c = \frac{1}{\sqrt{2}} - 1$. The value of $\tilde{f}_1(N)$, obtained averaging the minimum cut results over many samples of the system, should converge for large $N$ to the analytical value computed trough the population dynamic algorithm applied to Eq. (5.23). The data plotted in Fig. 5.2 show a very good agreement between experiments and predictions, although the convergence is slow due to the presence of subleading $O(\frac{1}{\sqrt{N}})$ finite size effects.

## 5.4  Conclusions

We argued for the existence of an anomalous $O(\sqrt{N})$ subleading correction to the thermodynamic average free energy $f_0$ of systems with a zero-mean external random field. This correction is limited to the ferromagnetic phase, since it is caused by the difference of free energy among the two pure states. If one is interested only in asymptotic quantities such as $f_0$, the slowing down of the convergence can be avoided in numerical simulations such as Markov Chain Monte Carlo methods choosing an initial condition uncorrelated to the realization of the disorder (as is indeed usually done), such that the dynamics gets trapped with equal probability in the state with lowest free energy or in the other one. Estimations of observables on systems of finite size through exact algorithms though are bound to follow a behaviour of the type $a(N) \sim a_0 + \frac{a_1}{\sqrt{N}}$, therefore the convergence is much slower then the usual $O(\frac{1}{N})$ behaviour one finds in the paramagnetic phase or in pure systems. On the other hand when the focus is on the finite size average value of some observable, particular care has to be taken for the definition used for the finite size free energy and for the method chosen to equilibrate the system.

Using a formalism inspired by some recent works [169] [170–172] we present a general framework to compute the coefficient $f_1$ of the $O(\sqrt{N})$ term in the average free energy. The computation has been carried out, using a variant of the replica trick, in two solvable mean field systems, the RFIM on the fully connected graph and on the RRG ensemble. The analytic results obtained, Eqs. (5.11)(5.15)(5.23), are found to be in strong agreement with the numerical simulations.

A different but equivalent approach to the problem, also based on the replica trick, can be taken using the techniques of Refs. [175, 176]. Instead of computing the large deviation function $\phi(n,m)$, where $n$ and $m$ replicas are constrained to be in the up and down state respectively, and expanding to small values of $n$ and $m$, the same conclusions could be obtained summing over all the saddle point contributions obtained from the partitioning of a the replicas in the two sets. We preferred the approach of Refs. [169] [170–172] because it is conceptually more clear and analytically less involving (although they share many similarities).

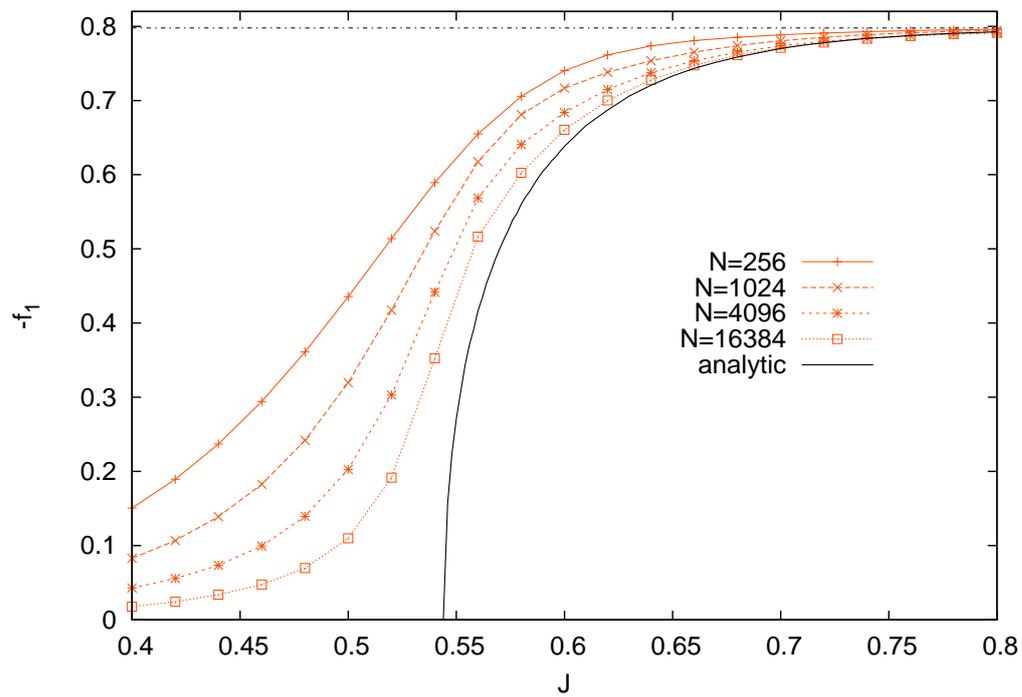

**Fig. 5.2.** The numerical estimates, obtained through the min-cut algorithm, of the coefficient of the $O(\sqrt{N})$ correction to the average free energy in the RFIM at zero temperature on Random Regular Graphs of connectivity $z = 4$, for many sizes of the systems (*orange lines*). The numerical curves (see Eq. 5.26 for their definition) extrapolate to the analytical prediction of Eqs. (5.11) and (5.23) for large $N$ (*black line*). The dashed black line is the asymptotic value $\sqrt{\frac{2}{\pi}}$.

While the formalism we have developed in Section 5.1 is completely general, the exact computation of the coefficient $\Delta^2$ (thus of $f_1$) can be achieved only in mean field models. In finite dimension one has to resort to a perturbative diagrammatic expansion of a replicated field theory. We did not take this path, but our numerical simulations with a min-cut algorithm, using the same procedure described in Section 5.3 for the RRG ensemble, show, qualitatively and also quantitatively, the scenario depicted in Fig. 5.2. Summing it up, as a general feature of models with zero-mean random external field, the average free energy density as a first finite size correction of order $O(\frac{1}{N})$ in the paramagnetic phase, and of order $O(\frac{1}{\sqrt{N}})$ in the ferromagnetic phase.

# Chapter 6

# The Euclidean Assignment Problem

## 6.1 Introduction and Main Results

In this Section we focus on a restricted version of the matching problem, the *complete bipartite matching problem*, also known as the *assignment problem*, that has a long standing tradition among the scholars [15, 18, 84]. In this problem we have two sets having the same cardinality $N$, let us call them $\mathcal{R}$ and $\mathcal{B}$, and we want to find a one-to-one correspondence between the elements of $\mathcal{R}$ and the elements of $\mathcal{B}$ in such a way that all the elements are paired and a certain global function of this *matching* (called the *cost function*) is minimized. An instance of the assignment problem is a $N \times N$ matrix $w$: each element $w_{ij}$ gives the partial cost of the assignment of the element $i \in \mathcal{B}$ to the element $j \in \mathcal{R}$. From a combinatorial point of view, an assignment is a permutation $\pi \in \mathcal{S}_N$, where $\mathcal{S}_N$ is the set of permutations of $N$ elements. Its cost is defined by

$$E_N[\pi; w] = \frac{1}{N} \sum_{i=1}^{N} w_{i\pi(i)}. \tag{6.1}$$

The optimization problem consists in finding the *optimal assignment* $\pi^*$, i.e., the assignment $\pi^*$ that satisfies the property $E_N[\pi^*, w] = \min_{\pi \in \mathcal{S}_N} E_N[\pi; w]$.

Some interesting questions arise when random instances of the problem are considered, that is, when the elements of the cost matrix $w$ are chosen accordingly to a certain probability law. We will discuss the properties of the system for different choices of the disorder, i.e., of the distribution of $w$. Let us consider the average optimal cost $E_N = \overline{E_N[\pi^*; w]}$, where $\overline{\bullet}$ denotes the expectation over the instances $w$ of the problem. If we choose the problem ensemble such that the matrix elements $w_{ij}$ are i.i.d. random variables, we obtain the so-called *random assignment problem*. This version of the problem was largely investigated in a set of papers in which both the distribution of the optimal weights in the large $N$ limit [18, 104] and the finite sizes corrections [95] were derived using different approaches. In this context, the celebrated replica approach, directly borrowed from the theory of spin glasses and disordered systems, proved to be an excellent tool to investigate the properties of these random optimization problems, and led to the celebrated formula $\lim_{N\to\infty} E_N = \frac{\pi^2}{6}$ under certain assumptions over the distribution of $w_{ij}$ [18]. Quite interestingly the average optimal cost $E_N$ for the random assignment problem is known exactly for every value of $N$. It is given by the simple formula $E_N = \sum_{k=1}^{N} \frac{1}{k^2}$, as has been first conjectured in [177] and independently proven in [105] and [178].

From an analytical point of view, a much more difficult problem arises when the elements of the cost matrix $w$ are correlated. This is indeed the case of the *Euclidean assignment problem*, also known as *Euclidean bipartite matching problem* (EBMP). The EBMP is an assignment problem in which a set of $N$ "blue" points $\mathcal{B} = \{\mathbf{b}_i\}_{i=1}^{N}$ and a set of $N$ "red" points $\mathcal{R} = \{\mathbf{r}_i\}_{i=1}^{N}$ are given on the hypercube $[0,1]^d$. Each point is supposed to be generated independently and uniformly at random in the hypercube. Periodic boundary conditions are imposed (in other words, to avoid scaling corrections due to border effects, we consider the sets of points on the torus $\mathsf{T}^d \equiv \mathbb{R}^d/\mathbb{Z}^d$). The cost of the matching between two points is then given by a function of their distance on the torus. We will generalize the Euclidean flat distance on the torus to a family of functions characterized by a cost exponent $p$, assuming that

$$w_{ij} = \|\mathbf{b}_i - \mathbf{r}_j\|^p, \tag{6.2}$$

where $\|\mathbf{b}_i - \mathbf{r}_j\| \equiv \sqrt{\sum_{k=1}^{d} \left(\min\left\{\left|b_i^k - r_j^k\right|, 1 - \left|b_i^k - r_j^k\right|\right\}\right)^2}$ is the Euclidean norm on the torus. Due to the underlying Euclidean structure, the elements of $w$ present very strong correlations. We shall denote with $E_N^{(p)}(d)$ the average cost of the optimal assignment between $N$ red points and $N$ blue points on $\mathsf{T}^d$ with cost exponent $p$, that is

$$E_N^{(p)}(d) \equiv \overline{E_N^{(p)}[\pi^*; \{\mathbf{r}_i, \mathbf{b}_i\}]}, \tag{6.3}$$

where the average is intended over the positions of the points and $\pi^*$ is the optimal permutation. In the following we will sometimes drop the dependence on $p$ and on $d$ of the optimal cost, and write simply $E_N$.

The scaling behaviour of the leading order of the optimal cost is well known for $p > 1$ and all values of $d$ and has been confirmed also by the investigations conducted by means of statistical physics methods. In fact, from a simple heuristic argument [58], we expect that, given a red point, the nearest blue points can be found approximately in a volume of order $O\left(\frac{1}{N}\right)$ around it: their distances from the red point is, for this reason, of order $N^{-\frac{1}{d}}$. Supposing that each red point is matched to one of its nearest blue points, the expected total cost scales as $E_N = O\left(N^{-\frac{p}{d}}\right)$ for large $N$. It turns out that this asymptotic estimation is correct only for $d \geq 3$ as it can be rigorously proved [179]. In a fundamental paper on this subject, published in 1984, [180] proved that for $d = 2$ a logarithmic correction appears, $E_N = O\left(\left(\frac{\ln N}{N}\right)^{\frac{p}{2}}\right)$. In dimension $d = 1$ instead, the divergence from the expected result is even greater, in fact it is informally known to the literature [96, 180] (even if to our knowledge nowhere formally stated), that $E_N = O\left(N^{-\frac{p}{2}}\right)$ in this case. We can resume the state of the art of knowledge regarding the asymptotic behaviour of the average optimal cost in the EBMP, with the following formula:

$$\beta_N^{(p)}(d) \equiv \frac{E_N^{(p)}(d)}{N^{-\frac{p}{d}}} = \begin{cases} O(N^{\frac{p}{2}}) & d = 1, \\ O\left((\log N)^{\frac{p}{2}}\right) & d = 2, \\ e_d^{(p)} + O(N^{-\gamma_d}) & d > 2. \end{cases} \quad (6.4)$$

The determination of the exponents $\gamma_d$ is one of the original contributions we present in this Section and it will be discussed in the following. The coefficients $e_d^{(p)}$ are not known to the literature and could not be derived using our approach either, even though we give accurate numerical estimates in Table 6.2. The rescaled average optimal cost, $\beta_N^{(p)}(d)$, is what a physicist would call the intensive energy, and will be used in the rest of the Section. As with $E_N$, we will sometimes drop the $p$ and $d$ dependence from it and write simply $\beta_N$.

The scenario depicted in Eq. (6.4) has to be compared with the one arising in the *Euclidean monopartite matching problem* (EMMP). [58] studied analytically this problem considering the correlations among the costs as perturbations around the case with random independent entries $w_{ij}$. In the EMMP there is a unique set of $2N$ points to be matched among themselves. It has been proven [181, 182] that, in the EMMP, the rescaled average optimal cost $\beta_N^{(p)}(d)$ has a finite limit and is a self-averaging quantity in every dimension $d$. The odd behaviour noticed in the bipartite case in low dimensions is due to the presence of differences, in small regions of space, between the number of red and blue points, that imply the presence of "long distance" pairings and the failure of arguments based on subadditivity [59]. Obviously, in the monoportite cases such problems do not exist, since a partial matching between the points in an arbitrary subregion of $[0,1]^d$ leaves only one point at most unpaired.

Moreover numeric and analytic arguments [59] show that in the EMMP the first subleading correction to the large $N$ limit of $\beta_N$ is of order $O(N^{-1})$ in any dimension. This assumptions though was also been improperly used in the EBMP to numerically extrapolate the value $e_d^{(p)} = \lim_{N \to \infty} \beta_N^{(p)}(d)$ for $d > 2$ in the case of flat distances [59] (i.e. $p = 1$). This led to some inaccurate estimations of $e_d^{(1)}$ that we address in Table 6.2. In fact in Section 6.3 we give numerical evidence that the appropriate value for the exponent $\gamma_d$ of the subleading correction to $\beta_N$ in dimension $d > 2$, as defined in Eq. (6.4), is

$$\gamma_d = \frac{d-2}{d}, \quad (6.5)$$

for any value of $p$. Notice that in the mean-field limit $d \to \infty$ one recovers the subleading scaling $O(N^{-1})$ of the random assignment problem [95, 156].

The main focus of this work is the EBMP with quadratic costs, i.e. the case $p = 2$ in Eq. (6.2). In Section 6.2, inspired by some considerations on the continuum equivalent of the matching problem, the so called Monge–Kantorovič problem, we present a powerful ansatz, Eq. (6.15), for the asymptotic dependence of the optimal cost from the realization of the disorder. After a careful treatment of the diverging quantities, through an appropriate renormalization procedure, we obtain a whole new set of analytic predictions for $\beta_N$. In fact we recover the whole scenario given in Eq. (6.4) for $p = 2$, deriving the proposed expression Eq. (6.5) for $\gamma_d$ as well. Moreover we refine the classification given in Eq. (6.4), with

$$\beta_N^{(2)}(d) \sim \begin{cases} \frac{1}{6}N + e_1^{(2)} & d = 1, \\ \frac{1}{2\pi} \ln N + e_2^{(2)} & d = 2, \\ e_d^{(2)} + \frac{\zeta_d(1)}{2\pi^2} N^{-\gamma_d} & d > 2, \end{cases} \qquad (6.6)$$

where $\zeta_d(x)$ is the Epstein zeta function. Here and in the following the symbol $\sim$ means that the term on the *l.h.s.* is asymptotically equal to the *r.h.s.* except for some additional term decaying faster than each term in the *r.h.s.* (e.g. $\beta_N^{(p)}(1) = \frac{1}{6}N + e_1^{(2)} + o(1)$). While the coefficients $e_d^{(2)}$ have to be determined numerically, we managed to obtain analytically the coefficients of the leading order expansion of $\beta_N$ for $d = 1, 2$ and of the subleading order for $d > 2$. In the following Sections we give a detailed derivation of these results.

## 6.2 A scaling hypothesis for the quadratic cost

### 6.2.1 The Monge–Kantorovič problem and Monge–Ampère Equation

Let us now briefly introduce the so called Monge–Kantorovič problem, as the conclusions of this paragraph have a crucial role in the following discussion. Given two measure densities $\rho_1\colon \mathsf{T}^d \to \mathbb{R}^+$ and $\rho_2\colon \mathsf{T}^d \to \mathbb{R}^+$, $\mathsf{T}^d = \mathbb{R}^d/\mathbb{Z}^d$ being the $d$-dimensional flat torus, $\int_{\mathsf{T}^d} \rho_1(\mathbf{x})\mathrm{d}^d x = \int_{\mathsf{T}^d} \rho_2(\mathbf{x})\mathrm{d}^d x = 1$, we define $\mathcal{M}$ as the set of measure preserving maps $\boldsymbol{\mu}\colon \mathsf{T}^d \to \mathsf{T}^d$, i.e. the set of all maps $\boldsymbol{\mu}$ such that:

$$\rho_1(\mathbf{x}) = \rho_2(\boldsymbol{\mu}(\mathbf{x})) \det \mathsf{J}_{\boldsymbol{\mu}}(\mathbf{x}) \qquad \forall \mathbf{x} \in \mathsf{T}^d, \qquad (6.7)$$

where $\mathsf{J}_{\boldsymbol{\mu}}(\mathbf{x})$ is the Jacobian matrix of $\boldsymbol{\mu}$, $(\mathsf{J}_{\boldsymbol{\mu}}(\mathbf{x}))_{ij} \equiv \frac{\partial \mu_i}{\partial x_j}(\mathbf{x})$. Given a transportation cost function $w\colon \mathsf{T}^d \times \mathsf{T}^d \to \mathbb{R}^+$, we introduce the cost functional

$$\mathsf{E}[\boldsymbol{\mu}; w] = \int_{\mathsf{T}^d} w(\mathbf{x}, \boldsymbol{\mu}(\mathbf{x}))\rho_1(\mathbf{x})\mathrm{d}^d x. \qquad (6.8)$$

We ask for the map $\mathbf{M} \in \mathcal{M}$ that minimizes the cost functional (6.8), i.e., such that $\mathsf{E}[\mathbf{M}; w] = \min_{\boldsymbol{\mu} \in \mathcal{M}} \mathsf{E}[\boldsymbol{\mu}; w]$. This problem is known in Measure Theory as the Monge transport problem [183, 184] and a lot of results have been obtained regarding the existence of the optimal map and its properties [185]. One of the most interesting cases is the quadratic one, in which the cost is given by the convex function $w(\mathbf{x}, \mathbf{y}) = \|\mathbf{x} - \mathbf{y}\|^2$, and we have to minimize the functional

$$\mathsf{E}^{(2)}[\boldsymbol{\mu}] = \int_{\mathsf{T}^d} \|\mathbf{x} - \boldsymbol{\mu}(\mathbf{x})\|^2 \rho_1(\mathbf{x}) \mathrm{d}^d x. \qquad (6.9)$$

In the case of quadratic cost it can be proved that the optimal map can be expressed as the gradient of a certain function $\varphi$ [184], i.e. $\mathbf{M}(\mathbf{x}) = \mathrm{grad}\, \varphi(\mathbf{x})$. Eq. (6.7) can be than rewritten in terms of $\varphi$, obtaining the so called *Monge–Ampère equation*

$$\rho_1(\mathbf{x}) = \rho_2(\mathrm{grad}\, \varphi(\mathbf{x})) \det \mathsf{Hess}\, \varphi(\mathbf{x}), \qquad (6.10)$$

where $(\mathsf{Hess}\, \varphi(\mathbf{x}))_{ij} = \frac{\partial^2 \varphi(\mathbf{x})}{\partial x_i \partial x_j}$ is the Hessian matrix of $\varphi$.

Suppose now that $\rho_1(\mathbf{x}) = 1 + \delta\rho_1(\mathbf{x})$ and $\rho_2(\mathbf{x}) = 1 + \delta\rho_2(\mathbf{x})$, being $|\delta\rho_1(\mathbf{x})| \ll 1$ and $|\delta\rho_2(\mathbf{x})| \ll 1 \; \forall \mathbf{x} \in \mathsf{T}^d$. We expect that, under these hypothesis, we can write $\mathbf{M}(\mathbf{x}) = \mathbf{x} + \mathbf{m}(\mathbf{x})$ with $\|\mathbf{m}(\mathbf{x})\| \ll 1 \; \forall \mathbf{x} \in \mathsf{T}^d$: in the first order approximation, $\det \mathsf{J}_\mathbf{M}(\mathbf{x}) \approx 1 + \operatorname{div} \mathbf{m}(\mathbf{x})$, so Eq. (6.7) becomes:

$$\operatorname{div} \mathbf{m}(\mathbf{x}) = \rho_1(\mathbf{x}) - \rho_2(\mathbf{x}) \equiv \delta\rho(\mathbf{x}). \tag{6.11}$$

In particular, if $w(\mathbf{x}, \mathbf{y})$ has the form as in Eq. (6.9) we can introduce $\mathbf{m}(\mathbf{x}) = \operatorname{grad} \phi(\mathbf{x})$, obtaining the simple Poisson equation

$$\Delta \phi = \delta\rho. \tag{6.12}$$

Denoting by $\delta\hat{\rho}_\mathbf{n} \equiv \int_{\mathsf{T}^d} \delta\rho(\mathbf{x}) \, \mathrm{e}^{-2\pi i \mathbf{n} \cdot \mathbf{x}} \, \mathrm{d}^d x$, in this case the total cost of the transport is given at the first order by

$$\mathsf{E}^{(2)}[\mathbf{M}] \approx \int_{\mathsf{T}^d} [\operatorname{grad} \phi(\mathbf{x})]^2 \, \mathrm{d}^d x = \sum_{\mathbf{n} \in \mathbb{Z}^d \setminus \{\mathbf{0}\}} \frac{|\delta\hat{\rho}_\mathbf{n}|^2}{4\pi^2 \|\mathbf{n}\|^2}. \tag{6.13}$$

Although the last equation has been derived under assumptions difficult to justify in the discrete and random version of the Monge–Kantorovič problem, that is in the EBMP, we will see how Eq. (6.13) retains its validity also in that case.

### 6.2.2 The scaling ansatz

Inspired by the previous considerations, we made an ansatz about the functional dependence of the optimal cost of the Euclidean bipartite matching problem with quadratic cost from the density of the two sets of points. The ansatz is simple, yet it is surprisingly predictive.

We denote with $\rho_\mathcal{B}(\mathbf{x}) \equiv \frac{1}{N} \sum_{i=1}^N \delta(\mathbf{x} - \mathbf{b}_i) \equiv \rho_1(\mathbf{x})$ and with $\rho_\mathcal{R}(\mathbf{x}) \equiv \frac{1}{N} \sum_{i=1}^N \delta(\mathbf{x} - \mathbf{r}_i) \equiv \rho_2(\mathbf{x})$ the random densities in $[0,1]^d$ of the $N$ $\mathcal{B}$-points and $\mathcal{R}$-points respectively. We suppose that periodic boundary conditions are imposed, so we work on the torus $\mathsf{T}^d$, as explained in the introduction. Let us call $\delta\rho(\mathbf{x}) \equiv \rho_\mathcal{B}(\mathbf{x}) - \rho_\mathcal{R}(\mathbf{x})$ the difference between the two densities and

$$\delta\hat{\rho}_\mathbf{n} \equiv \frac{1}{N} \sum_{i=1}^N \left( \mathrm{e}^{-2\pi i \mathbf{n} \cdot \mathbf{b}_i} - \mathrm{e}^{-2\pi i \mathbf{n} \cdot \mathbf{r}_i} \right) \qquad \mathbf{n} \in \mathbb{Z}^d, \tag{6.14}$$

its Fourier modes. Following the hint given by the continuous problem, Eq. (6.13), we introduce the following functional

$$\mathcal{E}_N[\delta\hat{\rho}] \equiv \sum_{\mathbf{n} \in \mathbb{Z}^d \setminus \{\mathbf{0}\}} \frac{|\delta\hat{\rho}_\mathbf{n}|^2}{4\pi^2 \|\mathbf{n}\|^2}. \tag{6.15}$$

Our hypothesis is that the functional $\mathcal{E}_N[\delta\hat{\rho}]$ at large $N$ captures the leading terms of the exact optimal cost $E_N^{(2)}[\pi^*; \{\mathbf{r}_i, \mathbf{b}_i\}]$, i.e. asymptotically $E_N^{(2)}[\pi^*; \{\mathbf{r}_i, \mathbf{b}_i\}] \sim \mathcal{E}_N[\delta\hat{\rho}]$, in the notation of Section 6.1. Note that we are using only (6.12) to evaluate the scaling of the optimal cost, without any reference to the optimality conditions itself. However, this is sufficient to reproduce the correct average behaviour: in fact, in the limit of validity of our linearisation of the Monge–Ampère equation, the solution of (6.12) on the torus is unique and therefore determines automatically the optimal map. It can be shown by direct calculation that $\overline{|\delta\hat\rho_{\mathbf n}|^2}=\frac{2}{N}$ for each $\mathbf n\neq \mathbf 0$. Therefore we have

$$E_N^{(2)}(d)\sim \overline{\mathcal E_N[\delta\rho]}=\frac{1}{2\pi^2 N}\sum_{\mathbf n\in\mathbb Z^d\setminus\{\mathbf 0\}}\frac{1}{\|\mathbf n\|^2}\Rightarrow \beta_N^{(2)}(d)\sim N^{\frac{2}{d}}\overline{\mathcal E_N[\delta\rho]}=\sum_{\mathbf n\in\mathbb Z^d\setminus\{\mathbf 0\}}\frac{N^{\frac{2-d}{d}}}{2\pi^2\|\mathbf n\|^2}. \tag{6.16}$$

For $d\geq 2$ the sum in the previous relation is divergent. However, by means of a proper regularisation of the sum, we can still extract useful informations on the scaling of $\beta_N$. For $d=2$ Eq. (6.16) provides, after the regularization procedure, the leading scaling behaviour with the correct prefactor, whilst for $d>2$ the procedure gives the leading scaling and the prefactor of the subleading behaviour. Sadly, in no case for $d>2$ the coefficient, which we name $e_d^{(2)}$, of the leading term in the $\beta_N$ expansion can be computed using our formalism.

### 6.2.3　Case $d=1$

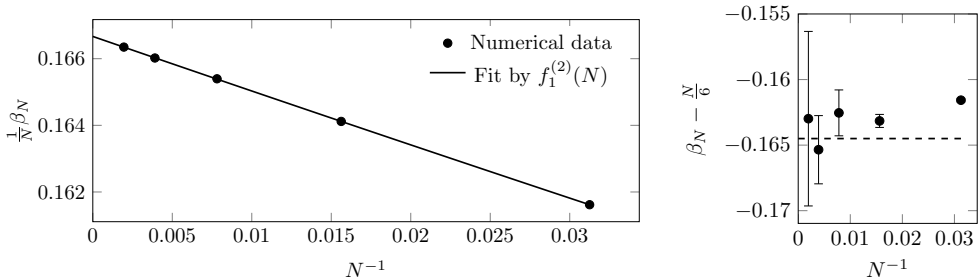

**Fig. 6.1.** Numerical simulations for $d=1$; the fit was performed using the three parameters fit function $f_1^{(2)}(N)=\alpha_1^{(2)}+\frac{e_1^{(2)}}{N}+\frac{c}{N^2}$.

Let us consider the EBMP in dimension $d=1$. In low dimensions, that is for $d=1,2$, the optimal cost is no longer a subadditive quantity and the fluctuations of the density of points are dominant [59]. Moreover for $d=1$ the optimal cost is also not self-averaging [59], while for $d=2$ this is still an open question.

In one dimension the structure of the optimal matching is known [59, 96]: for any $p>1$ and using open boundary conditions, the optimal matching is the one associating the $k$-th blue point to the $k$-th red point, ordering the points from left to right along the line. This consideration leads to the prediction $\beta_N^{(2)}(1)=O(N^{\frac{p}{2}})$ for the leading behaviour with generic cost exponent $p$. For closed periodic boundary conditions, the case we are considering, a similar scenario holds: enumerating the red points and the blue points clockwise or anticlockwise, the optimal matching is given by a cyclic permutation, with offset to be determined by the optimality condition.

The one-dimensional case constitutes the simplest application of our formula, Eq. (6.16), since this is the only where the sum is not divergent. We obtain straightforwardly

$$\beta_N^{(2)}(1)\sim \frac{N}{\pi^2}\sum_{n=1}^{+\infty}\frac{1}{n^2}=\frac{N}{6}. \tag{6.17}$$

This is indeed the exact asymptotic behaviour of $\beta_N$ [96], and it is a first validation of our very simple ansatz, Eq. (6.15): we were able to catch both the anomalous scaling $O(N)$ and its correct prefactor.

We checked the validity of Eq. (6.17) averaging the optimal cost of the EBMP given by an exact algorithm [107] for system sizes up to $N = 2048$. We found the numerical data for $\frac{1}{N}\beta_N$ to be well approximated by a three parameters fitting function of the form $f_1^{(2)}(N) = \alpha_1^{(2)} + \frac{e_1^{(2)}}{N} + \frac{c}{N^2}$, where an additional correction of higher order is included. From a least square fit we obtained the coefficient $\alpha_1^{(2)} = 0.166668(3)$, in perfect agreement with our analytical prediction (see Figure 6.1).

Once verified the validity of Eq. (6.17), we used it to extrapolate the subleading coefficient $e_1^{(2)}$, fixing $\alpha_1^{(2)} = \frac{1}{6}$ and using the fitting function $f_1^{(2)}(N)$ with two free parameters (see Figure 6.1 and Table 6.1).

### 6.2.4 Case $d = 2$

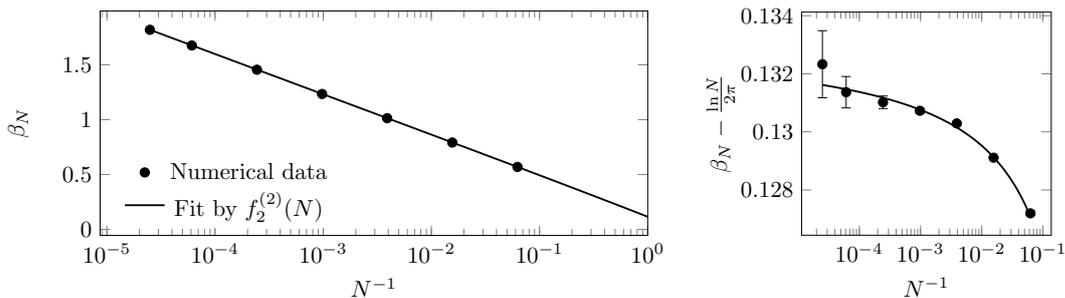

**Fig. 6.2.** Numerical simulations for $d = 2$. We fitted our numerical data for $\beta_N^{(2)}(2)$ using the function $f_2^{(2)}(N) = a \ln N + e_2^{(2)} + \frac{b}{\ln N}$. The $\frac{1}{\ln N}$ correction was suggested by the right hand plot.

As already stated in the introduction, in dimension two we have that $\beta_N = O(\ln N)$. In this paragraph we will show how to derive this scaling from our ansatz, Eq. (6.16), and we will also obtain the corresponding prefactor. The sum appearing in Eq. (6.16) diverges in dimension two and above, therefore we have to find a suitable regularization to give meaning to the expression. The regularization procedure presents some peculiarities at $d = 2$ from which the anomalous scaling arises.

We choose a regularizing smooth function $F(x)$ such that $F(0) = 1$ and $F(x) \to 0$ for $x \to +\infty$. The function has to decrease rapidly enough to make the series $\sum_{\mathbf{n} \in \mathbb{Z}^2 \setminus \{\mathbf{0}\}} \frac{1}{\|\mathbf{n}\|^2} F\left(\frac{2\pi\|\mathbf{n}\|}{2\pi\ell^{-1}}\right)$ converge: here we introduced a cut-off in momentum space, $2\pi\ell^{-1}$, where $\ell \equiv \frac{1}{\sqrt{N}}$ is the characteristic length for the system, being $\ell$ of the order of the average distance between a blue point and the nearest red point. Clearly $2\pi\ell^{-1} \to +\infty$ for $N \to +\infty$. Finally, let us denote by $\mathcal{N}_d(r) = \left|\{\mathbf{x} \in \mathbb{Z}^d \setminus \{\mathbf{0}\} | \|\mathbf{x}\| < r\}\right|$, the number of lattice points (excluded the origin) included in a ball of radius $r$ centred in the origin in dimension $d$. Then, for arbitrary

| | $d=1$ | $d=2$ | $d=3$ | $d=4$ | $d=5$ |
|---|---|---|---|---|---|
| $\frac{1}{2\pi^2}\zeta_d(1)$ | $\frac{1}{6}$ | – | $-0.45157\ldots$ | $-0.28091\ldots$ | $-0.21423\ldots$ |
| $\alpha_d^{(2)}$ | $0.166668(3)$ | – | $-0.4489(16)$ | $-0.282(4)$ | $-0.2139(13)$ |
| $e_d^{(2)}$ | $-0.1645(13)$ | $0.1332(5)$ | $0.66251(2)$ | $0.571284(6)$ | $0.584786(2)$ |

**Table 6.1.** Results of numerical simulations for $p = 2$. In the first line the analytical prediction for $\alpha_d^{(2)}$ for $d \neq 2$ is presented.

$a \in (0,1)$, we can write

$$\sum_{\mathbf{n}\in\mathbb{Z}^2\setminus\{\mathbf{0}\}} \frac{1}{\|\mathbf{n}\|^2} F\left(\frac{\|\mathbf{n}\|}{\sqrt{N}}\right) = \lim_{R\to\infty}\int_a^R \frac{1}{r^2} F\left(\frac{r}{\sqrt{N}}\right)\left[\frac{\partial \mathcal{N}_2(r)}{\partial r} - 2\pi r\right]\mathrm{d}r + 2\pi\int_a^\infty F\left(\frac{r}{\sqrt{N}}\right)\frac{\mathrm{d}r}{r}$$
$$\approx \lim_{R\to\infty}\int_a^R \frac{1}{r^2}\left[\frac{\partial \mathcal{N}_2(r)}{\partial r} - 2\pi r\right]\mathrm{d}r + 2\pi\int_{\frac{a}{\sqrt{N}}}^\infty \frac{F(r)}{r}\mathrm{d}r, \quad (6.18)$$

where we have isolated in the last term the singular part of the series. The first integral in the right hand side is well behaved: indeed, $\int_a^R \frac{1}{r^2}\left[\frac{\partial \mathcal{N}_2(r)}{\partial r} - 2\pi r\right]\mathrm{d}r = \left.\frac{\mathcal{N}_2(r)-\pi r^2}{r^2}\right|_a^R + \int_a^R \frac{\mathcal{N}_2(r)-\pi r^2}{2r^3}$. Both the first and the second term are finite in the $R \to \infty$ limit due to the fact that [186] $\mathcal{N}_2(r) - \pi r^2 \leq 1 + 2\sqrt{2}\pi r$. Therefore we have

$$\sum_{\mathbf{n}\in\mathbb{Z}^2\setminus\{\mathbf{0}\}} \frac{1}{\|\mathbf{n}\|^2} F\left(\frac{\|\mathbf{n}\|}{\sqrt{N}}\right) \approx \int_a^{+\infty} \frac{1}{2r^3}\left[\mathcal{N}_2(r) - \pi r^2\right]\mathrm{d}r + 2\pi \log\frac{\sqrt{N}}{a} + 2\pi\int_1^\infty \frac{F(r)}{r}\mathrm{d}r. \quad (6.19)$$

Eq. (6.16) for the case $d=2$ can then be rewritten as

$$\beta_N^{(2)}(2) \sim \frac{\ln N}{2\pi} + e_2^{(2)}. \quad (6.20)$$

where $e_2^{(2)}$ is some constant. To our knowledge the result $\lim_{N\to\infty}\frac{\beta_N^{(2)}(2)}{\ln N} = \frac{1}{2\pi}$ is new to the literature.

The validity of Eq. (6.20) has been confirmed by numerical simulation with system sizes up to $N = 4\cdot 10^4$. We found a three parameter function of the form $f_2^{(2)}(N) = a\ln N + e_2^{(2)} + \frac{b}{\ln N}$ to be the best suited to fit the data for $\beta_N$. From a least square fit we obtain the coefficient $2\pi a = 1.0004(6)$, in perfect agreement with our analytical prediction (see Figure 6.2). Once verified the validity of Eq. (6.20), we used it to extrapolate the subleading coefficient $e_2^{(2)}$, fixing $a = \frac{1}{2\pi}$ and fitting the other two parameters (see Figure 6.1 and Table 6.1).

### 6.2.5 Case $d > 2$

In dimensions greater than two the average optimal cost has the leading scaling $E_N^{(2)}(d) = O(N^{-\frac{2}{d}})$ that one could expect from very simple arguments [18], as already stated in the introduction. This is in fact the scaling obtained if each point is matched to one of its nearest points of different type, being their distance of order $O(N^{-\frac{1}{d}})$. Moreover it has been proven, using standard sub-additivity arguments, that $\beta_N$ is a self-averaging quantity and has a finite limit for $d > 2$ [59]. We will show that our ansatz Eq. (6.15) accounts for the subleading corrections to $\beta_N$.

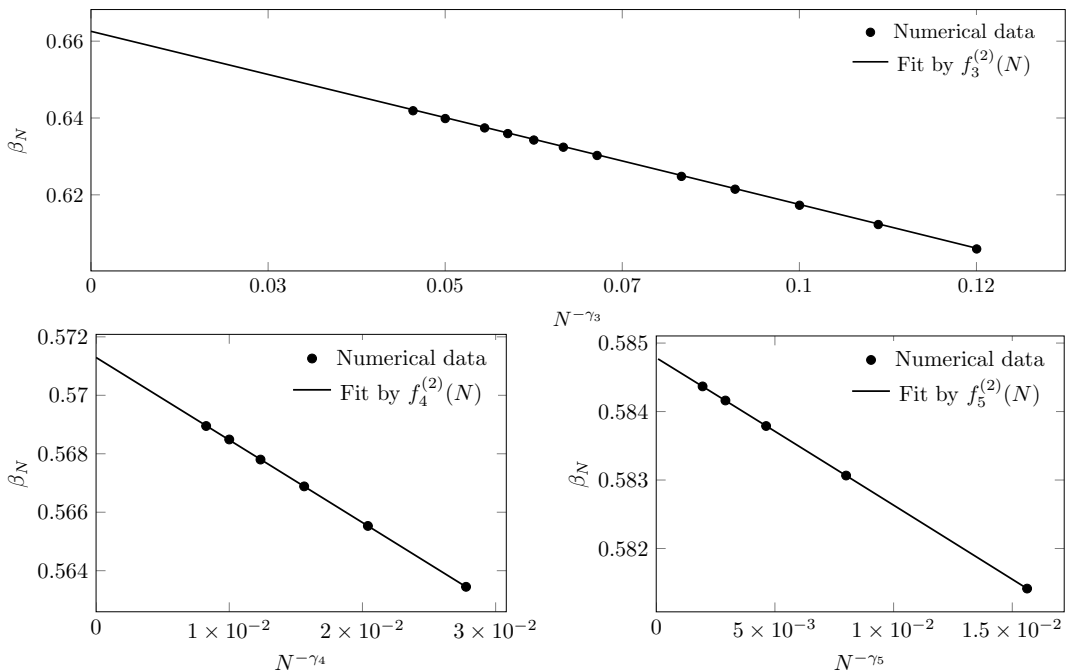

**Fig. 6.3.** Numerical simulations for $d > 2$. We fitted our numerical data for $\beta_N^{(2)}(d)$ using the function $f_d^{(2)}(N) = e_d^{(2)} + \alpha_d^{(2)} N^{-\gamma_d} + \frac{c}{N}$. The expected value for $\alpha_d^{(2)}$ was $\frac{1}{2\pi^2}\zeta_d(1)$.

For $d \geq 2$ the series $\sum_{\mathbf{n} \in \mathbb{Z}^d \setminus \{\mathbf{0}\}} \frac{1}{\|\mathbf{n}\|^2}$ is divergent. As in the previous paragraph, we use a sufficiently rapidly decaying function $F(x)$, with $\lim_{x \to \infty} F(x) = 0$ and $\lim_{x \to 0} F(x) = 1$, to regularize it. The characteristic length of the system is given by $\ell = \frac{1}{\sqrt[d]{N}}$. Denoting as before by $\mathcal{N}_d(r)$ the number of lattice points inside a sphere centred in the origin and of radius $r$, with the exclusion of the origin itself, we can write

$$\sum_{\mathbf{n} \in \mathbb{Z}^d \setminus \{\mathbf{0}\}} \frac{1}{\|\mathbf{n}\|^2} F\left(\frac{\|\mathbf{n}\|}{\sqrt[d]{N}}\right) = \int_0^{+\infty} \frac{1}{r^2} F\left(\frac{r}{\sqrt[d]{N}}\right) \left[\frac{\partial \mathcal{N}_d(r)}{\partial r} - S_d r^{d-1}\right] dr + N^{\frac{d-2}{d}} S_d \int_0^\infty F(r) r^{d-3} dr$$

$$\approx \int_0^{+\infty} \frac{1}{r^2} \left[\frac{\partial \mathcal{N}_d(r)}{\partial r} - S_d r^{d-1}\right] dr + N^{\frac{d-2}{d}} S_d \int_0^\infty F(r) r^{d-3} dr, \tag{6.21}$$

where $S_d = \frac{2\pi^{\frac{d}{2}}}{\Gamma(\frac{d}{2})}$ is the unit sphere surface in $d$ dimensions. The last term in Eq. (6.21) gives the leading order contribution to $\beta_N$, but in our formalism it cannot be explicitly computed since it depends on the choice of the regularizing function $F(x)$. We name the other integral on the right hand side as

$$\Sigma_d \equiv \int_0^{+\infty} \frac{1}{r^2} \left[\frac{\partial \mathcal{N}_d(r)}{\partial r} - S_d r^{d-1}\right] dr. \tag{6.22}$$

$\Sigma_d$ gives the first subleading correction to the leading scaling of $\beta_N$. Moreover it can be shown (see appendix C.1) that $\Sigma_d = \zeta_d(1)$, the analytic continuation to the point $s = 1$ of the function

$$\zeta_d(s) \equiv \sum_{\mathbf{n} \in \mathbb{Z}^d \setminus \{\mathbf{0}\}} \frac{1}{\|\mathbf{n}\|^{2s}} \qquad \text{for } \operatorname{Re} s > \frac{d}{2}. \tag{6.23}$$

The previous function is a particular case of a more general class of zeta functions, called *Epstein zeta functions*. Therefore we are able to compute analytically the subleading behaviour of $\beta_N$ for $d > 2$,

$$\beta_N^{(2)}(d) \sim e_d^{(2)} + \frac{\zeta_d(1)}{2\pi^2 N^{\gamma_d}}, \qquad \gamma_d \equiv \frac{d-2}{d}. \tag{6.24}$$

The expression for $\zeta_d(1)$ is given by Eq. (C.6), while the intensive costs $e_d^{(2)}$ have to be determined numerically. Note that for $d \to +\infty$ we recover the correct meanfield scaling behaviour already analysed by [59] for the random assignment problem, i.e. $\gamma_d \to 1$. However, for finite $d$, the scaling behaviour can be very different from the mean field one.

We verified the validity of Eq. (6.24) with numerical simulation on systems with sizes up to $N = 10648$ in dimension $d = 3$, $N = 14641$ in dimension $d = 4$ and $N = 32768$ in dimension $d = 5$. We used three parameter functions of the form $f_d^{(2)}(N) = e_d^{(2)} + \frac{\alpha_d^{(2)}}{N^{\gamma_d}} + \frac{c}{N}$ to fit our data for $\beta_N^{(2)}(d)$. The scaling exponents $\gamma_d$ are readily confirmed to be the right ones (see Figure 6.3) and the fitted coefficients $\alpha_d^{(2)}$ are in strong agreement with the analytical prediction $\frac{\zeta_d(1)}{2\pi^2}$ (Table 6.1). Then we fixed the $\alpha_d^{(2)} = \frac{\zeta_d(1)}{2\pi^2}$ in $f_d^{(2)}(N)$ to extrapolate the extensive coefficients $e_d^{(2)}$, reported in Table 6.1.

## 6.3 Results for generic $p$

The asymptotic form we proposed for the average optimal cost $E_N^{(2)}(d)$ in the EBMP with quadratic costs and periodic boundary conditions, Eq. (6.15), could not be extended to cover the case of generic cost exponent $p$. Nonetheless, our numerical simulations give strong evidence that, for $d > 2$ and any $p > 0$, $\beta_N$ has the asymptotic form

$$\beta_N^{(p)}(d) \sim e_d^{(p)} + \frac{\alpha_d^{(p)}}{N^{\gamma_d}}, \qquad \gamma_d \equiv \frac{d-2}{d}. \tag{6.25}$$

We thus find the same scaling exponent $\gamma_d$ analytically predicted only for the case $p = 2$. The non-trivial scaling exponent $\gamma_d$ differs from the mean-field exponent $\gamma_\infty = 1$ of the random link matching problem [95] and of the Eucliden monopartite matching [59]. The identification of the correct exponent $\gamma_d$ is crucial to the extrapolation of $e_d^{(p)}$ from numerical data. Since in the literature a mean-field like scaling has been considered [59], we report in Table 6.2 the values of $e_d^{(p)}$ and $\alpha_d^{(p)}$ for different values of $p$. They were obtained fitting $\beta_N$ with functions of the form $f_d^{(p)}(N) = e_d^{(p)} + \frac{\alpha_d^{(p)}}{N^{\gamma_d}} + \frac{c}{N}$. In Figure 6.4 we plot $\beta_N^{(1)}$, $\beta_N^{(3)}$ and $\beta_N^{(4)}$ for $d = 3, 4, 5$, along with the data already presented for $p = 2$ for comparison. The scaling exponent $\gamma_d = \frac{d-2}{d}$ is confirmed by our simulations. Generalizing the case $p = 2$, we therefore conjecture that the optimal cost as a function of the difference of the density of point $\delta\rho(\mathbf{x})$, for $|\delta\rho(\mathbf{x})| \ll 1$ and in dimension $d > 2$, can be approximated by

$$\mathcal{E}_N^{(p)}[\delta\hat{\rho}] \equiv A_d^{(p)} \sum_{\mathbf{n} \in \mathbb{Z}^d \setminus \{\mathbf{0}\}} \frac{|\delta\hat{\rho}_{\mathbf{n}}|^2}{4\pi^2 \|\mathbf{n}\|^2}. \tag{6.26}$$

where $A_d^{(p)}$ are unknown parameters. From last equation, the asymptotic form Eq. (6.25) for $\beta_N$ can be derived, but at odds with the case $p = 2$ where $A_d^{(2)} \equiv 1$ for any $d$, the lack of knowledge of the value of the parameters $A_d^{(p)}$ forbids the analytic prediction of the subleading coefficients $\alpha_d^{(p)}$. We also notice that Eq. (6.26) cannot be extended to dimensions $d = 1, 2$, since it is incompatible with the scaling scenario depicted in Eq. (6.4).

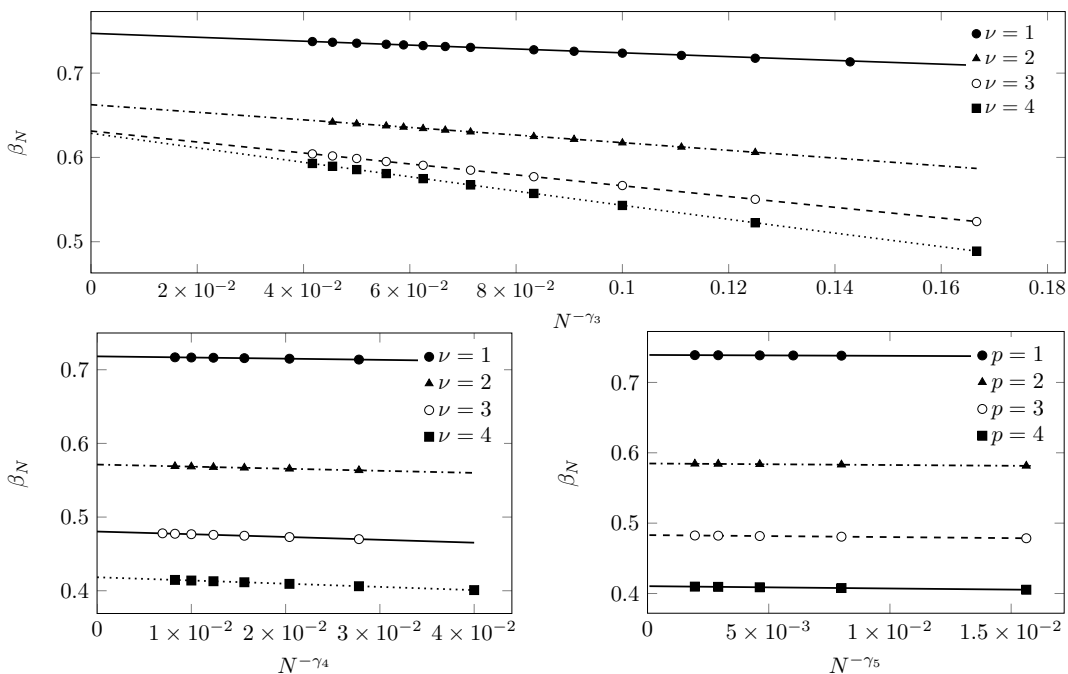

**Fig. 6.4.** Numerical simulations for $d > 2$. We fitted our numerical data for $\beta_N^{(p)}(d)$ using the function $f_d^{(p)}(N) = e_d^{(p)} + \alpha_d^{(p)} N^{-\gamma_d} + \frac{c}{N}$. We plotted the results obtained for $p = 1, 3, 4$. Fit results are presented in Table 6.2.

The ansatz given in Eq. (6.15) for $p = 2$ has been *a posteriori* confirmed by the correct prediction of both the exponents $\gamma_d$ and the subleading coefficients $\alpha_d^{(2)}$. On the other hand, for generic $p$, the ansatz (6.26) is only supported by the fact that it gives the correct exponents $\gamma_d$, which is the reason itself why it was introduced. Therefore we tried to verify the internal consistence of Eq. (6.26). In fact after some algebraic manipulations and averaging over the disorder one can derive

$$A_d^{(p)} \sim \frac{\|\mathbf{n}\|^2 \pi N^2}{2} \overline{\left(E_N^{(p)}[\pi^*; \{\mathbf{r}_i, \mathbf{b}_i\}] - E_N^{(p)}\right) |\delta\hat{\rho}_{\mathbf{n}}|^2}. \tag{6.27}$$

where we used the notation introduced in Eq. (6.3). We computed numerically the *r.h.s* of the previous equation for $d = 3$ and sizes up to $N = 10648$. The computation is numerically quite heavy since the density fluctuation are small. While the *l.h.s.* of equation (6.27) is independent of the mode **n** in the large $N$ limit, we observed a finite size effect that seemed to be best accounted for by the scaling form $A_3^{(p)}(\mathbf{n}, N) \sim A_3^{(p)} + b\frac{1}{\sqrt[3]{N}} + c\left(\frac{\|\mathbf{n}\|}{\sqrt[3]{N}}\right)^2$. Using this three parameters function to fit our whole set of data ($\|\mathbf{n}\| < 10$) at fixed $p$, we extrapolate the values of $A_3^{(p)}$ that we report in Table 6.3. In the same table we compare these extrapolations with the predictions steaming from Eq. (6.26), that is $A_d^{(p)} = \frac{\alpha_d^{(p)}}{\alpha_d^{(2)}} = \alpha_d^{(p)} \frac{2\pi}{\zeta_d(1)}$, where the coefficients $\alpha_d^{(p)}$ where computed in Table 6.2. The agreement between the two different sets of values is quantitatively and qualitatively good, even though we cannot definitively affirm the validity of Eq. (6.26) due to the complexity and imprecision of the procedure utilized to test it. A more sound verification would be to manually excite one of the modes through the addition a position dependent shift of the form $\epsilon\, \mathbf{n} \cos(2\pi \mathbf{n} \cdot \mathbf{x})$ to each randomly generated point in one of the sets. One should then observe a linear response of the totat cost to the variation of $\epsilon$ as predicted by Eq. (6.26).

|  | $p = 1$ | | |
|---|---|---|---|
|  | $d = 3$ | $d = 4$ | $d = 5$ |
| $\alpha_d^{(p)}$ | $-0.233(2)$ | $-0.152(17)$ | $-0.127(16)$ |
| $e_d^{(p)}$ | $0.7472(2)$ | $0.7181(2)$ | $0.73905(5)$ |

|  | $p = 3$ | | |
|---|---|---|---|
|  | $d = 3$ | $d = 4$ | $d = 5$ |
| $\alpha_d^{(p)}$ | $-0.652(3)$ | $-0.36(1)$ | $-0.28(4)$ |
| $e_d^{(p)}$ | $0.6313(2)$ | $0.4804(1)$ | $0.4830(1)$ |

|  | $p = 4$ | | |
|---|---|---|---|
|  | $d = 3$ | $d = 4$ | $d = 5$ |
| $\alpha_d^{(p)}$ | $-0.863(3)$ | $-0.44(2)$ | $-0.34(1)$ |
| $e_d^{(p)}$ | $0.6286(2)$ | $0.4183(2)$ | $0.41043(4)$ |

**Table 6.2.** Results of the fit of numerical data for $\beta_N^{(p)}(d)$ by a function of the form $f_d^{(p)}(N) = e_d^{(p)} + \frac{\alpha_d^{(p)}}{N^{\gamma_d}} + \frac{c}{N}$.

|  | $p = 1$ | $p = 2$ | $p = 3$ | $p = 4$ |
|---|---|---|---|---|
| $A_3^{(p)}$ from Eq. (6.26) | $0.516(5)$ | $1$ | $1.44(1)$ | $1.908(4)$ |
| $A_3^{(p)}$ from Eq. (6.27) | $0.51(3)$ | $0.99(2)$ | $1.46(3)$ | $1.96(2)$ |

**Table 6.3.** Values of $A_d^{(p)}$ for $d = 3$, extrapolated from Eqs. (6.26) and (6.27) respectively, as explained in the main text. The error in the second row are upscaled by a factor ten from those given by our fitting program (gnuplot), to assure the compatibility with the case $p = 2$.

## 6.4 Conclusions and perspectives

In the present work we proposed a simple form for the asymptotic behaviour of the average optimal cost, $E_N^{(2)}(d)$, in the Euclidean bipartite matching problem with quadratic costs and periodic boundary conditions. This ansatz, Eq. (6.15), contains no free parameters and leads to a strong set of predictions in every dimension, resumed in Eq. (6.6). The rescaled cost $\beta_N^{(2)}(d) \equiv \frac{E_N^{(2)}}{N^{-\frac{2}{d}}}$ for low dimensions is a diverging quantity in the EBMP, at odds with the monopartite case. We were able to prove that $\lim_{N\to\infty} \frac{\beta_N^{(2)}(1)}{N} = \frac{1}{6}$ and $\lim_{N\to\infty} \frac{\beta_N^{(2)}(2)}{\log N} = \frac{1}{2\pi}$ for $d=1$ and $d=2$ respectively. Above the critical dimension of the system, $d=2$, the rescaled cost $\beta_N^{(2)}(d)$ has a finite limit which is inaccessible to our formalism. We were able though to determine analytically both the subleading scaling, $O(N^{-\gamma_d})$ with $\gamma_d = \frac{d-2}{d}$, and its prefactor $\frac{\zeta_d(1)}{2\pi^2}$. All the above claims are overwhelmingly supported by numerical simulations.

Finally, we provided numerical evidences that, in dimension $d > 2$, the subleading scaling exponent $\gamma_d$ we predicted for the case of quadratic costs is the same for arbitrary cost exponent $p$. This led us to extend the ansatz proposed for quadratic costs, Eq. (6.15), to the general form Eq. (6.25) for $d > 2$. We tested numerically the validity of Eq. (6.25), obtaining good but not definitive results, therefore we proposed another numerical procedure that could give a firmer validation to the theory.

Although our scaling ansatz proved itself to be very powerful, as discussed above, a deeper theoretical investigation is required to derive analytically the limit of $\beta^{(p)}(d)$ at large $N$, not computable in our framework. Moreover, a rigorous justification of our simple ansatz is desirable, and could be inspired by the considerations we made in Section 6.2.1 on the Monge–Kantorovič problem.

# Part IV

# Beyond the Bethe approximation in finite dimension

# Chapter 7

# The Euclidean Matching Problem

We consider in this Chapter the Euclidean matching problem (EMP), not to be confused with the bipartite Euclidean matching problem we dealt in Chapter 6. Here there is no local density fluctuation among the points belonging to different sets, trivially since there is a unique set of points. As a consequence the source of the anomalous scaling behaviour addressed in Chapter 6 here is not present, and the leading cost is extensive in any dimension once rescaled according to a standard prescription.

We will perform a first principle calculation for the average cost in the EMP using the replica method, generalizing the computation of Parisi and Mézard [58]. Since the replicated action contains infinitely many diagrams, we will approximate the result considering only the terms corresponding to polygons, something akin to the HNC approximation for liquid systems.

We compute the contributions of polygons as a perturbation to the mean-field (Bethe) term. We will show that this loop contributions are formally equivalent to the terms appearing in the $O(1/N)$ correction to the total cost in the random-link matching problem, the one with independent random costs, as recently computed by Ratiéville and Parisi [95]. In Ref. [95] the finite size correction was computed according to the scheme we presented in Chapter 4, the only difference being that the authors proceeded through the diagonalization of the Hessian, therefore leaving uncovered the connection with the simple loops of the graphs. Here we establish this connection, relating the random-link model with the Euclidean model, where the terms in the action corresponding to loops appear explicitly since the beginning of the calculation.

Ultimately both the first finite size correction in the mean-field system and the polygons' contributions in the Euclidean model are due to the presence of simple loops of finite size in the interaction graph. The only difference is that the correlations among the cost of the edges in the loops are present only in the Euclidean system.

## 7.1  Replica calculation

Let's consider the monopartite euclidean matching problem with $N$ points uniformly distributed in the $[0, 1]^d$ box. The rescaled intensive energy $e_N$, according to the considerations of Chapter 6 is defined by

$$e_N = \frac{1}{N^{1-\frac{p}{d}}} \sum_{i<j} |x_i - x_j|^p \, n_{ij}^* \tag{7.1}$$

where $n_{ij}^*$ is the optimal matching and $p > 0$ is the cost exponent. Notice that we choose to work with $N$ points, as in Ref. [59], instead of the $2N$ points of Ref. [58]. This to compare our result with the more up to date numerical computations of Houdayer.

We will give the details of the calculation since they are non-trivial and where not specified in the original articles of Parisi and Mézard [18, 58]. We denote with $\overline{\bullet}$ the average over the distribution the joint distribution of the lengths $\ell_{ij} = |x_i - x_j|$ as induced by the uniformly distributed displacement of the points in the box. The replicated partition function for the model reads

$$\overline{Z^n} = \overline{\sum_{\{n_{ij}^a\}} \prod_{a=1}^n \prod_{i=1}^N \delta(\sum_j n_{ij}^a - 1) \prod_{i<j} e^{-\beta N^{\frac{p}{d}} l_{ij}^p \sum_a n_{ij}^a}} \tag{7.2}$$

$$= \int \overline{\prod_{i,a} \frac{d\lambda_i^a}{2\pi} e^{i\lambda_i^a} \prod_a \prod_{i<j} \left(1 + e^{-i(\lambda_i^a + \lambda_j^a) - \beta N^{\frac{p}{d}} l_{ij}^p}\right)} \tag{7.3}$$

$$\equiv \int \prod_{i,a} \frac{d\lambda_i^a}{2\pi} e^{i\lambda_i^a} \overline{\prod_{i<j} (1 + u_{ij})}, \tag{7.4}$$

$$\tag{7.5}$$

where in the last line we defined

$$u_{ij} = \sum_{r=1}^n e^{-r\beta N^{\frac{p}{d}} l_{ij}^p} \sum_{\alpha:|\alpha|=r} e^{-i \sum_{a\in\alpha}(\lambda_i^a + \lambda_j^a)}. \tag{7.6}$$

We can now expand the products of the $u_{ij}$ as a sum over all the subgraphs of the complete graph. If we call $K = \{(i,j) : i,j = 1, \ldots, N\}$ the set of the $N(N-1)/2$ edges, we then have

$$\overline{\prod_{i<j}(1 + u_{ij})} = 1 + \sum_{E \subset K} \overline{\prod_{\mathbf{e} \in E} u_{\mathbf{e}}}, \tag{7.7}$$

A scaling analysis of the lengths distribution shows that leading order contributions in $N$ are given by generalized loops, that is subgraphs with no dangling edges. We keep only the terms corresponding to simple loops, so that our approximate partition function takes the form

$$\overline{Z^n} \approx \int \prod_{i,a} \frac{d\lambda_i^a}{2\pi} e^{i\lambda_i^a} e^{\frac{1}{2} \sum_{i,j} \overline{u_{ij}} + \sum_{\ell \geq 3} \frac{1}{2\ell} \sum_{i_1,\ldots,i_\ell} \overline{u_{i_1 i_2} u_{i_2 i_3} \ldots u_{i_\ell i_1}}} \tag{7.8}$$

$$\tag{7.9}$$

Introducing the order parameters

$$Q_\alpha = \frac{1}{N} \sum_{i=1}^{N} e^{-i \sum_{a \in \alpha} \lambda_i^a} \qquad \alpha \subseteq \{1, \ldots, n\}, \ \alpha \neq \varnothing, \qquad (7.10)$$

and $\hat{Q}_\alpha$ as the associated Lagrange multiplier, we can write

$$\overline{Z^n} \approx \int \prod_{i,a} \frac{d\lambda_i^a}{2\pi} e^{i\lambda_i^a} \prod_\alpha d\hat{Q}_\alpha \, dQ_\alpha \ e^{-N i Q_\alpha \hat{Q}_\alpha + i \hat{Q}_\alpha \sum_i e^{-i \sum_{a \in \alpha} \lambda_i^a}}$$
$$\times e^{\frac{N}{2} \sum_{r=1}^n g_r \sum_{\alpha:|\alpha|=r} Q_\alpha^2 + N \sum_{\ell \geq 3} \frac{1}{2\ell} \sum_{r_1, \ldots, r_\ell} g_{r_1 \ldots r_\ell} \sum_{\boldsymbol{\alpha}:|\boldsymbol{\alpha}|=\mathbf{r}} Q_{\alpha_1 \cup \alpha_2} \cdots Q_{\alpha_\ell \cup \alpha_1}}. \quad (7.11)$$

Here the notation $|\boldsymbol{\alpha}| = \mathbf{r}$ stands for $\alpha_1 = r_1$, $\alpha_2 = r_2$, .... We have also defined the averaged quantities

$$g_r \equiv S_d \int dl \ l^{d-1} \ e^{-\beta l^p} = \frac{\Gamma(\frac{d}{p}) S_d}{(\beta r)^{\frac{d}{p}}}, \qquad (7.12)$$

$$g_{r_1 \ldots r_\ell} \equiv \int dP_\ell(l_1, \ldots, l_\ell) \ e^{-\beta \sum_{e=1}^{\ell} r_e l_e^p}, \qquad (7.13)$$

with the spherical shell surface given by $S_d = \frac{d\pi^{\frac{d}{2}}}{\Gamma(1+\frac{d}{2})}$. $P_\ell$ is the joint distributions of the lengths in the polygon of $\ell$ edges. Now that we have decoupled the sites we can perform the integration over the lambdas. We obtain

$$\overline{Z^n} \approx \int \prod_\alpha d\hat{Q}_\alpha \, dQ_\alpha \ e^{-\beta N S[Q, \hat{Q}]}. \qquad (7.14)$$

The action replicated free energy $S[Q, \hat{Q}]$ is defined by

$$S[Q, \hat{Q}] = S_{MF}[Q, \hat{Q}] + \sum_{\ell \geq 3} \frac{1}{2\ell} S_\ell[Q], \qquad (7.15)$$

with

$$-\beta S_{MF}[Q, \hat{Q}] = -\sum_\alpha i \hat{Q}_\alpha Q_\alpha + \frac{1}{2} \sum_{r=1}^n g_r \sum_{\alpha:|\alpha|=r} Q_\alpha^2$$
$$+ \log \left[ \int \prod_a \frac{d\lambda^a}{2\pi} e^{i \sum_a \lambda^a + \sum_\alpha i \hat{Q}_\alpha e^{-i \sum_{a \in \alpha} \lambda^a}} \right] \qquad (7.16)$$

and

$$-\beta S_\ell[Q] = \sum_{r_1, \ldots, r_\ell} g_{r_1 \ldots r_\ell} \sum_{\boldsymbol{\alpha}:|\boldsymbol{\alpha}|=\mathbf{r}} Q_{\alpha_1 \cup \alpha_2} \chi_{\alpha_\ell \cap \alpha_1 = \varnothing} \cdots Q_{\alpha_\ell \cup \alpha_1} \chi_{\alpha_\ell \cap \alpha_1 = \varnothing}, \qquad (7.17)$$

where $\chi$ is the indicator function.

We will now compute the partition function (7.14) through saddle point evaluation.

## 7.2  RS saddle point

We restrict the saddle point analysis to the mean field action. Subsequently we will consider the terms $S_\ell$ as a perturbation to the the saddle point. Therefore the saddle point equations read

$$i\hat{Q}_\alpha = g_{r(\alpha)} Q_\alpha \tag{7.18}$$

$$Q_\alpha = \langle\langle e^{-i\sum_{a\in\alpha}\lambda^a} \rangle\rangle \tag{7.19}$$

Here $r(\alpha)$ is the cardinality of the set $\alpha$, $r(\alpha) = |\alpha|$ and the expectation $\langle\langle \bullet \rangle\rangle$ is over the (normalized) integrand in the logarithm of Eq. (7.16)

In order to compute the saddle point we make the Replica Symmetric assumption

$$Q_\alpha \equiv Q_{r(\alpha)}, \qquad i\hat{Q}_\alpha \equiv i\hat{Q}_{r(\alpha)}. \tag{7.20}$$

At the leading order in small $n$, after some manipulations [19], the replicated action is given by

$$-\beta S_{MF} \sim n\bigg\{ -\sum_{r\geq 1}\frac{(-1)^{r-1}}{r} i\hat{Q}_r Q_r + \frac{1}{2}\sum_{r\geq 1}\frac{(-1)^{r-1}}{r} g_r Q_r^2 \\ - \int \mathrm{d}u \left(e^{-e^u} - e^{-\sum_r \frac{(-1)^{r-1}}{r!} i\hat{Q}_r e^{ru}}\right)\bigg\} \tag{7.21}$$

It is now convenient to parametrize the variable $Q_r, \hat{Q}_r$ with the function

$$G(u) \equiv \sum_{r\geq 1} \frac{(-1)^{r-1}}{r!} i\hat{Q}_r\, e^{\beta r u}. \tag{7.22}$$

In the previous definition we inserted a factor $\beta$ in the exponent to have a good zero temperature limit, as we shall later see. From last two equation it follows immediately

$$S_{MF} = n\bigg\{ \frac{1}{2}\int \mathrm{d}u\, G(u)\, e^{-G(u)} - \int \mathrm{d}u \left(e^{-e^{\beta u}} - e^{-G(u)}\right)\bigg\} \tag{7.23}$$

To write last equation we also exploited the saddle point equations

$$i\hat{Q}_r = g_r\, Q_r \tag{7.24}$$

and

$$Q_r = \beta \int \mathrm{d}u\, \frac{e^{\beta r u}}{r!} r\, e^{-G(u)} \tag{7.25}$$

From last two equation and the definition of $G(u)$ given in Eq. (7.22), we can write a saddle-point self consistent equation for $G(u)$:

$$G(u) = \frac{S_d}{p}\int \mathrm{d}v \int_0^{+\infty} \mathrm{d}l\, l^s\, \frac{\partial}{\partial v} I(\beta(u+v-l))\, e^{-G(v)} \tag{7.26}$$

with

$$I(x) = \sum_{r=1}^{+\infty} \frac{(-1)^{r-1}}{(r!)^2} e^{rx} = 1 - J_0\left(2e^{\frac{x}{2}}\right) \quad (7.27)$$

In Eq. (7.26) we also introduced for convenience the coefficient $s$, defined as

$$s = \frac{d}{p} - 1. \quad (7.28)$$

## 7.3 Corrections from polygons

Using the RS assumption and treating the term $S_\ell$ as a first order perturbation to the mean field saddle point, the correlation terms in the action due to the polygonal graphs can be written as

$$S_\ell = -\frac{1}{\beta} \sum_{\alpha_1,\ldots,\alpha_\ell} \int \prod_{i=1}^{\ell} d^d x_i \ \prod_{i=1}^{\ell} e^{-\beta r(\alpha_i)|x_i - x_{i+1}|^p} Q_{r(\alpha_i)+r(\alpha_{i+1})} 1_{\alpha_i \cap \alpha_{i+1} = \varnothing}. \quad (7.29)$$

. We see that in the spatial part we have a convolution, therefore we can apply a Fourier transforms to deal with it. We define the function

$$\tilde{g}_r(k) \equiv S_d \int dl \ l^{d-1} \ e^{-\beta r \ l^p} \ {}_0F_1\left(\frac{d}{2}, -\frac{1}{4}k^2 l^2\right) \quad (7.30)$$

in order to write in the loop free energy in the compact form

$$\begin{aligned} S_\ell &= -\frac{1}{\beta} \frac{S_d}{(2\pi)^d} \int dk \ k^{d-1} \sum_{\alpha_1\ldots\alpha_L} \prod_{i=1}^{\ell} Q_{r_i+r_{i+1}} 1_{\alpha_i \cap \alpha_{i+1} = \varnothing} \ \tilde{g}_{r_i}(k), \\ &= -\frac{1}{\beta} \frac{S_d}{(2\pi)^d} \int dk \ k^{d-1} \ \mathrm{Tr}\left[T^\ell(k)\right] \end{aligned} \quad (7.31)$$

In the last line we have introduced the operator

$$T_{\alpha\alpha'}(k) = Q_{r+r'} 1_{\alpha \cap \alpha' = \varnothing} \sqrt{\tilde{g}_r(k)\tilde{g}_{r'}(k)}, \quad (7.32)$$

which is the analogue for the Euclidean matching problem of the operator of Eq. (4.22) we encounter in Chapter 4 when dealing with diluted graphs. An import difference here, is the presence of spatial correlations in the couplings on top of the simple loop topology, due to the Euclidean structure of the ambient space. From now on we shall omit the dependence on $k$. Following Ref. [156] we decompose the matrix $T$ in the irreducible representations of the replica permutation group. The method is similar to the spectral formalism we employed in Chapter 3 for the replicated Ising transfer matrix. The restriction of $T$ to the irreducible subspaces $D^{(q)}$, $q = 0, 1, \ldots$, is given, in the mall $n$ limit, by the matrices

$$N_{rr'}^{(q)} = (-1)^{r'} \frac{(r+r'-1)!(r'-1)!}{(r-1)!(r'-q)!(r'+q-1)!} \ Q_{r+r'} \ \sqrt{\tilde{g}_r \tilde{g}_{r'}} \quad (7.33)$$

Still following Ref. [156] we can apply a transformation to the matrices $N^{(q)}$ that does not alter the eigenvalues. We then consider the matrices

$$M^{(q)}_{rr'} = (-1)^{r'+q}\frac{(r+r'+2q-1)!}{(r+2q-1)!\,r'!}\,Q_{r+r'+2q}\,\tilde{g}_{r'+q} \tag{7.34}$$

The eigenvalue equation for $M^{(q)}$ reads

$$\begin{aligned}\lambda c_r &= \sum_{r'} M^{(q)}_{rr'} c_{r'} = (-1)^q \frac{1}{(r+2q-1)!}\sum_{r'}(-1)^{r'}\frac{(r+r'+2q-1)!}{r'!}\,Q_{r+r'+2q}\,\tilde{g}_{r'+q} c_{r'}\\ &= (-1)^q\beta\int dv\,\frac{e^{\beta(r+q)v}}{(r+2q-1)!}\sum_{r'}(-1)^{r'}\frac{e^{\beta(r'+q)v-G(v)}}{r'!}\,\tilde{g}_{r'+q} c_{r'}\end{aligned} \tag{7.35}$$

We can then turn last eigenvalue equation in an integral eigenvalue equation noticing that if $\{c_r\}$ satisfies last equation then

$$f(u) = \sum_r (-1)^r \frac{e^{\beta(r+q)u-\frac{G(u)}{2}}}{r!}\,c_r\,\tilde{g}_{r+q} \tag{7.36}$$

satisfies

$$\lambda f(u) = (-1)^q \int dv\, A^{(q)}(u,v) f(v) \tag{7.37}$$

with

$$A^{(q)}(u,v) = \beta\, e^{-\frac{G(u)}{2}-\frac{G(v)}{2}+\beta q(u+v)}\sum_r \frac{(-1)^r e^{\beta r(u+v)}}{(r+2q-1)!\,r!}\,\tilde{g}_{r+q}(k) \tag{7.38}$$

We notice that $A^{(q)}$ depends implicitly on the mode $k$ through $\tilde{g}_{r+q}(k)$ defined in Eq. (7.30).

Finally the contribute in the action of the polygons of length $\ell$ can be expressed in terms of the traces of the operator $A^{(q)}$ as

$$S_\ell = -\frac{1}{\beta}\frac{S_d}{(2\pi)^d}\sum_{q\geq 0} d_q \int dk\, k^{d-1}\,\mathrm{Tr}\left((-1)^q A^{(q)}_k\right)^\ell. \tag{7.39}$$

In last equation, as expected from the discussion of Chapter 3, the Sectors' degeneracies are given by

$$d_q = \binom{n}{q} - \binom{n}{q-1} \tag{7.40}$$

and in particular $d_0 = n$ and $d_1 = n-1$.

## 7.4 The zero temperature limit

### 7.4.1 Mean Field term

Let us consider the limit $\beta \uparrow \infty$ for the mean field part of the action. For the function I(x) defined in Eq. (7.27) we have [18]

$$\lim_{\beta\to\infty} I(\beta x) = \theta(x). \tag{7.41}$$

The argument leading to the above equation involves the representation of the sum in $I(x)$ as a path in the complex plane. The saddle point equation (7.26) becomes

$$G(u) = \frac{S_d}{p} \int \mathrm{d}v \; (u+v)^s \, \theta(u+v) \, e^{-G(u)}. \tag{7.42}$$

We can then compute the average cost of the optimal matching in the mean field approximation through the formula

$$e_{MF} \equiv \lim_{\beta \to \infty} \lim_{n \to 0} \frac{S_{MF}}{n} \tag{7.43}$$

$$= \frac{1}{2} \int \mathrm{d}u \; G(u) \, e^{-G(u)} - \int \mathrm{d}u \; \left(\theta(-u) - e^{-G(u)}\right) \tag{7.44}$$

We now turn to the computation of the zero temperature limit for the polygonal corrections to the free energy.

### 7.4.2 Longitudinal and Anomalous Sectors

Before considering the $T \downarrow 0$ (i.e. $\beta \uparrow \infty$) limit for the average free energy, some considerations regarding the Longitudinal and Anomalous sectors, represented by the matrices $N^0_{rr'}$ and $N^1_{rr'}$, in the decomposition of $T_{\alpha\alpha'}$. At finite $n$ the two representations read [95, 156]

$$N^{(0)}_{rr'}(n) = \binom{n-r}{r'} Q_{r+r'} \sqrt{\tilde{g}_r \tilde{g}_{r'}} \qquad r, r' = 1, \ldots, n \tag{7.45}$$

whose $n$ eigenvalues have to be counted multiplicity $d_0 = 1$, and

$$N^{(1)}_{rr'}(n) = \binom{n-r}{r'} Q_{r+r'} \sqrt{\tilde{g}_r \tilde{g}_{r'}} \frac{r'}{r'-n} \qquad r, r' = 1, \ldots, n-1, \tag{7.46}$$

which has $n-1$ eigenvalues, each one contributing with a multiplicity $d_1 = n-1$. As it can be readily seen the two matrices have the same limit at $n = 0$. At odds with the replicated Ising transfer matrix considered in Chapter 3 though, the dimension of $N^{(0)}$ is $n$ and not $n+1$, and for $n \downarrow 0$ all its eigenvalues coincides with the one of $N^{(1)}$, and the total multiplicity then becomes $n$, as expected. In the Ising RTM case instead, the first eigenvalue $\lambda = 1$ of the Longitudinal Sector does not appear in the Anomalous Sector.

The correspondent continuous operators, according to definition (7.38) and after some manipulations, differ only by an overall minus sign, and are given by

$$A^{(0/1)}(u,v;k) = (-1)^{0/1} \frac{S_d}{p} e^{-\frac{G(u)}{2} - \frac{G(v)}{2}} \int_0^\infty \mathrm{d}l \; l^s \; {}_0F_1\left(\frac{d}{2}, -\frac{1}{4}k^2 l^{\frac{2}{p}}\right) \frac{\partial}{\partial v} J_0(e^{\beta(u+v-l)}). \tag{7.47}$$

The operators $A^{(0/1)}(u,v;k)$ are thus well defined in the zero temperature limit, where we obtain

$$A^{(0/1)}(u,v;k) = (-1)^{1/0} \, e^{-\frac{G(u)}{2} - \frac{G(v)}{2}} \Theta_k(u+v). \tag{7.48}$$

with
$$\Theta_k(h) \equiv \frac{S_d}{p} h^s \, \theta(h) \, {}_0F_1\left(\frac{d}{2}, -\frac{1}{4}k^2 \, h^{\frac{2}{p}}\right). \tag{7.49}$$

Therefore eigenvalues of $A^{(0/1)}$ are finite and, due to the factor $\frac{1}{\beta}$ in Eq. (7.39), do not contribute to the zero temperature limit of $S_\ell$.

A finite contribution is given by the first order correction to the eigenvalues for small $n$ though. Moving alone with the computation, in order to account correctly for all the terms in the small $n$ limit, the eigenvalue shifts can be computed using the matrix

$$\begin{aligned}\Delta_{rr'} &\equiv \lim_{n\to 0} \frac{1}{n}\left[N^{(0)}_{rr'}(n) - N^{(1)}_{rr'}(n)\right](t) \\ &= \frac{(-1)^{r'+1}}{r'}\frac{r+r'-1}{(r-1)!r'!}Q_{r+r'}\sqrt{\tilde{g}_r(k)\tilde{g}_{r'}(k)},\end{aligned} \tag{7.50}$$

along with ordinary perturbation theory. The contribution to the loop correction $S_\ell$ given by the degeneracy between the first two sectors is then given by

$$\Gamma_\ell = -\lim_{\beta\to\infty} \frac{S_d}{(2\pi)^d} \int_d dk \, k^{d-1} \sum_\lambda \ell \frac{\Delta_\lambda}{\beta} \lambda^{\ell-1} \tag{7.51}$$

where $\Delta_\lambda$ is defined in term of the left and right eigenvectors eigenvectors, $\langle L_\lambda|$ and $|R_\lambda\rangle$, of $N^{(0)}(n=0)$ as
$$\Delta_\lambda \equiv \langle L_\lambda|\Delta|R_\lambda\rangle. \tag{7.52}$$

While we expect the eigenvalues $\lambda$ of $N^{(0)}$ to be finite in the large $\beta$ limit, as discussed in the next paragraph. Moreover, according to result for the random link case of Ref. [95], we also expect $\Delta_\lambda = O(\beta)$ for large $\beta$, therefore we have a non-null contribution $\Gamma_\ell$ from the first two Sectors to the zero temperature free energy.

### 7.4.3 Sectors $q \geq 2$

Following the same procedure of Parisi and Ratiéville [95] we compute the $\beta \uparrow \infty$ limit of the operator $A^{(q)}$ defined in Eq. (7.38). We will see that the limit is non trivial only if $q$ is appropriately scaled with $\beta$. In fact we can rewrite the operator $A^{(q)}$ as

$$A^{(q)}(u,v) = \beta \, e^{-\frac{G(u)}{2}-\frac{G(v)}{2}} \frac{i}{2\pi}\frac{S_d}{p} \int_C dz \int_0^\infty dl \, e^{S(z,l,q,\beta)} \, l^s \, {}_0F_1\left(\frac{d}{2}, -\frac{1}{4}k^2 l^{\frac{2}{p}}\right) \tag{7.53}$$

with
$$S(z,\ell,q,\beta) = \beta q(u+v-l) - z + \frac{e^{\beta(u+v-l)}}{z} - 2q\log(-z) \tag{7.54}$$

For large $\beta$ the exponential term in the integrand becomes concentrated on the point given by the saddle point conditions
$$\frac{\partial S}{\partial z} = 0; \qquad \frac{\partial S}{\partial l} = 0, \tag{7.55}$$

while the other terms in the integrand do not depend on $\beta$. The saddle point conditions give
$$z^* = -q; \qquad l^* = u + v - 2\frac{\log(q)}{\beta} \tag{7.56}$$

It is easy to see that in the summation over $q$ the only terms giving a non zero contribution are of order $\log q \sim \beta$. Fixing $\beta$ to $\beta = \log(q)/t$ we obtain $l^* = u+v-2t$.

The computation of the saddle point contribution and of the Gaussian fluctuation for the integral in Eq. (7.53) is formally equivalent to the one performed in the appendices of Ref. [95]. In the defining

$$H_{k,t}(u,v) \equiv \lim_{\beta \to \infty} A^{(q=e^{\beta t})}(u,v), \tag{7.57}$$

we obtain

$$H_{k,t}(u,v) = e^{-\frac{G(u)}{2} - \frac{G(v)}{2}} \Theta_k(u+v-2t) \tag{7.58}$$

with

$$\Theta_k(h) \equiv \frac{S_d}{p} h^s \, \theta(h) \, _0F_1\left(\frac{d}{2}, -\frac{1}{4}k^2 h^{\frac{2}{p}}\right). \tag{7.59}$$

We notice that $\Theta_0(u+v)$ corresponds to the kernel of the integral operator of the saddle point Eq. (7.42). Some considerations are now in order to proceed further. Since in Eq. (7.39) we have an alternating sign in $q$, we split the contribution in odds and even sectors. As showed in Ref. [95] the sectors $q = 0$ and $q = 1$ do not contribute to the zero temperature limit. For large $\beta$ and small $n$ we obtain, the contribution to $S_\ell$ from the irreducible sectors with $q \geq 2$ is given by

$$-\frac{n}{\beta} \frac{S_d}{(2\pi)^d} \int \mathrm{d}k \, k^{d-1} \left[ -\sum_{\substack{q>1 \\ q \text{ even}}} \frac{2q-1}{q(q-1)} \operatorname{Tr}\left(A_k^{(q)}\right)^\ell + (-1)^\ell \sum_{\substack{q>1 \\ q \text{ odd}}} \frac{2q-1}{q(q-1)} \operatorname{Tr}\left(A_k^{(q)}\right)^\ell \right] \tag{7.60}$$

Moreover, still for large $\beta$ the sum over the sectors $q$ turns into an integral :

$$\frac{1}{\beta} \sum_{q \text{ odd}} \frac{2}{q} \quad \overset{q=e^{\beta t}}{\longrightarrow} \quad \int_0^{+\infty} \mathrm{d}t \tag{7.61}$$

Finally, using the zero temperature limit of the operator $A^{(q)}$ given in Eq. (7.58), we arrive to compute the contribution to the energy of small simple loops from the sectors $q \geq 2$:

$$\Omega_\ell = \begin{cases} \frac{2S_d}{(2\pi)^d} \int_0^\infty \mathrm{d}t \, \mathrm{d}k \, k^{d-1} \operatorname{Tr} H_{k,t}^\ell & \ell \text{ odd} \\ 0 & \ell \text{ even} \end{cases} \tag{7.62}$$

### 7.4.4 Wrapping it up

The total average energy $e$ in our approximation is thus given by

$$e \approx e_{MF} + \sum_{\ell \geq 3} \frac{1}{2\ell} e_\ell \tag{7.63}$$

where

$$e_\ell \equiv \lim_{\beta \to \infty} \lim_{n \to 0} \frac{S_\ell}{n} = \Gamma_\ell + \Omega_\ell \tag{7.64}$$

contains the contribution $\Gamma_\ell$ from the degeneracy of the Longitudinal and Anomalous sectors

$$\Gamma_\ell = -\lim_{\beta \to \infty} \frac{S_d}{(2\pi)^d} \int_d dk \ k^{d-1} \sum_\lambda \ell \frac{\Delta_\lambda}{\beta} \lambda^{\ell-1}, \tag{7.65}$$

and the contribution form higher sectors

$$\Omega_\ell = \begin{cases} \frac{2S_d}{(2\pi)^d} \int_0^\infty dt\, dk \ k^{d-1} \operatorname{Tr} H_{k,t}^\ell & \ell \text{ odd} \\ 0 & \ell \text{ even} \end{cases} \tag{7.66}$$

The operator $H_{k,t}$ has eigenvalue $\lambda = 1$ for $k = t = 0$. We show it taking the derivative of saddle point equation (7.42) for $G(u)$ and integrating by parts, obtaining the relation

$$G'(u) = \int dv \ \Theta_0(u+v) \ G'(v) \, e^{-G(v)}. \tag{7.67}$$

Then it is easy to prove that the function

$$h(u) = G'(u) \, e^{-\frac{G(u)}{2}} \tag{7.68}$$

is eigenfunction of $H_{0,0}$ with eigenvalue $\lambda = 1$.

We performed numerical simulations to compute the value of the eigenvalues of $H_{k,t}$. First we discretize the zero temperature saddle point equation (7.42) for $G(u)$. Then we discretize the operator $H_{k,t}$ and compute its eigenvalues with the linear algebra library Armadillo [187]. The result is shown in Fig. 7.1. $H_{k,t}$ has a discrete spectrum, its eigenvalues continuously decreasing in both $k$ and $t$. We focus the analysis on the largest eigenvalue of $H_{k,t}$, since it dominates the trace of $H_{k,t}^\ell$ for large $\ell$. The largest eigenvalue $\lambda(k,t)$ is decreasing in $t$ and $k$ and, moreover for $k = t = 0$ we have $\lambda(0,0) = 1$, with eigenfunction $h(u)$ as already stated. For large $\ell$ the dominant contribution in the integration over $k$ and $t$ is given by the region of small $t$ and $k$. In that region the leading behaviour is given by

$$\lambda(k,t) \sim e^{-at-bk^2}. \tag{7.69}$$

as it can be seen in Fig. 7.1.

Using the knowledge of the $k = t = 0$ eigenfunction $h(u)$ defined in Eq. (7.68), we can use perturbation theory to determine the coefficient $a$ and $b$ of Eq. (7.69). Le us define the cavity message distribution

$$P(u) = G'(u) \, e^{-G(u)} \tag{7.70}$$

Perturbation theory in $t$ gives

$$a = -2\frac{S_d}{p} \frac{\int du\, dv \ P(u)\,(u+v)^s \theta(u+v)\, P'(v)}{\int du\ h^2(u)} \tag{7.71}$$

where in the second line we have used Eq. (7.67) along with an integration by parts.

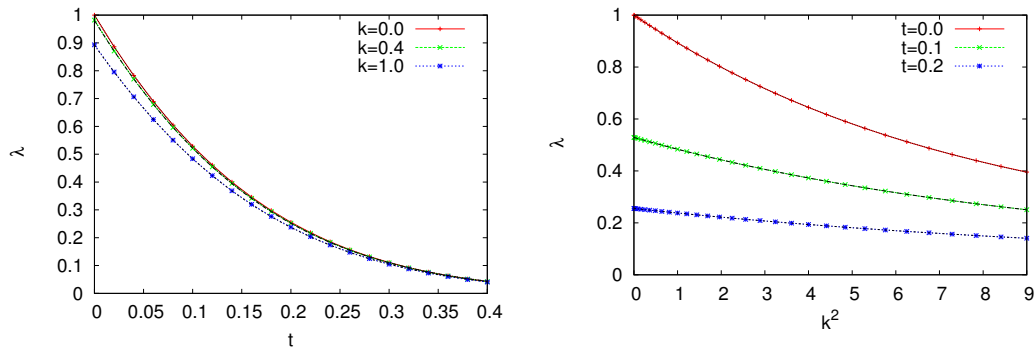

**Fig. 7.1.** Maximum eigenvalue of the operator $H_{k,t}$ of Eq. (7.58) as a function of $t$ (*left*) and $k$ (*right*). Black lines are numerical fit of the form $\lambda = \exp(A - Bx - Cx^2)$.

Using the analytic formula for $G(u)$

$$G(u) = \ln\left(1 + e^{\frac{S_d}{d}u}\right) \qquad for \ \ d = p, \tag{7.72}$$

equation (7.71) gives

$$a = \frac{4}{3} \qquad for \ \ d = p. \tag{7.73}$$

This result is confirmed by numerical simulations.

The perturbation in $k$ to $H_{t,k}$ is obtained expanding to the first order in $k^2$ the hypergeometric function in definition of $\tilde{g}_r(k)$ in Eq. (7.22). We have

$$b = \frac{1}{6}\frac{S_d}{p}\frac{\int du\, dv\ P(u)(u+v)^{s+\frac{2}{p}}\theta(u+v)P(v)}{\int du\ h^2(u)} \tag{7.74}$$

Some further progress can be made for $p = 1, 2$. In fact a few integrations by parts, along with $\int G'e^{-G} = \int P = 1$, yield

$$b = \frac{s+1}{6\int du\ G'^2(u)\,e^{-G(u)}} \qquad for \ \ p = 2, \tag{7.75}$$

and

$$b = \frac{(s+2)(s+1)}{6}\frac{\int du\ G(u)e^{-G(u)}}{\int du\ G'^2(u)\,e^{-G(u)}} \qquad for \ \ p = 1, \tag{7.76}$$

In the analytically solvable case $d = p$ we have

$$b = \begin{cases} \frac{\pi^2}{36} & p = d = 1 \\ \frac{1}{6} & p = d = 2 \end{cases} \tag{7.77}$$

Once we assume the leading dependence on $k^2$ and $t$ of the eigenvalues $\lambda(k,t)$, we can perform separately the two integrations in $\Omega_\ell$:

$$\int dt\ e^{-at\ell} = \frac{1}{\ell}\frac{1}{a}, \tag{7.78}$$

$$\int dk\ k^{d-1}e^{-bk^2\ell} = \frac{1}{\ell^{\frac{d}{2}}}\frac{\Gamma(\frac{d}{2})}{2b^{\frac{d}{2}}}. \tag{7.79}$$

Using these results to approximate $\Omega_\ell$ we obtain

$$\sum_{\ell \geq 3} \Omega_\ell \approx \frac{S_d}{(2\pi)^d} \frac{\Gamma(\frac{d}{2})}{2ab^{\frac{d}{2}}} \sum_{\substack{\ell \geq 3 \\ \ell \text{ odd}}} \frac{1}{\ell^{2+\frac{d}{2}}} \tag{7.80}$$

The sum over $\ell$ can be expressed in term of the generalized Rienmann zeta function $\zeta(x,y)$, thus giving

$$\sum_{\ell \geq 3} \Omega_\ell \approx \frac{S_d}{(2\pi)^d} \frac{\Gamma(\frac{d}{2})}{2ab^{\frac{d}{2}}} \frac{1}{2^{2+\frac{d}{2}}} \zeta\left(2+\frac{d}{2}, \frac{3}{2}\right) \tag{7.81}$$

A similar approximation can be done for the terms $\Gamma_\ell$, at the present point we did not make the computation though.

## 7.5 Finite size corrections

### 7.5.1 Connection with the random link problem

Let us take a step back from the Euclidean Matching Problem and consider the random link model, where the elements $w_{ij}$ of the cost matrix are extracted independently at random according to a certain distribution (after a proper rescaling of the costs with $N$) $\rho(w) = c\, w^\gamma$, where $c, \gamma > 0$ are some coefficients. According to Parisi and Mézard [156], and as in the calculations we performed in Chapter 4 on diluted graphs, there are two terms contributing to the $O\left(1/N\right)$ correction to the average cost. The first one, $\Delta F^1$, comes from the $1/N$ correction to the action itself. The second one, $\Delta F^2$, is due to the Gaussian fluctuations around the saddle point. This second correction, at finite temperature, can be written, using our notation, as

$$\Delta F^2 = \lim_{n \to 0} \frac{1}{2\beta n} \ln \det(\mathbb{I} - \mathsf{T}) \tag{7.82}$$

with

$$\mathsf{T}_{\alpha\alpha'} = Q_{r(\alpha)+r(\alpha')}\, 1_{\alpha \cap \alpha' = \varnothing}\, \sqrt{g_{r(\alpha)} g_{r(\alpha')}} - \frac{1}{2} Q_{r(\alpha)} Q_{r(\alpha')}\, \sqrt{g_{r(\alpha)} g_{r(\alpha')}}, \tag{7.83}$$

and

$$g_r = \int \mathrm{d}w\; \rho(w)\, e^{\beta r w}. \tag{7.84}$$

These last formulas differs from the ones presented in Refs. [156] and [95] due a different notation: their $Q_r$ is rescaled by a factor $g_r$ and the consider a matching among $2N$ points instead than $N$. At this point we see that the matrix $\mathsf{T}$ corresponds exactly to our transfer matrix for the polygonal correction $T(k)$ defined in Eq. (7.32), except for wo reasons. One is the dependence on the mode $k$. This is natural since in the random link model we loose the spatial correlation of the costs in the polygon and the zero mode encodes all the relevant information. The second is the additional term $Q_{r(\alpha)} Q_{r(\alpha')}$. In Ref. [95] this is shown to contribute only in the Anomalous and Longitudinal sectors, and it has the effect of exactly counterbalancing the same contributions of the term $Q_{r(\alpha)+r(\alpha')} 1_{\alpha \cap \alpha'}$, therefore $O(1/N)$ Gaussian fluctuations in the random link problem are given only by the Sectors $q \geq 2$ and contribute positively to the free energy.

Expanding the logarithm in Eq. (7.82), as we did in Chapter 4 with the Hessian in random graphs, we obtain

$$\Delta F^2 = \lim_{n \to 0} -\frac{1}{\beta n} \sum_{\ell \geq 1} \frac{1}{2\ell} \operatorname{Tr} \mathsf{T}^\ell \qquad (7.85)$$

The structure of $O(1/N)$ finite size correction, $\Delta F^1 + \Delta F^2$, in the random link problem, is similar to the one we found in Chapter 4 for Ising models on Erdős-Rényi graphs. In fact, as clearly stated in Ref. [60], each nodes in the complete edge has only a finite number of low cost edges relevant for the optimal matches. Therefore the complete graph is well approximated with a Poissonian random graph with finite connectivity. The $\ell = 2$ term in Eq. (7.85) is cancelled out by a corresponding term in $\Delta F_1$. Therefore the total finite size correction $\Delta F = \Delta F^1 + \Delta F^2$ is ultimately due to two factors. The first, corresponding to the term $\ell = 1$ in Eq. (7.85), is due to the fluctuation on the number of edges in the Poissonian graph. The second, corresponding to the terms $\ell \geq 3$, is given by the presence of simple loops of finite length. Last proposition is confirmed by previous arguments and comparison of Eq. (7.85) and Eqs. (7.15, 7.31) for the Euclidean matching problem.

We have thus established the connection between finite size correction in the mean field matching model, *i.e.* the random link problem, and the polygonal correction to the Bethe approximation of the leading cost in the Euclidean version of the problem.

### 7.5.2 Anomalous scaling behaviours

Let us consider again the random link problem. We will show how we can derive an anomalous scaling for the second order correction to the thermodynamic free energy. We express the average optimal cost as

$$e_{\mathrm{RL}}(N) = e_\infty + e_1 \frac{1}{N} + o\left(\frac{1}{N}\right). \qquad (7.86)$$

For $\rho(w) \sim 1$, corresponding to the case $d = p$ in the Euclidean matching, $e_\infty = \frac{\pi^2}{12}$ [18]. For simplicity we will stick to this case, the argument is general though. Following the discussion of previous paragraph we approximate the $e_1$ with the contribution given by loops. According to Ref. [95] we have

$$e_1 \approx \sum_{\substack{\ell \geq 3 \\ \ell \text{ odd}}} \frac{1}{2\ell} \int \mathrm{d}t \ \operatorname{Tr} H_t^\ell, \qquad (7.87)$$

where
$$H_t(u, v) = e^{-\frac{G(u)}{2} - \frac{G(v)}{2}} \ \theta(u + v - 2t) \qquad (7.88)$$

Once again we stress the equivalence between last equation and the operator $H_{k,t}$ of Eq. (7.58). We approximate traces with powers of the leading eigenvalue of $H_t$ we arrive to the expression, as we did in Section 7.4. Since for large $\ell$ the integral is dominated by the region of small $t$, we can use the asymptotic form $\lambda(t) \sim e^{-at}$. Integration in $t$ in Eq. (7.87) than yields

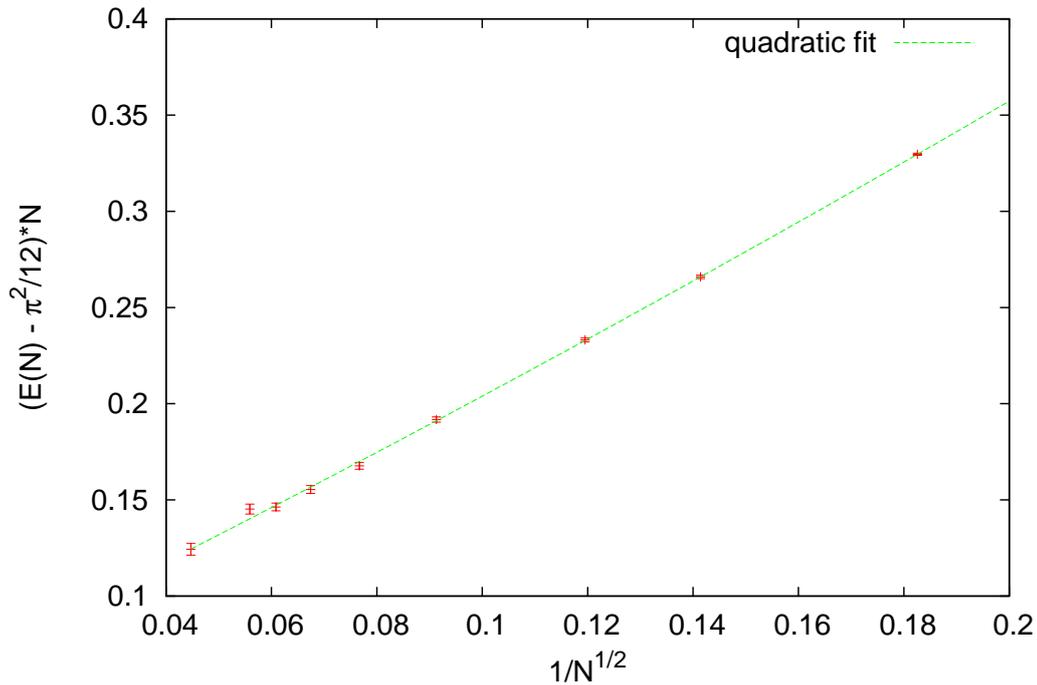

**Fig. 7.2.** Average optimal cost in the monopartite random link matching as a fuction of $N$. The subsubleading scaling has an anomalous $O(N^{-\frac{3}{2}})$ behaviour.

$$e_1 \approx c \sum_{\substack{\ell \geq 3 \\ \ell\, odd}} \frac{1}{\ell^2}, \tag{7.89}$$

where $c$ is some constant.

Since the probability for a random path of length $\ell$ in the complete graph to intersect itself is $O(\ell^2/N)$ for large $\ell$ and $N$, therefore we expect a correction to last formula that can be as expressed in terms of a regularizing function, $\sum_{\substack{\ell \geq 3 \\ \ell\, odd}} \frac{1}{\ell^2} f(\ell^2/N)$. Equivalently we can introduce a cutoff $\Lambda = \sqrt{N}$ on the series, and write

$$c \sum_{\substack{\ell \geq 3 \\ \ell\, odd}}^{\Lambda} \frac{1}{\ell^2} = e_1 + O\left(\frac{1}{\sqrt{N}}\right). \tag{7.90}$$

Therefore the asymptotic expansion of the average optimal cost in the random link problem takes the form

$$e_{\text{RL}}(N) = e_\infty + e_1 \frac{1}{N} + e_{\frac{3}{2}} \frac{1}{N^{\frac{3}{2}}} + o\left(\frac{1}{N^{\frac{3}{2}}}\right). \tag{7.91}$$

We have thus individuated a subleading correction with a non trivial fractional exponent. This result is confirmed by numerical simulations as shown in Figure 7.2. Fit of numerical data give $e_1 = 0.063(3)$ and $e_{\frac{3}{2}} = 1.35(6)$ for exponentially distributed costs, whereas the leading value, as computed in Ref. [18], is $e_\infty = \frac{\pi^2}{12}$.

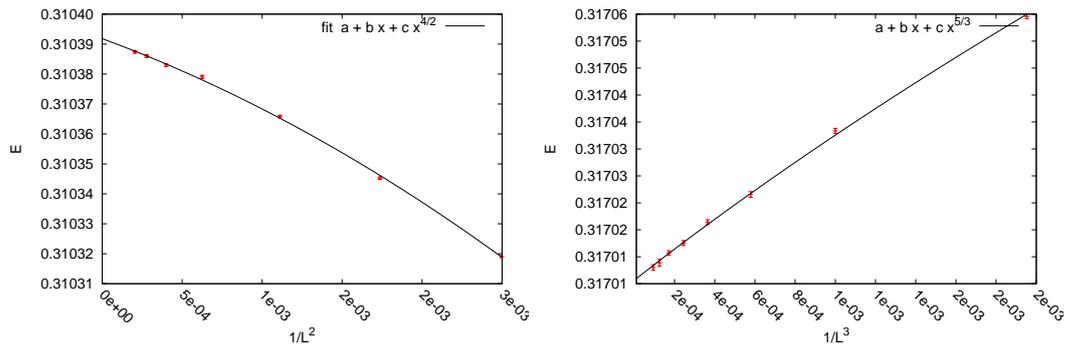

**Fig. 7.3.** Average optimal cost in the Euclidean matching problem as a function of $L = \sqrt[d]{N}$ for $d = 2$ (*left*) and $d = 3$ (*right*). Numerical data do not allow for a definitive validation of the subsubleading scaling behaviour.

A similar argument can be used to compute a finite size correction in the Euclidean matching. In this case the loops contribute to the leading order energy approximately with

$$c \sum_{\substack{\ell \geq 3 \\ \ell \text{ odd}}} \frac{1}{\ell^{2+\frac{d}{2}}}. \tag{7.92}$$

We now impose a cutoff $\Lambda$. Some simple considerations will allow us to find its scaling with $N$. In fact from Eq. 7.30 we see that dimensionally $[k] = [\ell^{-2}]$. At finite size the integration over $k$ in Eq. (7.31) is replaced by a sum over the Fourier modes corresponding to the box $[0, L]^d$, with $L = N^{\frac{1}{d}}$ since we rescaled the original unit box. Therefore the cutoff $\Lambda$ has to scale as $\Lambda \sim k_{min}^{-2} \sim L^2$, and we have

$$c \sum_{\substack{\ell \geq 3 \\ \ell \text{ odd}}}^{\Lambda} \frac{1}{\ell^{2+\frac{d}{2}}} = c \sum_{\substack{\ell \geq 3 \\ \ell \text{ odd}}} \frac{1}{\ell^{2+\frac{d}{2}}} + O\left(\frac{1}{L^{d+2}}\right). \tag{7.93}$$

According to this argument the optimal cost in the Euclidean matching problem should scale as

$$e_d(N) = e_d + \Delta_d \frac{1}{L^{d+2}} + o\left(\frac{1}{L^{d+2}}\right). \tag{7.94}$$

Unfortunately numerical simulations contradicts this scenario. In dimension $d = 3$ and with flat cost exponent, $p = 1$, we averaged the optimal cost over $10^6$ to $10^7$ samples for each size of the box $L = 8, 10, 12, \ldots, 22$ and with $N = L^d$ points. The results are clearly incompatible with the scaling $1/L^5$ from of Eq. (7.94). It turns out our data is compatible both with a fitting function of the form $e_3(N) \approx a + b/L^3 + c/L^5$ and one with the form $e_d(N) \approx a + b/L^c$. This results are presented in Fig. 7.3 and Table 7.1.

The $O(1/L^{d+2})$ correction we have highlighted is overshadowed by a slower correction coming from some other mechanism, possibly $O(1/N)$. In any case the presence of non-exponential correction to the leading order energy is per se an unusual feature for a finite dimensional system, denoting a critical behaviour.

| **d = 3** | $a$ | $b$ | c |
|---|---|---|---|
| $e(N) \approx a + b/L^3 + c/L^5$ | 0.317010(1) | 0.033(2) | -0.5(1) |
| $e(N) \approx a + b/L^c$ | 0.317009(1) | 0.009(2) | 2.5(1) |

**Table 7.1.** Fitted parameters from numerical simulations on the EMP with $N = L^d$ points and flat distances ($p = 1$).

## 7.6 Conclusions

We performed a detailed replica calculations for the average cost in the Euclidean matching problem. We expanded upon the work of Parisi and Mézard [58], which considered only correlation in the costs of links forming triangles, including all the cost correlations on simple loops. The approach we took is based on the diagonalization of a replicated transfer matrices in the irreducible subspaces of the replica permutation group. This approach is the same used in Refs. [13] and [95] to compute the $O(1/N)$ finite size correction to the random link matching problem trough the diagonalization of the Hessian. For the leading order in $N$ of the EMP our computations gives the result

$$e_d \approx e_{\text{MF}} + \sum_{\ell \geq 3} \frac{1}{2\ell} \Gamma_\ell + \sum_{\substack{\ell \geq 3 \\ \ell \text{ odd}}} \frac{1}{2\ell} \Omega_\ell, \tag{7.95}$$

where $e_{\text{MF}}$ is the mean field term computed in the random link (i.e. independent weights) approximation, $\Gamma_\ell$ is the contribution from the Longitudinal and Anomalous Sectors degeneracy in the loops of length $\ell$, and $\Omega_\ell$ id the contribution from the other Sectors, which is null for even lengths. Additionally we showed how to efficiently compute the terms $\Omega_\ell$ and $\Gamma_\ell$, although some analytical progress can still be done for $\Gamma_\ell$. Finally we proved the equivalence, up to a simple Fourier mode dependence, between the terms $\Gamma_\ell$, and the terms appearing in the $1/N$ computation in the random link problem as computed by Parisi and Ratieville [95].

The work presented in this Chapter is incomplete, although the great part of the expected results have been achieved. A more in-depth analysis of the term $\Gamma_\ell$ is needed, and also analytic results have to be confirmed by numerical simulations. We will address these deficiencies in the following months.

In the long term perspective, we expect that the terms in the Euclidean replicated action which are of higher order in the in the number of cycles are in one-to-one correspondence to higher order terms in the $O(1/N)$ expansion of the random link matching problem.

The results of this Chapter nicely fit with the discussion of the $O(1/N)$ finite size corrections in diluted random graphs of Chapter 4, characterizing simple loops contributions to the free energy, and with we present in the next Chapter on the large $M$ expansion.

# Chapter 8

# The Large M Expansion

## 8.1 Introduction

Analytical investigation of finite dimensional disordered systems is a difficult challenge for physicist, scholars relying on uncontrolled renormalization schemes [188–190] and facing diverging (already at the first order) perturbative theories [191].

Fully-connected mean field models would constitute a solid starting point for perturbative expansions, but they could fail to capture qualitative feature of finite dimensional systems [192]. Another class of mean field models, possessing some of the feature of real systems, is that of diluted random graphs. As already discussed in Chapter 2, diluted graphs, being locally tree-like in the thermodynamic limit, can be investigated through the use of Cavity Method, which we will also call Bethe approximation (although generally this name indicates the Replica Symmetric Cavity Method).

Many attempts to perform systematic expansion around the Bethe approximation where produced in recent years, namely those of Parisi and Slanina [193], Chertkov and Chernyak [167,168], Montanari and Rizzo [157,194] (can only compute corrections to marginals), Mori and Tanaka [195,196]. While some of this attempts are very promising [167,196], they where conceived to be applied to single realizations of the disorder, therefore it is difficult to see how to use them to compute disorder averages. The other schemes are plagued by infrared divergences [193] or are algorithmically oriented and limited to the computation of marginals and not free energies [157].

In this Chapter we contribute to the literature developing a formalism to systematically perform a perturbative expansion around the Bethe free energy for disorder system in finite dimension, which we call large $M$ expansion. We will employ a combinatorial technique to interpolate between an arbitrary finite graph and a locally tree-like graph infinite graph, for which the Bethe approximation holds true.

We will make use of both the Replica and the Cavity Method in our computation. The replica computation is similar to the one performed in Chapter 4, but it is slightly more involved, since we have also to address an $U(1)$ symmetry giving some null modes. This work is at an early stage, therefore the results we presents are incomplete. Some additional effort is required in the future to polish the formalism and explore its applications. We will see though that the work we reported in Chapter 4 on the the finite size correction in diluted random graph is a valuable

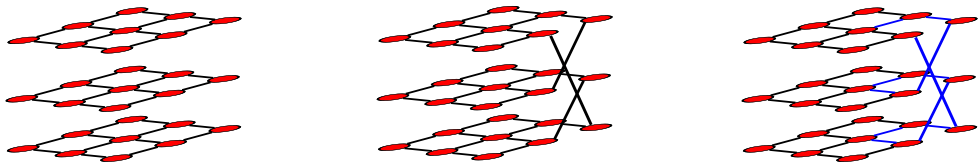

**Fig. 8.1.** Pictorial representation of the $M$ layer construction. The rightmost figure outlines a simple loop in the $M$-graph, corresponding to a non-backtracking closed path in the original lattice.

guide for the large $M$ expansion.

## 8.2 The model

Here we introduce the general idea of our large $M$ expansion. Let $G = (V, F, E)$ be an arbitrary factor graph, with $V$ and $F$ its variable and factor nodes set respectively, and $E$ its edge set. To each node $i \in V$ we associate a variable $x_i$ taking values in a set $X$ having finite or infinite cardinality. To each factor $r \in F$ we associate a random interaction $\psi(x_i : i \in \partial r\,;\, J_r)$, where $J_r$ is some parameter encoding the realization of the disorder. The partition function of the system is given by

$$Z(\mathbf{J}) = \sum_{\mathbf{x}} \prod_{r \in F} \psi(x_i : i \in \partial r\,;\, J_r) \tag{8.1}$$

We then create $M$ copies of our system. We will refer to the different copies as layers and label them with the Greek letters $\alpha, beta = 1, \ldots, M$. Each variable of the new system is then labelled by a couple $(i, \alpha)$, with $i \in V, \alpha \in [M]$. We consider for each layer $\alpha$ a different realization of the disorder $\boldsymbol{J}_\alpha$. The partition function of the whole system is trivially

$$Z(\mathbf{J}_1) \times \cdots \times Z(\mathbf{J}_M) = \sum_{\mathbf{x}_1 \ldots \mathbf{x}_M} \prod_{\alpha=1}^{M} \prod_{r \in F} \psi(x_{i\alpha} : i \in \partial r\,;\, J_{r\alpha}). \tag{8.2}$$

Now we introduce the main ingredient of the construction. For each link $(r, i) \in E$ define a permutation $\pi_r i$ of the labels $\alpha = 1, \ldots, M$ and rewire the links according to these permutations. The partition function of the new system is thus given by

$$Z_M(\{\mathbf{J}_\alpha\}, \boldsymbol{\pi}) = \sum_{\{\mathbf{x}_\alpha\}} \prod_{\alpha} \prod_{r} \psi(x_{i\beta} : i \in \partial r\,, \beta = \pi_{ri}(\alpha)\,;\, J_{r\alpha}). \tag{8.3}$$

We refer to the $M$ graph as to the factor graph underlying the partition function $Z_M$. The free energy of the model is defined as

$$\mathcal{F}_M \equiv -\frac{1}{\beta M} \mathbb{E}_{\boldsymbol{J}} \mathbb{E}_{\boldsymbol{\pi}} \log Z_M(\{\mathbf{J}_\alpha\}, \boldsymbol{\pi}) \tag{8.4}$$

Alternative definitions can be considered for $\mathcal{F}_M$, exchanging quenched averages with annealed averages or sampling the same disorder for each of the $M$ layers.

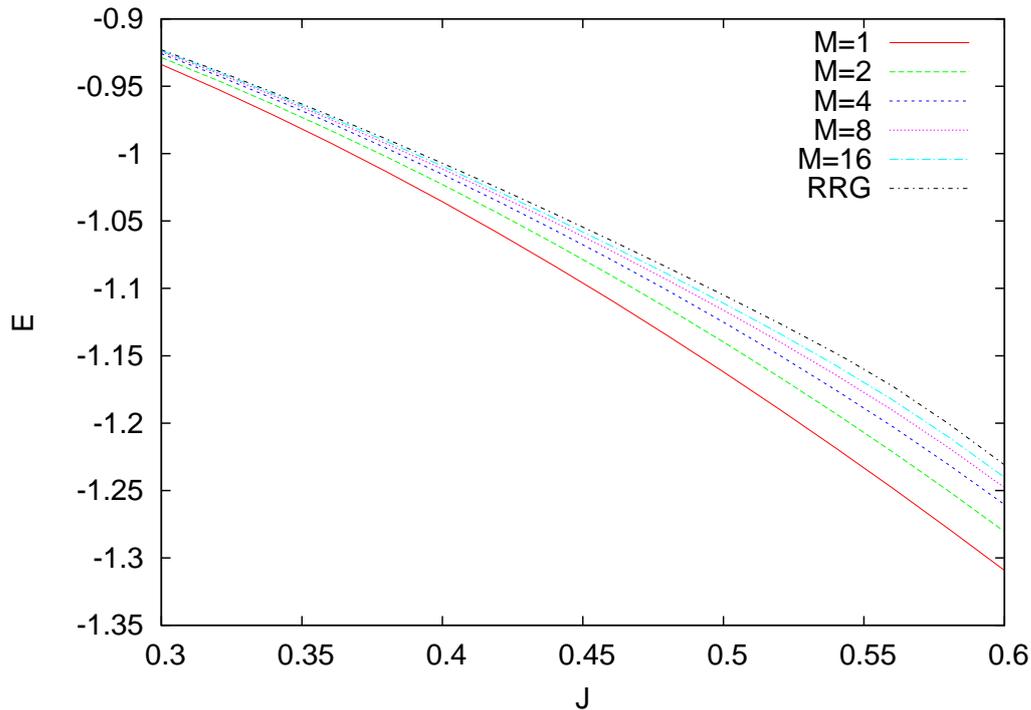

**Fig. 8.2.** Average energy for the RFIM at zero temperature on the square lattice $d = 2$ and $L = 16$. The values extrapolate to the energy of an infinite random regular graph with connectivity $z = 4$.

Setting $M = 1$ in $\mathcal{F}_M$ we trivially obtain the average free energy $\mathcal{F}$ of the original model. More thought is required to understand the $M \uparrow \infty$ limit.

The infinite computational tree of the graph is sort of "unfolded".

We resume the previous considerations in

$$\mathcal{F}_1 = \mathcal{F}, \qquad \lim_{M \to +\infty} \mathcal{F}_M = \mathcal{F}_{\text{Bethe}}. \qquad (8.5)$$

We now assume that $\mathcal{F}_M$ has a regular $1/M$ expansion around the point $M = +\infty$, to write

$$\mathcal{F}_M = \mathcal{F}_{\text{Bethe}} + \sum_{k=1}^{+\infty} A_k \frac{1}{M^k}. \qquad (8.6)$$

We have thus produce a perturbative expansion around the Bethe free energy for any given factor graph (see Fig 8.2). Notice that we did not take a large graph limit in order to produce it. Generally the limit $N \uparrow \infty$ and $M \uparrow \infty$ do not commute.

## 8.3 Replica Calculation

Here we discuss ho to apply the replica method to compute the large $M$ expansion for the Ising model on an arbitrary graph $G = (V, E)$. The Hamiltonian appearing in the random partition function $Z_M$ is given by

$$\mathcal{H}_M[\boldsymbol{\sigma}] = -\sum_{\alpha,\beta=1}^{M} \sum_{(i,j)} C_{ij}^{\alpha\beta} J_{ij}^{\alpha\beta} \sigma_{i\alpha}\sigma_{j\beta} - \sum_{\alpha}\sum_{i} h_i^{\alpha} \sigma_{i\alpha} \qquad (8.7)$$

with *i.i.d.* couplings $J_{ij}^{\alpha\beta}$ and fields $h_i^{\alpha}$. They elements of the adjacency matrix $C_{ij}^{\alpha\beta}$ of the $M$ are induced by the uniformly chosen permutation of the links. Their joint probability distribution is factorized on each edge $(i,j)$, and is given by

$$\mathcal{P}(\mathbf{C}) = \prod_{(i,j)} P(\mathbf{C}_{ij}) \qquad (8.8)$$

with

$$P(\mathbf{C}) = \frac{1}{M!} \sum_{\{C^{\alpha\beta}=0,1\}} \prod_{\alpha} \delta\left(\sum_{\beta} C^{\alpha\beta} = 1\right) \prod_{\beta} \delta\left(\sum_{\alpha} C^{\alpha\beta} = 1\right) \qquad (8.9)$$

We notice that these are exactly the constrains defining a bipartite matching. In order to perform the quenched average over coupling, fields and the realization of the $M$ graph, we have to replicate the system. The replicated Hamiltonian is then given by $\sum_{a=1}^{n} \mathcal{H}_M[\boldsymbol{\sigma}^a]$ and the replicated system will contain $n \times N \times M$ spins $\sigma_{i\alpha}^a$. As usual we will often suppress the replica index from now on. The computation of the replicated partition function is very similar to the one performed for the *RRG* in Appendix B.1, therefore here we will skip some of the passages. After introducing the Lagrange multiplicators $\lambda_{i\to j}^{\alpha}$ to open the Kronecker delta in $\mathcal{P}(\mathbf{C})$ we obtain

$$\overline{Z_M^n} = \frac{1}{(M!)^{|E|}} \sum_{\sigma} \int \prod_{(i\to j)} \prod_{\alpha} \left[\frac{d\lambda_{i\to j}^{\alpha}}{2\pi} e^{-i\lambda_{i\to j}^{\alpha}}\right] \prod_{i,\alpha} e^{H(\sigma_{i\alpha})}$$
$$\times \prod_{(i,j)} \prod_{\alpha,\beta} e^{\log\left(1 + e^{i\lambda_{i\to j}^{\alpha}} U(\sigma_{i\alpha},\sigma_{j\beta}) e^{i\lambda_{j\to i}^{\beta}}\right)}. \qquad (8.10)$$

The functions $U(\sigma,\tau)$ and $H(\sigma)$ are defined by

$$U(\sigma,\tau) = \mathbb{E}_J \, e^{J \sum_{a=1}^{n} \sigma^a \tau^a} \qquad (8.11)$$

$$e^{H(\sigma)} = \mathbb{E}_h \, e^{h \sum_{\alpha=1}^{n} \sigma^{\alpha}} \qquad (8.12)$$

As we will later check we can expand the logarithm just to the first order around one. In fact higher powers are suppressed by the $\lambda$ integrations. Therefore we have

$$\overline{Z_M^n} = \frac{1}{(M!)^{|E|}} \sum_{\sigma} \int \prod_{(i\to j)} \prod_{\alpha} \left[\frac{d\lambda_{i\to j}^{\alpha}}{2\pi} e^{-i\lambda_{i\to j}^{\alpha}}\right] \prod_{i,\alpha} e^{H(\sigma_{i\alpha})}$$
$$\times \prod_{(i,j)} e^{\sum_{\alpha,\beta} e^{i\lambda_{i\to j}^{\alpha}} U(\sigma_{i\alpha},\sigma_{j\beta}) e^{i\lambda_{j\to i}^{\beta}}} \qquad (8.13)$$

Defining for each directed link the function $\rho_{i\to j}(\sigma)$ through

$$\hat{\rho}_{i\to j}(\sigma) = \sum_{\alpha} e^{i\lambda_{i\to j}^{\alpha}} \delta(\sigma - \sigma_{i\alpha}), \qquad (8.14)$$

using the relation

$$e^{\int uUv} = e^{\int \frac{1}{4}(u+v)U(u+v) - \frac{1}{4}\int(u-v)U(u-v)}$$
$$= \int \frac{\mathrm{d}x\,\mathrm{d}y}{\pi \det U} e^{-\int xU^{-1}x - \int yU^{-1}y + \int u(x+iy) + \int v(x-iy)} \quad (8.15)$$

and integrating out the lambdas we obtain

$$\overline{Z_M^n} = \frac{1}{(M!)^{|E|}} \sum_\sigma \int \prod_{(i,j)} \frac{\mathrm{d}\hat{\rho}_{i\to j}\mathrm{d}\hat{\rho}_{j\to i}}{\pi^{2n} \det U} e^{-\int \hat{\rho}_{i\to j}U^{-1}\hat{\rho}_{j\to i}}$$
$$\times \prod_a \hat{\rho}_{i\to j}(\sigma_{i\alpha}) \prod_\beta \hat{\rho}_{j\to i}(\sigma_{j\beta}) \prod_{i,\alpha} e^{H(\sigma_{i\alpha})} \quad (8.16)$$

The integration variable $\hat{\rho}_{i\to j}$ are complex and such that $\hat{\rho}_{i\to j} = \hat{\rho}^*_{j\to i}$. We can now sum over the spin configurations. We make the change of variables $\hat{\rho}_{i\to j}(\sigma) = \sqrt{M}\int U(\sigma,\tau)\rho_{j\to i}(\tau)$. Notice that we inverted the site indexes, to obtain more meaningful message passing equations at the saddle point, as we will later show. The equation for the replicated partition function now reads

$$\overline{Z_M^n} = \left[\frac{M^M M^{2n} \det U}{\pi^{2n} M!}\right]^{|E|} \int \prod_{(i,j)} [\mathrm{d}\rho_{i\to j}\,\mathrm{d}\rho_{j\to i}]\, e^{-MS[\rho]}, \quad (8.17)$$

with the action $S[\rho]$ defined by

$$S[\rho] = \sum_{(i,j)} \int \mathrm{d}\sigma\mathrm{d}\tau\, \rho_{i\to j}(\sigma)\, U(\sigma,\tau)\, \rho_{j\to i}(\tau) - \sum_i \log \int \mathrm{d}\sigma\, e^{H(\sigma)} \prod_{k\in\partial i}[\int U\rho_{k\to i}](\sigma). \quad (8.18)$$

We notice that the action is invariant under a complex rotation of the fields

$$\rho_{i\to j}(\sigma) \to \rho_{i\to j}(\sigma)e^{i\theta_{i\to j}}, \quad (8.19)$$

with $\theta_{i\to j} = \theta_{j\to i}$, since the two field a re complex conjugate. In order to integrate over the orbits we can introduce for each edge the constraints

$$2\pi \int \mathrm{d}r_{ij}\, r_{ij}\, \delta\left(\sum_\sigma \rho_{i\to j}(\sigma) - r_{ij}\right) \delta\left(\sum_\sigma \rho_{j\to i}(\sigma) - r_{ij}\right) \quad (8.20)$$

rescaling the fields $\rho$ by a factor $r_{ij}$ we can integrate out the $r_{ij}$. In fact after the rescaling the $r$ dependent part for each edge is given by

$$2\pi \int \mathrm{d}r_{ij}\, r_{ij}(r_{ij}^2)^{2n-1}\, e^{-Mr_{ij}^2 \int \rho_{i\to j}U\rho_{j\to i}}\, r_{ij}^{2M} = \frac{\pi\Gamma(M+2^n)}{M^{M+2^n}} e^{-(M+2^n)\log \int \rho_{i\to j}U\rho_{j\to i}} \quad (8.21)$$

At this point we finally arrive to

$$\overline{Z_M^n} = \left[\frac{\pi\Gamma(M+2^n)\det U}{\pi^{2n} M!}\right]^{|E|} \int \mathrm{d}\rho\, \delta\left(\sum \rho - 1\right) e^{-MS[\rho]}. \quad (8.22)$$

The integration measure and the delta function in last equation stand for

$$\mathrm{d}\rho\, \delta\left(\sum \rho - 1\right) \equiv \prod_{(i,j)} \left[\mathrm{d}\rho_{i\to j}\mathrm{d}\rho_{j\to i}\, \delta(\sum_\sigma \rho_{i\to j}(\sigma) - 1)\, \delta(\sum_\sigma \rho_{j\to i}(\sigma) - 1)\right]. \quad (8.23)$$

The action $S[\rho]$ is defined by

$$S[\rho] = \left(1 + \frac{2^n}{M}\right) \sum_{(i,j)} \log \int d\sigma d\tau \, \rho_{i \to j}(\sigma) \, U(\sigma, \tau) \, \rho_{j \to i}(\tau)$$
$$- \sum_i \log \int d\sigma \, e^{H(\sigma)} \prod_{k \in \partial i} [\int U \rho_{k \to i}](\sigma) \quad (8.24)$$

The learned reader will recognize at this point that $S[\rho]$ has the form of the Bethe free entropy. we will later see that at the saddle point the function $\rho_{j \to i}(\sigma)$ will correspond to a distribution of cavity fields $h_{j \to i}$.

## 8.4 RS Saddle Point

We proceed to the saddle point evaluation of the action, subject to the constraints $\delta(\sum \rho - 1)$. For the time being we ignore the constraints, we will later see that there exist a set of $\rho_{i \to j}$ satisfying the unconstrained saddle point equation that is compatible with them. Let us define

$$\hat{\rho}_{i \to j}(\sigma) \equiv \int d\tau \, U(\sigma, \tau) \rho_{i \to j}(\tau) \quad (8.25)$$

The saddle point conditions for $M \to \infty$, $\frac{\partial S}{\partial \rho_{j \to i}(\sigma)} = 0$, give

$$\frac{\int U \rho_{i \to j}(\sigma)}{\int \rho_{i \to j} U \rho_{j \to i}} - \frac{\int U e^{H(\sigma)} \prod_{k \in \partial i \setminus j} \hat{\rho}_{k \to i}(\sigma)}{\int e^H \prod_{k \in \partial i} \hat{\rho}_{k \to i}} = 0 \quad (8.26)$$

which can be otherwise stated as

$$\frac{\rho_{i \to j}(\sigma)}{\int \rho_{i \to j} U \rho_{j \to i}} = \frac{e^{H(\sigma)} \prod_{k \in \partial i \setminus j} \hat{\rho}_{k \to i}(\sigma)}{\int e^H \prod_{k \in \partial i} \hat{\rho}_{k \to i}}. \quad (8.27)$$

We see that given a solution $\{\rho_{j \to i}\}$ of last equations, we can multiply each function $\rho_{j \to i}$ for a different arbitrary constant $c_{j \to i}$ and obtain a new solution. This scale invariance is destroyed thanks to the constraints $\sum_\sigma \rho_{j \to i}(\sigma)$ in Eq. (8.22), which, by the same scaling invariance property, can always be met by the unconstrained saddle point. Here we consider the Replica Symmetric approximation. In this case the value of $\rho_{i \to j}(\sigma)$ depends only on the sum of the $\sum_{a=1}^n \sigma^a$. As usual we can parametrize the order parameter at the saddle point with a function $P(h)$:

$$\rho_{i \to j}(\sigma) = \int dh \, P_{i \to j}(h) \frac{e^{\beta h \sum_a \sigma^a}}{(2 \cosh \beta h)^n} \quad (8.28)$$

For each value of $n$ the normalization condition implies $\int dh \, P_{i \to j}(h) = 1$. In the $n \to 0$ limit the saddle point equations (8.27) become the distributional cavity equation

$$P_{i \to j}(h) = \mathbb{E}_H \int \prod_{k \in \partial i \setminus j} dQ_{k \to i}(u_k) \, \delta\Big(h - (H + \sum_{k \in \partial i \setminus j} u_k)\Big)$$
$$Q_{i \to j}(u) = \mathbb{E}_J \int dP_{i \to j}(h) \, \delta\left(u - \tilde{u}(J, h)\right) \quad (8.29)$$

Inserting previous two equation in the action (8.24), we finally obtain

$$\lim_{M \to \infty} \mathcal{F}_M = \mathcal{F}_{\text{Bethe}} \tag{8.30}$$

with

$$-\beta \mathcal{F}_{\text{Bethe}} = \sum_{(i,j)} \phi_{ij} + \sum_i (1 - |\partial i|) \phi_i. \tag{8.31}$$

The edge and site free entropies are defined by

$$\phi_{ij} = \ll \ln\left[4 \cosh(h_{i \to j}) \cosh(h_{j \to i}) \cosh(J_{ij})(1 + \tanh(h_{i \to j}) \tanh(h_{j \to i}) \tanh(J_{ij}))\right] \gg \tag{8.32}$$

and

$$\phi_i = \ll \ln\left[2 \cosh(H_i + \sum_{k \in \partial i} u_{k \to i})\right] \gg \tag{8.33}$$

## 8.5 The 1/M term

The computation in the replica formalism of the $1/M$ term, that is the Gaussian contribution to the action, poses some technical difficulties that we did not manage to overcome yet. The presence of the constraints in the partition function 8.22, ultimately due to the gauge invariance, kill the corresponding directions in the functional space, therefore we have to change accordingly the base in order to compute only the relevant fluctuations. The presence of a Gauge invariance is common to all the problems involving bipartite matching, therefore was not encountered in the ER and RRG finite size corrections of Chapter 4. This problem was encountered also in Ref. [156] and [95] when computing the $1/N$ correction to the random link matching problem. In that case though, they could bypass the problem not including the longitudinal sector contribution in the diagonalization of the Hessian, since that sector is exactly the one suppressed by the constraints. Since we don't want to proceed to the direct diagonalization of the Hessian but we want to retain our physical interpretation for the correction to the free energy in terms of closed and open chains, that road cannot be taken. The approach we plan to take exploits the formal limit

$$\delta\left(\sum_\sigma \rho_{i \to j}(\sigma) - 1\right) = \lim_{\epsilon \to 0} \frac{1}{2\pi\epsilon} e^{-\frac{1}{2\epsilon}\left(\sum_\sigma \rho_{i \to j}(\sigma) - 1\right)^2}. \tag{8.34}$$

Therefore we add some extra Gaussian terms to the action to suppress longitudinal fluctuations. Note that using the integral representation of $\delta\left(\sum_\sigma \rho_{i \to j}(\sigma) - 1\right)$ using a Lagrange multiplier would give null modes in the Hessian as well.

We did not complete yet the replica calculation of the Hessian, therefore, for the time being, we leverage the result obtained in the diluted graphs to obtain the expression for the $1/M$ term we where looking for.

We assume that we deal with a regular lattice in dimension $d$ with periodic boundary conditions. In this case the solution to the saddle point equation is homogeneous,

$$\rho_{i \to j}(\sigma) \equiv \rho(\sigma) \qquad P_{i \to j}(h) \equiv P(h), \tag{8.35}$$

and the asymptotic free energy is exactly the same of RRGs with connectivity $2d$. Writing the expansion of the free energy as

$$\mathcal{F}_M = \mathcal{F}_{\text{Bethe}} + \Delta F \frac{1}{M} + o\left(\frac{1}{M}\right) \tag{8.36}$$

the first sub-leading order term in the expansion has to take the form

$$\Delta F = \sum_{\ell \geq 3} \mathcal{N}_\ell \, \Delta \phi_\ell \tag{8.37}$$

The coefficients $\Delta \phi_\ell$ are exactly the same, qualitatively and quantitatively, as those computed in random regular graphs in Chapter 4.

$$\Delta \phi_\ell = \phi_\ell^c - \ell(\phi_\ell^o - \phi_{\ell-1}^o). \tag{8.38}$$

The coefficient $\mathcal{N}_\ell$, accounting for the average number of loops in the graph, is not $\frac{(2d-1)^\ell}{2\ell}$, as one would expect in the RRG ensemble. In fact one convince himself that the average number of loops at the order $O(1/M)$ in the $M-$graph is given by the number of non-backtracking paths of length $\ell$, both simple and non simple, on the original lattice. If we define the $2|E| \times 2|E|$ non-backtracking matrix $B$, indexed by the directed edges $i \to j$ of the lattice, as

$$B_{k \to l, j \to i} = \delta_{lj}(1 - \delta_{ik}), \tag{8.39}$$

then the coefficient $N_\ell$ is given by

$$\mathcal{N}_\ell = \frac{1}{2\ell} \operatorname{Tr} B^\ell \tag{8.40}$$

In order to compute Eq. (8.37) we approximate the free energy shift $\Delta \phi_\ell$ with the highest eigenvalue in the relevant sector of the replicated transfer matrix (see Chapter 3), that is

$$\Delta \phi_\ell \sim \begin{cases} c_1 \ell \lambda_1^\ell & \text{for the RFIM} \\ c_2 \lambda_2^\ell & \text{for the spin-glass} \end{cases} \tag{8.41}$$

With this approximation, it turns out that the series in Eq. (8.37) can be exactly resummed, since we can compute the generating function for the number of non-backtracking walks, as we show in the next paragraph.

## 8.6 Non-Backtracking Random Walks

Let us first establish some notations for the number of non-backtackig walks in the infinite lattice $Z^d$:

$$\begin{aligned} b_\ell(x) &\equiv \# \text{ NBWs of length } \ell \text{ from 0 to } x \\ \hat{b}_\ell(k) &\equiv \sum_{x \in Z^d} b_\ell(x) \, e^{ikx} \qquad k \in [-\pi, \pi]^d \\ \hat{B}_z(k) &\equiv \sum_{\ell \geq 0} \hat{b}_\ell(k) \, z^\ell \end{aligned} \tag{8.42}$$

Notice that since we are interested in closed paths we have to compute $b_\ell(0)$. We will use a result obtained in Ref. [197] for $\hat{B}_z(k)$, the generating function of the Fourier transform of $b_\ell(x)$. The authors proved the relation

$$\hat{B}_z(k) = \frac{1 - z^2}{1 + (2d-1)z^2 - 2z \sum_{\iota=1}^{d} \cos k_\iota}. \tag{8.43}$$

We now show how to compute the generating function for $\mathcal{N}_\ell$, that we call $A_\ell$. This is related to $\hat{B}_z(k)$ by

$$A(z) = \int \frac{\mathrm{d}^d k}{(2\pi)^d} \hat{B}_z(k) \tag{8.44}$$

Using last two equations, some simple manipulations give

$$\begin{aligned} A(z) &= \int \frac{\mathrm{d}^d k}{(2\pi)^d} (1-z^2) \int_0^{+\infty} \mathrm{d}t \, e^{-t\left(1+(2d-1)z^2 - 2z \sum_{\iota=1}^d \cos k_\iota\right)} \\ &= (1-z^2) \int_0^{+\infty} \mathrm{d}t \, e^{-t\left(1+(2d-1)z^2\right)} [I_0(2zt)]^d \end{aligned} \tag{8.45}$$

where $I_\alpha(x)$ is the modified Bessel function of the first kind of order $\alpha$. In two dimensions we have

$$A(z\,;\,d=2) = (1-z^2) \frac{2}{\pi} \frac{K\left(\frac{16z^2}{(3z^2+1)^2}\right)}{3z^2 + 1} \tag{8.46}$$

where $K(x)$ is the complete elliptic integral of the first kind.

More generally for every dimension $d$ we have

$$A(z) \to \begin{cases} 1 & \text{for } z \to 0 \\ +\infty & \text{for } z \to \frac{1}{2d-1} \end{cases} \tag{8.47}$$

The analytic knowledge of the generating function $A(z)$ through Eq. (8.45), along with the results from the replicated transfer matrix theory of Chapter 3 that yield Eq. (8.41), allow to approximate the infinite sum of Eq. (8.37) to a high degree of precision.

## 8.7 Conclusions

In this Chapter we proposed a new perturbative approach around the Bethe approximation for finite dimensional disordered systems. The expansion parameter is the number $M$ of layers in a auxiliary random lattice models. For $M = +\infty$ we obtain an infinite RRG graph, to which all the machineries of Cavity and Replica Methods apply. The first contribution to the the free energy in a $1/M$ expansion is computed, although the replica computations poses some problems yet to be addressed. The result for the coefficient $\Delta F$ of the $O(1/M)$ term

$$\Delta F = \sum_{\ell \geq 3} \mathcal{N}_\ell \, \Delta \phi_\ell, \tag{8.48}$$

appears to be sound, motivated by a cavity argument and by the results of Chapter 4. Hopefully we will soon address the difficulties in the replica computation and investigate the analytical and numerical consequences of our approach.

# Part V

# Conclusions

# Conclusions

Along this thesis I discussed many different topics, strictly or loosely related. Here I resume and discuss briefly the results we achieved, trying to underline their common ground, and I invite the reader to refer to the conclusions of the various chapters for more specific discussions.

One research line bridges the work on finite size corrections in diluted random graphs of Chapter 4 and Refs. [2] [3], with the large $M$ expansion around the Bethe approximation that we proposed in Chapter 8. We can express the result we derived for the coefficient $\Delta f$ of the $1/N$ and the $1/M$ terms in diluted random graphs and lattice models respectively through the unique formula

$$\Delta f = \sum_{\ell \geq 3} \mathcal{N}_\ell \, \Delta \phi_\ell. \tag{8.49}$$

This formula relates the first correction to the Bethe free energy to the presence of simple (i.e. non-intersecting) loops. The factor $\phi_\ell$ here is the free energy difference due to the addition of a closed chain of length $\ell$ to an infinite tree. The coefficient $\mathcal{N}_\ell$ is a combinatorial factor accounting for the average number of simple loops of length $\ell$. It takes value $\tilde{z}^\ell/(2\ell)$, with $\tilde{z}$ the average residual degree, while in the large $M$ expansion for finite dimensional lattices is the number of closed non-backtracking paths of length $\ell$ in the lattice. The infinite sum in last equation can be computed with high degree of precision thanks to the replicated transfer matrix formalism we developed in Chapter 3 and Ref. [1]. A short term goal is to fix some issues, due to gauge symmetries, in the replica computation for the large $M$ expansion, thus supporting with the replica formalism the result obtained through the probabilistic argument. In mid/long term perspective it would be desirable to find a general rule to express the free energy shift of a loopy structure to an infinite graph. An accurate comparison with Loop Calculus of Cherniak and Chertkov [167] is also needed. Finally, the large $M$ formalism could help to shed some light on phenomena, such as the breaking of dimensional reduction in the RFIM [56], normally not captured by perturbative theories [57].

Another line of research is dedicated to the Euclidean Matching Problem, both in its monopartite (Chapter 7) and bipartite versions (Chapter 6 and Ref. [5]). As we discuss in depth in Chapter 6, bipartiteness as deep consequences for the optimal matching and the scaling behaviour of its cost, therefore the two problems require separate analysis.

In the bipartite case, through a continuous approximation for this discrete optimization problem, we propose a (stochastic) Poisson equation relating a transport field and the density differences among the two set of points. The average optimal cost can then be expressed as an infinite series, diverging for $d \geq 2$, that we manage to regularize with an analytic continuation. In the end, in the case of quadratic costs, the following original results:

$$e_d(N) \sim \begin{cases} \frac{1}{6}N + e_1 & d = 1, \\ \frac{1}{2\pi} \ln N + e_2 & d = 2, \\ e_d + \frac{\zeta_d(1)}{2\pi^2} N^{-\gamma_d} & d > 2, \end{cases} \qquad (8.50)$$

with $\gamma_d = \frac{d-2}{d}$. We see that the effects of set-to-set fluctuations captured by our approximation is dominant for low dimension, while for $d > 2$ gives an anomalous subleading scaling. Noticeably in our formalism the intensive $O(1)$ term can not be predicted in any dimension. We hope that similar techniques could be applied to other optimization problems as well.

The approach to the monopartite problem in Chapter 7 is completely different. It is based on a replica computation that goes beyond the simple mean field approximation including the correlation among the weights of the link in simple loops. The expression we obtain is similar to Eq. (8.49). In fact, in relation to the $O(1/N)$ and $O(1/M)$ expansion discussed above, analyzing the $O(1/N)$ finite size in the random link matching problem we managed to show a one-to-one correspondence between the terms appearing in the fully-connected model and the ones in Euclidean model, ultimately relating all to the presence of small simple loops. This Chapter represent another brick establishing the connection between finite size corrections in mean field models and perturbative expansions in finite dimensional systems. The perspective here as well is to compute the contribution of more complex loopy structures and to extract from the series the scaling of finite size corrections.

Throughout these works the results for the Replicated Transfer Matrix of Chapter 3 and Ref. [1] proved to be valuable conceptual, analytical and numerical tools. It is desirable to extend the formalism to the 1RSB scenario and to p-spin models.

# Appendices

# Appendix A

# Appendix to Chapter 4.2

## A.1 Coefficients $a_L$

The coefficients $a_L$ can be determined by computing recursively all the functions $S_L$. A little bit of thought should convince oneself that $S_L$ is given by the probability of the event $E_L \equiv \{\text{L consecutive links are present}\}$, by splitting it in at least two smaller events $E_{L_1}$ and $E_{L_2}$ with $L_1 + L_2 = L$. Since $p_L = \text{Prob}[E_L]$ it should be clear that $q_L \equiv p_L - S_L$ is like a "connected" probability to obtain the $L$ links from a unique structure. It is not hard to derive a recursive equation for the functions $S_L$, valid for any $L \geq 2$:

$$S_L = q_1 p_{L-1} + q_2 p_{L-2} + \cdots + q_{L-1} p_1 , \tag{A.1}$$

from which we get, for any $L \geq 1$,

$$p_L(1 + q_0) = \sum_{k=0}^{L} q_k p_{L-k} , \tag{A.2}$$

where $q_0 \equiv 0$ and $p_0 = 1$ thanks to the fact that $p_L = z^L[1 - L(L+1)/2N]$. The above equation can be easily solved by introducing the generating functions:

$$\begin{aligned} p(x) &\equiv \sum_{k=0}^{\infty} p_k x^k , \\ q(x) &\equiv \sum_{k=0}^{\infty} q_k x^k , \end{aligned} \tag{A.3}$$

that must satisfy

$$(1 + q_0)[p(x) - p_0] = q(x)p(x) - q_0 p_0 \implies q(x) = p_0 - \frac{1}{p(x)} . \tag{A.4}$$

Keeping only terms up to order $1/N$ the result is

$$q(x) = zx - \frac{1}{N}\frac{zx}{1 - zx} , \tag{A.5}$$

implying
$$q_1 = z\left(1 - \frac{1}{N}\right) \quad q_k = -\frac{z^k}{N} \quad \text{for } k \geq 2 \ . \tag{A.6}$$

Rewriting eq.(A.2) as
$$q_L = \sum_{k=L}^{\infty}(k - L + 1)a_k + \sum_{k=L+1}^{\infty} c_k \ , \tag{A.7}$$

we can obtain
$$q_L - q_{L+1} = \sum_{k=L}^{\infty} a_k + c_{L+1} \implies \sum_{k=L}^{\infty} a_k = q_L = -\frac{z^L}{N} \quad \forall L \geq 2 \ , \tag{A.8}$$

by noticing that $c_L = -q_L$ for $L \geq 3$. Moreover, for $L = 1$ we have
$$\sum_{k=1}^{\infty} a_k = q_1 - q_2 = z + \frac{z^2 - z}{N} \ . \tag{A.9}$$

In conclusion the coefficients $a_L$ are given by the following expressions
$$\begin{aligned} a_1 &= z + \frac{1}{N}(2z^2 - z) \ , \\ a_L &= \frac{1}{N}(z^{L+1} - z^L) \ \text{ for } L \geq 2 \ . \end{aligned} \tag{A.10}$$

## A.2   Combinatorics of $\text{Tr}\left(T^L\right)$

Here we prove eq. (4.23), relating $\text{Tr}\left(T^L\right)$ in the small $n$ limit to the free energies of open and closed cavity chains. Let's rewrite our $2^n \times 2^n$ matrix as:
$$\begin{aligned} T(\sigma, \sigma') =& z\, \mathbb{E}\Bigg\{\frac{1}{[2\text{ch}(\beta h)]^n}\bigg[e^{J \sum_a \sigma_a \sigma'_a + h \sum_a \sigma'_a} \\ & - \frac{1}{[2\text{ch}(\beta h')]^n}\left(\int d\tau\, e^{J \sum_a \sigma_a \tau_a + h \sum_a \tau_a}\right) e^{h' \sum_a \sigma'_a}\bigg]\Bigg\}. \end{aligned} \tag{A.11}$$

where expectation is taken over the coupling $J$ and the cavity fields $h, h'$, which are distributed according to the solution of eq. (4.16). We immediatly note that the factor $[2\text{ch}(\beta h)]^n$ reduces to $1 + n \log 2\text{ch}(\beta h) + o(n)$, in the small $n$ limit, allowing to rewrite $T(\sigma, \sigma')$, with $o(n)$ accuracy, as:
$$\begin{aligned} T(\sigma, \sigma') =& z\, \mathbb{E}\left[e^{J \sum_a \sigma_a \sigma'_a + h \sum_a \sigma'_a} - \left(\int d\tau\, e^{J \sum_a \sigma_a \tau_a + h \sum_a \tau_a}\right) e^{h' \sum_a \sigma'_a}\right] \\ & + nz\mathbb{E}_h \log 2\text{ch}(\beta h) + o(n) \ . \end{aligned} \tag{A.12}$$

We recognize that the term $\mathbb{E}[e^{J \sum_a \sigma_a \sigma'_a + h \sum_a \sigma'_a}]$ is the replicated transfer matrix of a one-dimensional chain, and so when we take the trace of $T(\sigma, \sigma')$ we simply get:
$$\text{Tr}[T] = -n\beta z[\phi_1^c - \left(\phi_1^a + \beta^{-1}\mathbb{E}_h \log 2\text{ch}(\beta h)\right)] + o(n) \ . \tag{A.13}$$

When computing the trace of $T^L$, for $L \geq 2$, the term $nz\mathbb{E}_h \log 2\mathrm{ch}(\beta h)$ in eq. (A.12) gives only contributions of order $o(n)$ and thus can be completely neglected in the following calculation. Let's call $A = \mathbb{E}[e^{J\sum_a \sigma_a \sigma'_a + h\sum_a \sigma'_a}]$ and $B = \mathbb{E}[(\int d\tau \, e^{J\sum_a \sigma_a \tau_a + h\sum_a \tau_a}) e^{h\sum_a \sigma'_a}]$. The product $T^L$ is formed of a linear combination of all the possible products of $L$ matrices chosen between $A$ and $B$, therefore we now consider the traces of such products. A simple inspection shows immediately the $\mathrm{Tr}\left(A^L\right)$ is nothing else that the replicated partition function of a cavity loop, that is a closed chain of length $L$ embedded in a locally tree-like random graph. Consider instead a term with one insertion of the matrix $B$, $\mathrm{Tr}\, A \ldots ABA \ldots$. Since $B$ is factorized, its insertion prevents the closure of the chain and we obtain the replicated partition function of an open cavity chain of length $L$. Generalizing the argument we can see that the trace of a product containing $k$ matrices $B$ yields the product of $k$ replicated partition functions of open chains, whose total lengths adds up to $L$. Since in the $n \downarrow 0$ limit products of partition functions become the sum of free energies, we can write

$$\frac{\partial}{\partial n} \mathrm{Tr}\left(T^L\right) = -\beta z^L \Big[\, \phi_L^c + \sum_{l=1}^L b_l \, \phi_l^a \,\Big] + o(1) \,, \qquad (\mathrm{A}.14)$$

where the coefficients $b_l$ have to be determined. It is easy to see that $b_L = -L$ and $b_{L-1} = L$, while a simple combinatoric argument gives the remaining coefficients. We can construct an open chain of length $l < L-1$ in the first $l+1$ positions of the product and than multiply for the $L$ possible ways of obtaining the same trace. So we consider products of the form $BA^{l-1}B \times \{2^{L-l-1}$ *different combinations of A and B*$\}$. Taking into account the number of insertions of $B$ in the last $L-l-1$ positions we obtain

$$b_l = L \times \sum_{k=0}^{L-l-1} (-1)^k \binom{L-l-1}{k} = 0 \qquad for \quad l < L-1 \,, \qquad (\mathrm{A}.15)$$

which immediately yields eq. (4.23).

# Appendix B

# Appendix to Chapter 4.3

## B.1 Integral representation of $[Z^n]_{\text{av}}$

The average of the replicated partition function of the model reads:

$$[Z^n]_{\text{av}} = \left[\sum_{\{\sigma\}} \left(\prod_{i<j} \exp\left(\beta J_{ij} C_{ij} \sum_{a=1}^n \sigma_i^a \sigma_j^a\right) \prod_i \exp\left(\beta H_i \sum_{a=1}^n \sigma_i^a\right)\right)\right]_{\text{av}}, \quad \text{(B.1)}$$

where the sum has to be taken over the replicated spins $\{\sigma_i^a\}$ for $i = 1, \ldots, N$ and $a = 1, \ldots, n$. The average $[\ \cdot\ ]_{\text{av}}$ has to be performed over the RRG ensemble, the couplings $J_{ij}$ and the random fields $H_i$. The graph ensemble is defined by the probability of sampling one of its element, having adjacency matrix $C_{ij}$. This is given by

$$\mathcal{P}(C) = \frac{1}{\mathcal{N}} \prod_{i<j} \left[\left(1 - \frac{c}{N}\right)\delta(C_{ij}) + \frac{c}{N}\delta(C_{ij} - 1)\right] \prod_i \delta\left(\sum_{j\neq i} C_{ij} - c\right). \quad \text{(B.2)}$$

Here the variable $c$ is the connectivity of the nodes, the weighting factors $1 - \frac{c}{N}$ and $\frac{c}{N}$ have been chosen for convenience and $\mathcal{N}$ is a normalization factor. Let us define

$$\begin{aligned} U(\sigma_i, \sigma_j) &= \mathbb{E}_J\left(e^{\beta J \sum_a \sigma_i^a \sigma_j^a}\right), \\ B(\sigma_i) &= \log \mathbb{E}_H\left(e^{\beta H \sum_a \sigma_i^a}\right). \end{aligned} \quad \text{(B.3)}$$

The arguments of the functions $U(\sigma_i, \sigma_j)$ and $B(\sigma_i)$ indicate the replicated spins $\sigma_i \equiv (\sigma_i^1, \ldots, \sigma_i^n)$. When it can cause confusion, the replica label $a$ will be explicitly written.

After averaging over the graph ensemble [198], Eq. (B.1) takes the following form:

$$[Z^n]_{\text{av}} = \frac{1}{\mathcal{N}} \sum_{\{\sigma\}} \int \left(\prod_i \mathrm{d}\lambda_i\ e^{-i\lambda_i c + B(\sigma_i)}\right) \exp\left\{\sum_{i<j} \log\left[1 + \frac{c}{N}\left(e^{i\lambda_i} U(\sigma_i, \sigma_j) e^{i\lambda_j} - 1\right)\right]\right\}, \quad \text{(B.4)}$$

where we used the integral representations of the Kronecker $\delta$-functions appearing in $\mathcal{P}(C)$:

$$\prod_i \delta\left(\sum_j C_{ij} - c\right) = \int_0^{2\pi} \prod_i \frac{\mathrm{d}\lambda_i}{2\pi} \, e^{i\lambda_i \left(\sum_{j\neq i} C_{ij} - c\right)}. \tag{B.5}$$

We expand the argument of the exponential in Eq. (B.4) to obtain

$$\begin{aligned}
\sum_{i<j} \log\left[1 + \frac{c}{N}\left(e^{i\lambda_i} U(\sigma_i, \sigma_j) e^{i\lambda_j} - 1\right)\right] &= \left(\frac{c}{2N} + \frac{c^2}{2N^2}\right)\left[\sum_{ij} e^{i\lambda_i} U(\sigma_i, \sigma_j) e^{i\lambda_j}\right] \\
&\quad - \frac{c}{2N}\left[\sum_i e^{2i\lambda_i} U(\sigma_i, \sigma_i)\right] \\
&\quad - \frac{c^2}{4N^2}\left[\sum_{ij} e^{2i\lambda_i} U^2(\sigma_i, \sigma_j) e^{2i\lambda_j}\right] \\
&\quad + A(N,c) + O\left(\frac{1}{N}\right),
\end{aligned} \tag{B.6}$$

where the constant $A(N,c)$ is given by

$$A(N,c) = -\frac{cN}{2} + \frac{c}{2} - \frac{c^2}{4}. \tag{B.7}$$

In order to compute the sum over the spin variables in Eq. (B.4), we need to decouple the sites. Site factorization can be achieved by introducing two functions $\rho_1(\sigma)$ and $\rho_2(\sigma)$, defined as

$$\begin{aligned}
\rho_1(\sigma) &= \frac{1}{N} \sum_i e^{i\lambda_i} \prod_a \delta(\sigma^a - \sigma_i^a), \\
\rho_2(\sigma) &= \frac{1}{N} \sum_i e^{2i\lambda_i} \prod_a \delta(\sigma^a - \sigma_i^a),
\end{aligned} \tag{B.8}$$

if we are interest only in the first correction in $\frac{1}{N}$. To obtain higher orders in the expansion we would need up to $c$ different functions $\rho_k(\sigma)$, as will be clear from what follows. Using $\rho_1(\sigma)$ and $\rho_2(\sigma)$, along with the expansion (B.6), Eq. (B.4) becomes

$$\begin{aligned}
[Z^n]_{\mathrm{av}} \sim \exp\bigg\{&\frac{Nc}{2}\left(1 + \frac{c}{N}\right)\int \mathrm{d}\sigma\,\mathrm{d}\tau\,\rho_1(\sigma) U(\sigma,\tau)\rho_1(\tau) \\
&+ N\log\int\frac{\mathrm{d}\lambda}{2\pi}\,\mathrm{d}\sigma\,\exp\left[-ic\lambda + B(\sigma) - \frac{c}{2N}e^{2i\lambda} U(\sigma,\sigma)\right] \\
&- \frac{c^2}{4}\int \mathrm{d}\sigma\,\mathrm{d}\tau\,\rho_2(\sigma) U^2(\sigma,\tau)\rho_2(\tau) + A(N,c) - \log\mathcal{N}\bigg\},
\end{aligned} \tag{B.9}$$

up to the order $O(1)$. In the previous equation the notation $\int \mathrm{d}\sigma$ stands for:

$$\int \mathrm{d}\sigma \equiv \prod_{a=1}^n \sum_{\sigma^a=\pm 1}. \tag{B.10}$$

We note also that the function $U(\sigma, \sigma)$, evaluated on the same first and second arguments, is independent on $\sigma$. To lighten the notation we then define:

$$U(\sigma, \sigma) = \mathbb{E}_J e^{n\beta J} \equiv U_0. \tag{B.11}$$

Proceeding in the calculation, we introduce two $\delta$-functionals to enforce the definitions of $\rho_1(\sigma)$ and $\rho_2(\sigma)$:

$$1 = \int \mathcal{D}\rho_k \, \delta\left[\rho_k(\sigma) - \frac{1}{N}\sum_i e^{ik\lambda_i} \prod_a \delta(\sigma^a - \sigma_i^a)\right] \quad \text{for} \quad k = 1, 2. \tag{B.12}$$

The functional measure $\mathcal{D}\rho$ is defined as $\mathcal{D}\rho \equiv \prod_{\sigma \in \mathbb{R}^n} d\rho(\sigma)$. Moreover, we use the following integral representation of the $\delta$-functional:

$$\delta[\rho] = \int \mathcal{D}\hat{\rho} \, e^{-\int d\sigma \, \rho(\sigma)\hat{\rho}(\sigma)}, \tag{B.13}$$

where $\mathcal{D}\hat{\rho} \equiv \prod_{\sigma \in \mathbb{R}^n} \frac{d\hat{\rho}(\sigma)}{2\pi}$, to rewrite the replicated partition function (B.9) as

$$
\begin{aligned}
[Z^n]_{\text{av}} \sim \int \left(\prod_{k=1}^{2} \mathcal{D}\rho_k \mathcal{D}\hat{\rho}_k\right) \exp\Bigg\{ &- N \int d\sigma \, \rho_1(\sigma)\hat{\rho}_1(\sigma) \\
&+ \frac{Nc}{2}\left(1 + \frac{c}{N}\right) \int d\sigma \, d\tau \, \rho_1(\sigma) U(\sigma, \tau) \rho_1(\tau) \\
&- \int d\sigma \, \rho_2(\sigma)\hat{\rho}_2(\sigma) - \frac{c^2}{4} \int d\sigma \, d\tau \, \rho_2(\sigma) U^2(\sigma, \tau) \rho_2(\tau) \\
&+ N \log \int d\sigma \, e^{B(\sigma)} \int \frac{d\lambda}{2\pi} \, \exp\left[\hat{\rho}_1(\sigma)e^{i\lambda} - ic\lambda + \frac{e^{2i\lambda}}{N}\left(\hat{\rho}_2(\sigma) - \frac{cU_0}{2}\right)\right] \\
&+ A(N, c) - \log \mathcal{N} \Bigg\}.
\end{aligned}
\tag{B.14}
$$

We now can carry out the $\lambda$ integration, expanding the exponential and obtaining

$$
\int_0^{2\pi} \frac{d\lambda}{2\pi} \, \exp\left[\hat{\rho}_1(\sigma)e^{i\lambda} - ic\lambda + \frac{e^{2i\lambda}}{N}\left(\hat{\rho}_2(\sigma) - \frac{cU_0}{2}\right)\right] \\
= \sum_{m=0}^{c/2} \frac{1}{m!(c-2m)!} \frac{\hat{\rho}_1(\sigma)^{c-2m}}{N^m} \left(\hat{\rho}_2(\sigma) - \frac{cU_0}{2}\right)^m.
\tag{B.15}
$$

In the sum on the r.h.s. of Eq. (B.15) we retain only the leading terms, corresponding to $m = 0$ and $m = 1$. The partition function (B.14) now reads

$$[Z^n]_{\text{av}} \sim \int \left(\prod_{k=1}^{2} \mathcal{D}\rho_k \mathcal{D}\hat{\rho}_k\right) \exp\left\{-N \int d\sigma\, \rho_1(\sigma)\hat{\rho}_1(\sigma)\right.$$
$$+ \frac{Nc}{2}\left(1 + \frac{c}{N}\right)\int d\sigma\, d\tau\, \rho_1(\sigma)U(\sigma,\tau)\rho_1(\tau) - \int d\sigma\, \rho_2(\sigma)\hat{\rho}_2(\sigma)$$
$$- \frac{c^2}{4}\int d\sigma\, d\tau\, \rho_2(\sigma)U^2(\sigma,\tau)\rho_2(\tau) + N\log\int d\sigma\, e^{B(\sigma)}\hat{\rho}_1(\sigma)^c$$
$$\left. + c(c-1)\frac{\int d\sigma e^{B(\sigma)}\hat{\rho}_1(\sigma)^{c-2}\left(\hat{\rho}_2(\sigma) - \frac{cU_0}{2}\right)}{\int d\sigma\, e^{B(\sigma)}\hat{\rho}_1(\sigma)^c} + A(N,c) - \log\mathcal{N} - N\log c!\right\}.$$
(B.16)

At this point it is natural to define a new field $r(\sigma)$ as

$$r(\sigma) = c^2 \frac{\int d\sigma\, e^{B(\sigma)}\hat{\rho}_1(\sigma)^{c-2}}{\int d\sigma\, e^{B(\sigma)}\hat{\rho}_1(\sigma)^c}, \qquad (B.17)$$

and to observe that the integral over $\hat{\rho}_2(\sigma)$ gives

$$\int \mathcal{D}\hat{\rho}_2 \exp\left\{-\int d\sigma\hat{\rho}_2(\sigma)\left[\rho_2(\sigma) - \frac{c-1}{c}r(\sigma)\right]\right\} = \delta\left[\rho_2 - \frac{c-1}{c}r\right]. \qquad (B.18)$$

Integrating out also $\rho_2$, we obtain

$$[Z^n]_{\text{av}} \sim \int \mathcal{D}\rho_1 \mathcal{D}\hat{\rho}_1 \exp\left\{-N\int d\sigma\, \rho_1(\sigma)\hat{\rho}_1(\sigma) + \frac{Nc}{2}\left(1 + \frac{c}{N}\right)\int d\sigma\, d\tau\, \rho_1(\sigma)U(\sigma,\tau)\rho_1(\tau)\right.$$
$$+ N\log\int d\sigma\, e^{B(\sigma)}\hat{\rho}_1(\sigma)^c - \frac{(c-1)^2}{4}\int d\sigma\, d\tau\, r(\sigma)U^2(\sigma,\tau)r(\tau)$$
$$\left. - \frac{(c-1)U_0}{2}\int d\sigma\, r(\sigma) + A(N,c) - \log\mathcal{N} - N\log c!\right\}.$$

The integral over $\rho_1$ is Gaussian and can be performed explicitly and we get:

$$[Z^n]_{\text{av}} \sim [\det(cU)]^{-1/2} \int \mathcal{D}\hat{\rho}_1 \exp\left\{-\frac{Nc}{2}\int d\sigma\, d\tau\, \hat{\rho}_1(\sigma)U^{-1}(\sigma,\tau)\hat{\rho}_1(\tau)\right.$$
$$+ N\log\int d\sigma\, e^{B(\sigma)}\hat{\rho}_1(\sigma)^c + \frac{1}{2}\int d\sigma\, d\tau\, \hat{\rho}_1(\sigma)U^{-1}(\sigma,\tau)\hat{\rho}_1(\tau)$$
$$- \frac{(c-1)^2}{4}\int d\sigma\, d\tau\, r(\sigma)U^2(\sigma,\tau)r(\tau) - \frac{(c-1)U_0}{2}\int d\sigma\, r(\sigma)$$
$$\left. + A(N,c) - \log\mathcal{N} - N\log c!\right\}.$$
(B.19)

We make the following change of variables, introducing at last the order parameter $\rho(\sigma)$ through

$$\hat{\rho}_1(\sigma) = c\int d\sigma\, U(\sigma,\tau)\rho(\tau), \qquad (B.20)$$

which redefines also the field $r(\sigma)$ as

$$r(\sigma) = \frac{e^{B(\sigma)}\left[\int \mathrm{d}\tau\ U(\sigma,\tau)\rho(\tau)\right]^{c-2}}{\int \mathrm{d}\sigma\ e^{B(\sigma)}\left[\int \mathrm{d}\tau\ U(\sigma,\tau)\rho(\tau)\right]^{c}}. \tag{B.21}$$

The computation of the factor $\log \mathcal{N}$ can be done among the same lines of the preceding derivation, and gives [198]:

$$\log \mathcal{N} \sim N(c \log c - \log c! - c) + \frac{c}{2} + \frac{1}{4} - \frac{\log 2}{2}. \tag{B.22}$$

Calling for brevity $\mathcal{A}(N,c)$ the quantity

$$\begin{aligned}\mathcal{A}(N,c) &= A(N,c) + Nc \log c - N \log c! - \log \mathcal{N} \\ &= \frac{Nc}{2} - \frac{c^2+1}{4} + \frac{\log 2}{2},\end{aligned} \tag{B.23}$$

we obtain the final expression of the $[Z^n]_{\mathrm{av}}$ up to the order $O(1)$, that is

$$[Z^n]_{\mathrm{av}} \sim [\det(cU)]^{1/2} e^{\mathcal{A}(N,c)} \int \mathcal{D}\rho\ e^{-N\mathcal{S}_0[\rho] - \mathcal{S}_1[\rho]}. \tag{B.24}$$

Here the integration measure is given by $\mathcal{D}\hat{\rho} \equiv \prod_{\sigma \in \mathbb{R}^n} \frac{\mathrm{d}\rho(\sigma)}{\sqrt{2\pi}}$, and the functionals $\mathcal{S}_0[\rho]$ and $\mathcal{S}_1[\rho]$ read

$$\begin{aligned}\mathcal{S}_0[\rho] &= \frac{c}{2}\int \mathrm{d}\sigma\,\mathrm{d}\tau\ \rho(\sigma)U(\sigma,\tau)\rho(\tau) - \log \int \mathrm{d}\sigma\ e^{B(\sigma)}\left[\int \mathrm{d}\tau\ U(\sigma,\tau)\rho(\tau)\right]^c, \\ \mathcal{S}_1[\rho] &= -\frac{c^2}{2}\int \mathrm{d}\sigma\,\mathrm{d}\tau\ \rho(\sigma)U(\sigma,\tau)\rho(\tau) + \frac{(c-1)^2}{4}\int \mathrm{d}\sigma\,\mathrm{d}\tau\ r(\sigma)U^2(\sigma,\tau)r(\tau) \\ &\quad + \frac{(c-1)U_0}{2}\int \mathrm{d}\sigma\ r(\sigma).\end{aligned} \tag{B.25}$$

## B.2 Computing the finite size corrections

There are two sources for the $1/N$ finite size corrections to the thermodynamic free energy density. The first contribution comes from the subleading part of the replicated action $\mathcal{S}_1[\rho]$, evaluated in the saddle point solution $\rho_*$, given in eq. (4.57). The second one stems from the Gaussian integral obtained by expanding the leading action $\mathcal{S}_0[\rho]$ around the saddle point. This is given by

$$[\det(cU)]^{1/2} \int \mathcal{D}\chi\ e^{-\frac{1}{2}\int \chi\, \partial^2 \mathcal{S}_0^*\, \chi} = e^{-\frac{1}{2}\log\det(\mathbb{I} - \Sigma)}, \tag{B.26}$$

where $\chi(\sigma)$ is the rescaled fluctuation of $\rho(\sigma)$ around the saddle point and we omitted the dependence on the replicated spins in the exponent of the l.h.s.. The matrix $\Sigma(\sigma,\tau)$ is defined as

$$\begin{aligned}\Sigma(\sigma,\tau) &= (c-1)T(\sigma,\tau) - c\left[\int \mathrm{d}\sigma'\ U(\sigma,\sigma')\rho_*(\sigma')\right]\rho_*(\tau), \\ T(\sigma,\tau) &= U(\sigma,\tau)r_*(\tau).\end{aligned} \tag{B.27}$$

Summing up all the contributions, the averaged replicated partition function $[Z^n]_{\text{av}}$ up to order $O(1)$ becomes:

$$[Z^n]_{\text{av}} \sim \exp\left[\mathcal{A}(N,c) - N\mathcal{S}_0[\rho_*] - \mathcal{S}_1[\rho_*] + \frac{1}{2}\sum_{\ell=1}^{\infty}\frac{1}{\ell}\operatorname{Tr}\Sigma^\ell\right]. \tag{B.28}$$

The introduction of the auxiliary matrix $T(\sigma,\tau)$ will simplify a lot the computation of the trace appearing in Eq. (B.28). The crux is to observe that $\rho_*(\sigma)$ is a left eigenvector of $T(\sigma,\tau)$ with eigenvalue 1, i.e.

$$\int \mathrm{d}\sigma\; \rho_*(\sigma)T(\sigma,\tau) = \rho_*(\tau). \tag{B.29}$$

This property can be verified by acting on the left with $T(\sigma,\tau)$ on the saddle point equation (4.57).

It is useful to define also an auxiliary function $\hat{\rho}_*(\sigma)$ as follows:

$$\hat{\rho}_*(\sigma) = \int \mathrm{d}\tau\; U(\sigma,\tau)\rho_*(\tau). \tag{B.30}$$

As a consequence of the saddle point equation (4.57), the function $\hat{\rho}_*(\sigma)$ has the following interesting property:

$$\int \mathrm{d}\sigma\; \hat{\rho}_*(\sigma)\rho_*(\sigma) = 1. \tag{B.31}$$

Using the definition of $T(\sigma,\tau)$ and $\hat{\rho}_*(\sigma)$, the matrix $\Sigma(\sigma,\tau)$ can be cast in a simpler form, that reads

$$\Sigma(\sigma,\tau) = (c-1)T(\sigma,\tau) - c\hat{\rho}_*(\sigma)\rho_*(\tau). \tag{B.32}$$

The two matrices $T(\sigma,\tau)$ and $\hat{\rho}_*(\sigma)\rho_*(\tau)$ in the r.h.s of Eq. (B.32) they do commute with each other, as can be checked by inspection. Therefore, the trace of the $\ell$-th power of the matrix $\Sigma$ can be written as

$$\operatorname{Tr}\Sigma^\ell = \sum_{k=0}^{\ell}\binom{\ell}{k}(c-1)^{\ell-k}(-c)^k \operatorname{Tr}\left[(\hat{\rho}_*\rho_*)^k T^{\ell-k}\right]. \tag{B.33}$$

Observing that in all the terms of the sum, but the one corresponding to $k=0$, the matrix $T$ is multiplied on the left by its left eigenvector with unitary eigenvalue, we easily get the following result:

$$\operatorname{Tr}\left[(\hat{\rho}_*\rho_*)^k T^{\ell-k}\right] = \begin{cases}\operatorname{Tr} T^\ell & \text{for}\; k=0 \\ 1 & \text{for}\; k\neq 0.\end{cases} \tag{B.34}$$

We can now immediately evaluate Eq. (B.33) and we find:

$$\operatorname{Tr}\Sigma^\ell = (c-1)^\ell\left[\operatorname{Tr} T^\ell - 1\right] + (-1)^\ell. \tag{B.35}$$

Inserting Eq. (B.35) into Eq. (B.28) we get

$$[Z^n]_{\text{av}} \sim \exp\left[-N\mathcal{S}_0[\rho_*] - \mathcal{S}_1[\rho_*] + \frac{1}{2}\sum_{\ell=1}^{\infty}\frac{(c-1)^\ell}{\ell}\left[\operatorname{Tr} T^\ell - 1\right] + \mathcal{A}(N,c) - \frac{\log 2}{2}\right]. \tag{B.36}$$

The term $\mathcal{S}_1[\rho_*]$ can be expressed using the matrix $T$ in the following way:

$$\mathcal{S}_1[\rho_*] = \frac{c^2+1}{4} + \frac{(c-1)^2}{4}\left[\operatorname{Tr} T^2 - 1\right] + \frac{c-1}{2}\left[\operatorname{Tr} T - 1\right]. \quad (B.37)$$

Therefore we see that the terms $\ell = 1$ and $\ell = 2$ in the sum in Eq. (B.36) cancel out with the terms coming from $\mathcal{S}_1[\rho_*]$. Moreover, using the definiton of $\mathcal{A}(N,c)$, and the noting that $\mathcal{S}_0[\rho_*]$ equals

$$\mathcal{S}_0[\rho_*] = \frac{c}{2} - \log \int d\sigma \; e^{B(\sigma)} \left[\int d\tau \; U(\sigma,\tau)\rho_*(\tau)\right]^c, \quad (B.38)$$

we finally obtain

$$[Z^n]_{\mathrm{av}} \sim \exp\left\{N \log \int d\sigma \; e^{B(\sigma)}\left[\int d\tau \; U(\sigma,\tau)\rho_*(\tau)\right]^c + \frac{1}{2}\sum_{\ell=3}^{\infty} \frac{(c-1)^\ell}{\ell}\left[\operatorname{Tr} T^\ell - 1\right]\right\}. \quad (B.39)$$

We split the free energy density $f(N)$ into the sum of the leading term plus the $1/N$ correction:

$$f(N) = f_0 + \frac{1}{N}f_1 + o(1/N). \quad (B.40)$$

The quantity $f_1$ is given by

$$f_1 = -\frac{1}{\beta}\lim_{n\to 0}\sum_{\ell=3}^{\infty}\frac{(c-1)^\ell}{2\ell}\partial_n \operatorname{Tr} T^\ell. \quad (B.41)$$

## B.3 Evaluating $\operatorname{Tr} T^\ell$

The matrix $T(\sigma,\tau)$ is defined as $T(\sigma,\tau) = U(\sigma,\tau)r(\tau)$. In the replica-symmetric regime, we can parametrize the field $r(\sigma)$ as

$$r(\sigma) = \int dr \; R_n(r)\frac{e^{\beta r \sum_{a=1}^n \sigma^a}}{[2\cosh(\beta r)]^n}, \quad (B.42)$$

where the density $R_n(r)$ is non-negative and normalized to 1 in the limit $n \to 0$. In order to compute the $O(1/N)$ correction to the free energy, we need also to compute its normalization up to order $O(n)$. Inserting the parametrization (B.42) in the equation defining $r(\sigma)$, i.e. Eq. (B.21), and considering also the $n$-dependence of the distribution $P_n(h)$ parametrizing $\rho(\sigma)$, we obtain

$$\int dr \; R_n(r) = 1 - n\,\mathbb{E}_{J,r,u}\log\left[\frac{\cosh(\beta J)\cosh(\beta r + \beta u)}{\cosh(\beta r)\cosh(\beta u)}\right] + O(n^2). \quad (B.43)$$

The random variable $u$ is called a cavity bias and is drawn from the distribution

$$Q(u) = \mathbb{E}_J \int dh \; P(h)\,\delta[u - \hat{u}(\beta,J,h)], \quad (B.44)$$

with $P(h)$ solution of Eq. (4.59). In Eq. (B.43) the random variable $r$ is distributed as $R(r) = \lim_{n\to 0} R_n(r)$, solution to Eq. (B.21).

With this considerations in mind, the matrix $T(\sigma,\tau)$ can be written, for small $n$, as

$$T(\sigma,\tau) = \mathbb{E}_{J,r}\left[\prod_{a=1}^{n}\exp\left(\beta J\sigma_a\tau_a + \beta r\tau_a\right)\right] \\ - n\,\mathbb{E}_{J,r,u}\log\left[\frac{2\cosh(\beta J)\cosh(\beta r + \beta u)}{\cosh(\beta u)}\right] + O(n^2). \quad (B.45)$$

The first term in Eq.(B.45) is the replicated transfer matrix of a 1-dimensional disordered Ising chain with random couplings $J$ and random fields $r$. Let's call $T_n(\sigma,\tau)$ this first term.

The second term in Eq. (B.45) is proportional to the thermodynamic free-energy density $\phi$ of an Ising chain with random couplings $J$ and random fields $r$ [138] [1], explicitly:

$$\mathbb{E}_{J,r,u}\log\left[\frac{2\cosh(\beta J)\cosh(\beta r + \beta u)}{\cosh(\beta u)}\right] = -\beta\phi. \quad (B.46)$$

The full matrix $T(\sigma,\tau)$, in the limit $n\to 0$, then becomes:

$$T(\sigma,\tau) = T_n(\sigma,\tau) + n\beta\phi + O(n^2). \quad (B.47)$$

Taking the trace $\operatorname{Tr}\left(T^\ell\right)$ we find

$$\operatorname{Tr} T^\ell = \operatorname{Tr} T_n^\ell + n\ell\beta\phi + o(n^2). \quad (B.48)$$

Now we observe that

$$\lim_{n\to 0}\partial_n \operatorname{Tr} T_n^\ell = -\beta\phi_\ell^c, \quad (B.49)$$

where $\phi_\ell^c$ is the free energy of a closed chain (loop) of length $\ell$, receiving a field $r$ on each of its vertex. Eventually taking the derivative and then the limit $n\to 0$ of the full trace $\operatorname{Tr} T^\ell$, we get

$$\lim_{n\to 0}\partial_n \operatorname{Tr} T^\ell = -\beta\left(\phi_\ell^c - \ell\phi\right) \equiv \Delta\phi_\ell. \quad (B.50)$$

Coming back to the equation (B.41) for $f_1$, and substituting the previous result (B.50), we finally obtain the formula given in the main text:

$$f_1 = \sum_{\ell=3}^{\infty}\frac{(c-1)^\ell}{2\ell}\Delta\phi_\ell. \quad (B.51)$$

# Appendix C

# Appendix to Chapter 6

## C.1 Evaluation of $\Sigma_d$

In this appendix we will show that, in dimension $d > 2$, the function

$$\zeta_d(s) = \sum_{\mathbf{n} \in \mathbb{Z}^d \setminus \{\mathbf{0}\}} \frac{1}{\|\mathbf{n}\|^{2s}}, \tag{C.1}$$

a particular Epstein zeta function defined for $\operatorname{Re} s > \frac{d}{2}$, analytically continued to the point $s = 1$, takes the value of $\Sigma_d$ defined in Eq. (6.22), i.e. $\zeta_d(1) = \Sigma_d$. Then we will derive an easily computable representation of the analytic continuation of $\zeta_d(s)$ which was already presented in the literature [180, 199].

Let us fix $d > 2$ and $a > 0$. Then, for $\operatorname{Re} s > \frac{d}{2}$, we can rewrite Eq. (C.1) as

$$\begin{aligned} \zeta_d(s) &= \lim_{R \to +\infty} \left[ \sum_{\mathbf{n} \in \mathbb{Z}^d \setminus \{\mathbf{0}\}}^{\|\mathbf{n}\| \leq R} \frac{1}{\|\mathbf{n}\|^{2s}} - S_d \int_a^R \frac{r^{d-1}}{r^{2s}} \mathrm{d}r + S_d \int_a^R \frac{r^{d-1}}{r^{2s}} \mathrm{d}r \right] \\ &= \lim_{R \to +\infty} \left[ \sum_{\mathbf{n} \in \mathbb{Z}^d \setminus \{\mathbf{0}\}}^{\|\mathbf{n}\| \leq R} \frac{1}{\|\mathbf{n}\|^{2s}} - S_d \int_a^R \frac{r^{d-1}}{r^{2s}} \mathrm{d}r \right] + S_d \frac{a^{d-2s}}{2s - d}. \end{aligned} \tag{C.2}$$

Assuming that the limit in the last equation exists also for $\operatorname{Re} s < \frac{d}{2}$, we have isolated the singular term. The analytic continuation of $\zeta_d(s)$ then reads

$$\zeta_d(s) = \lim_{R \to +\infty} \left[ \sum_{\mathbf{n} \in \mathbb{Z}^d \setminus \{\mathbf{0}\}}^{\|\mathbf{n}\| \leq R} \frac{1}{\|\mathbf{n}\|^{2s}} - S_d \int_0^R \frac{r^{d-1}}{r^{2s}} \mathrm{d}r \right] \qquad \text{for } \operatorname{Re} s < \frac{d}{2}, \tag{C.3}$$

where to limit $a \to 0$ has been taken. Note that, comparing the previous equation with Eq. (6.22), $\zeta_d(1) \equiv \Sigma_d$. On the other hand, for $\operatorname{Re} s > \frac{d}{2}$

$$\begin{aligned} \zeta_d(s) &= \sum_{\mathbf{n} \in \mathbb{Z}^d \setminus \{\mathbf{0}\}} \frac{1}{\Gamma(s)} \int_0^{+\infty} z^{s-1} \, e^{-\|\mathbf{n}\|^2 z} \, \mathrm{d}z \\ &= \frac{\pi^s}{\Gamma(s)} \int_0^{+\infty} z^{s-1} \left( \Theta^d(z) - 1 \right) \mathrm{d}z, \end{aligned} \tag{C.4}$$

where $\Theta(z)$ is defined by $\Theta(z) \equiv \sum_{n=-\infty}^{+\infty} e^{-\pi n^2 z} \equiv \vartheta(0; iz)$, where $\vartheta(\tau; z)$ is the Jacobi theta function. Noticing that for $z \ll 1$ asymptotically we have $\Theta(z) \sim \frac{1}{\sqrt{z}}$, while $\Theta(z) \sim 1 + 2\,e^{-z}$ for $z \gg 1$, we can isolate the singular parts of $\zeta_d(s)$ writing

$$\zeta_d(s) = \frac{\pi^s}{\Gamma(s)} \left[ \frac{2}{2s-d} - \frac{1}{s} + \int_1^{+\infty} z^{s-1} \left( \Theta^d(z) - 1 \right) \mathrm{d}z + \int_0^1 z^{s-1} \left( \Theta^d(z) - \frac{1}{z^{\frac{d}{2}}} \right) \mathrm{d}z \right]. \tag{C.5}$$

Last expression can be readily continued to the region $\operatorname{Re} s < \frac{d}{2}$. Using the property $\sqrt{t}\,\Theta(t) = \Theta(t^{-1})$, we can write

$$\Sigma_d = \zeta_d(1) = \pi \left[ \frac{2}{2-d} - 1 + \int_1^{+\infty} \left( 1 + z^{\frac{d}{2}-2} \right) \left( \Theta^d(z) - 1 \right) \mathrm{d}z \right]. \tag{C.6}$$

# Appendix D

# Appendix to Chapter 8

## D.1 Normalization Factor

Let's compute the total number $\mathcal{N}$ of realizations of $G^M$. Simple combinatorics gives $\mathcal{N} = (M!)^{|E|}$. We will rederive this results using delta functions and Lagrange multipliers. We have

$$\mathcal{N} = \sum_{\{n\}} \prod_{(i \to j)} \prod_a \delta\left(\sum_b n_{ij}^{ab} = 1\right). \tag{D.1}$$

Using Lagrange multipliers to expand the Kronecker deltas we obtain

$$\begin{aligned}
\mathcal{N} &= \int \prod_{(i \to j)} \prod_a \frac{d\lambda_{i \to j}^a}{2\pi} e^{-i\lambda_{i \to j}^a} \times \prod_{(i,j)} \prod_{ab} (1 + e^{i\lambda_{i \to j}^a + i\lambda_{j \to i}^b}) \\
&= \int \prod_{(i \to j)} \prod_a \frac{d\lambda_{i \to j}^a}{2\pi} e^{-i\sum_{i \to j} \sum_a \lambda_{i \to j}^a + \sum_{(i,j)} (\sum_a e^{i\lambda_{i \to j}^a})(\sum_b e^{i\lambda_{j \to i}^b})} \\
&= \left[\int \prod_a \frac{d\lambda_{1 \to 2}^a}{2\pi} \frac{d\lambda_{2 \to 1}^a}{2\pi} \ e^{-i\sum_a \lambda_{1 \to 2}^a - i\sum_a \lambda_{2 \to 1}^a + (\sum_a e^{i\lambda_{1 \to 2}^a})(\sum_a e^{i\lambda_{2 \to 1}^a})}\right]^{|E|}
\end{aligned} \tag{D.2}$$

where the second passage will be a posteriori justified. Using the relation

$$e^{AB} = e^{\frac{(A+B)^2 - (A-B)^2}{4}} = \int \frac{dx\,dy}{\pi} e^{-x^2 - y^2 + A(x+iy) + B(x-iy)} \tag{D.3}$$

e poi integrando su $\lambda$ si ottiene

$$\begin{aligned}
\mathcal{N} &= \left[\int \frac{dx\,dy}{\pi} e^{-x^2 - y^2}(x+iy)^M(x-iy)^M\right]^{|E|} \\
&= \left[\int \frac{dx\,dy}{\pi} e^{-(x^2+y^2)}(x^2+y^2)^M\right]^{|E|} \\
&= \Gamma(M+1)^{|E|}
\end{aligned} \tag{D.4}$$

where we obtained the expected result in last equation passing through polar coordinates.

# Publications